%% file: main.tex
\documentclass[
11pt, 
english, 
singlespacing, 
parskip, 
headsepline, 
dvipsnames 
]{MastersDoctoralThesis} 

\usepackage[utf8]{inputenc} 
\usepackage[T1]{fontenc} 
\usepackage{multirow}
\usepackage{mathpazo} 

\usepackage{mathtools}

\usepackage[backend=bibtex,style=phys,natbib=true]{biblatex} 
\usepackage[normalem]{ulem} 

\usepackage[autostyle=true]{csquotes} 

\makeatletter
\def\blfootnote{\gdef\@thefnmark{}\@footnotetext}
\makeatother

\usepackage{epigraph}

\usepackage{mathtools}
\usepackage{lettrine}
\usepackage{amsmath,amssymb}
\allowdisplaybreaks
\usepackage{dsfont}
\usepackage{subcaption}
\usepackage{ stmaryrd }
\usepackage{pdfpages}
\usepackage{enumerate}

\usepackage{imakeidx} 
\usepackage{graphicx} 
\usepackage{float} 
\usepackage{tensor} 

\usepackage{dsfont} 
\usepackage{lmodern} 
\usepackage{cancel} 
\usepackage{bm} 
\usepackage{tikz} 
\usetikzlibrary{calc,trees,positioning,arrows,chains,shapes.geometric,%
	decorations.pathreplacing,decorations.pathmorphing,shapes,%
	matrix,shapes.symbols}
\usetikzlibrary{calc,angles,positioning,intersections}
\usepackage{psvectorian} 

\usepackage{ulem} 
\usepackage{ifthen}
\usepackage[pagestyles,extramarks]{titlesec} 
\usepackage{comment} 
\usepackage{tikz-cd} 
\usepackage{fontawesome} 
\usepackage[most]{tcolorbox} 

\usepackage{afterpage} 
\usepackage{bbold} 
\usepackage[export]{adjustbox}

\usepackage{marginnote}

\usepackage[font=small,labelfont=bf,
justification=justified,
format=plain]{caption} 

\usepackage[framemethod=tikz]{mdframed} 
\newmdenv[
topline=false,
bottomline=false,
skipabove=\topsep,
skipbelow=\topsep,
leftmargin=-10pt,
rightmargin=0pt,
innertopmargin=0pt,
innerbottommargin=0pt
]{example}

\allowdisplaybreaks

\newcommand{\ud}{\mathrm{d}}

\thesistitle{Classical and quantum aspects of perturbations in Primordial Universe} 
\supervisor{Przemys\l aw   \textsc{Ma\l kiewicz}} 
\examiner{} 
\degree{Doctor of Philosophy} 
\author{Alice \textsc{Boldrin}} 
\addresses{} 

\subject{Physical Sciences} 
\university{\href{http://www.ncbj.gov.pl}{National Centre for Nuclear Research}} 
\department{\href{}{Department of Fundamental Research \\ Theoretical Physics Division}} 
\group{{}} 
\faculty{\href{}{Factulty of Physics}} 

\AtBeginDocument{
\hypersetup{pdftitle=\ttitle} 
\hypersetup{pdfauthor=\authorname} 
\hypersetup{pdfkeywords=\keywordnames} 
\hypersetup{linkcolor=RoyalBlue}
}

\graphicspath{{pictures/}} 

\usepackage{titlesec}
\titleformat{\chapter}[display]
{\normalfont\Large\raggedleft}
{\MakeUppercase{\chaptertitlename}%
	\rlap{ \resizebox{!}{1.5cm}{\thechapter} \rule{5cm}{1.5cm}}}
{10pt}{\Huge}
\titlespacing*{\chapter}{0pt}{30pt}{20pt}

\newcommand{\dd}{\mathrm{d}} \newcommand{\ex}{\mathrm{e}}

\newcommand{\be}{\begin{equation}} \newcommand{\ee}{\end{equation}}
\newcommand{\ba}{\begin{eqnarray}} \newcommand{\ea}{\end{eqnarray}}

\newcommand{\Tr}{\text{Tr}}

\usetikzlibrary{calc}
\newcommand*{\boxcolor}{orange}
\makeatletter
\renewcommand{\boxed}[1]{\textcolor{\boxcolor}{%
		\tikz[baseline={([yshift=-1ex]current bounding box.center)}] \node [rectangle, minimum width=1ex,rounded corners,draw] {\normalcolor\m@th$\displaystyle#1$};}}
\makeatother

\begin{document}

\frontmatter 

\pagestyle{plain} 


\begin{titlepage}
\begin{center}

{\scshape\LARGE \univname\par}\vspace{1.5cm} 
\textsc{\Large Doctoral Thesis}\\[0.5cm] 

\HRule \\[0.4cm] 
{\huge \bfseries \ttitle\par}\vspace{0.4cm} 
\HRule \\[1.5cm] 
 
\begin{minipage}[t]{0.4\textwidth}
\begin{flushleft} \large
\emph{Author:}\\
\href{https://google.com}{\authorname} 
\end{flushleft}
\end{minipage}
\begin{minipage}[t]{0.4\textwidth}
\begin{flushright} \large
\emph{Supervisor:} \\
\href{https://www.ncbj.gov.pl/}{\supname}\\ 
\emph{Auxiliary supervisor:} \\
\href{https://www.ncbj.gov.pl/}{Patrick \textsc{Peter}}\\ 
\end{flushright}
\end{minipage}\\[2cm]
 
\vspace{-1 cm}
\large \textit{A thesis submitted in fulfillment of the requirements\\ for the degree of \degreename}\\[0.3cm] 
\textit{in the}\\[0.4cm]
\groupname
\deptname\\[-2cm] 
\includegraphics[width=0.5\textwidth]{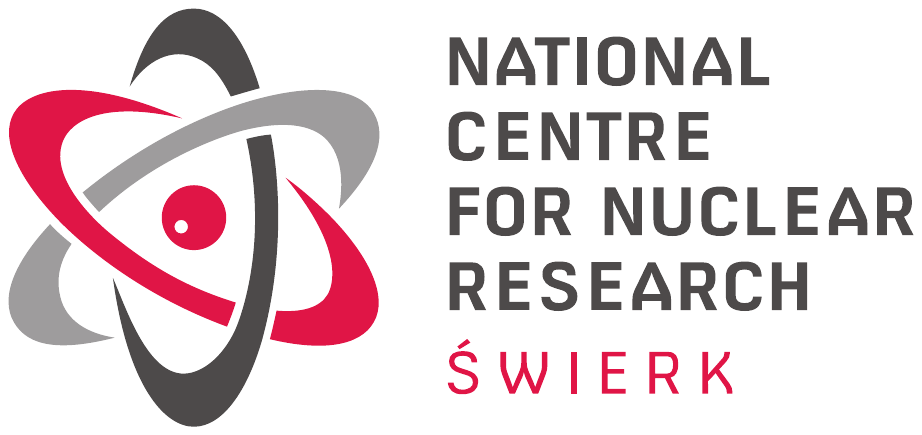}\\[-3cm]
\vfill
{\large \today} 
 
\end{center}
\end{titlepage}


\begin{declaration}
\addchaptertocentry{\authorshipname} 
\noindent I, \authorname, declare that this thesis titled, \enquote{\ttitle} and the work presented in it are my own. I confirm that:

\begin{itemize} 
\item This work was done wholly or mainly while in candidature for a research degree at the National Centre for Nuclear Research.
\item Where any part of this thesis has previously been submitted for a degree or any other qualification at the National Centre for Nuclear Research or any other institution, this has been clearly stated.
\item Where I have consulted the published work of others, this is always clearly attributed.
\item Where I have quoted from the work of others, the source is always given. With the exception of such quotations, this thesis is entirely my own work.
\item I have acknowledged all main sources of help.
\item Where the thesis is based on work done by myself jointly with others, I have made clear exactly what was done by others and what I have contributed myself.\\
\end{itemize}
 
\noindent Signed:\\
\rule[0.5em]{25em}{0.5pt} 
 
\noindent Date:\\
\rule[0.5em]{25em}{0.5pt} 
\end{declaration}

\cleardoublepage


\vspace*{0.2\textheight}

\noindent\enquote{\itshape She generally gave herself very good advice, (though she very seldom followed it).}\bigbreak

\hfill  Lewis Carroll, Alice’s Adventures in Wonderland / Through the Looking-Glass  

\begin{abstract}
\addchaptertocentry{\abstractname} 
The quest for a comprehensive description of the initial stages of our Universe leads through the understanding of quantum gravity. In this work, our aim is to obtain a Hamiltonian formulation suitable for canonical quantization. Moreover, we assume that the early Universe can be described with fewer initial symmetries, thus we abandon the isotropy assumption and instead explore anisotropic universes, beginning with the simplest one, namely Bianchi I.

 The presence of small initial fluctuations in the early universe can be well described by perturbations around a homogeneous background. General relativity (GR) is a constrained system, and we apply the so-called Dirac procedure for constrained systems to derive a gauge-invariant Hamiltonian formulation suitable for quantization. In this work, we present how this procedure can be extended to a generic background and its relation to the Kucha\v r decomposition. Subsequently, we apply this formulation to a Bianchi I universe, obtaining new and  interesting results on the gauge-invariant representation of matter and geometry perturbations. Contrary to the Friedman-Lemaître-Robertson-Walker (FLRW) case, in which all the modes decouple, in Bianchi I we see that scalar and tensor modes do not decouple. 

We show that new types of gauge-fixing conditions exit in this case. For instance, a gravitational wave can be encoded into scalar modes, by introducing a new gauge which is not valid in FLRW.

Furthermore, we make a first step towards a consistent and unified quantization of the composite system made of a background mode and perturbation modes. Specifically, we study tensor modes in a FLRW universe. We focus on the relation between the choice of internal time of the universe and the quantum evolution it undergoes.
Our results indicate that the time reparametrization invariance in general relativity affects the quantum evolution of the background and perturbation modes. However, in the classical limit, i.e. for a large universe, the dynamics becomes unique. Thus, the predictive power of the theory is maintained.
\end{abstract}

\begin{extraAbstract}
\addchaptertocentry{\extraAbstractname}

Droga do kompleksowego opisu początkowych etapów naszego Wszechświata prowadzi przez zrozumienie grawitacji kwantowej. W tej pracy naszym celem jest uzyskanie sformułowania Hamiltonowskiego odpowiedniego do kanonicznej kwantyzacji. Zakładamy ponadto, że wczesny Wszechświat można opisać z mniejszą liczbą początkowych symetrii, dlatego rezygnujemy z założenia izotropii i zamiast tego badamy anizotropowe wszechświaty, zaczynając od najprostszego, mianowicie Bianchi I.

Obecność małych początkowych fluktuacji we wczesnym Wszechświecie można dobrze opisać za pomocą perturbacji wokół jednorodnego tła. Ogólna teoria względności (OTW) jest systemem z więzami, i stosujemy tzw. procedurę Diraca dla systemów z więzami, aby wyprowadzić sformułowanie Hamiltonowskie niezmiennicze względem cechowania, odpowiednie do kwantyzacji. W tej pracy pokazujemy, jak tę procedurę można rozszerzyć na dowolne tło oraz jej związek z dekompozycją Kuchařa. Następnie stosujemy to sformułowanie do wszechświata Bianchi I, uzyskując nowe i interesujące wyniki dotyczące reprezentacji niezmienniczej względem cechowania dla perturbacji materii i geometrii. W przeciwieństwie do przypadku Friedman-Lemaître-Robertson-Walker (FLRW), w którym wszystkie mody się wszechświata typu
odsprzęgają się, w Bianchi I widzimy, że mody skalarne i tensorowe się nie odsprzęgają się.

Pokazujemy, że w tym przypadku istnieją nowe typy warunków uniezgadniania cechowania. Na przykład fala grawitacyjna może być zakodowana w modach skalarnych, poprzez wprowadzenie nowego cechowania, które nie jest ważne w FLRW.

Ponadto, wykonujemy pierwszy krok w kierunku spójnej i całkowitej kwantyzacji złożonego systemu składającego się z modu tła i modów perturbacyjnych. W szczególności badamy mody tensorowe we wszechświecie FLRW. Skupiamy się na związku między wyborem wewnętrznego czasu Wszechświata a jego kwantową ewolucją. Nasze wyniki wskazują, że niezmienność względem przekształceń czasu w ogólnej teorii względności wpływa na kwantową ewolucję modu tła i modów perturbacyjnych. Jednak w klasycznym limicie, tj. dla dużego wszechświata, dynamika staje się unikalna. Tym samym moc predykcyjna teorii jest zachowana.
\end{extraAbstract}

\begin{acknowledgements}
\addchaptertocentry{\acknowledgementname} 

I would like to thank my supervisor Przemek for his guidance and for making my PhD the most enjoyable time of my studies, and for allowing me to fulfil my dream of working on the time problem.  I also want to thank my second supervisor, Patrick, for his help during my stay in Paris, and for the long talks we had during our meetings which were very interesting and from which I learned a lot. 

I also want to thank Monika, who patiently tried hard to teach me Polish (and its horrible numbers) and who has always been of great support. Her wisdom has helped make my life into what it is now.

Moreover, I want to thank my friend and former office mate, Anatolii, for his great sense of humour, for allowing me to share my cooking recipes with him, and for the enjoyable discussions we had both during and after office hours. I am especially grateful to him for introducing me to his wife, Diana, who has since become a great friend. I also wish to thank Ubaldo for the nice moments spent together, and whose friendship made this journey more fun.

For my stay in Tartu I would like to thank Adam for suggesting me for the position, and especially Maria. I truly enjoyed the time spent together both in the office and outside.

I also want to thank my bachelor supervisor, Prof. Ansoldi, who has taught me a lot and who has played a big part in achieving my dream.

A big thank goes also to Irene, who, despite the distance, has always been a great friend ready to listen to my complains or useless ideas. 

I would not have been able to apply for a PhD without the support of my parents, my sister and my dog, who have always been my number one supporters and helped me through ups and down during this long journey.

A special thanks goes to my partner, with whom I shared all my studies, who helped me immensely, believing I could do it when I did not, and who stood by me regardless. But my biggest thank you goes to our son, Levi, who made the writing of my thesis harder (if not impossible) but immensely more pleasant.

Lastly, I want to thank Anna, who may not be able to enjoy the end of this journey with me, but whose presence and teachings throughout my life have been essential in allowing me to achieve everything I have always wanted.
 

\end{acknowledgements}


\tableofcontents 

\listoffigures 

\dedicatory{To Anna and Levi, who sadly never met.} 

\mainmatter 

\pagestyle{thesis} 

\include{Chapters/Introduction}

\include{Chapters/Chapter1}
\include{Chapters/Chapter2}
\include{Chapters/Chapter3}

\include{Chapters/Chapter4}

\include{Chapters/Chapter5}


\printbibliography[heading=bibintoc]



\appendix 

\include{Appendices/AppendixBianchi}

\include{Appendices/Appendix-time}

\end{document}

%% file: Chapters/Introduction.tex

\chapter*{Introduction} 
\addcontentsline{toc}{chapter}{Introduction}
\markboth{Introduction}{}

\label{Introduction} 

The quest for quantum gravity aiming to deepen our understanding of spacetime, overcome the current limitations, and achieve a unified description of the universe, has been ongoing for more than 80 years \cite{Bronstein:2012zz, Rovelliqg, Misner:1972js}. 
Approaching quantum gravity means deepening our understanding of quantum mechanics (QM) and general relativity (GR). Quantum mechanics is very successful at explaining atomic-scale phenomena but ignores gravity's effects predicted by general relativity.  These theories have revolutionized our understanding of the universe, yet they turn out very incompatible in attempts to describe such extreme phenomena as black holes or the origin of the universe. A quantum gravity (QG) theory might help us go beyond the limits of GR marked by singularities within black holes \cite{Husain:2004yz} and the puzzle of information loss in black holes \cite{Hawking:2005kf, Kiefer:2020cbu}, as well as the initial singularity in the early universe. By filling this gap, QG promises us an enhanced understanding of spacetime and of the fundamental nature of gravity.

The aim of this thesis is to introduce some necessary tools such as a Hamiltonian description of a cosmological system for a more complete approach towards quantization.
We study the early universe, through cosmological perturbation theory, which in turn allows us to study the small geometry and matter fluctuations present in the early universe. These inhomogeneities eventually grew into the large-scale structures we observe today, such as galaxies, clusters, and cosmic filaments \cite{Calzetta:1995ys}. In particular, perturbation theory enables us to explain the slight temperature fluctuations in the cosmic microwave background (CMB) across the sky \cite{Sachs:1967er}. These fluctuations carry information about the universe's early conditions, including its composition, geometry, and history of expansion.
Most importantly, we are then able to make predictions about observable quantities, such as the CMB power spectrum, galaxy clustering, and gravitational lensing. By comparing these predictions with observational data from experiments like the Planck satellite \cite{Planck:2018nkj}, we can test and refine our cosmological models, shedding light on the initial state of the universe and its evolution. It is also important to 
constantly refine and expand our theoretical framework to make maximal use of forthcoming data from experiments, such as LiteBIRD, CMB S4 and Cosmic Origins Explorer (CORE) \cite{Matsumura:2013aja, LiteBIRD:2024twk, CMB-S4:2020lpa, CORE:2017krr}.

The approach we take involves formulating a Hamiltonian description of GR in perturbation theory around generic homogeneous backgrounds. 
The fact that GR is a constrained theory also holds in perturbation theory with the difference that now, the latter is not a totally constrained system, as it includes both constraints and a non-vanishing part.
 As such, we are able to apply the Dirac method for constrained systems \cite{Diraclec, regge, Henneaux:1992ig}, which consists of identifying the primary constraints, constructing the total Hamiltonian by adding these constraints to the initial Hamiltonian, and deriving the secondary constraints through the Poisson brackets by imposing the consistency of the constraints. This process is repeated iteratively until all constraints are accounted for, thereby allowing for a consistent Hamiltonian formulation.
Once we find the Hamiltonian formulation, by using the Dirac observables, which are gauge invariant variables commuting with the constraints, we obtain a physical description of our Hamiltonian without constraints. 
This procedure enables to separate the dynamical degrees of freedom from the unphysical ones that can be used to reconstruct the spacetime. More simply, the obtained gauge-invariant Hamiltonian written in terms of the Dirac observables allows us to explore different gauges with little effort but still keeping the physical description invariant.

Once this formalism is established, we can apply it to specific backgrounds. Previously, it has been used for the Friedman-Lemaître-Robertson-Walker (FLRW) background \cite{Malkiewicz_2019}, which describes a homogeneous and isotropic universe. However, in this work, we apply it to an anisotropic universe described by the Bianchi I background, which is the simplest anisotropic cosmological models.

Although current observations indicate that the Universe today is well described by isotropic and homogeneous backgrounds \cite{Planck:2018nkj}, this has been disputed (see, e.g. \cite{Jones:2023ncn}) and may not have always been the case. It is widely believed that as we approach the initial singularity, the conditions of the early universe could have been different, because of unstable symmetries. The absence of symmetries would have played a crucial role in the dynamics during those early stages.

By relaxing the strict symmetry assumptions, we aim to achieve a more nuanced and realistic description of the primordial universe. The Bianchi I universe, which allows for anisotropic expansion, is a good starting point for exploring these early conditions. Understanding anisotropic effects is not only theoretically interesting but may also be essential for constructing a more comprehensive picture of the universe's evolution from its inception to its current state \cite{Uzan}. The quantization of the Hamiltonian formalism will serve as a starting point to deepen our understanding on this very topic. 

In order to better understand how to quantize the Hamiltonian in Bianchi I, we decided to simplify our system. We quantize a FLRW background mode and perturbations (tensor) modes in such a way as to obtain a bounce scenario, avoiding the initial singularity and introducing clear quantum effects into the model. By using the Dirac observables  and their property of being non-dynamical quantities, we were able to tackle one of the big conceptual problems encountered when trying to merge QM and GR, i.e. the time problem \cite{Isham:1992ms}. This issue has been widely discussed, yet with no widely accepted solution. Although many papers have been published on this topic, our novel approach focus on the trajectories of both the background and perturbations in a semi-classical approximation leading to new and interesting results. In particular, we find that the time problem is relevant only around the bounce and disappears once we approach a classical domain. Hence, the theory is predictive for this regime, where the classical approximation is valid.

The outline of this thesis is the following: In Ch. \ref{Chapter1} we discuss the theory behind constrained and gauge systems, in Ch. \ref{Chapter2} we present our method for a generic background and in Ch. \ref{Chapter3} we apply the method to a Bianchi I background. Finally, in Ch. \ref{Chapter4}, we tackle the time problem using the Dirac observables.

%% file: Chapters/Chapter1.tex

\chapter{Constrained systems} 

\label{Chapter1} 

In this chapter we are going to introduce the mathematical background needed for the understanding of the Dirac method, which is the fundamental tool used in the other chapters. 
Some of the material and examples in the following sections have been written following my personal notes of the lectures on "Advanced General Relativity" given by Prof. J. Lewandowski at the University of Warsaw in the summer semester of 2020.

\section{ADM formalism }\label{SecADM} 

\begin{figure}[H]
	\centering
	\begin{subfigure}{0.35\textwidth}
		\centering
		\includegraphics[width=\linewidth]{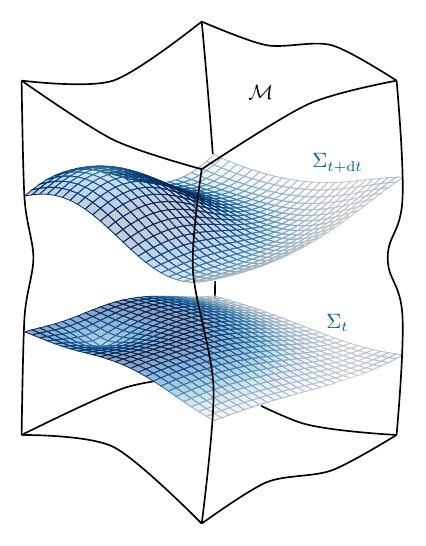}
		\caption{}
		\label{fol}
	\end{subfigure}%
	\begin{subfigure}{0.35\textwidth}
		\centering
		\includegraphics[width=\linewidth]{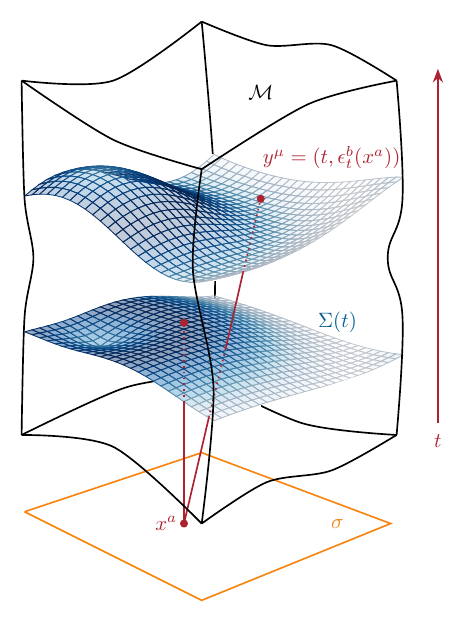}
		\caption{ }
		\label{fol-emb}
	\end{subfigure}
	\caption{Representation of 4-dimensional spacetime, with its evolution described by the evolution of a fixed 3-dimensional hypersurface.
		}
	\label{fig:side-by-side}
	\end{figure}

The ADM formalism \cite{Arnowitt:1962hi} splits spacetime into a family of \textbf{space-like 3-dimensional hyper-surfaces} $\{\Sigma_t\}_{t{\in \mathbb{R}}}$, each of which is defined at an instant of time $t$, globally defined. More formally, a generic hypersurface $\Sigma \in \{\Sigma_t\}_{t{\in \mathbb{R}}}$ is said to be a co-dimension $1$ (i.e. dimension $D$) space-like subsurface in a $(D+1)$-dimensional pseudo-Riemannian manifold\footnote{We denote the metric $2$-form by $g$, that is $g=g_{\mu\nu}  \mathrm{d}x^\nu  \mathrm{d}x^\mu$. We work with signature $+2$.} $(\mathcal{M},g)$. The coordinates on the hypersurfaces $\Sigma$ will be denoted by $x^a$, while the coordinates in the manifold $\mathcal{M}$ will be $y^\mu$. We use Greek indices for $(D+1)$-dimensional quantities, and Latin indices for $D$-dimensional quantities. 
	\begin{figure}[H]
	\includegraphics[width=0.4\textwidth,center]{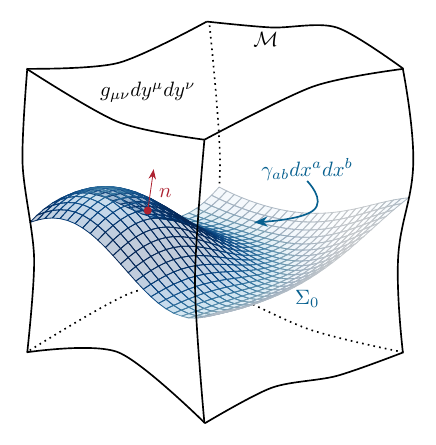}
	\caption{Representation of the manifold $\mathcal{M}$ together with a vector belonging to the vector field $n$ normal to the subsurface $\Sigma_0$ .}
	\label{admgeom1}
	\end{figure}
Since, by definition, these subsurfaces are chosen to be \textbf{non-null}, the intrinsic geometry $\gamma$ of $\Sigma$, which is the restriction of the total metric $g$ in the surface $\Sigma$, is non-degenerate. In other words, the rank of the matrix form of $\gamma$ is the highest possible. This is important since a non-null surface allows us to define a vector field $n=n^\mu\partial_\mu$ normal to the surface $\Sigma$, which is also \textbf{not} tangent to $\Sigma$ itself (Fig. \ref{admgeom1}). The normalization of $n$ at each point can be chosen to be\footnote{The sign of the normalization of $n$ depends on the choice of the metric signature. In our work, considering signature $+2$, we have $n^\mu n_\mu=-1$.} $n^\mu n_\mu=\pm 1$. Physically speaking, the existence of $n$ ensures that a time-like or light-like path in this formalism consists of a succession of points each one in a different hyper-surface, e.g. $\Sigma_t$, $\Sigma_{t+\ud t}$, and so on (Fig. \ref{fol}).  Alternatively, the dynamics can be seen as each sheet $\Sigma(t)$ representing a $t$-dependent embedding $\epsilon_t$ of a standard co-dimension $1$ surface $\sigma$ (Fig. \ref{fol-emb}). From this point of view, all the points $y^\mu$ of $\mathcal{M}$ (the spacetime manifold) are the images (under the embeddings) of the fixed point $x^a$ on the standard surface $\sigma$.
The two approaches presented above are equivalent and each one of them is better suited to different contexts. Unless otherwise stated, we will use the first approach.

\begin{figure}[t]
	\includegraphics[width=0.4\textwidth,center]{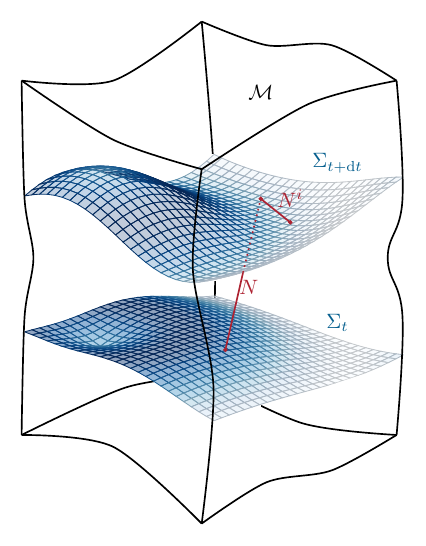}
	\caption{Representation of the role of the lapse and shift function in the definition of coordinates system in the 4-dimensional spacetime.}
	\label{admgeom}
\end{figure}

The ADM formalism involves splitting spacetime into space-like hypersurfaces, or sheets, each at a specific time coordinate. This approach allows for a clear separation of spatial and temporal components.
This implies that the usual quantities defined in the manifold $\mathcal M$ should also be split in \textbf{space-like} and \textbf{time-like} components. 
It is  thus instructive to start by looking at the metric whose purpose is indeed to describe the geometry of the spacetime. 
We need to introduce two ad hoc quantities, the so called \textbf{lapse function} and \textbf{shift vector} (Fig. \ref{admgeom}). The first one is the temporal component of the normal vector describing how time evolves from one sheet to the another, i.e. it describes how much time has passed for an observer travelling between two neighbouring hypersurfaces. The latter is a 3-dimensional vector which describes the shift from a point $x^i$ in the sheet $\Sigma_t$ to a point $x^i+\ud x^i$ in the sheet $\Sigma_{t+\ud t}$. The lapse and shift can be mathematically defined as
\begin{equation}
	\begin{aligned}
	\ud\tau&=N \ud t;\\
	x^i_{\Sigma_{t+\ud t}}&=x^i_{\Sigma_{t}}-N^i\ud t.
	\end{aligned}
\end{equation}
where $\tau$ is the proper time. With the introduction of the lapse and shift, the metric in this formalism can be defined as follows
\begin{equation}\label{dsadm}
	\ud s^2=-\left(N\ud t\right)^2+\gamma_{ij}\left(\ud x^i+N^i\ud t\right)\left(\ud x^j+N^j\ud t\right).
\end{equation} 
We can now study the dynamics in this formalism. To do so, as per usual, we just need to vary the action $S=\int \ud^{D+1} x~ \mathcal{L}$ with respect to the metric and study the conditions for which $\delta S=0$.
For simplicity, in what follows we consider the case in which $D+1=4$. The Lagrangian density, up to some constant\footnote{Considering $G=c=1$, one could usually introduce in the definition of the Lagrangian a factor $\frac{1}{16\pi}$ \cite{gravitation}.}, is 
\begin{equation}
	\mathcal{L}=\sqrt{-g}\, \prescript{(4)}{}{R},
\end{equation}
where $\prescript{(4)}{}{R}$ is the Ricci scalar in the spacetime manifold.
The first step is to write the curvature of the spacetime manifold $\mathcal{M}$ in terms of the curvature of the $3$-dimensional hypersurface $\Sigma$ and the \textbf{extrinsic curvature}\footnote {The extrinsic curvature represents the curvature of an embedded subspace, i.e. in our case it is the curvature of the hypersurface $\Sigma_t$ embedded in  the manifold $\mathcal{M}$.} $K_{ij}$. The extrinsic curvature is defined as 
\begin{equation}\label{extrcurv}
	K_{ik} =
	\frac{1}{2N}\left(N_{i;k}+N_{k;i}-\dot{\gamma}_{ik}\right),
\end{equation}
where $N_{i;k}$ is defined by $N_{i;k}=\gamma_{ji}D_kN^j$, and where $D_k$ is the 3-dimensional covariant derivative calculated from $\gamma_{ij}$. The quantity $\dot{\gamma}_{ik}$ is the Lie derivative of the spatial metric with respect to the time $t$, i.e. $\dot{\gamma}_{ik}=\frac{\partial \gamma_{ik}}{\partial t}$. The lapse and shift, which are non-physical coordinates, are necessary in Eq. \eqref{extrcurv} to account for the non-orthogonality of the time flow, as the hypersurfaces are curved differently. Thus, through Eq. \eqref{extrcurv}, we can consider the variation of the 3-metric $\gamma_{ij}$ from one sheet to another.
After some computations \cite{Wald,gravitation} we find
one of the so called \textbf{Gauss-Codazzi  equations} 
\begin{equation}\label{GCeq}
	^{(4)}\tensor{R}{^m_{ijk}}
	={^{(3)}}\tensor{R}{_{ijk}^m}
	+(n^\mu n_\mu)^{-1} \left(K_{ij}\tensor*{K}{_k}{^m}
	-K_{ik}\tensor*{K}{_j}{^m}\right),
\end{equation}
where ${^{(3)}}\tensor{R}{_{ijk}^m}$ represents the intrinsic curvature of the hypersurface $\Sigma$, which, along with the extrinsic curvature, describes the curvature of the 4-geometry.
 Notice that, when the extrinsic curvature vanishes, the curvature of the 3-geometry equates the one in the 4-geometry.
Contracting the indices of the Riemann tensor we can now obtain the Ricci scalar 
\begin{equation}\label{ricciadm}
{^{(4)}R}=
{^{(3)}R}+(n^\mu n_\mu)
	\left[
	\Tr {K}^2-\left(\Tr  K\right)^2
	\right]
	+2(n^\mu n_\mu)
		^{(4)}\tensor{R}{^i_{jij}}
,
\end{equation}
where $\Tr {K}=\gamma^{ij}K_{ij}=K^j_j$ and $ \Tr{K}^2=\tensor*{K}{_j}{^m}\tensor*{K}{_m}{^j}$.
As previously mentioned, we want to derive the equations of motion from $\delta S=0$,
allowing us to ignore total divergences in the Lagrangian. Therefore, the ADM Lagrangian is given by
\begin{equation}\label{ADMlag}
	\mathcal{L}_{_{ \text ADM}}=\sqrt{-g}\, 
	\left[
	\prescript{(3)}{}{R}
	+(n\cdot n)
\left(\left(\Tr  K\right)^2
-\Tr  K^2\right)
	\right].
\end{equation}	
We can finally write the action as\footnote{The coefficient in front of the integral comes from the use of natural units.}
\begin{equation}\label{S}
	S = \int \mathcal{L}  \, \mathrm{d}^4x
	= 
	\frac{1}{16\pi}\int
	\sqrt{-\gamma}
	 \left[^{(3)}R
	+(n^\mu n_\mu)\left(\left(\Tr  K\right)^2
	-\Tr  K^2\right)\right]N
	\, \mathrm{d}^4x.
\end{equation}
The 4-metric density can be written using the 3-metric density and lapse function, that is, we have $\sqrt{-g}=\sqrt{-\gamma}~N$ by using \eqref{dsadm}. For the sake of this section's clarity, we are ignoring the matter contribution to the action.
We integrate Eq. \eqref{S} by parts and ignore the surface terms, which, although they do not contribute to the bulk dynamics, can still impose relevant boundary conditions. After performing the necessary computations \cite{gravitation}, we obtain the following result:
\begin{equation}\label{varS}
S  =  
	\frac{1}{16\pi}
	\int 
	\left[ 
	\pi^{ij} \dot{\gamma}_{ij}
	-N \mathcal{H}_0 \left(\pi^{ij},\gamma_{ij}\right)
	-N_i \mathcal{H}^i \left(\pi^{ij},\gamma_{ij}\right) 
	\right]\mathrm{d}^4 x,
\end{equation}
where 
\begin{equation}
	\mathcal{H}_0(\pi^{ij},\gamma_{ij})=\gamma^{-1/2}
	\left[
	\Tr \pi^2-\frac{1}{2} \left(\Tr \pi^2\right)
	\right],
\end{equation} 
is called the \textbf{superhamiltonian}, whereas
\begin{equation}
	\mathcal{H}^i(\pi^{ij},\gamma_{ij})=-2 \tensor{\pi}{^{ij}_{;k}} \quad
\end{equation}
is called the \textbf{supermomentum}.
The quantity 
\begin{equation}\label{admomento}
\pi^{ij}=\gamma^{1/2}\left(g^{ij}\Tr{K}-K^{ij}\right)
\end{equation}
 is the \textbf{canonical momentum} which accounts for the extrinsic curvature contribution, and it is conjugate to the spatial metric $\gamma_{ij}$.
Notice that the integrand in Eq.~\eqref{varS} has the same form as the inverse Legendre transform of $\mathcal{L}$.
The \textbf{Hamiltonian} is thus given by
\begin{equation}\label{hamgra}
	H_{\text{gravity}}
	=
	\int (N\mathcal{H}_0+N^i \mathcal{H}_i)\mathrm{d}^3 x.
\end{equation}
Eq. \eqref{hamgra} can be written in a more compact way defining a vector $C_\mu=(\mathcal{H}_0,\mathcal{H}_i)$, hence obtaining
\begin{equation}\label{hamgracon}
	H_{\text{gravity}}=\int \ud^3x N^\mu C_\mu.
\end{equation} 

According to the variational principle, $\delta S=0$ under infinitesimal variations $\delta \gamma$ and $\delta \pi$ around the physical solution, which represents the extremum of the action $S$.
As is known, variables with no time derivative in the equation of motion (eom) are considered \textbf{auxiliary variables}. In Eq. \eqref{varS} we see that both the lapse $N$ and shift $N^i$ are auxiliary variables and take the form of \textbf{Lagrange multipliers}.

In particular, if we vary the action with respect to the lapse function $N$, we get that $\mathcal{H}_0=0$, analogously by varying the shift vector $N^i$ we get $\mathcal{H}_i=0$. This results implies that the superhamiltonian $\mathcal{H}_0$ and supermomentum $\mathcal{H}^i$ are \textbf{constraints} of our system. This known result arises due to the general covariance of General Relativity, i.e., the freedom to choose the spacetime coordinates.




\section{Constraints and gauges}\label{constraint}

The study of constrained systems has been of great interest for a long time due to the relevance that these systems have in physics.
 This is very much so in General Relativity since, as seen in Sec. \ref{SecADM}, the \textbf{gravitational Hamiltonian} is a constraint of the theory. In this section, we will give a mathematical description of constraints and gauges within the context of phase space. Our main goal is to offer readers a geometric intuition, useful to better understand the \textbf{Dirac method} for constrained systems. We will then focus on the concept of \textbf{gauge transformations} which usually arise in constrained systems, and conclude with a discussion of physical observables, known as \textbf{Dirac observables}, that can be derived from the theory.

\subsection{Geometrical description of constraints}\label{geomconst}

\begin{figure}[t]
	\includegraphics[width=0.5\textwidth,center]{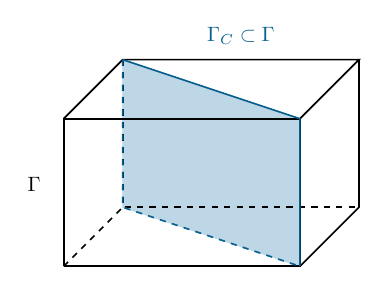}
	\caption{Representation of the constraint surface $\Gamma_C$ on the \\ manifold $\Gamma$.}
\label{ConstGamma}
\end{figure}

We start by considering a phase space described by a \textbf{symplectic manifold} $(\Gamma, \,\Omega)$, where $\Gamma$ is an even dimensional manifold  with an associated symplectic 2-form $\Omega$, i.e. a closed 2-form, where closed means $\ud\Omega=0$. We are interested in understanding what happens when the system is constrained to lie in a subset $\Gamma_C$ of the phase space (see Fig. \ref{ConstGamma}), identified by the condition $C=0$, where $C:\Gamma\rightarrow\mathbb{R}\,|\, C\in C^{\infty}(\Gamma)$.
By imposing the condition $C=0$, we are eliminating one of the dimensions of our initial manifold $\Gamma$ obtaining $\Gamma_C$, a co-dimension 1 manifold. By definition, we thus have that the space $(\Gamma_C,\Omega_C)$ is not symplectic\footnote{It is important to have a symplectic manifold for the Liouville's theorem to hold.}.
This also means that the associated 2-form $\Omega_C$ might be degenerate, where a form is said to be non-degenerate iff $X\lrcorner \, \Omega= X_\nu \Omega_{\mu\nu} = 0 \,\, \Leftrightarrow \,\, X=0$. In this context, the symbol $\lrcorner$ denotes the contraction of a vector field ($X$ in this case) with a $n$-form ($\Omega$ where $n=2$). 
We notice that  $\Omega_C$ is degenerate with respect to the vector field $\delta_C$ defined by $\delta_C \lrcorner \, \Omega = -\ud C$, where $\delta_C$ is a vector tangent to $\Gamma_C$. Indeed, one can show that the vector field $\delta_C$ lies in the $C=0$ plane, since\footnote{Intuitively, one can think of $\ud C=\partial_\mu C \ud x^\mu$, where $x^\mu$ are generic coordinates in $\Gamma$. The quantity $\delta_C \lrcorner \, \ud C$ can be understood as the scalar product between the vector field $\delta_C$ and the gradient of $C$.}
\begin{equation}\label{deg1}
\delta_C \lrcorner \, \ud C = -\delta_C\lrcorner \,(\delta_C\lrcorner \, \Omega) = 0.
\end{equation} 
Eq. \eqref{deg1} holds because $\Omega$ is antisymmetric, that is $\delta_C\lrcorner \,(\delta_C'\lrcorner \, \Omega) = -\delta_C'\lrcorner \,(\delta_C\lrcorner \, \Omega)$.
From the linear property of the pullback obtained from $\psi: \Gamma_C \rightarrow \Gamma$, it is easy to see that\footnote{From Eq. \eqref{deg1} we know that the $1$-form $\ud C=-\delta_C\lrcorner \, \Omega$ can be interpreted as being "orthogonal" to $\Gamma_C$, so its pullback to $\Gamma_C$ is zero.}
\begin{align}
\delta_C\lrcorner \, \Omega_C = \delta_C\lrcorner \, \psi^*\Omega = \psi^*(\delta_C\lrcorner \, \Omega)
=
0,
\end{align}
which proves the degeneracy of $\Omega_C$ along the direction $\delta_C$.
From Eq. \eqref{deg1} we conclude that moving along the directions generated by the vector field $\delta_C$, our system does not change, i.e. $\delta_C$ generates a \textbf{gauge transformation}.

\begin{figure}[t]
\includegraphics[width=0.5\textwidth,center]{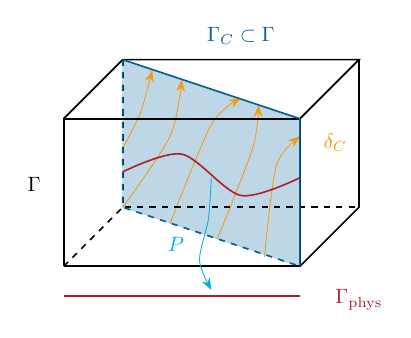}
\caption{Gauge orbits generated by $\delta_C$ in the constraint surface $\Gamma_C$. They give rise to a natural projection to a lower dimension physical phase space $\Gamma_{\text{phys}}$. Notice that the gauge-fixing surface, defined by the curved red line, crosses the gauge orbits once and only once.
	}
\label{gaugeorb}
\end{figure}

At this point, given that the subspace $\Gamma_C$ is not physical, we need to find the \textbf{physical phase space} $\Gamma_{\text{phys}}$. Recalling that $\Gamma_C$ is of odd dimension, it is natural to proceed by getting rid of another dimension to be able to work with an even dimensional space.
Thus, a degree of freedom (dof) varying along $\delta_C$ seems to be a natural choice since it represents a redundant dof anyway. We define the physical phase space $\Gamma_{\text{phys}}$ by projecting-out, through a projection operator $P$, the redundant dof by setting a \textbf{gauge-fixing surface}, which is a line crossing all the gauge orbits (see Fig. \ref{gaugeorb}). Once the physical phase space is obtained, we need to find the $2$-form $\Omega_{\text{phys}}$ associated with it. This can be done considering a theorem, which states that if there exists a vector field $\delta_C$ such that 
\begin{align}
\mathcal{L}_{\delta_C}\Omega_C = 0,
\,
\text{and}
\quad
\delta_C\lrcorner \,\Omega_C = 0,
\end{align}
where $\mathcal{L}_{\delta_C}$ is the Lie derivative with respect to $\delta_C$,
then there exists a $2$-form $\Omega_{\text{phys}}$ over $\Gamma_{\text{phys}}$ such that $\Omega_C = P^*\Omega_{\text{phys}}$, where $P$ is a "projection operator"\footnote{Notice that we are actually \textbf{embedding} the gauge-fixing surface into the physical phase space.}. In order to apply this theorem we need to show that the hypothesis $\mathcal{L}_{\delta_C}\Omega_C = 0$ is satisfied. This means that we need to prove that the vector field $\delta_C$, which defines the \textbf{gauge orbits}, generates a symmetry
\begin{equation}
	\mathcal{L}_{\delta_C}\Omega=\delta_C\lrcorner \ud \Omega+\ud (\underbrace{\delta_C\lrcorner \Omega}_{-\ud C})
	=\ud(-\ud C)=0,
\end{equation}
where we used the fact that $\Omega$ is a closed $2$-form and also $\ud^2=0$ by definition.
  In this way we obtain the \textbf{symplectic space} $(\Gamma_{\text{phys}},\,\Omega_{\text{phys}})$ describing the physical system, which only contains all the possible configurations of the physical system which are not gauge equivalent\footnote{Indeed from Fig.\ref{gaugeorb} we see that the gauge fixing surface is cutting through all the gauge orbits only once.}.
Let us consider some examples to better understand the role of $\Gamma$, $\Gamma_C$ and $\Gamma_{\text{phys}}$.
\bigskip
\begin{example}
\textbf{Example 1}

Let $\Gamma=\{(q^1,...,q^n;p_1,...,p_n)\}$ be our initial phase space with the associated symplectic form $\Omega=\sum_{i=1}^n \mathrm{d}p_i \wedge \mathrm{d}q^i$. The constraint is given by $C=p_n$, that is we are removing the $n$-th momentum component.
Applying the constraint $C$ we find that the submanifold $\Gamma_C$ is given by
\begin{equation*}
\Gamma_C=\{(q^1,...,q^n;p_1,...,0)\}.
\end{equation*}
The gauge generator $\delta_C$ can be found considering the condition $\delta_C\lrcorner \, \Omega= -\mathrm{d}C$, which in this case reads $\delta_C\lrcorner \, \Omega
=
\delta_C\lrcorner
\sum_{i=1}^n \mathrm{d}p_i \wedge \mathrm{d}q^i = -\mathrm{d}p_n$. This means that 
\begin{equation*}
\delta_C=\frac{\partial}{\partial q^n}.
\end{equation*}
Therefore, in this example, the gauge orbits are generated by a vector field ($\delta_C$) directed along the $q_n$ axis. In other words, we are free to choose any $q_n$ since it only represents a redundancy of our system.
Using the projector $P$ we find that the physical phase space is 
\begin{equation*}
\{ (q^1, \dots,q^{n-1}, q^n ; p_1, \dots, p_{n-1}, 0) \}
\xrightarrow{P}
\underbracket{\{ (q^1, \dots, q^{n-1} ; p_1, \dots, p_{n-1}) \}}_{\Gamma_{\text{phys}}}.
\end{equation*}
It is easy to see that in this case, the physical symplectic form $\Omega_{phys}$ is given by
\begin{equation*}
\Omega_{\text{phys}} = \sum_{i=1}^{n-1} \mathrm{d}p_i \wedge \mathrm{d}q^i.
\end{equation*}
Hence, we have that, in this example, the constraint $C=p_n$ eliminates two dof: $(q^n,p_n)$.
\end{example}
\bigskip
We now want to study a new system, similar to \textbf{Example 1}, but with two constraints.
\bigskip
\begin{example}
	\textbf{Example 2}
	
	As in \textbf{Example 1}, we consider $\Gamma=\{(q^1,...,q^n;p_1,...,p_n)\}$ and $\Omega=\sum_{i=1}^n \mathrm{d}p_i \wedge \mathrm{d}q^i$. The constraints of the system are now $C_1=p_n$ and $C_2=p_{n-1}$. Therefore we are removing $2$ dof from our initial manifold $\Gamma$. Notice that, in this case, the submanifold $\Gamma_C$ is even dimensional. Nonetheless it does not define a symplectic manifold because $\Omega_C$ had two degenerate directions.
	Proceeding as before, we find the submanifold $\Gamma_C$ to be
	\begin{equation*}
		\Gamma_C=\{(q^1,...,q^n;p_1,...,p_{n-2},0,0)\}.
	\end{equation*}
	The gauge orbits are now generated by
	\begin{equation*}
		\delta_{C_1}=\frac{\partial}{\partial q^n},
		\quad
		\text{and}
		\quad
		\delta_{C_2}=\frac{\partial}{\partial q^{n-1}}
	\end{equation*}
	Finally, the physical phase space and the symplectic form are given by
	\begin{equation*}
		\Gamma_{\text{phys}}=
			\{ (q^1, \dots, q^{n-2} ; p_1, \dots, p_{n-2}) \},
\quad
		\Omega_{\text{phys}} = \sum_{i=1}^{n-2} \mathrm{d}p_i \wedge \mathrm{d}q^i.
	\end{equation*}
	It is important to notice that the Poisson brackets between the two constraints are
	\begin{equation*}
\{p_{n-1}, p_{n-2}\}=0.
	\end{equation*}
	This condition defines the \textbf{first class} constraints which will be discussed in detail in Sec. \ref{secdirac}.
\end{example}
\bigskip
Lastly, we look at an example in which the two constraints behave differently with respect to \textbf{Example 2}.
\bigskip
\begin{example}
\textbf{Example 3}

Let us consider the same initial phase space manifold as in \textbf{Example 1 } and \textbf{Example 2}. The constraints of our system in this case are $p_n=q^n=0$. Therefore the reduction of the degrees of freedom will be different, because in this example we do not have the residual freedom to move along any direction. The submanifold $\Gamma_C$ and the symplectic form associated are defined as follows
\begin{equation*}
\Gamma_C = \{ (q^1, \dots, q^{n-1}, 0 ; p_1, \dots, p_{n-1}, 0) \}
\quad
\text{and}
\quad
\Omega_C = \sum_{i=1}^{n-1} \mathrm{d}p_i \wedge \mathrm{d}q^i.
\end{equation*}
Since there are no degenerate directions along which we can move, we simply have $\Gamma_C = \Gamma_{\text{phys}}$ and $\Omega_C = \Omega_{\text{phys}}$ because we cannot further project along any non-trivial vector field.
Once again, we look at the Poisson brackets between the two constraints, which in this case reads
\begin{equation*}
\{q_n,p_n\}=1.
\end{equation*}
Therefore the Poisson brackets are non vanishing, this relation defines the so called \textbf{second class} constraints, which will also be discussed in Sec. \ref{secdirac}.
\end{example}
\bigskip
In this section, we formally presented the constraints and the associated gauge freedom. In the following section we will see how constrained systems can be approached using the so called Dirac method.
\subsection{Dirac method}\label{secdirac}

We will now present a method to treat constrained Hamiltonian systems, which was introduced by Dirac \cite{Dirac1950}. Intuitively speaking, as shown in Fig. \ref{ConstGamma}, in a constrained system, once the constraints are solved, a reduced phase space is obtained. The system's dynamics might not be well described by the Poisson brackets in the constraints surface $\Gamma_C$.  In order to deal with this issue we need to introduce a new set of brackets that will take the constraints into consideration, these are usually referred to as Dirac brackets. In what follows we will give a more formal representation of the method, in particular when working with a Hamiltonian system.

We are interested in studying the constraints in a Hamiltonian theory \cite{Diraclec}.  In the general framework we start from the action integral $S=\int \mathcal{L}(q^i,\dot{q}^i)\,\ud t$, where $\mathcal{L}(q^i,\dot{q}^i)$ is the Lagrangian of the system, and where $q$ and $\dot{q}$ are respectively the canonical positions and canonical velocities with $i=1,\dots,N$. 
We consider no explicit $t$-dependence of the Lagrangian, which we also assume to be singular, i.e. $\mathcal{L}$ satisfies the \textit{necessary} and \textit{sufficient} condition \cite{regge} 
\begin{equation}\label{hessian}
	\text{Det}(\bm{H})=	\text{Det}\left(\frac{\partial \mathcal{L}}{\partial \dot{q}^j \partial \dot{q}^i}\right)=0,
\end{equation}
where $\bm{H}=\frac{\partial \mathcal{L}}{\partial \dot{q}^j \partial \dot{q}^i}$ is called \textbf{Hessian}. This condition ensures the existence of gauge 
degrees of freedom, which always imply a constrained Hamiltonian system 
\cite{Henneaux:1992ig}.

 To see why this is so, we notice that, for the action $S$ to be stationary, i.e. for its variation to vanish, the Euler-Lagrange equations
\begin{equation}\label{EL}
	\frac{\mathrm{d}}{\mathrm{d}t}
	\left(
	\frac{\partial\mathcal{L}}{\partial \dot{q}^i}
	\right)
	=
	\frac{\partial\mathcal{L}}{\partial {q}^i}
\end{equation}
must be satisfied, where $i=1,...,N$.
Solving Eq. \eqref{EL} for the accelerations $\ddot{q}^i$, we obtain a second order differential equation for the coordinates $q^i$, which is
\begin{equation}\label{eomq}
	\ddot{q}^i \frac{\partial^2\mathcal{L}}{\partial \dot{q}^i \partial \dot{q}^j}
	=
	\frac{\partial\mathcal{L}}{ \partial {q}^i}
	-
	\dot{q}^i
	\frac{\partial^2\mathcal{L}}{\partial {q}^i \partial \dot{q}^j}.
\end{equation}
Notice that the accelerations $\ddot{q}^i$ cannot be entirely determined from the canonical coordinates $q^i$ and the velocities $\dot{q}^i$ if the Hessian is non-invertible. This means that the solution of Eq. \eqref{eomq}, $q^i$, can be defined using arbitrary functions of time  which constitute gauge degrees of freedom.
More precisely, when switching to the Hamiltonian formalism, the canonical variables change from $(q^i,\dot{q}^j)$ in the Lagrangian to $(q^i,p_i)$, where
\begin{equation}\label{phamlag}
	p_i=\frac{\partial\mathcal{L}}{\partial \dot{q}^i}.
\end{equation}
The condition \eqref{hessian} implies the presence of \textbf{primary constraints} \cite{Anderson:1951ta}
\begin{equation}\label{constphi}
	\phi^m(q,p)=0,\quad m=1,\dots M,
\end{equation}
whose name is due to the fact that it refers to relations between $q$ and $p$ which holds regardless of the use of the equations of motion.

Notice that the condition in Eq. \eqref{constphi} does not mean that the constraints vanish on the phase space $\Gamma$. Indeed, the constraints have non-zero Poisson bracket with the canonical variables, as can be seen considering \textbf{Example 1} presented in Sec. \ref{constraint}, where $\phi=p_n$ and $\{q^i,p_n\}=\delta^i_n$. Two functions $F$ and $G$ are said to be \textbf{weakly equal} if they coincide in the submanifold $\Gamma_C$ where $\phi^m=0$.  Mathematically we can write \cite{Henneaux:1992ig}
\begin{equation}
	F\approx G\Leftrightarrow F-G= c_m(q,p)\phi^m.
\end{equation}
In the same way, if two functions are equal in the whole phase space $\Gamma$, it is defined as a \textbf{strong equality}.
Therefore it is easy to see that the presence of the constraints implies some sort of redundancy in our system since each quantity can be defined up to a constraint. Indeed, the  Hamiltonian obtained from the initial Lagrangian through Eq. \eqref{phamlag} is not unique since it now depends on the gauge fixing choice, as we will see later in more details. 
\begin{figure}[t]
	\includegraphics[width=0.5\textwidth,center]{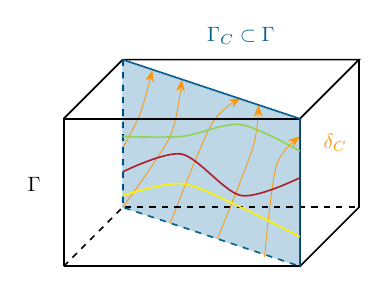}
	\caption{Different gauge-fixing surfaces (yellow, red and green curves) are shown crossing the gauge orbits $\delta C$. }
\label{diffgauges}
\end{figure}

As pictured in Fig. \ref{diffgauges}, we can choose several gauge-fixing surfaces, each of which crossing the gauge orbits in a different way, which tells us that the physical phase space $\Gamma_{\text{phys}}$ is not unique. Another way to see this, is to consider the mapping between the $(q,\dot{q})$-space and $(q,p)$-space. The map that bring $(q,\dot{q})$ to $(q,p)$ is not injective, which means that more than one pair of $(q,\dot{q})$ are mapped to the same  $(q,p)$.

This is what makes the Hamiltonian interesting in this context. Indeed, the Hamiltonian, which is defined as the {Legendre transformation} of the Lagrangian, $\mathcal{H}=\dot{q}^ip_i-\mathcal{L}$, is $\dot{q}$-dependent only through $p$ defined in Eq. \eqref{phamlag}. We have that the Hamiltonian is well defined only in the constraint surface, since otherwise it would not be uniquely defined, i.e.
\begin{equation}\label{dirham}
\tilde{\mathcal{H}}=	\mathcal{H}+c_m(q,p)\phi^m,
\end{equation}
where $c_m$ are some functions of $p$ and $q$ and $\phi^m$ are constraints. Notice that in the physical phase space we cannot distinguish between $\tilde{\mathcal{H}}$ and $\mathcal{H}$ since the additional term is zero in the constraints surface.
Let us consider a system with $N$ degrees of freedom and $M$ constraints, where $M\leq2N$.
It can be shown (Theorem 1.2 in \cite{Henneaux:1992ig}) that the Hamilton equations are then given by
\begin{equation}\label{hameq}
\begin{aligned}
	\dot{q}^i&
	=
	\frac{\partial \mathcal{H}}{\partial p_i}
	+u_m \frac{\partial \phi^m}{\partial p_i},
	\\
	-\dot{p}_i&
	=
	\frac{\partial \mathcal{H}}{\partial q_i}
	+u_m \frac{\partial \phi^m}{\partial q_i},
\end{aligned}
\end{equation}
where $u_m$ are the coordinates of the preimage of $p$ of the transformation function defined by Eq. \eqref{phamlag}. Moreover, the $u_m$ are introduced to make the Legendre transformation invertible, which can be done by using Eq. \eqref{hameq} and Eq. \eqref{constphi}.
It is interesting to notice that Eq. \eqref{hameq} can be derived from the variation of the action $\delta S=\delta \int^{t_1}_{t_0}\left(\dot{q}^ip_i-\mathcal{H}-u_m \phi^m\right)\ud t$, where the coefficients $u_m$ assume the role of Lagrange multipliers.
Considering Eq. \eqref{dirham} and  Eq. \eqref{hameq}, the evolution for a generic function $g$, can be described using the Poisson brackets
$
\{f,g\}=\frac{\partial f}{\partial q_i}\frac{\partial g}{\partial p^i}
-
\frac{\partial f}{\partial p_i}\frac{\partial g}{\partial q^i}
$
as
\begin{equation}\label{dotg}
\dot{g}=
\frac{\partial g}{\partial q_i}\dot{q}_i
+ \frac{\partial g}{\partial p_i} \dot{p}_i
=
\{g,\mathcal{H}\}
+u_m\{g,\phi^m\}.
\end{equation}
Recalling that the coefficients $u_m$ are \textbf{not} functions of $p$ and $q$ but of $q$ and $\dot{q}$, their Poisson brackets are not well defined. However we can still use all the properties (linearity, antisymmetry, product law and Jacobi identity) to prove that, knowing that $\phi^m\approx 0$, we can write the equation of motion for $g$ as \cite{Diraclec}
\begin{equation}
	\dot{g}
	=
	\{g,\mathcal{H}
	+u_m\phi^m\}.
\end{equation}
Thus we can define the \textbf{total Hamiltonian} as
\begin{equation}\label{totHg}
\mathrm{H}_{\text{tot}}
	=
\mathcal{H}
	+u_m\phi^m.
\end{equation}
It is important to notice the difference between Eq. \eqref{dirham} and Eq. \eqref{totHg}. The first one is a transformation which leaves the system unchanged, while the second one is a formulation of the Hamiltonian which gives us the equations of motion equivalent to the ones obtained from the Lagrangian (Euler-Lagrange equations).
Using the total Hamiltonian \eqref{totHg} we can also compute the dynamics of the constraints $\phi^m$, simply by considering Eq. \eqref{dotg} for the special case in which $g=\phi^n$. Self-consistency requires
\begin{equation}\label{phieom}
\{\phi^n,\mathcal{H}\}+u_m \{\phi^n,\phi^m\}=\dot{\phi}^n\approx 0,
\end{equation}
where  the index $m$ runs though all the possible constraints of the system.
The above equations give rise to different possibilities:
\begin{itemize}
	\item[\textit{a})] If we have $0=0$ it means that we have no new information and the primary constraints directly satisfy the identity;
	
	\item [\textit{b})] If the LHS is \textbf{not} identically equal to zero, we can have two additional possibilities:
	\begin{itemize}
	\item[\textit{i)}] It can depend on the $u_m$'s, in which case it turns into a system of equations in which the $u_m$'s are the unknown which we need to find;
	
	\item[\textit{ii)}] It is independent from the $u_m$'s and from the primary constraints. 
	 This means that we find ourself with an equation similar to Eq. \eqref{constphi}.
	Thus we need to introduce another constraint which is usually called \cite{Diraclec, Henneaux:1992ig, regge} \textbf{secondary constraint} $\chi(q,p)$. Notice that these constraints, contrary to the primary ones, make use of the equations of motion. Once these constraints are found, the procedure is the same as the one presented before, where now the consistency equation is defined as
	\begin{equation}\label{chieom}
		\{\chi^n,\mathcal{H}\}+u_m \{\chi^n,\phi^m\}=\dot{\chi}^n\approx 0.
	\end{equation}
		\end{itemize}
\end{itemize}
Point \textit{a}) is trivial, and point \textit{ii}) bring us back to the two other possibilities since the study of the secondary constraints proceed analogously to the one for the primary constraints. Therefore we are left studying point  \textit{i}). Indeed, the study of Eq. \eqref{phieom} can give us further information on the coefficients $u_m$ \cite{Henneaux:1992ig,Diraclec}. In particular we notice that Eq. \eqref{phieom} is of the form $A+u_m B=0$ which is a system of linear equations in the unknowns $u_m$. First we can solve the system as it is, obtaining the so called \textbf{general inohomeneous} solution ($U_m$). If we now solve the equation assuming the term $A=0$, we obtain the so called \textbf{associated homogeneous} solution ($V_m$). Therefore once added to the general inhomogeneous one, we find a complete general solution for $u_m$. The full set of solutions for the coefficients $u_m$ is given by
\begin{equation}\label{umtot}
u_m=U_m+v_aV^a_m,
\end{equation}
where $a$ is an index for all the possible homogeneous solutions and $v_a$ is a set of arbitrary coefficients. Using Eq. \eqref{umtot} we can write the total Hamiltonian \eqref{totHg} in terms of the $U_m$'s and $V_m$'s
\begin{equation}\label{totHg-UV}
	\begin{aligned}
	\mathrm{H}_{\text{tot}}
	&=
	\underset{\mathcal{H}'}{\boxed{\mathcal{H}
	+U_m\phi^m}}
	+v_aV^a_m\phi^m
	\\
	&=
	\mathcal{H}'
	+v_a
	\underset{\phi^a}{\boxed{V^a_m\phi^m}}
		\\
	&=
	\mathcal{H}'
	+v_a\phi^a.
	\end{aligned}
\end{equation}
From this we get that the eom can now be defined as
\begin{equation}
	\dot{g}\approx\{g,	\mathrm{H}_{\text{tot}}\}.
\end{equation}
It might seem like Eqs. \eqref{dirham}, \eqref{totHg} and \eqref{totHg-UV} are trivially the same. This is however \textbf{not} the case, and it gives us a chance to summarize and clarify what we obtained up until now.
In Eq. \eqref{dirham}, as explained, we defined a Hamiltonian transformation that leaves the dynamics invariant, with the coefficients $c_m$ depend on $(q,p)$. We then introduced a different Hamiltonian transformation such that the eom are the same as those obtained from the Lagrangian. This can be obtained by using the total Hamiltonian \eqref{totHg} where the coefficients $u_m$ are functions of $(q,\dot{q})$.  It is important to note that the coefficients $c_m$ and $u_m$ must satisfy certain consistency equations, meaning they cannot be chosen arbitrarily.

However, using the constraint's evolution, we are able to separate the coefficients $u_m$ in two parts, an arbitrary one and one dependent of the consistency equations. Therefore we are able to find a set of \textbf{arbitrary} coefficients $v_a$ with which to define a new total Hamiltonian \eqref{totHg-UV}. This defines a formulation of the Hamiltonian which makes the presence of \textbf{gauge freedom} manifest, as well as their generators. Indeed, the arbitrariness of the $v_a$'s implies that they can also be
chosen as being  arbitrary \textbf{time dependent} function while still satisfying the  consistency requirements \cite{Diraclec}. In particular it implies that, although for fixed $(q,p)$ we can define the dynamics of the system, the inverse is not true, i.e. from the dynamics we can not infer the value of all the phase space variables, which means that not all the $(q,p)$ are physical. This can explicitly be shown \cite{Henneaux:1992ig} by choosing $v_a(t_1)= \tilde{v}_a(t_1)$ and $v_a(t_2)\neq \tilde{v}_a(t_2)$, such that for $t_1<t<t_2$ we have $\delta v_a=(v_a-\tilde{v}_a)$, which means that there will be two possible evolutions of a dynamical variable $R$ at any time $t$, such that: 
\begin{equation}\label{gaugecont}
\delta \dot{R}=\delta v_a \{R,\phi^a\}.
\end{equation}

In this section we introduced primary and secondary constraints. However, these separation is not of fundamental importance. Instead, it is useful to introduce a new classification of the constraints. In the literature \cite{Diraclec}, a variable is said to be \textbf{first} or \textbf{second} class when it satisfies the following properties:
\begin{itemize}
\item [\textit{i})]	A \textbf{first class} dynamical variable $R$ has zero Poisson brackets with the constraints $\phi^m$, i.e. 
	\begin{equation}\label{firstclass}
		\{R,\phi^m\}\approx 0.
	\end{equation}
	It is important to notice that the Poisson brackets of two \textbf{first class} variables are still \textbf{first class}. To prove that we can use the Jacobi identity $\{\{A,B\},C\}=\{A,\{B,C\}\}-\{B,\{A,C\}\}$ for two first class variables $R$ and $F$ such that $\{R,\phi^m\}=r^m_n\phi^n$ and $\{F,\phi^m\}=f^m_n\phi^n$. Therefore we have
	\begin{equation}
		\begin{aligned}
			\{\{R,F\},\phi^m\}&=\{R,\{F,\phi^m\}\}-\{F,\{R,\phi^m\}\}
			\\
			&=
			\{R,f^m_n\phi^n\}-\{F,r^m_n\phi^n\}
			\\
			&=
			\{R,f^m_n\}\phi^n+f^m_n\{R,\phi^n\}
			-
			\{F,r^m_n\}\phi^n-r^m_n\{F,\phi^n\}
			\\
			&=
			\{R,f^m_n\}\phi^n+f^m_nr^m_n\phi^n
			-
			\{F,r^m_n\}\phi^n-r^m_nf^m_n\phi^n
			\\
			&\approx 0.
		\end{aligned}
	\end{equation}
	
\item[\textit{ii})]	A \textbf{second class} dynamical variable $R$ satisfies
	\begin{equation}\label{secondclass}
		\{R,\phi^m\} \not\approx 0,
	\end{equation}
	that is a variable whose Poisson brackets are not weakly zero for\textbf{ at least }one of the constraints.
	\end{itemize}
 It is important to notice that both $\mathcal{H}'$ and $\phi^a$ in Eq. \eqref{totHg-UV} are \textbf{first class}, and we will now prove it. Let us start with the constraints $\phi^a$, whose interesting feature is that, by definition, they are the only independent weakly vanishing quantities.
Considering the definition of $\phi^a=V^a_m \phi^m$ if we take the Poisson bracket with another constraint $\phi^n$, we obtain
\begin{equation}
	\begin{aligned}
		\{V^a_m \phi^m,\phi^n\}
		&=
		V^a_m \{\phi^m,\phi^n\}
		+
		\{V^a_m ,\phi^n\}\phi^m
		\\
		&
		\approx
		V^a_m 	\{\phi^m,\phi^n\}
		=0,
	\end{aligned}
\end{equation}
where the last equality comes from the definition of the coefficients $V^a_m$ which are defined to satisfy the homogeneous equation $V_m\{\phi^m,\phi^n\}=0$.
To prove that also $\mathcal{H}'$ is first class we use the definition $\mathcal{H}'=\mathcal{H}+U_m\phi^m$, from which we obtain
\begin{equation}
	\begin{aligned}
		\{\mathcal{H}',\phi^n\}
		&=
		\{\mathcal{H}+U_m\phi^m,\phi^n\}
		\\
		&
		=
		\{\mathcal{H},\phi^n\}
		+
		U_m\{\phi^m,\phi^n\}
		+
		\{U_m,\phi^n\} \phi^m
		\\
		&
		\approx
		\{\mathcal{H},\phi^n\}
		+
		U_m\{\phi^m,\phi^n\}
			\\
		&
		=0,
	\end{aligned}
\end{equation}
where the last step is due to the fact that the coefficients $U_m$ are defined as the solution of the inhomogeneous equation. 
Therefore Eq. \eqref{totHg-UV} gives the total Hamiltonian in terms of a first order Hamiltonian and first order primary constraints $\phi^a$.

 We are interested in removing \textbf{redundant} variables from the formulation of a constrained system, in particular this means that we need to introduce new brackets defined on the constraints surface, since Poisson brackets are not necessarily well defined in this context. To do so we start by arranging our initial constraints in linear combinations that are \textbf{first class}, i.e. $V^a_m\phi^m$. The remaining independent combinations are the \textbf{second class} constraints.
  Following Dirac's notation, we denote these second class constraints as $\chi_s$ where $s=1,...,S$ with $S$ the number of remaining second class constraints. It can be proven \cite{Diraclec} that the determinant of the matrix formed by the Poisson brackets taken between these constraints, is \textbf{strongly} non vanishing, i.e. these second class constraints form a \textit{nonsingular} $S\times S$ matrix of commutations, i.e. $C_{ij}=\{\chi_i,\chi_j\}$ . Notice that, a key part of the proof, is the assumption that the dimension of the second class vector is the smallest possible. Due to the antisymmetry of the Poisson brackets we have that $C_{ij}$ is also antisymmetric, and therefore \textit{even dimensional}\footnote{It is known that the determinant of an odd-dimensional antisymmetric matrix must vanish.}. Thank to this matrix we are now able to define a new set of brackets, called \textbf{Dirac brackets} which allow to eliminate \textit{second class} constraints.
Since $\text{Det}(C_{ij})\neq 0$ we know that the matrix is invertible. We define a new set of brackets $\{A,B\}_D$ between two quantities $A$ and $B$ by
\begin{equation}\label{dirbra2}
	\{A,B\}_D=\{A,B\}-\{A,\chi_i\}(C_{ij})^{-1}\{\chi_j,B\}.
\end{equation}
It is possible to prove \cite{Diraclec} that the Dirac brackets satisfy the same properties of the Poisson brackets. 
An important feature is that the new brackets between a general variable $A$ and any second class constraint vanish, i.e.
\begin{equation}\label{dirbra}
	\{A,\chi_s\}_D
	=
	\{A,\chi_s\}
	-\{A,\chi_i\}
	\underset{\delta^i_s}{\boxed{(C_{ij})^{-1}\{\chi_j,\chi_s\}}}
	=0.
\end{equation}
Therefore, substituting all the Poisson brackets with Dirac brackets allows to reduce the number of variables by solving the second class constraints.

\begin{figure}[t]
	\centering
	\begin{subfigure}{0.35\textwidth}
		\centering
		\includegraphics[width=1.3\textwidth]{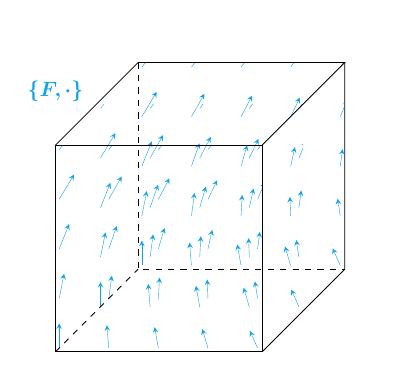}
		\caption{}
		\label{vecfildF}
	\end{subfigure}%
	\begin{subfigure}{0.35\textwidth}
		\centering
		\includegraphics[width=1.2\textwidth]{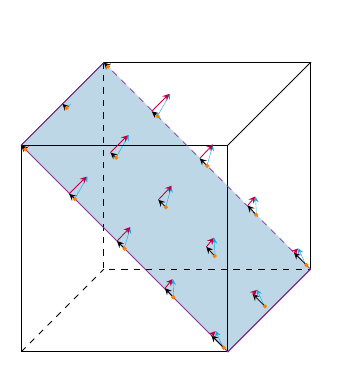}
		\caption{ }
		\label{vecfildFconst}
	\end{subfigure}
	\caption{Representation of the dynamics between subsurfaces, each at a fixed time $t$.}
	\label{FIGdiracbrack}
\end{figure}

We now want to give a geometrical intuition of the Dirac brackets.
Let us consider the vector field given by $\vec{v}=\{F, \cdot\}$ for a generic function $F(q,p)$ of the phase space  (See. Fig. (\ref{vecfildF})).
Notice that, if $\vec{v}(g)=\{F,g\}=0$, then $g$ is constant along the trajectories defined by the vector field $\vec{v}$. Therefore $\vec{v}_D=\{F,\cdot\}_D$ defines a vector field tangent to the surface $\chi^m=\text{const}$, where $\chi^m$ are second class constraints. Indeed, we know that, by definition, the Dirac brackets between $F$ and second class constraints are $\{F,\chi^m\}_D=0$, which means that $\chi^m$ are constant along the vector field $\vec{v}_D$. In Fig.( \ref{vecfildFconst}) only one of these surfaces is pictured.

\subsubsection{Dirac observables}
Considering what we discussed until now, each constraint can be categorized as either first-class or second-class, and, in turn, they can be further classified as primary or secondary constraints. The classification into primary and secondary constraints is not important once we go from the Lagrangian to the Hamiltonian formalism as it is mainly related to the Lagrangian. 
Since the first class constraints can modify the dynamics of a constraint system without affecting its physical properties, they are \textbf{generators of gauge transformations}, i.e. transformations which leave the physical system invariant as shown in Eq. \eqref{gaugecont}.

By definition GR is a \textbf{diffeomorphism invariant} theory, which reflects the fact that each observer can choose a different reference frame. The diffeomorphisms of a spacetime $(\mathcal{M},g)$ form the group $\text{Diff}(\mathcal{M})$. 
More precisely, the freedom of choosing the slicing of spacetime is generated by the \textbf{scalar Hamiltonian} $\mathcal{H}_0$, while the freedom of choosing the coordinates in each slices is generated by the \textbf{vector Hamiltonian} $\mathcal{H}_i$.
 This gauge freedom, although useful, can be a problem if we aim to eventually quantise our theory since, as we know, in quantum mechanics physical variables are defined by operators. 
Thus, there is the need to find some gauge invariant variables which represents our observables \cite{Bergmann:1961wa}. Generally speaking these variables, in the context of constrained systems, are known as \textbf{Dirac observables} $D$, which are defined as those variables which commute\footnote{These commutation is intended with Poisson brackets.} with the total Hamiltonian \eqref{totHg-UV}. In particular, we can define \cite{Giesel:2008zz} the so called \textbf{strong} Dirac observable as
\begin{equation}\label{Dirstrong}
\{D,C_m\}=0, \forall m
\end{equation}
where $C_m$ are the constraints of the system.
We can also define \textbf{weak} Dirac observables as
\begin{equation}\label{Dirweak}
	\{D,C_m\}\approx0, \forall m.
\end{equation}
Therefore the strong Dirac observables are quantities that strongly commute with the constraints, while the weak Dirac observables commute only on the constraints surface.
In our specific case in which the system's constraint is the Hamiltonian, the definition of Dirac observables implies a conditions on their evolution. In particular, in full GR, the Dirac observables are \textbf{constants of motion}. This is a known problem since all the observables of our theory would be non dynamical. This issue has been addressed several times in the literature (see e.g. \cite{Rovelli:1990ph}).

An important property of the Dirac observables is that they form a\textbf{ closed algebra}, i.e.
\begin{equation}\label{Dirclosal}
	\begin{aligned}
\{
\{ D_i~, D_j\},
C_m
\}
&=
\{
 D_i D_j,
 C_m
\}
-
\{
 D_j D_i
,
C_m
\}
\\
&
=
 D_i\{
 D_j
,
C_m
\}
+
\{
 D_i
,
C_m
\}
 D_j
\\
&
-
 D_j\{
 D_i
,
C_m
\}
-
\{
 D_j
,
C_m
\}
 D_i
\\
&
\approx0,
\end{aligned}
\end{equation}
where the last weak equality is substituted by a strong equality in case of strong Dirac observables. Eq. \eqref{Dirclosal} shows that, by definition of Dirac observables, i.e. a quantity which commutes with the constraints, the Poisson bracket of two Dirac observables commutes with the constraints. This means that $\{D_i,D_j\}\propto D_k$ which proves the algebra is closed.
This makes them good candidates for quantization, see for example \cite{Rovelli-an-hyp}. Moreover, the algebra is still closed when computed using the Poisson Brackets, as the Dirac observables commutes with the second class constraints making the second term in the RHS of Eq. \eqref{dirbra2} zero,
\begin{equation}\label{Dalgebra}
	\{ D_i, D_j\}_D\approx\{ D_i, D_j\}.
\end{equation}
It is also straightforward to see that the definition of Dirac observables does not depend on the particular choice of the representatives as they all must differ by a constraint and
\begin{equation}
	\{ D_i+{\alpha}^{m} \mathcal{H}_{m}, D_j\}\approx\{ D_i, D_j\},
\end{equation}
for any ${\alpha}^{m}$.

The number of (independent) Dirac observables is equivalent to the number of reduced variables in the constraints parametrizing any gauge-fixing surface. Using these observables we can write the Hamiltonian in a \textbf{gauge-invariant} way, see Ch. \ref{Chapter2} for details.

In the explicit case in which the second class constraints are defined by the gauge-fixing functions and the first class constraints given by the Hamiltonian, defining the invertible matrix $C_{ij}=\{g_i,\mathcal{H}_j\}$, we have that any physical variable in the gauge-fixing surface must be equal to a Dirac observable modulo a combination of constraints and gauge-fixing conditions, see \ref{gid}. This implies that there must exist $D_i$ such that
\begin{equation}\label{Dirphys}
	v^{\text{phys}}_i= D_i+{\beta}^m_i C_m,
\end{equation}
for some ${\beta}^{m}$.  Eq. \eqref{Dirphys} is equivalent to the mapping
\begin{equation}\label{iso}
	\mathcal{D}\ni D_i\mapsto  v_i\approx  D_i\big|_{\delta c_{\mu}= 0}~,
\end{equation}
and implies that there exists a \textbf{canonical isomorphism} between the physical variables in any gauge-fixing surface and the Dirac observables due to the invariance of the Dirac brackets under second class constraints, i.e.
\begin{equation}\label{vphyD}
	\{v^{\text{phys}}_i~,v^{\text{phys}}_j\}_D\approx\{ D_i~, D_j\}_D\bigg|_{g_i=0}.
\end{equation}
Since this mapping is a canonical isomorphism this means that from the physical space $\Gamma_{\text{phys}}$ we can pull-back to the space of Dirac observables, which we call Dirac space $\mathcal{D}$, see Fig. \ref{fig:Dirac}.
This means that the Dirac observables parametrize a\textbf{ unique phase space }where the dynamics are generated by a\textbf{ unique Hamiltonian} that is a function of the Dirac observables.
Note that if we choose the representatives of the Dirac observables in such a way that they commute with a given set of gauge-fixing conditions,
\begin{equation}\label{spD}
	\{\delta D_i,g_{j}\}=0,
\end{equation} 
for all $j$'s, then the Dirac bracket \eqref{dirbra} can be equivalently expressed as 
\begin{equation}
	\{A,B\}_D=\{A,\delta D_i\}\{\delta D_i,\delta D_j\}^{-1}\{\delta D_j,B\},
\end{equation}
for any $A$ and $B$.
The above formula shows how any variable inserted into the Dirac bracket is first unambiguously associated with a Dirac observable \eqref{Dirweak} which coincides with that variable in a given gauge-fixing surface. Next, the resulting observable is computed in accordance with the commutation rule of Eq. \eqref{Dalgebra} and yields a Dirac observable that is again given in the representation (\ref{spD}).

\section{Summary}

In this chapter, we introduced the concept of constraints and constraint surface. 
In particular, we discussed how by solving the constraints, i.e., going from the full 
phase-space to the subspace defined by imposing the constraints, we are left with 
some redundant degrees of freedom which represent the gauge freedom of our system. 
We also introduced different classifications of constraints, like the definition of 
primary and secondary constraints associated with the equations of motion of our system.

Additionally, we provided the more useful definitions of first and second-class 
constraints, which do or do not commute with the remaining constraints, respectively. 
This latter distinction is more fundamental, as it is related to the nature of the 
symmetries involved in our system and plays an important role in the quantization. 
Moreover, through the introduction of second-class constraints, we were able to introduce 
the Dirac bracket, which acts as a generalization of the Poisson brackets in constrained 
systems.

Finally, we introduced the concept of Dirac observables, which are defined as quantities 
which (weakly) commute with the constraints and, as such, in general relativity they are 
not dynamical. These observables are good candidates for quantization as they are gauge-invariant 
quantities and their algebra is closed. They will play an essential role in the physical 
interpretation of the results obtained in Ch. \ref{Chapter3}.

%% file: Chapters/Chapter2.tex

\chapter{Constrained systems in perturbation theory} 

\label{Chapter2} 

In Ch. \ref{Chapter1}, we introduced the theory of constrained Hamiltonian systems in the ADM formalism, which can be studied using the Dirac method as presented in Sec. \ref{secdirac}. In the current chapter, we will use the notions presented in Ch. \ref{Chapter1}, focusing on the theory of perturbations around any spatially homogeneous spacetime. 
Cosmological Perturbation Theory (CPT) is widely accepted as a model for understanding the behaviour of gravity on large cosmological scales and at early times.
It allows to describe primordial density fluctuations which gave rise to the large-scale structures we observe today, such as galaxies and galaxy clusters. As such, it plays an important role  in the development of quantum theories that aim to explain the origin of the primordial structure of the universe. 

In this chapter we will  keep our considerations general without specifying any particular background model. We will then study the complete Hamiltonian formalism, focusing on the gauge-independent description of cosmological perturbations, as well as on the issues of gauge-fixing, gauge transformations and spacetime reconstruction by the use of the Dirac method. Next, we will introduce a new parametrization of the spacetime based on the Kucha\v r decomposition, which will provide us a better understanding of the gauge-fixing procedure and spacetime reconstruction. 
This procedure paves an ideal route towards gravity quantization, and the approach to cosmological perturbation theory follows essentially the same path. However, there are also differences between the non-perturbative and perturbative approaches that should be noted. The most striking difference is that, in non-perturbative canonical gravity, Dirac observables are constants of motion, as the Hamiltonian itself is a constraint. Consequently, the dynamical variables cannot be expressed exclusively in terms of Dirac observables, and an extra variable, the internal clock, is needed. It is assumed that the internal clock commutes with all the Dirac observables. This assumption makes the symplectic structure of the physical phase space depend on the choice of internal clock, giving rise to the so-called multiple choice time problem. \cite{kuchar1988,Hajicek:1999ti,Malkiewicz:2014fja,Malkiewicz:2015fqa,Alexander:2012tq}. On the other hand, in the perturbative approach the second-order Hamiltonian is not a constraint, and the respective Dirac observables are, in general, dynamical. The reason for this to happen is that the first-order gauge transformations keep the background ``time" fixed whereas the true dynamics of perturbations occurs in the evolving background. Hence, the multiple choice problem is confined to zeroth order. Nonetheless, if one is to quantize both the background and the perturbations, the choice of the background clock has to be made and it will affect the dynamics of perturbations \cite{Malkiewicz2020}.

The Hamiltonian formulation in CPT has been derived before in simple background spacetime such as  the Friedmann universe \cite{Mukhanov:1990me,Langlois:1994ec, Dapor_2013, Malkiewicz_2019, Artigas_2022, Domenech:2017ems} or the Bianchi Type I model \cite{Uzan,Agullo2020, boldrin2021dirac} (See Ch. \ref{Chapter3}).

The following work has been presented in \cite{Boldrin:2022vcp}.

\section{Hamiltonian formalism} 

We start this chapter by introducing the Hamiltonian formalism in perturbation theory. We will first present its form up to $n$-th order of perturbation from which we will then restrict our attention to the Hamiltonian up to second order.

\subsection{Hamiltonian up to $n$-th order of perturbation}

In this section we aim to present a general form of the perturbed Hamiltonian up to the $n$-th order of perturbation, subsequently we are going to restrict to the case $n=2$.
To understand the form of the perturbed Hamiltonian, we need to first consider its full form. In particular, from Eq.\eqref{hamgra} we know that the Hamiltonian is \textbf{linear} in both the lapse function and shift vector.  This means that in the perturbative expansion of $H$, there can not be more than one derivative with respect to $N^\mu$. Furthermore we can assume the Hamiltonian to be at most second order in the momenta (this allows us to obtain the canonical kinetic term), which implies that we can not have more than the second derivative with respect to the momentum $p$.

To obtain a perturbative expansion of $H$ we need to write its Taylor expansion. We start by writing the expansion of the variables $N^\mu$, $p$ and $q$.
The expansion will be in the form $\epsilon^{n}(\beta_n+\delta^{(n)}\beta)$, where $\epsilon$ is a bookkeeping parameter ultimately set to 1,
 while $\beta_n$ is the \textbf{homogeneous part} of the contribution and $\delta^{(n)}\beta$ is the \textbf{inhomogeneous }one.
So we have
\begin{align}
	N^\mu&=N^\mu(\epsilon)=
	N^\mu_0
	+\epsilon
	\left(
	N^\mu_1+\delta^{(1)}N^\mu
	\right)
	+\epsilon^2
	\left(
	N^\mu_2+\delta^{(2)}N^\mu
	\right)
	+...;\label{expN}\\
	p^i&=p^i(\epsilon)=
	p^i_0
	+\epsilon
	\left(p^i_1+\delta^{(1)}p^i
	\right)
	+\epsilon^2
	\left(p^i_2+\delta^{(2)}p^i
	\right)
	+...;\label{expp}\\
	q_j&=q_j(\epsilon)=
	{q_j}_0
	+\epsilon
	\left({q_j}_1+\delta^{(1)}q_j
	\right)
	+\epsilon^2
	\left({q_j}_2+\delta^{(2)}q_j
	\right)
	+...;\label{expq}
\end{align}
As previously mentioned, the two variables $N^\mu$ and $p$ are perturbed up to first and second order respectively, while the position $q$ has no order restriction. The total Taylor expansion of the Hamiltonian reads 
\begin{equation}
\begin{aligned}\label{HexpNpq}
	H(N,p^i,q_j)=&
	H_0
	+
	\sum_{\alpha=1}^{\infty}
	\frac{1}{\alpha!}
	\frac{\partial^{\alpha} H}{\partial^{\alpha} q}	\delta^{(\alpha)} q
	+
	\sum_{\alpha=1}^{\infty}
	\frac{1}{\alpha!}
	\frac{\partial^{\alpha} H}{\partial^{\alpha} q}	
	\frac{\partial H}{\partial N}	
	\delta N\delta^{(\alpha)} q
		\\
	&
	+
	\sum_{\alpha=1}^{\infty}
	\frac{1}{\alpha!}
	\frac{\partial^{\alpha} H}{\partial^{\alpha} q}	
	\frac{\partial H}{\partial p}
	\delta^{(\alpha)} q	\delta p	
	+
	\sum_{\alpha=1}^{\infty}
	\frac{1}{\alpha!}
	\frac{\partial^{\alpha} H}{\partial^{\alpha} q}	
	\frac{\partial H}{\partial N\partial p}
	\delta^{(\alpha)} q\delta N\delta p
		\\
	&
	+
	\frac{1}{2}
	\sum_{\alpha=1}^{\infty}
	\frac{1}{\alpha!}
	\frac{\partial^{\alpha} H}{\partial^{\alpha} q}	
	\frac{\partial^2 H}{\partial p\partial p}
	\delta^{(\alpha)} q	\delta^2 p
	+
	\frac{1}{2}
	\sum_{\alpha=1}^{\infty}
	\frac{1}{\alpha!}
	\frac{\partial^{\alpha} H}{\partial^{\alpha} q}	
	\frac{\partial^3 H}{\partial N\partial p\partial p}	
	\delta^{(\alpha)} q
	\delta N\delta^2 p
\end{aligned}
\end{equation}
where we are using the notation $	\frac{\partial^{\alpha} H}{\partial^{\alpha} q}	=\frac{\partial^{\alpha} H}{\partial q^{i_1}...\partial q^{i_\alpha}}$, where the sum over $\alpha$ is a placeholder for $\sum_{n=1}^{\infty} \sum_{\alpha=1}^{n}$, and we are using Eqs. \eqref{expN}, \eqref{expp}, \eqref{expq}.
 The \textbf{general form} of the Hamiltonian at $2n$-th order is given by
\begin{equation}\label{genham16}
	\begin{aligned}
	\mathds{H}^{(2n)}=
	&
	\Psi^{(0)}_{ij}(p_n^i+\delta^{(n)}p^i)(p_n^j+\delta^{(n)}p^j)
	+
	\Phi^{(0)}_{ij}(q_{i,n}+\delta^{(n)}q_i)(q_{j,n}+\delta^{(n)}q_j)\\
	&
	+
	\Theta^{(0)}_{ij}(q_{i,n}+\delta^{(n)}q_i)(p_n^j+\delta^{(n)}p^j)
	+
	\Pi^{(n)}_{i}(p_n^i+\delta^{(n)}p^i)\\
	&
	+
	\Delta^{(n)}_{i}(q_{i,n}+\delta^{(n)}q_i)
	+C^{(n)}(N_n+\delta^{(n)}N)
	\end{aligned}
\end{equation}
where $	\Psi^{(0)}_{ij}$, $\Phi^{(0)}_{ij}$, $\Theta^{(0)}_{ij}$, are time-dependent background coefficients and $\Pi^{(n)}_{i}$, $\Delta^{(n)}_{i}$ depends on lower order variables as well as background ones, while $C^{(n)}$ is a term which contributes only to the constraint equations.
The Hamiltonian \eqref{genham16} is obtained from the expansion of the Hamilton equations in $\epsilon$, order by order.

Remember that $N$ is a vector containing both the shift vector and lapse function.
\subsection{ADM Hamiltonian up to $2$nd order of perturbation}\label{ADMpert}

Let us now consider the Hamiltonian in the ADM formalism given in Eq. \eqref{hamgra}
\begin{equation}\label{hamtemspa}
	H_{\text{gravity}}=\int \left(N\mathcal{H}_0+N_i\mathcal{H}^i\right)\ud^3x,
\end{equation}
where \cite{Wald,gravitation}
\begin{equation}\label{hamtemp}
	\mathcal{H}_0=
	\sqrt{q}\left[-{^3}R
	+q^{-1}(\pi^i_j\pi^j_i
	-\frac{1}{2}\pi^2)\right]
\end{equation}
and
\begin{equation}\label{hampspa}
	\mathcal{H}^i=-2\sqrt{q}D_j\left(\frac{\pi^{ij}}{\sqrt{q}}
	\right).
\end{equation}
The covariant derivative is defined as
\begin{equation}\label{covariantderivativegen}
	D_{j}\pi^{il}=\partial_j \pi^{il}+ \tensor{\Gamma}{^i_{kj}}\pi^{kl}+ \tensor{\Gamma}{^l_{kj}}\pi^{ik}.
\end{equation}
In the following we will perturb the Hamiltonian \eqref{hamtemspa} up to \textbf{second order}. We consider a spatially homogeneous background with fixed spatial coordinates, such that the background shift vector components $N^i$ vanish \cite{Bojowald:2006tm}, while the background lapse is left unspecified. Also, the diffeomorphism constraints of the background model $\mathcal{H}^{i(0)}=0$ must vanish trivially if the metric does not explicitly depend on spatial coordinates. The total Hamiltonian will have the form
\begin{equation}\label{Htot}
	\mathbb{H}=\int \left( N\mathcal{H}^{(0)}+N\mathcal{H}^{(2)}+\delta N^\mu\delta \mathcal{H}_\mu\right)\ud^3x,
\end{equation}
where $N$ is the \textbf{zeroth order} lapse function, $\delta N^\mu=(\delta N, \delta N^i)$ are the \textbf{first order} lapse and shift functions, $\mathcal{H}^{(0)}$ and $\mathcal{H}^{(2)}$ are respectively the \textbf{zeroth order} constraint and the \textbf{second order }scalar Hamiltonian, and $\delta \mathcal{H}_\mu$ are linearized scalar and diffeomorphism \textbf{constraints}. {Notice that there are no linear terms in the perturbations as
	they all average to zero when integrated over all space.} This Hamiltonian is a function of the homogeneous three-metric $\overline{q}_{ij}$ and three-momentum  $\overline{\pi}^{ij}$, and the pure inhomogeneous perturbations of the \textbf{three-metric } and \textbf{three-momentum } given by
\begin{equation}\label{3mpert}
\delta q_{ij}=q_{ij}-\overline{q}_{ij},
\\ ~~
 	\delta{\pi}^{ij}={\pi}^{ij}-\overline{\pi}^{ij}.
 \end{equation}
The total canonical structure can be shown to be the sum of the \textbf{homogeneous} and \textbf{inhomogeneous} canonical structures.

The interpretation of the terms in Eq. \eqref{Htot} is as follows. The zeroth-order constraint $\mathcal{H}^{(0)}$ generates time transformations in the homogeneous background spacetime while keeping the inhomogeneous fields fixed. The first-order constraints $\delta\mathcal{H}_\mu$ generate linearized transformations of the inhomogeneous spacetime while keeping the homogeneous background fixed. The second-order Hamiltonian $\mathcal{H}^{(2)}$ generates the dynamics of perturbations that must occur simultaneously with the dynamics of the homogeneous background generated by $\mathcal{H}^{(0)}$.

\begin{figure}[t]
	\includegraphics[width=0.8\textwidth,center]{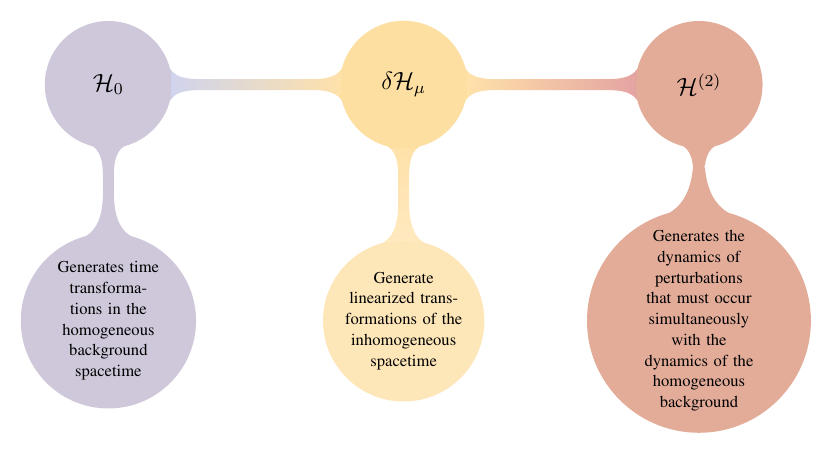}
	\caption{Schematization of the role of the order of perturbation terms of the Hamiltonian up to second order.}
	\label{mindmap}
\end{figure}

In what follows, we will calculate the terms of the perturbed Hamiltonian within the framework of \textbf{linear metric perturbation}, extending the computation up to the second order.
\subsubsection{$0$th order}

We start by computing the form of the Hamiltonian at\textbf{ zeroth order}.
In particular we want to find
\begin{equation}\label{hamzerotot}
	\mathds{H}^{(0)}=N{\mathcal{H}_0}^{(0)}+N_i{\mathcal{H}^i}^{(0)}.	
\end{equation}
The components of Eq. \eqref{hamtemp} and Eq. \eqref{hampspa} depend on the specific metric chosen. In \cite{Malkiewicz_2019}, an application to the Friedman-Lemaître-Robertson-Walker metric is shown. In Chapter \ref{Chapter3}, we will explore the application to an anisotropic model, the \textbf{Bianchi I} universe, which represents one of the original contributions of this thesis.
From Eq.  \eqref{hamtemp} we have
\begin{equation}\label{hamtemp0}
	{\mathcal{H}_0}^{(0)}=
	\overline q^{-1/2}
	\left(
	\tensor{\overline{\pi}}{^i_j}
	\tensor{\overline{\pi}}{^j_i}
	-\frac{1}{2}
	\overline{\pi}^2
	\right)-\sqrt{q}\,\,{}^3\overline{R},
\end{equation}
where all overlined quantities are considered at the background level. The Einstein summation convention is assumed throughout.
For completeness, we will now explicitly write the zeroth-order components of Eq. \eqref{hamtemp0}.
The trace of the conjugate momentum square can be written 
\begin{equation}
\begin{aligned}\label{pipi0}
	\overline q^{-1/2}
	\tensor{\overline\pi}{^i_j}
	\tensor{\overline \pi}{^j_i}
	&=
\overline q^{-1/2}
	{\overline \pi^{ik}}
	{\overline q_{kj}}
	{\overline \pi^{jn}}
	{\overline q_{ni}}
\end{aligned}
\end{equation}
The trace square of the conjugate momenta reads
\begin{equation}
\begin{aligned}\label{pi20}
	\overline \pi^2&
	={\tensor{\overline\pi}{^i_i}}
	{\tensor{\overline\pi}{^j_j}}
	={\overline\pi^{ik}}
	{\overline q_{ki}}
	{\overline\pi^{jl}}
	{\overline q_{lj}}
\end{aligned}
\end{equation}
The \textbf{Ricci scalar} in Eq. \eqref{hamtemp0} is defined as 
\begin{equation}\label{Ricci}
	{^3}R={^3}R_{ih}q^{ih} = 
	q^{ih}\partial_{l} \tensor{\Gamma}{^{l}_{hi}} 
	- 
	q^{ih}\partial_{h} \tensor{\Gamma}{^{k}_{k i}} 
	-q^{ih}\tensor{\Gamma}{^{m}_{l i}}\tensor{\Gamma}{^{l}_{hm}}
	+q^{ih}\tensor{\Gamma}{^{m}_{hi}}\tensor{\Gamma}{^{l}_{lm}}.
\end{equation}
The \textbf{Christoffel symbols} are defined by
\begin{equation}\label{Chris}
	\tensor{\Gamma}{^{i}_{mn}} = \frac{1}{2} q^{ik}(q_{k m, n} + q_{k n, m} - q_{mn, k}) ,
\end{equation}
where $q_{km,n}$ denotes the spatial derivative with respect to $q_n$, i.e. $q_{km,n}=\frac{\partial q_{km}}{\partial q^n}$.
Note that Eq. \eqref{Chris} contains only spatial derivatives of the metric. For any homogeneous background metric, which we will assume moving forward, all spatial derivatives are zero. Therefore, the zeroth-order term of the Ricci scalar vanishes. 


Let us now make some considerations for the \textbf{vector part} of the background Hamiltonian,
\begin{equation}\label{hampspa0}
	{\mathcal{H}^i}^{(0)}=-2\sqrt{\overline q}D_j{\left(\frac{\pi^{ij}}{\sqrt{q}}
		\right)}^{(0)}.
\end{equation}
Since the Christoffel symbols vanish when computed with a homogeneous background metric, the covariant derivative can be substituted with a normal derivative at zeroth order, $D_j\rightarrow \partial_j$. Analogously, the conjugate momenta at zeroth order would only be time dependent, which would make the total contribution of Eq.\eqref{hampspa0}, null. 

\subsubsection{$1$st order}

We want to compute $\delta^{(1)}\mathcal{H}^i$ and $\delta^{(1)}\mathcal{H}^0$.
Starting with the \textbf{vector component}, we want to obtain 
\begin{equation}\label{Hi1-inizio}
	\delta^{(1)}\mathcal{H}^i
	=
		-2
		\left(\sqrt{q}\right)^{(0)}
		\left[D_j
		\left(\frac{\pi^{ij}}{\sqrt{q}}
		\right)
		\right]^{(1)}
	-2(\sqrt{q})^{(1)}
	\left[
	D_j\bigg(\frac{\pi^{ij}}{\sqrt{q}}\bigg)
	\right]^{(0)}
		\end{equation}
		The second term on the RHS of Eq. \eqref{Hi1-inizio} vanishes, thus we simply need to compute the spatial derivative of the conjugate momenta $\pi^{ij}$ at the background. 
We are interested in \textbf{linear perturbation} of the metric, hence there will not be second order perturbations of the metric but just products of first order ones. In other words, in all our further perturbations,  we will have equations proportional to $\delta q_{ij}$ and $\delta \pi^{nm}$ or their products (for second order perturbations). 
The \textbf{first term} on the RHS of Eq. \eqref{Hi1-inizio} is obtained by computing the perturbation of both the covariant derivative $D_j$ and the conjugate momenta $\pi_{ij}$, that is
		\begin{equation}\label{Dpi-first}
			\begin{aligned}
\left[D_j
\left(\frac{\pi^{ij}}{\sqrt{q}}
\right)
\right]^{(1)}
	&
	=
		\overline q^{-1/2}\partial_j {\delta\pi^{ij}}
		+ \boxed{q^{-1/2}\tensor{\Gamma}{^i_{kj}}^{(0)}{\delta \pi^{kj}}}
			\\
		&
		+\boxed{\overline q^{-1/2} \tensor{\Gamma}{^j_{kj}}^{(0)}{\delta\pi^{ik}}}
			+\overline{\pi}^{ij}\partial_j(q^{-1/2})^{(1)}
				\\
			&
		+\overline q^{-1/2}
		\tensor{\Gamma}{^i_{kj}}^{(1)}\overline{\pi}^{kj}
		+ \overline q^{-1/2}\tensor{\Gamma}{^j_{kj}}^{(1)}\overline{\pi}^{ik},
	\end{aligned}
\end{equation}
where the two boxed terms vanish.
The first order terms $\tensor{\Gamma}{^j_{kj}}^{(1)}$ and $(q^{-1/2})^{(1)}$ are easily computed following the same reasoning showed for the zeroth order, obtaining
\begin{equation}
	\begin{aligned}\label{Gamma1-b}
		\tensor{\Gamma}{^{i}_{kj}}^{(1)}
		&=
		\frac{1}{2} {q^{il}}
		\left(
		\delta q_{l k, j} 
		+\delta  {q_{l j, k}}
		-\delta  {q_{kj, l}}
		\right),
	\end{aligned}
\end{equation}
where we used the perturbation of the contravariant form of the metric, i.e. 
\begin{equation}\label{delta-q-contrav}
		\delta q^{ik}=\delta q_{lj} \overline q^{lk}\overline q^{ij},
\end{equation}
as well as
	\begin{equation}\label{qprimo}
		(\sqrt{q})^{(1)}
		=
		\frac{1}{2}
		\overline q^{1/2}
		\overline q^{nm}\delta q_{nm}.
\end{equation}
In a similar fashion to Eq. \eqref{qprimo}, we have
\begin{equation}\label{qprimo-inverse}
	\left( q^{-1/2}\right) ^{(1)}
=
	-\frac{1}{2}\overline q^{-1/2} \overline q^{ij}\delta q_{ij}.
\end{equation}

Let us now compute the first order for the \textbf{scalar part} of the Hamiltonian, i.e. 

\begin{equation}\label{H0-1}
	\begin{aligned}
	\delta^{(1)}{\mathcal{H}_0}
	&
	=
	\left[-\sqrt{q}\,\,\,{}^3{R}
	+q^{-1/2}
	\left({\pi^i_j}{\pi^j_i}
	-
	\frac{1}{2}{\pi^2}
	\right)
	\right]^{(1)}
	\\
	&
	=
-{	\sqrt{q}^{(0)}\,\,\,{^3}{R^{(1)}}}
-\boxed{			\sqrt{q}^{(1)}\,\,\,{^3}{R^{(0)}}}
\\
&
	+\left(q^{-1/2}\right)^{(1)}
	\left(\overline{\pi}^i_j\overline{\pi}^j_i
	-
	{\frac{1}{2}{\pi^2}^{(0)}}
	\right)
	\\&
	+
	\left(q^{-1/2}\right)^{(0)}
	\left[2\delta\pi^i_j\overline{\pi}^j_i
	-
	\frac{1}2\left(\pi^2\right)^{(1)}
	\right],
	\end{aligned}
\end{equation}
where, as before, the boxed term vanishes. Notice that in the last line the term $2\delta\pi^i_j\overline{\pi}^j_i$ is multiplied by a factor $2$ taking into account the symmetry.

From the definition of the\textbf{ Ricci scalar} in Eq. \eqref{Ricci} we know that we need to compute products of first and zeroth order for both the metric and the Christoffel symbols. Once again, the equation can be largely simplified for a homogeneous background metric, obtaining
\begin{equation}\label{Ricci-1}
	\begin{aligned}
	{^3}R^{(1)} 
	&= 
	\left(q^{ih}\partial_{l} \tensor{\Gamma}{^{l}_{hi}} 
	- 
	q^{ih}\partial_{h} \tensor{\Gamma}{^{k}_{k i}} 
	-q^{ih}\tensor{\Gamma}{^{m}_{l i}}\tensor{\Gamma}{^{l}_{hm}}
	+q^{ih}\tensor{\Gamma}{^{m}_{hi}}\tensor{\Gamma}{^{l}_{lm}}
	\right)^{(1)}
	\\
	&
	=
	\overline	q^{ih}\left(\partial_{l} \tensor{\Gamma}{^{l}_{hi}} \right)^{(1)}
- 
\overline q^{ih} \left(\partial_{h}\tensor{\Gamma}{^{k}_{k i}} \right)^{(1)}
	\end{aligned}
\end{equation}
As we can see, in order to compute the first order of the Ricci scalar we just need to use Eq. \eqref{delta-q-contrav} and Eq. \eqref{Gamma1-b} for different sets of indices.  
Lastly, we need the first order of the trace of the square and the square of the trace of the conjugate momentum.
The \textbf{trace of the conjugate momentum square} is
\begin{equation}
	\begin{aligned}\label{pipi1}
	\overline	q^{-1/2}{\tensor{\delta \pi}{^i_j}}
		{\tensor{\overline\pi}{^j_i}}
		=
\overline		q^{-1/2}
		\delta{\pi}^{ik}\overline{q}_{kj}
		\overline{\pi}^{jn}\overline{q}_{ni}
		+
	\overline		q^{-1/2}
		\overline{\pi}^{ik}
		\delta{q_{kj}}
		\overline{\pi}^{jn}
		\overline{q}_{ni},
		\end{aligned}
\end{equation}
and the \textbf{square of the trace} reads
\begin{equation}
	\begin{aligned}\label{pi21}
		\delta{\pi^2}
		=
		2\left(
		{\delta\pi^{ik}}
		{\overline q_{ki}}
		{\overline\pi^{jl}}
		{\overline q_{lj}}
		+
		{\overline\pi^{ik}}
		{\delta q_{ki}}
		{\overline\pi^{jl}}
		{\overline q_{lj}}
		\right),
	\end{aligned}
\end{equation}
where the factor $2$ accounts for symmetrization.

The steps showed up to this point allow one to write the Hamiltonian at first order.

In the same manner, we now proceed to show how we can obtain the second order perturbation.

\subsubsection{$2$nd order}
At second order, we only need to compute the perturbation of the scalar Hamiltonian \eqref{hamtemp}. This is so because the Hamiltonian \eqref{hamtemspa} at second order does not include a term such as $N_i \left(\mathcal{H}^i\right)^{(2)}$, where $N_i=0$ as explained below Eq. \eqref{covariantderivativegen}.

In the following, we will briefly showcase how to compute the Hamiltonian at second order. The steps will follow from the previous order; however, it is interesting to note how this initially tedious computation can be later simplified by introducing the scalar-vector-tensor (SVT) decomposition. 

 Once again, since we are expanding up to first order in the dynamics, we are only considering terms such as $\delta q \delta q$ or $\delta \pi \delta \pi$, and ignoring terms like $\delta^{(2 )}q$ or $\delta^{(2)} \pi$.

The second order perturbed scalar Hamiltonian is
\begin{equation}\label{H0-2}
	\begin{aligned}
	{\delta^{(2)}{\mathcal{H}_0}}=&
	\left[-\sqrt{q}\,\,{^3}{R}
	+q^{-1/2}
	\left({{\pi^i_j}{\pi^j_i}}
	-
\frac{1}{2}{\pi^2}\right)
	\right]^{(2)}
	\\
	=&
	-\left(\sqrt{q}\right)^{(0)}{\,\,{^3}R}^{(2)}
	-\left(\sqrt{q}\right)^{(1)}{\,\,{^3}R}^{(1)}
	\\
	&
	+\left(q^{-1/2}\right)^{(0)}
	\left(
	{\delta \pi^i_j}{\delta\pi^j_i}
	-
	\frac{1}{2}{\delta\pi^i_i}{\delta\pi^j_j}
	+2
	{\delta^{(2)}\pi^i_j}{\overline\pi^j_i}
	-
	{{\delta ^{(2)}\pi^i_i}}{\overline\pi^j_j}
	\right)
	\\&
	+\left(q^{-1/2}\right)^{(1)}
	\left(2{\delta\pi^i_j}{\overline\pi^j_i}
	-
	{{\delta\pi^i_i}{\overline\pi^j_j}}\right)
	\\&
	+\left(q^{-1/2}\right)^{(2)}
	\left({{\overline\pi^i_j}{\overline\pi^j_i}}
	-
	{\frac{1}{2}{\overline\pi^2}}\right)
	\end{aligned}
\end{equation}
Most of the terms in Eq.\eqref{H0-2} are products of the previously shown equations \eqref{Gamma1-b}, \eqref{qprimo-inverse}, \eqref{Ricci},  \eqref{pi21}, \eqref{pipi1}; therefore we are left with the task of computing the second order of the Ricci scalar, the metric and conjugate momentum.
  At first order we know that $\delta q=q q^{ij}\delta q_{ij}$, therefore we find
\begin{equation}\label{q-sqrt-2}
	\delta^{(2)} q^{-1/2}
	=
	\frac{1}{4} \overline q^{-1/2}(\overline q^{ij}
	\delta q_{ij})^2+\frac{1}{2}  \overline q^{-1/2}\delta q^{ij}\delta q_{ij}, 
\end{equation}
where we used Eq. \eqref{qprimo-inverse}.
The second order in perturbation of the conjugate momentum, using Eq. \eqref{pi21}, is given by
\begin{equation}\label{delta2piij}
	{\pi^i_j}^{(2)}=(\pi^{ik}q_{jk})^{(2)}=\delta\pi^{ik}\delta q_{jk}.
\end{equation}
Finally, the last term needed to compute Eq. \eqref{H0-2} is the Ricci scalar
\begin{equation}\label{Ricci-2}
	\begin{aligned}
	{{^3}R}^{(2)}
	=&
	{{q^{ih}}^{(0)}\partial_{l} (\tensor{\Gamma}{^{l}_{hi}})^{(2)} }
	- 
	{q^{ih}}^{(0)}\partial_{h} (\tensor{\Gamma}{^{k}_{k i}})^{(2)}
	+
	{{q^{ih}}^{(1)}\partial_{l} (\tensor{\Gamma}{^{l}_{hi}})^{(1)} }
	- 
	{{q^{ih}}^{(1)}\partial_{h} (\tensor{\Gamma}{^{k}_{k i}})^{(1)}}
	\\
	&
	-
	{(q^{ih})^{(0)}(\tensor{\Gamma}{^{m}_{li}})^{(1)}(\tensor{\Gamma}{^{l}_{hm}})^{(1)}}
	+
	{(q^{ih})^{(0)}(\tensor{\Gamma}{^{m}_{hi}})^{(1)}(\tensor{\Gamma}{^{l}_{lm}})^{(1)}},
	\end{aligned}
\end{equation}
where Eq. \eqref{Gamma1-b} and \eqref{delta-q-contrav} are needed to obtain the final form of the second order. We also need to compute the second order perturbation of the Christoffel symbols, i.e.
\begin{equation}\label{Gamma2}
	\begin{aligned}
	{\tensor{\Gamma}{^{l}_{hi}}}^{(2)}
	=
	\frac{1}{2} \delta{q^{lk}}(\delta q_{k h, i}+ \delta q_{k i, h}- \delta q_{hi, k}),
	\end{aligned}
\end{equation}
while the Christoffel symbols with contracted indices can be simplified as
\begin{equation}\label{Gammacont2}
	{\tensor{\Gamma}{^k_{ik}}}^{(2)}=
	\frac{1}{2}\delta {q^{kh}}\delta {q_{{h k},i}}.
\end{equation}

Once Eqs. \eqref{pipi0}, \eqref{qprimo}, \eqref{qprimo-inverse}, \eqref{Ricci-1},  \eqref{pi21}, \eqref{q-sqrt-2}, \eqref{delta2piij} and \eqref{Ricci-2} are substituted into Eq. \eqref{H0-2}, after some straightforward but tedious computations, we obtain 
\begin{align}\label{delta2Hsecondpart}
		{\delta^{(2)}{\mathcal{H}_0}}
	=&
		\frac{1}{2}
		q^{1/2}
	\bigg(
	\delta{q^{ih}}  q^{lk}
	{\delta q_{k h, il}}
	-
	\delta{q^{ih}} 
	q^{kl}
	\delta q_{{l k},ih}
	-2
	q^{ln}\delta {q^{hm}} \delta {q_{hm, nl}}
	+2
	q^{ih}
	q^{mk}
	(q^{ln}
	\delta q_{{n l}}) \delta {q_{hi, km}}\bigg)
	\nonumber
	\\
	&
	+q^{-1/2}
	\bigg[
	{
		\left(\delta\pi^{ik}q_{jk}+\pi^{ik}\delta q_{jk}\right)
		\left(\delta\pi^{jl}q_{il}+\pi^{jl}\delta q_{il}\right)
	}\nonumber
			\\
	&
	-\frac{1}{2}
	{\left(\delta\pi^{ik}q_{ik}+\pi^{ik}\delta q_{ik}\right)
		\left(\delta\pi^{jl}q_{jl}+\pi^{jl}\delta q_{jl}\right)}
	+
	{2\delta\pi^{ik}\delta q_{jk}\pi^{jl} q_{il}}
	-
	{\delta\pi^{ik}\delta q_{ik}\pi^{jl} q_{jl}}
	\bigg]\nonumber
	\\&
	-\frac{1}{2}
	q^{-1/2}
	q^{mn}\delta q_{mn}
	\bigg[
	{	2\left(\delta\pi^{ik}q_{jk}+\pi^{ik}\delta q_{jk}\right)
		\pi^{jl}q_{il}}
	-
	{\left(\delta\pi^{ik}q_{ik}+\pi^{ik}\delta q_{ik}\right)
		\pi^{jl}q_{jl}}
	\bigg]\nonumber
	\\&
	+\frac{1}{2} \bigg[
	\frac{1}{4}q^{-1/2}\left(q^{ij}\delta q_{ij}\right)^2+\frac{1}{2} q^{-1/2}\delta q^{ij}\delta q_{ij} 
	\bigg]
	\left({\pi^{ik}q_{jk}\pi^{jl}q_{il}}
	-
	{\frac{1}{2}\pi^{ik}q_{ik}\pi^{jl}q_{jl}}
	\right),
\end{align}
where the coefficients for each term follow from the expansion in Eq. \eqref{HexpNpq} and we omitted  the bar over background quantities.

Notice that for the above computations a universe with no matter content is assumed. Were there to be matter, the related Hamiltonian should be perturbed too. In Ch. \ref{Chapter3} we will apply this formalism to a Bianchi I universe filled with a scalar field as matter content.

\subsection{Hamiltonian constraints and constraint algebra}\label{ham-const}

As discussed in Chapter \ref{Chapter1}, the Hamiltonian in GR is a constraint, therefore we expect this property to persist in perturbation theory. The existence of constraints within our theory gives rise to gauge freedom. In the context of CPT, the gauge-fixing must happen at each order of perturbation. At zeroth order our constraint, defined by $\mathcal{H}_0$, provides one degree of freedom to be chosen which corresponds to the freedom in fixing our time. We can set our time to be any parameter suitable to describe the dynamics of our system, like the position of a moving particle in a mechanical system. 



The Hamiltonian \eqref{Htot} defines a gauge system if the constraints are truncated at first order. This is due to the fact that the constraints $\delta \mathcal{H}_\mu$ are first-class at first order, which means that they (weakly) commute at first order. 
 Specifically, at each spatial point, the algebra of the linearized constraints reads 
\begin{align}\label{comms0}
	\begin{split}
		\{\delta \mathcal{H}_i,\delta \mathcal{H}_j\}\approx0,&~~\{\delta \mathcal{H}_j,\delta \mathcal{H}_0\}\approx0,\end{split}
\end{align}
where we used the weak equality since the commutators can be proportional to the zeroth order Hamiltonian which is a constraint.
We note that the linearized constraints commute strongly at first order and thus the group of gauge transformations that they generate for each spatial point must be abelian. This is true independently from any particular choice of background spacetime model. Moreover, Eq. \eqref{comms0}, holds for perturbation theory around any homogenous background, since the homogeneous and inhomogeneous variables commute with each other.

The constraints are dynamically stable on the constraint surface, namely,
\begin{equation}
	\label{comms}
	\begin{gathered}
		\delta\dot{\mathcal{H}}_0
		=
		\left\{
		\int (\mathcal{H}^{(0)}
		+\mathcal{H}^{(2)})
		,\delta \mathcal{H}_0(x)
		\right\}
		=-\delta \mathcal{H}^i_{~,i}(x)\approx 0,
		\\
			\delta\dot{\mathcal{H}}_i
		=
		\left\{
		\int (\mathcal{H}^{(0)}+\mathcal{H}^{(2)})
		,\delta \mathcal{H}_i(x)
		\right\}
		= 0\end{gathered}
\end{equation}
where $\delta \mathcal{H}^i(x)=\overline{q}^{ij}\delta \mathcal{H}_j(x)$, and the dot represents a derivative with respect to $t$.

Eqs \eqref{comms0} and \eqref{comms} are a linearized version of the algebra of hypersurface deformations of canonical relativity \cite{10.2307/100497}. The full deformation algebra, and hence its linearization, are universal in the sense that they do not depend on any particular theory of gravity \cite{HOJMAN197688}.  It is worth noting that the algebra can abelianize naturally for spherically symmetric hypersurface deformations \cite{PhysRevD.105.026017,universe8030184}. 

\section{Dirac method in CPT}\label{sec-Dirac-metodo}
The Dirac method introduced in Sec. \ref{secdirac}, provides a powerful tool for setting valid gauges and for reconstructing the spacetime metric from gauge-invariant variables in the Hamiltonian formalism. 
In the following we will apply the Dirac method to CPT, while still maintaining a generic background. This will give us an idea of the practical and useful applications of this method. For an explicit application see \cite{Malkiewicz_2019} and Ch. \ref{Chapter3}.

\subsection{Gauxe-fixing and reduced Hamiltonian}
In order to remove the gauge freedom generated by solving the constraints $\delta\mathcal{H}_\mu$, we choose 4 gauge-fixing\footnote{There are as many gauge-fixing conditions as constraints.} conditions, denoted by $\delta c_{\mu}=0$, such that the commutation relations between the gauge-fixing functions and the linear constraints form an invertible matrix, that is,
\begin{equation}\label{det}
	\text{Det}\{\delta c_{\nu},\delta\mathcal{H}_{\mu}\}\neq 0.
\end{equation}
This condition is essential for transitioning from a constrained system to an unconstrained one. In particular, the invertibility of the matrix ensures that the chosen gauge-fixing conditions remove all and only the unphysical degrees of freedom, establishing a consistent and reduced description of the system.
The physical variables describing the system are obtained from the reduction of the formalism by solving both the gauge-fixing conditions $\delta c_{\mu}=0$ and the constraints $\delta \mathcal{H}_\mu$. More specifically the set of 12 ADM perturbation variables $(\delta q_{ij},\delta \pi^{ij})$, after solving the gauge-fixing conditions and the constraints, are replaced by a reduced set of 4 physical variables denoted by $(\delta q^{\text{phys}}_{I},\delta \pi_{\text{phys}}^{I})$, where $I=1,2$. This reduction is explicitly performed in Ch. \ref{Chapter3} for different gauges.
The physical variables can be chosen in such a way as to form a canonical coordinate system on the submanifold in the kinematical phase space, on which the gauge-fixing functions and the constraints vanish. We call this submanifold the physical phase space. The canonical structure of the \textbf{physical phase space} is encoded in the Dirac bracket, 
\begin{equation}\label{DB}
	\{A,B\}_D=\{A,B\} -\{A,\delta \phi_{\mu}\}\{\delta \phi_{\mu},\delta \phi_{\nu}\}^{-1}\{\delta \phi_{\nu},B\},
\end{equation}
where $\delta\phi_{\mu}\in (\delta \mathcal{H}_1,\dots,\delta \mathcal{H}_4,\delta c_{1},\dots,\delta c_{4})$ and where $\{\delta \phi_{\mu},\delta \phi_{\nu}\}^{-1}$ is the inverse matrix with regards to the indices. Note that the Dirac bracket depends on the choice of the gauge-fixing conditions $(\delta c_{1},\dots,\delta c_{4})$.  
Writing the Hamiltonian in terms of the new reduced variables gives the  \textbf{reduced Hamiltonian}:
\begin{equation}\label{redH}
	\left(N\mathcal{H}^{(2)}+\delta N^\mu\delta \mathcal{H}_\mu\right)\Big|_{\delta c_{\mu}=0=\delta\mathcal{H}_{\mu}}=N\mathcal{H}^{(2)}_{\text{red}}(\delta q^{\text{phys}}_I,\delta \pi_{\text{phys}}^I).
\end{equation}
As a consequence we also find the reduced Hamilton equations defined in terms of the reduced set of variables, the reduced Hamiltonian, and the Dirac bracket:
\begin{equation}\label{redHE}
	\begin{aligned}
		\frac{\ud}{\ud t}\delta {q}^{\text{phys}}_{I}=\{\delta q^{\text{phys}}_{I},\int \big( N\mathcal{H}^{(0)}+N\mathcal{H}^{(2)}_{\text{red}}\big)\ud^3x\}_D,\\
		\frac{\ud}{\ud t}\delta {\pi}_{\text{phys}}^{I}=\{\delta \pi_{\text{phys}}^{I},\int \big( N\mathcal{H}^{(0)}+N\mathcal{H}^{(2)}_{\text{red}}\big)\ud^3x\}_D.
		\end{aligned}
\end{equation}
Note that the term $\int N\mathcal{H}^{(0)}$ generates the dynamics of the background coefficients, which are typically present in the definitions of $\delta q^{\text{phys}}_{I}$ and $\delta \pi_{\text{phys}}^{I}$.


\subsection{Gauge-invariant Hamiltonian and Dirac observables}\label{gid}
\begin{figure}[t]
	\begin{center}
	\includegraphics[width=0.8\textwidth]{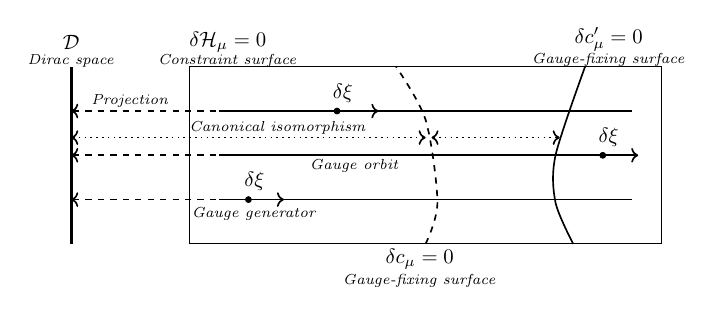}
	\end{center}
	\caption{Illustration of the key concepts involved in the Dirac procedure: the constraint surface, the gauge-fixing surface, the gauge orbit, the Dirac space and the canonical isomorphism between different gauge-fixing surfaces.}
	\label{fig:Dirac}
\end{figure}

The reduced Hamiltonian is obtained by fixing the gauge, this means that one particular choice of gauge must be made. 
Although obtained from a particular choice of gauge-fixing conditions, the reduced Hamiltonian and the physical variables in fact encode the gauge-independent dynamics of the model.
 This can be shown with the help of the \textbf{Dirac observables} presented in Sec. \ref{secdirac}. In the context of perturbation theory, considering the constraints $C_\mu=\delta \mathcal{H}_\mu$, the definition of the Dirac observables in Eq. \eqref{Dirweak}, now denoted by $\delta D_I$, becomes:
\begin{equation}\label{DDO}
	\{\delta D_I,\delta \mathcal{H}_\mu\}\approx 0~~\textrm{for all}~\mu.
\end{equation}
As discussed, by definition, Dirac observables commute with the 4 constraints $\delta\mathcal{H}_\mu$ and are understood as functions on the constraint surface. Hence, the number of independent Dirac observables must be equal to the number of \textbf{reduced} ADM perturbation variables, which is equivalent to the number of \textbf{physical variables} $(\delta q^{\text{phys}}_I,\delta \pi_{\text{phys}}^I)$. The Dirac observables provide a parametrization of the space of the gauge orbits in the constraint surface whereas the physical variables provide a parametrization of a particular gauge-fixing surface that crosses each gauge orbit once and only once as depicted in Fig. \ref{fig:Dirac}. 
Therefore, as shown in Sec. \ref{secdirac} for the full GR case, there exists a \textbf{one-to-one correspondence} between the Dirac observables and the physical variables. Specifically, for any Dirac observable $\delta D_I$ there must exist one and only one physical variable $\delta O^{\text{phys}}_I$ such that Eq. \eqref{Dirphys} with the new notation becomes
\begin{equation}\label{diracbra}
	\delta D_I+\xi_I^\mu\delta c_{\mu}+\zeta_I^\mu\delta\mathcal{H}_\mu=\delta O^{\text{phys}}_I(\delta q^{\text{phys}},\delta \pi_{\text{phys}}),
\end{equation}
for some background coefficients $\xi_I^\mu$ and $\zeta_I^\mu$. Notice that,  in CPT, Eq. \eqref{Dalgebra}, up to first order in perturbation, now reads
\begin{equation}\label{dirac-poisson}
\{\delta D_I,\delta D_J\}=\{\delta D_I,\delta D_J\}_D.
\end{equation}
Analogously, substituting Eq. \eqref{diracbra} into \eqref{dirac-poisson} we find again Eq. \eqref{vphyD}, i.e. a canonical isomorphism between Dirac observables and physical variables: 
\begin{equation}
	\begin{split}
		\{\delta D_I,\delta D_J\}
		&=\{\delta D_I,\delta D_J\}_D
		\\
		&=\{\delta D_I+\xi_I^\mu\delta c_{\mu}+\zeta_I^\mu\delta\mathcal{H}_\mu,\delta D_J+\xi_J^\mu\delta c_{\mu}+\zeta_J^\mu\delta\mathcal{H}_\mu\}_D\\
		&=\{\delta O^{\text{phys}}_I,\delta O^{\text{phys}}_J\}_D.
\end{split}
\end{equation}
Hence, we are able to define gauge-invariant Hamilton equation using the reduced Hamiltonian and the Dirac bracket, as
\begin{equation}\label{Dham}
	\begin{aligned}
		\frac{\ud}{\ud t}\delta D_I=\{\delta D_I,\int \big( N\mathcal{H}^{(0)}+N\mathcal{H}^{(2)}_{\text{red}}\big)\ud^3x\}_D,
		\end{aligned}
\end{equation}
where the reduced Hamiltonian is now understood as a function of the Dirac observables, $\mathcal{H}^{(2)}_{\text{red}}=\mathcal{H}^{(2)}_{\text{red}}(\delta D)$.
The presence of the zeroth-order constraint in \eqref{Dham}, expressed as $\{\delta D_I, \int N\mathcal{H}^{(0)}\mathrm{d}^3x\}$, arises from the fact that the Dirac observables $\delta D_I$ are linear combinations of the ADM perturbation variables $(\delta q_{ab}, \delta \pi^{ab})$ with time-dependent background coefficients. The dynamics of these 
coefficients must also be taken into account. 
 In the following Section we see how this comes into play in our theory.

\subsection{Physical Hamiltonian}
With the Dirac observables $\delta D_I$ and the physical variables $(\delta q^{\text{phys}}_{I},\delta \pi_{\text{phys}}^{I})$ at our disposal, it is advantageous to rewrite the Hamiltonian as a function of either of these variables. 
Both sets of variables are given by time-dependent combinations of the initial ADM perturbation variables. Hence, the new parametrizations are obtained through time-dependent canonical transformations. We shall use the Dirac observables $\delta D_I$ as \textbf{basic canonical variables}.
As mentioned, those variables are linear combinations of time-dependent background variables, as such, when performing a time dependent canonical transformation, the initial Hamiltonian written in the new coordinates acquires new terms responsible for the dynamics of the time-dependent coefficients.
When computing the symplectic form, the additional term from this transformation, called the \textbf{extra Hamiltonian density} $\mathcal{H}^{(2)}_{\text{ext}}$, is added. The new Hamiltonian density, in terms of the new canonical variables, is the sum of the reduced Hamiltonian and the extra Hamiltonian.
 \begin{equation}\label{physham}
	\mathcal{H}^{(2)}_{\text{phys}}=\mathcal{H}^{(2)}_{\text{red}}+\mathcal{H}^{(2)}_{\text{ext}}.
\end{equation}
	 This is referred to as the \textbf{physical Hamiltonian density}. With this choice of canonical variables, the Hamilton equations \eqref{Dham} are modified to read 
\begin{equation}\label{hphys}
		\frac{\ud}{\ud t}\delta D_I=
		\left\{
		\delta D_I,\int N\mathcal{H}^{(2)}_{\text{phys}}\ud^3x
		\right\}_D
\end{equation}
Indeed, with such a choice of canonical variables the dynamics of the perturbations is now given purely by the \textbf{second-order physical Hamiltonian}. The Dirac bracket  \begin{equation}\label{dirbraD}
	\{A,B\}_D=\{A,\delta D_J\}\{\delta D_I,\delta D_J\}^{-1}\{\delta D_I,B\}
	\end{equation}
	can now be expressed in terms of  the Poisson bracket and the Dirac observables instead of the gauge-fixing functions. 
	Although the gauge-fixing functions $\delta C_\mu$ are not explicitly defined in the formulation of the Dirac brackets \eqref{dirbraD}, there is an implicit dependence based on the assumption that the Dirac observables \eqref{DDO} commute with the gauge-fixing functions, i.e., $\{\delta D_I,\delta c_{\mu}\}_D=0$. This can always be achieved given the ambiguous definition of $\delta D_I$, i.e. Eq. \eqref{diracbra}.

\subsection{Spacetime reconstruction}\label{ISR}
We have seen that the dynamics can be obtained from a gauge-invariant description, i.e., we can write the Hamilton equations in terms of the physical Hamiltonian. The\textbf{ physical interpretation} of the obtained dynamics depends on the gauge-fixing conditions. Thus, the physical interpretation can be considered independently from the dynamical equations. 

Eq.  \eqref{det} implies  the existence of a \textbf{one-to-one map} between the values of the {gauge-fixing functions}, the {constraint functions} and the Dirac observables on one hand, and the values of the ADM perturbation variables on the other, i.e.
\begin{equation*}
	(\delta\mathcal{H}_\mu,\delta c_{\mu},\delta D_I)\leftrightarrow (\delta q_{ab},\delta \pi^{ab}).
\end{equation*}
As a consequence, the geometry of the spatial leaf in terms of the ADM  perturbation variables,  is unambiguously determined by fixing $\delta\mathcal{H}_\mu=0$ and $\delta c_{\mu}=0$, from the values of $\delta D_I$.


To have a full description of the entire spacetime, it is crucial not only to understand the geometry of the spatial slices but also to determine the values of the first-order lapse function and the shift vector.This is what we refer to as \textbf{spacetime reconstruction}, i.e. $(\delta c_\mu,\delta D_I)\rightarrow (\delta q_{ab},\delta \pi^{ab},N,N^i)$. In order to obtain the values of the lapse and shift we use the fact that the gauge-fixing dynamics must be preserved in the kinematical phase-space. Mathematically this means that the Poisson brackets between the gauge-fixing conditions and the total Hamiltonian must vanish, i.e. 
\begin{equation}\label{stabilityEq}
\{\delta c_{\nu},\mathbb{H}\}\approx0.
\end{equation}
 Substituting the integrand of Eq. \eqref{Htot} and solving for $\delta N^\mu$ we obtain
\begin{equation}\label{coneq}
\frac{\delta N^\mu}{N}\approx-\{\delta c_{\nu},\delta \mathcal{H}_\mu\}^{-1}\left(\{\delta c_{\nu},\mathcal{H}^{(0)}\}+\{\delta c_{\nu},\mathcal{H}^{(2)}\}\right).
\end{equation}
The above equation is physically meaningful only in the constraint surface, that is, it holds weakly.

Finally, with Eq. \eqref{coneq}  and Eq. \eqref{physham}, we conclude the overview of the Dirac method in perturbation theory. This method allows us to obtain a gauge-invariant reduced formulation of the Hamiltonian as well as the spacetime reconstruction. However the spacetime reconstruction can only be obtained using the full Hamiltonian, which can be cumbersome as the relation between the full and reduced Hamiltonian is hidden. Moreover, as we will see in the next section, using the Dirac method it is not clear how to define the so called partial gauge-fixing (see Sec. \ref{PGF}).

As such, it is interesting and useful to introduce a new parametrization of the kinematical phase space, which can simplify the spacetime reconstruction and the choice of gauge-fixing conditions.

\section{Kucha\v r decomposition}\label{kuchar}
In the present section we revisit the procedure outlined above by means of the so-called Kucha\v r decomposition that is a special\textbf{ parametrization of the kinematical phase space} with constraints encoded into canonical variables \cite{doi:10.1063/1.1666050, Hajicek:1999ht}. The existence of such a parametrization should become obvious as we proceed, nevertheless a general proof, valid beyond perturbation theory, can be found in \cite{Maskawa:1976hw}. As we will see in Sec. \ref{GT-kuchar} using the Kucha\v r decomposition, the gauge transformations form an \textbf{abelian group of translations}, $\mathbf{G}=\mathbb{R}^n$ (at each spacetime point). The space of all valid gauges becomes explicit, and the distinction between complete and partial gauge-fixing is very clear and analogous to electrodynamics.
\subsection{Motivations}
Recalling briefly the common issue that leads us to use the Dirac method for constrained systems, in GR we start with $12$ initial degrees of freedom, then we solve $4$ constraints and we apply $4$ gauge-fixing conditions which leave us with $4$\textbf{ physical degrees of freedom}, i.e. two pairs of conjugate physical variables. 
In the previous section we treated the problem of the redundant degrees of freedom applying what we called \textbf{Dirac method}. Although, as explained, the method gives good results by providing the physical Hamiltonian expressed in terms of the Dirac observables, the issue with this method is that 
true dynamical variables are mixed with the variables defining the spacetime coordinates \cite{doi:10.1063/1.1666050}.
This mixing would constitute a problem
in the quantum case in which the separation between variables of different types is essential for their quantization.

\begin{figure}[t]
	\begin{center}
		\includegraphics[width=0.8\textwidth]{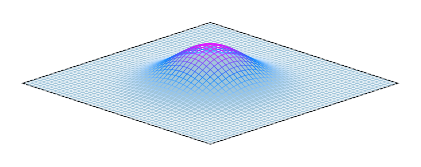}
	\end{center}
	\caption{Thin sandwich of two spatial slices, one pictured in green and the other with a white grid. The time bubble is denoted by a colour gradient. It shows how the time is accounted for even when the spacetime slicing choice is left arbitrary.}
	\label{fig:bubble}
\end{figure}

The solution proposed by Kucha\^r in \cite{doi:10.1063/1.1666050} is to perform a canonical transformation which separates the true dynamical variables and then solve the constraints.
 In particular, this decomposition is  defined by splitting the 12 ADM variables into three sets of coordinates. Two of these sets are \textbf{internal variables} as they come from the\textbf{ geometrical properties }of the hypersurfaces, while the third and last set is given by variables defining the\textbf{ true dynamical degrees of freedom} of the system. The important step which makes this decomposition interesting is the arbitrariness of the spacetime slicing. In the ADM formalism, usually each slice has a fixed coordinates system as the time change is usually defined as the change from one slice $\Sigma_t$ to another $\Sigma_{t+\ud t}$, which means that the variable $t$ is assigned to the entire hypersurface.
In this new decomposition the time can be locally changed, leading to two hypersurfaces which are only locally different. The time change will thus create a sort of\textbf{ bubble deformation }(see Fig. \ref{fig:bubble}), as Kucha\^ r calls it in his paper  \cite{doi:10.1063/1.1666050}, and from which the new formalism takes the name of \textbf{bubble-time canonical formalism}.


\subsection{Decomposition - how to build the new formalism} \label{sec-decomp}
 
We now introduce the new variables. To do so we will closely follow the arguments presented in \cite{doi:10.1063/1.1666050}. As previously explained, 4 of the ADM variables $(q_{ij},\pi^{ij})$, are non-dynamical variables and they are given by the first-class constraints of the system. They will represent the \textbf{first group }of canonical variables for the kinematical phase space in this new parametrization. The \textbf{second group} is given by the variables canonically conjugate to the first ones. As explained in Sec. \ref{geomconst}, this is the set of variables playing the role of the gauge-fixing conditions, and, as for the variables of the first group, those are non-dynamical too.
The \textbf{third group} is given by the Dirac observables  which, by definition, are gauge-invariant variables undergoing the physical dynamics of the system.
Let us now apply this formalism to redefine the system described in Sec. \ref{sec-Dirac-metodo}. More explicitly, the procedure will be as follows:
\begin{enumerate}
\item We introduce new canonical variables in the kinematical (ADM) phase space.
As mentioned, we need to define \textbf{two sets of canonical pairs}, the first one given by \textbf{the constraints} $\delta \mathcal{H}_\mu$  which, at first order, strongly commute between themselves \eqref{comms0}, and the second one given by the 4 gauge-fixing functions $\delta C^\mu$. We now have the set of canonical pairs $(\delta \mathcal{H}_\mu,\delta C^\mu)$.

\item We define the so-called \textbf{ strong Dirac observables} $\delta D_{I}$ that  are uniquely determined by the requirement that they strongly commute with the constraint functions and the gauge-fixing functions, see Eq. \eqref{Dirstrong}. 
Their Poisson algebra is closed and they form canonical pairs which we shall denote by $(\delta Q_I,\delta P^I)$. 

\item After finding the variables for the first, second and third group, we have that the new set of canonical variables in the kinematical phase space are:
\[(\delta \mathcal{H}_\mu,\delta C^{\mu},\delta Q_I,\delta P^I).\]
\end{enumerate}

The algebra of these new variables, up to first order, is
\begin{equation}
	\label{PbD}\{\delta\mathcal{H}_\mu(x),\delta C^{\nu}(y)\}
	=
	\delta_{\mu}^{~\nu}\delta^3(x-y),
	~\{\delta Q_I(x),\delta P^J(y)\}
	=\delta_{I}^{~J}\delta^3(x-y),
	\end{equation}
with all the remaining basic commutation relations vanishing. This new parametrization arises from the canonical transformation 
\begin{equation}\label{Kmap}
	\mathbb{R}^{12}\ni (\delta q_{ab},\delta \pi^{ab})\mapsto (\delta \mathcal{H}_\mu,\delta C^{\mu},\delta Q_I,\delta P^I)\in\mathbb{R}^{12}.
	\end{equation}
This transformation is time-dependent as the Kucha\v r variables are linear combinations of the ADM perturbation variables with time-dependent zeroth-order coefficients. Hence the Hamiltonian in the new parametrization $\mathbb{H}_K$, will be
\[\mathbb{H}\rightarrow\mathbb{H}_K=\mathbb{H}+\mathbb{K},\]
where $\mathbb{K}$ is the extra Hamiltonian needed to compensate for the dynamics of the zeroth-order coefficients present in the definition of the new canonical variables.
According to the theorem in \cite{Landau1976Mechanics} we know that Poisson brackets are invariant under canonical transformations. Therefore the Poisson brackets expressed in terms of the ADM variables and the Kucha\v r variables are equivalent.

In the Kucha\v r parametrization the \textbf{total Hamiltonian }is given by 
\begin{equation}\label{HK}
	\mathbb{H}_K=\int
	\left[
	N\mathcal{H}^{(0)}+N(\mathcal{H}^{(2)}+\mathcal{K}^{(2)})+\delta N^\mu\delta \mathcal{H}_\mu
	\right]\ud^3x,
\end{equation}
where $\int N\mathcal{K}^{(2)}\ud^3x=\mathbb{K}$. It generates the following \textbf{Hamilton equations}:
\begin{equation}
	\label{eom}
	\begin{aligned}
		\delta \dot{Q}_I&=N\frac{\partial(\mathcal{H}^{(2)}+\mathcal{K}^{(2)})}{\partial \delta{P}^I},~~\delta \dot{P}^I=-N\frac{\partial(\mathcal{H}^{(2)}+\mathcal{K}^{(2)})}{\partial \delta{Q}_I},\\
		\delta \dot{\mathcal{H}}_\mu&=N\frac{\partial(\mathcal{H}^{(2)}+\mathcal{K}^{(2)})}{\partial \delta{C}^\mu},~~\delta \dot{C}^\mu=-N\frac{\partial(\mathcal{H}^{(2)}+\mathcal{K}^{(2)})}{\partial {\delta\mathcal{H}}_\mu}
		-\delta N^\mu.
	\end{aligned}
\end{equation}
It is interesting to note how, through logical steps and consistency checks within the theory, it is possible to simplify the form of the total Hamiltonian $\mathbb{H}_K$ using Hamilton equations \eqref{eom}. In the following steps, we will outline the logical procedures for this reduction.
\begin{itemize}
\item[$\hookrightarrow$] The constraints' dynamics is conserved in the constraints, surface, this means that 
\[
\delta \dot{\mathcal{H}}_\mu=N\frac{\partial(\mathcal{H}^{(2)}+\mathcal{K}^{(2)})}{\partial \delta{C}^\mu}
\approx0.
\]
Therefore, the terms $\propto \delta{C}^\nu\delta{C}^\mu$, $\propto \delta Q_I\delta{C}^\mu$, and $\propto \delta P^I\delta{C}^\mu$ must not be present in the total Hamiltonian $\mathbb{H}_K$, as their presence would result in non-vanishing terms in $\delta\dot{\mathcal{H}}_\mu$.
 Notice that the absence of terms such as $\propto \delta Q_I\delta{C}^\mu$ and $\propto \delta P^I\delta{C}^\mu$ can also be understood by examining the dynamics of Dirac observables, which are expected to be independent of the gauge choice.
\item[$\hookrightarrow$]  Notice that the Hamiltonian density $\mathcal{H}^{(2)}+\mathcal{K}^{(2)}$ must be weakly equal to the Hamiltonian density $\mathcal{H}^{(2)}_{\text{phys}}$ of Eq. \eqref{physham}, more specifically, $\mathcal{H}^{(2)}\approx\mathcal{H}^{(2)}_{\text{red}}$ and $\mathcal{K}^{(2)}\approx\mathcal{H}^{(2)}_{\text{ext}}$. 
	\end{itemize}
The total Hamiltonian $\mathbb{H}_K$ is the sum of a \textbf{physical part}  \eqref{physham} and a \textbf{weakly vanishing part}. Thus, in the Kucha\v r parametrization the total Hamiltonian $\mathbb{H}_K$ reads:
\begin{equation}\label{KHw}
	\small{
	\mathbb{H}_K
	=
	N
	\int
	\bigg[
	\underbrace{
		\mathcal{H}^{(2)}_{\text{phys}}(\delta Q_I,\delta P^I)
		}_\text{physical part}
		+\underbrace{
			(
			\lambda_{1}^{\mu I}
		\delta{Q}_I
		+\lambda_{2I}^{\mu}\delta{P}^I
		+\lambda_{3}^{\mu\nu}\delta\mathcal{H}_\nu
		+\lambda_{4\nu}^\mu\delta C^\nu
		+\frac{\delta N^\mu}{N}
		)\delta \mathcal{H}_\mu
		}_\text{weakly vanishing part}
	\bigg]
	\ud^3x,
}
\end{equation}
where, in general, the zeroth-order coefficients $\lambda_1$, $\lambda_2$ and $\lambda_3$ depend on the particular choice of gauge-fixing functions $\delta C^{\mu}$. Computing Eq. \eqref{coneq} obtained from Eq.\eqref{eom} we obtain an equation for the lapse function and shift vector which depends on  $\lambda_1$ and $\lambda_2$ as
\begin{equation}\label{lapseshiftfull}
	\frac{\delta N^\mu}{N}\approx -\lambda_{1}^{\mu I}\delta{Q}_I-\lambda_{2I}^{\mu}\delta{P}^I.
\end{equation}
Given that the lapse and shift are gauge-dependent quantities, while the Dirac observables $\delta Q_I$ and $\delta P^I$ are not by definition, it is evident that $\lambda_1$ and $\lambda_2$ must be \textbf{gauge-dependent} quantities. In contrast, the value of $\lambda_4$ is \textbf{gauge-invariant}, meaning it does not depend on the particular gauge-fixing functions used. Indeed, from Eq. \eqref{KHw} we have that
\begin{equation}\label{lambda4}
	\lambda_{4\nu}^\mu\delta \mathcal{H}_\mu=\{\delta\mathcal{H}_\nu,\mathbb{H}_K\}=\{\delta\mathcal{H}_\nu,\int (\mathcal{H}^{(0)}+\mathcal{H}^{(2)})\},
\end{equation}
that is, the matrix $\lambda_4$ is fixed unambiguously by the algebra of hypersurface deformations \eqref{comms}.

The value of $\lambda_3$ is irrelevant for the physical content of the theory. 

\subsection{Gauge transformations}	\label{GT-kuchar}
The variables defined by the Kuchař parametrization depend on the gauge-fixing conditions $\delta C_\mu$; therefore, for each choice of gauge, a new reparametrization is defined.

We define a new set of gauge-fixing functions with $\delta \tilde{C}^{\mu}$.
The full \textbf{gauge transformation} is given by the canonical map $\mathbf{G}$:
\begin{equation}\label{Gmap}
	\mathbf{G}:~(\delta \mathcal{H}_\mu,\delta C^{\mu},\delta Q_I,\delta P^I)\mapsto (\delta \tilde{\mathcal{H}}_\mu,\delta \tilde{C}^{\mu},\delta \tilde{Q}_I,\delta \tilde{P}^I).
	\end{equation} 
 Notice that the constraint functions are preserved in the transformation, i.e. $\delta \tilde{\mathcal{H}}_\mu=\delta \mathcal{H}_\mu$.

We assume the new gauge-fixing functions $\delta \tilde{C}^{\mu}$ to be canonically conjugate to the constraints $\delta{\mathcal{H}}_\mu$, i.e., $\{\delta\mathcal{H}_\mu(x),\delta \tilde{C}^{\nu}(y)\}=\delta_{\mu}^{~\nu}\delta^3(x-y)$. This assumption is justified, as any set of gauge-fixing functions, $\delta {C}^{\mu}_{\text{ini}}$, can be transformed into the \textbf{canonical form}, $\delta {C}_{\text{can}}$, as follows:
\begin{equation}\label{can}
	\delta {C}_{\text{can}}^{\mu}(x)=
	\int M^\mu_{~\nu}(x,y)\delta {C}^{\nu}_{\text{ini}}(y)\ud^3y,
\end{equation}
where $M^{\mu}_{~\nu}(x,y)=\{\delta\mathcal{H}_\mu(x),\delta {C}^{\nu}_{\text{ini}}(y)\}^{-1}$.
Therefore, the difference between any two sets of canonical gauge-fixing functions should satisfy
\begin{equation*}
	\{\delta\mathcal{H}_\nu,\delta \tilde{C}^{\mu}-\delta {C}^{\mu}\}=0,
\end{equation*}
from which we find that the most general form of gauge-fixing functions can be written as
\begin{equation}\label{newgauge}
	\delta \tilde{C}^{\mu}=
	\delta C^\mu 
	+\alpha^\mu_{~I}\delta P^I
	+\beta^{\mu I} \delta Q_{I}
	+\gamma^{\mu\nu}
	 \delta\mathcal{H}_\nu.
\end{equation}
The parameters $\alpha^\mu_{~I}$, $\beta^{\mu I}$ and $\gamma^{\mu\nu}$ are \textbf{background-dependent} with the index $I$ running from $1$ to half of the number of basic Dirac observables for a given system. Thus, in the vacuum case $I\in\{1,2\}$ labels two polarization modes of the gravitational wave. The first two parameters $\alpha^\mu_{~I}$ and $\beta^{\mu I}$ are $4\times 2$ matrices, whereas $\gamma^{\mu\nu}$ is a $4\times 4$ matrix. Since gauge-fixing conditions are physically relevant only in the constraint surface, where $ \delta\mathcal{H}_\nu=0$, it follows that the only independent parameters involved in the gauge transformation \eqref{newgauge} must be $\alpha^\mu_{~I}$ and $\beta^{\mu I}$. In other words,\textit{ the space of gauge-fixing conditions for any fixed label $\mu$ is the affine space of dimension equal to the number of Dirac observables in the system}. The background-dependent parameters can in principle depend on time; their explicit formulation can be obtained with the following formulas
\begin{equation}\label{alphabeta}
	\begin{aligned}
\alpha^\mu_{~I}
&=
\left\{\delta Q_I,\int(\delta \tilde{C}^\mu
-
\delta C^\mu)
\right\},
\\
\beta^{\mu I} 
&=
\left\{\int(\delta \tilde{C}^\mu
-
\delta C^\mu)
,\delta P^I
\right\}.
\end{aligned}
\end{equation}
The change in the class of gauge-fixing functions implies that the \textbf{symplectic form},
\begin{equation}
	\Omega=
	\ud\delta Q_I\wedge \ud\delta P^I
	+ \ud \delta \mathcal{H}_\mu \wedge  \ud\delta C^\mu,
\end{equation}
can now be reformulated in an equivalent form for any $(\alpha, \beta)$ as
\begin{equation}
	\begin{aligned}
	\Omega
	&	
	=	
	\ud\left(\delta Q_I{-}{\alpha^\mu_{~I}\delta\mathcal{H}_\mu}\right)
	\wedge 
	\ud\left(\delta P^I{+}{\beta^{\nu I}\delta\mathcal{H}_\nu}\right)
		\\&
		+
	\ud \delta\mathcal{H}_\mu
	\wedge
	\ud
	\left[
	\delta {C}^\mu+\alpha^\mu_{~I}\delta P^I
	+\beta^{\mu I} \delta Q_{I}
	+
	\frac{1}{2}\left(
	\alpha^\mu_{~I}
	\beta^{\nu I}
	-
	\alpha^\nu_{~I}
	\beta^{\mu I}	
	\right)
	\delta\mathcal{H}_\nu
	\right] 
	\\&
	+
	\ud t
	\wedge 
	\ud\left[
	\dot{\alpha}^\mu_{~I} \delta\mathcal{H}_\mu\delta P^I
	+\dot{\beta}^{\mu I}\delta\mathcal{H}_\mu \delta Q_{I}+\frac{1}{2}
	\left(
	\dot{\alpha}^\mu_{~I}
	\beta^{\nu I}
	-
	\alpha^\mu_{~I}
	\dot{\beta}^{\nu I}
	\right)
	\delta\mathcal{H}_\mu
	\delta\mathcal{H}_\nu
	\right].
	\end{aligned}
\end{equation}
Thus, the new\textbf{ Kucha\v r variables }are
\begin{equation}\label{gtr}
	\begin{aligned}
		\delta\tilde{Q}_I&= \delta Q_I-\alpha^\mu_{~I}\delta\mathcal{H}_\mu,\\
		\delta\tilde{P}^I&=
		\delta P^I+\beta^{\mu I}\delta\mathcal{H}_\mu,\\
		\delta\tilde{\mathcal{H}}_\mu&=\delta\mathcal{H}_\mu,\\
		\delta \tilde{C}^\mu
		&=
		\delta C^\mu
		+\alpha^\mu_{~I}\delta P^I
		+\beta^{\mu I} \delta Q_{I}
		+
		\underset{\gamma^{\mu\nu}}
		{\boxed{\frac{1}{2}	\left(
			\alpha^\mu_{~I}
			\beta^{\nu I}
			-
			\alpha^\nu_{~I}
			\beta^{\mu I}
			\right)}}
		\delta\mathcal{H}_\nu,
	\end{aligned}
\end{equation}
Notice that the matrix $\gamma^{\mu\nu}$ is determined entirely by the parameters $\alpha^\mu_{~I}$ and $\beta^{\mu I}$, thus completely defining a gauge transformation.
The \textbf{extra Hamiltonian} density coming from the time-dependent coordinates transformation is given by:
\begin{equation}\label{Kden}
	\mathcal{K}^{(2)}=-	\left[\dot{\alpha}^\mu_{~I} \delta\mathcal{H}_\mu\delta P^I
	+\dot{\beta}^{\mu I}\delta\mathcal{H}_\mu \delta Q_{I}+\frac{1}{2}
	\left(
	\dot{\alpha}^\mu_{~I}
	\beta^{\nu I}
	-
	\alpha^\mu_{~I}
	\dot{\beta}^{\nu I}
	\right)
	\delta\mathcal{H}_\mu
	\delta\mathcal{H}_\nu\right].
\end{equation}
By substituting Eq. \eqref{Kden} into Eq. \eqref{HK}, we obtain the \textbf{new Hamiltonian} that we shall denote by $\mathbb{H}_{\tilde{K}}$.  We note that the extra Hamiltonian is weakly zero as it must be in order for the dynamical equations \eqref{eom} for the Dirac observables to be preserved in the constraint surface.
 The gauge transformation does not affect the definition of the Dirac observables, as their definition in Eq. \eqref{Kden} has the parameters $\alpha^\mu_{~I}$ and $\beta^{\mu I}$ multiplied by the constraints $\delta\mathcal{H}_\mu$, which is zero in the constraint surface. Nevertheless, the gauge transformation does modify the Dirac observables in their extension beyond the constraint surface, which, however, is not physically relevant.

From Eqs. \eqref{gtr}, we can see that the local space of gauge-fixing conditions is defined by the parameters $\alpha^\mu_{~I}$ and $\beta^{\mu I}$. This defines a space without an origin, as our system is invariant under gauge parametrization, meaning no gauge is preferred over another. Therefore, we conclude that the local space of gauge-fixing conditions is an \textbf{affine space} of dimension $n$, and the local gauge group is the space of displacement vectors in this affine space, $\mathbf{G}=\mathbb{R}^n$, where $n$ is the number of $\alpha^\mu_{~I}$'s and $\beta^{\mu I}$, i.e. the number of Dirac observables times the number of gauge-fixing conditions. Hence, the group of canonical transformations \eqref{Gmap} is \textbf{abelian}.
\begin{equation}
	\mathbf{G}_{\alpha,\beta}\circ\mathbf{G}_{\alpha',\beta'}=\mathbf{G}_{\alpha+\alpha',\beta+\beta'}.
\end{equation}
The \textbf{physical part} of the Hamiltonian \eqref{KHw}, is transformed according to the following replacement: 
\begin{equation*}
	\mathcal{H}^{(2)}_{\text{phys}}(\delta Q_I,\delta P^I)
	\rightarrow 
	\mathcal{H}^{(2)}_{\text{phys}}(\delta \tilde{Q}_I,\delta \tilde{P}^I),
\end{equation*}
i.e., as we would expect, it is \textbf{gauge-invariant}. The transformation of the \textbf{weakly vanishing} part of the Hamiltonian \eqref{KHw} together with Eqs. \eqref{gtr} must also take into consideration the transformation of the $\lambda$'s coefficient, which go as follows:

\begin{equation}\label{ltr}
	\begin{aligned}
		\bm{\lambda_{1}^{\mu I}}
		\rightarrow 
		\tilde{\lambda}_{1}^{\mu I}
		&=
		\lambda_{1}^{\mu I}
		-\dot{\beta}^{\mu I}
		-\lambda_{4\nu}^\mu\beta^{\nu I}
		-\frac{\partial^2\mathcal{H}^{(2)}_{\text{phys}} }{\partial \delta Q_I\partial \delta P^J}\beta^{\mu J}
		+\frac{\partial^2\mathcal{H}^{(2)}_{\text{phys}} }{\partial \delta Q_I\partial \delta Q_J}\alpha^\mu_{~J},
		\\
		\\
		\bm{\lambda_{2I}^{\mu}}
		\rightarrow 
		\tilde{\lambda}_{2I}^{\mu}
		&=
		\lambda_{2I}^{\mu}
		-\dot{\alpha}^\mu_{~I}
		-\lambda_{4\nu}^\mu\alpha^\nu_{~I}
		+\frac{\partial^2\mathcal{H}^{(2)}_{\text{phys}} }{\partial \delta P^I\partial \delta Q_J}\alpha^\mu_{~J}
		-\frac{\partial^2\mathcal{H}^{(2)}_{\text{phys}} }{\partial \delta P^I\partial \delta P^J}\beta^{J\mu},
		\\
		\\
		\bm{\lambda_{3}^{\mu\nu}}
		\rightarrow 
		\tilde{\lambda}_{3}^{\mu\nu}
		&=
		\lambda_{3}^{\mu\nu}
		+\frac{1}{2}
		\left(
		\dot{\alpha}^\mu_{~I}\beta^{\nu I}
		-\alpha^\mu_{~I}\dot{\beta}^{\nu I}
		\right)
		+\frac{1}{2}\lambda_{4\kappa}^\mu
		\left(
		\alpha^\kappa_{~I}\beta^{\nu I}
		-\alpha^\nu_{~I}\beta^{\kappa I}	
		\right)
		\\
		&
		+\lambda_{1}^{\mu I}\alpha^\nu_{~I}
		-\lambda_{2I}^{\mu}\beta^{\nu I}
		-\frac{\partial^2\mathcal{H}^{(2)}_{\text{phys}} }{\partial \delta Q_I\partial \delta P^J}\alpha^\mu_{~I}\beta^{\nu J}
				\\
		&
		+\frac{1}{2}\frac{\partial^2\mathcal{H}^{(2)}_{\text{phys}} }{\partial \delta Q_I\partial \delta Q_J}\alpha^\mu_{~I}\alpha^\nu_{~J}
		+\frac{1}{2}\frac{\partial^2\mathcal{H}^{(2)}_{\text{phys}} }{\partial \delta P^I\partial \delta P^J}\beta^{\mu I}\beta^{\nu J},
		\\
		\\
		\bm{\lambda_{4\nu}^\mu}
		\rightarrow 
		\tilde{\lambda}_{4\nu}^\mu
		&=
		\lambda_{4\nu}^\mu,
\end{aligned}
\end{equation}

i.e., $\lambda_{1}$, $\lambda_{2}$ and $\lambda_{3}$ are \textbf{gauge-dependent}, whereas $\lambda_{4}$ is \textbf{gauge-invariant}, see end of Sec. \ref{sec-decomp}.

\subsection{Spacetime reconstruction}

As done in Sec. \ref{ISR} for the Dirac method, we will discuss the spacetime reconstruction in the Kucha\v r parametrization.

The \textbf{gauge stability condition} $\delta \dot{C}^\nu= 0$ is a dynamical equation, defined as the Poisson bracket between the gauge-fixing condition and the Hamiltonian. Therefore, it is important to specify the particular parametrization used in the definition of the Poisson bracket to ensure the appropriate Hamiltonian is employed. We denote the Poisson bracket in the Kuchař parametrization as $\{\cdot,\cdot\}_K$. Thus, the gauge stability condition reads:
\begin{equation}\label{cdot}
	\{\delta C^\nu, \mathbb{H}_{K}\}_K=0.
\end{equation}
Making use of Eq. \eqref{HK} and Eq. \eqref{cdot}, we find a formula for the computation of the \textbf{lapse} and \textbf{shift} functions:
\begin{equation}\label{deltaN}
	\frac{\delta N^\mu}{N}
	=
	-\frac{\partial(\mathcal{H}^{(2)}+\mathcal{K}^{(2)})}{\partial {\delta\mathcal{H}}_\mu}.
\end{equation}
The above formula involves only the \textbf{weakly vanishing} part of the Hamiltonian \eqref{KHw} as the lapse and shift are \textbf{pure gauge-dependent} quantities. Notice, however, that the difference between  the lapse and shift computed in two different gauges depends only on the \textbf{gauge-independent} part of the Hamiltonian  \eqref{KHw}, which simplifies the task of spacetime reconstruction. Indeed, after substituting  Eqs. \eqref{ltr} into Eq. \eqref{lapseshiftfull} we find that in the constraint surface the different values of $\delta{N}^\mu$ computed in different gauges reads
\begin{equation}\label{lapseshift}
	\begin{aligned}
		\frac{\delta\tilde{N}^\mu}{N}\bigg|_{\delta\tilde{C}^{\mu}=0}
		- \frac{\delta{N}^\mu}{N}\bigg|_{\delta{C}^{\mu}=0}
		&
		\approx 
		\left(
		\lambda_{4\nu}^\mu\beta^{\nu I}
		+\dot{\beta}^{\mu I}
		+\frac{\partial^2\mathcal{H}^{(2)}_{\text{phys}} }{\partial \delta Q_I\partial \delta P^J}\beta^{\mu J}
		-\frac{\partial^2\mathcal{H}^{(2)}_{\text{phys}} }{\partial \delta Q_I\partial \delta Q_J}\alpha^\mu_{~J}
		\right)\delta{Q}_I
		\\
		&
		+\left(\lambda_{4\nu}^\mu\alpha^\nu_{~I}
		+\dot{\alpha}^\mu_{~I}
		-\frac{\partial^2\mathcal{H}^{(2)}_{\text{phys}} }{\partial \delta P^I\partial \delta Q_J}\alpha^\mu_{~J}
		+\frac{\partial^2\mathcal{H}^{(2)}_{\text{phys}} }{\partial \delta P^I\partial \delta P^J}\beta^{\mu J}\right)\delta{P}^I.
\end{aligned}
\end{equation}
From Eq. \eqref{lapseshift} it is straightforward to see that the difference between the lapse and shift in any two gauges is completely determined by the \textbf{physical part} of the Hamiltonian $\mathcal{H}^{(2)}_{\text{phys}}$ and the \textbf{gauge-invariant }coefficient $\lambda_{4}$. As explained at the end of Sec. \ref{sec-decomp}, the value of $\lambda_{4}$ can easily be obtained from the algebra of the hypersurface deformations \cite{HOJMAN197688}. Once the lapse and shift has been found for one gauge-fixing condition, they can easily be found for any other gauge without using the full Hamiltonian.

Let us now see how the \textbf{three-geometries} transform under the gauge transformations. Consider the following linear map between the Kucha\v r and the ADM variables:
\begin{equation}
	\label{reconM}
	\left(
	\begin{array}{c}
		\delta \mathcal{H}_\mu
		\\ 
		\delta C^{\mu}
		\\ 
		\delta Q_I
		\\
		 \delta P^I
		 \end{array}
	\right)
	 =
	 \mathbf{M}
	 \left(
	 \begin{array}{c}
	 	\delta q_{ab}
	 	\\ \\  
	 	\delta \pi^{ab}
	 	\end{array}
	 	\right),
\end{equation}
where $\mathbf{M}$ is a matrix of the background coefficients computed for the preferred gauge-fixing functions $\delta C^{\mu}$. Then the\textbf{ physical three-surface} is obtained from the vanishing of $\delta C^{\mu}$:
\begin{align}
	\left(\begin{array}{c}\delta q_{ab}\\ \\ \delta \pi^{ab}\end{array}\right)=\mathbf{M}^{-1}\left(\begin{array}{c}0\\ 0\\ \delta Q_I\\ \delta P^I\end{array}\right).
\end{align}
From Eq. \eqref{gtr} we find that the physical three-surface in any gauge $\delta \tilde{C}^{\mu}=0$ reads
\begin{align}\label{ab3g}
	\left(\begin{array}{c}\delta \tilde{q}_{ab}\\ \\ \delta \tilde{\pi}^{ab}\end{array}\right)=\mathbf{M}^{-1}\left(\begin{array}{c}0\\ -\alpha^\mu_{~I}\delta P^I
		-\beta^{\mu I} \delta Q_{I}\\ \delta Q_I\\ \delta P^I\end{array}\right),
\end{align}
and is a linear function of the coefficients $\alpha^\mu_{~I}$ and $\beta^{\mu I}$.

Thus, we found a map allowing us to change from one parametrization to the other in any possible gauge. See \ref{canism} for an application in the context of a Bianchi I background.

\subsection{Partial gauge-fixing}\label{PGF}

To conclude this section on the Kucha\v r parametrization, we explore the possibility of fixing a gauge in other ways than by explicitly setting the gauge-fixing conditions $\delta C^{\mu}=0$. For instance, we may impose the \textbf{synchronous gauge} \cite{Landau:1975pou,Malik:2008im}, which consists of specifying the spacetime coordinate system by means of conditions on the lapse and shift functions. More generally, we may replace only some of the gauge-fixing conditions with conditions on the lapse and shift functions. Nevertheless, we find it sufficient to restrict our attention to the case of 4 conditions on the lapse and shift functions. We shall call this method "\textbf{partial gauge-fixing}" to distinguish it from the method used in the previous subsection. 

Let us first observe an interesting analogy with the well-known gauge theory in electrodynamics (see e.g. \cite{Heitler}, \cite{Weinberg:1995mt}).
 In electrodynamics, the\textbf{ Coulomb gauge}, $\nabla\vec{A}=0$, is an example of a gauge-fixing condition on the kinematical phase space made of the spatial components of the four-potential and their conjugate momenta $(\vec{A},\vec{\pi})$. On the other hand, the \textbf{Lorenz gauge}, $\partial_\mu A^\mu=0$, is an example of a partial gauge-fixing condition on the temporal component of the four potential $A^{0}$. The latter plays a role of the Lagrange multiplier analogously to the lapse and shift in gravity, and multiplies the only constraint of electrodynamics, the \textbf{Gauss constraint}. As we will see below, the partial gauge-fixing in the present theory respects a limited amount of covariance, which is a clear counterpart of the Lorentz-invariance of the Lorenz gauge in electrodynamics.

\begin{figure}[t]
	\begin{center}
		\includegraphics[width=0.5\textwidth]{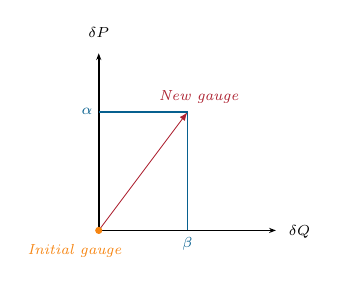}
	\end{center}
	\caption{\small{A displacement vector in the space of gauge-fixing conditions determines a new gauge via a shift from the initial gauge at the point of origin. }}
	\label{fig:gauge}
\end{figure}

Let us first study the subspace of gauge transformations that preserve the lapse and shift functions, i.e., $\frac{\delta\tilde{N}^\mu}{N}\big|_{\delta\tilde{C}^{\mu}=0}- \frac{\delta{N}^\mu}{N}\big|_{\delta{C}^{\mu}=0}=0$. This will determine the residual gauge freedom associated with this method of gauge-fixing. Using Eq. \eqref{lapseshift}, we find the ambiguity in the choice of the respective gauge-fixing conditions, expressed here in terms of $\alpha^\mu_{~I}$ and $\beta^{\mu I}$. These parameters satisfy the following dynamical equations (for each perturbation mode, $\vec{k}$):
\begin{equation}\label{kernel}
	\begin{aligned}
		\dot{\alpha}^\mu_{~I} &=-
		\beta^{\mu J}
		\frac{\partial^2\mathcal{H}^{(2)}_{\text{phys}} }{\partial \delta P^J\partial \delta P^I}
		+\alpha^\mu_{~J}
		\frac{\partial^2\mathcal{H}^{(2)}_{\text{phys}} }{\partial \delta Q_J\partial \delta P^I}-\lambda_{4\nu}^\mu\alpha^\nu_{~I},\\
		\dot{\beta}^{\mu I} &=
		-\beta^{\mu J}
		\frac{\partial^2\mathcal{H}^{(2)}_{\text{phys}} }{\partial \delta P^J\partial \delta Q_I}
		+\alpha^\mu_{~J}
		\frac{\partial^2\mathcal{H}^{(2)}_{\text{phys}} }{\partial \delta Q_J\partial \delta Q_I}-\lambda_{4\nu}^\mu\beta^{\nu I},
	\end{aligned}
\end{equation}
where the second-order partial derivatives yield the background coefficients of the physical Hamiltonian. Note that the solution of Eq. \eqref{kernel} does not depend on the particular choice of the lapse and shift. Once $\alpha^\mu_{~I}$ and $\beta^{\mu I}$ are set at an initial time for all $I={1,2}$ and $\mu={0,1,2,3}$, a unique solution $t\mapsto (\alpha^\mu_{~I}(t),\beta^{\mu I}(t))$ exists. Hence, at the initial time $t_0$ we have complete freedom in defining the gauge-fixing functions,
\begin{align}
	\delta \tilde{C}^{\mu}(t_0)\approx \delta C^\mu(t_0) +\alpha^\mu_{~I}(t_0)\delta P^I+\beta^{\mu I} (t_0)\delta Q_{I},
\end{align}
where $\delta C^\mu(t_0)$ lies at the point of origin of the gauge frame and $\delta \tilde{C}^{\mu}(t_0)$ are arbitrary gauge-fixing functions. Once the choice is made, Eq. \eqref{kernel} determines the gauge-fixing functions at all other times. There is a very clear spacetime picture associated with this ambiguity (see Fig. \ref{fig:gauge}): once $\delta \tilde{C}^{\mu}(t_0)$ are chosen, the initial three-surface with coordinates on it is fixed. If the initial values of the gauge-invariant variables $(\delta Q_{I}(t_0), \delta P^I(t_0))$ are known, then the initial three-surface may be reconstructed explicitly in terms of the ADM perturbation variables.

Furthermore, the evolution of the three-surface with its coordinates is uniquely determined via the evolution of $\delta \tilde{C}^{\mu}(t)$ and the independent evolution of the gauge-invariant variables $(\delta Q_{I}(t), \delta P^I(t))$. Hence the full spacetime geometry is reconstructed. Note the very important feature that the spacetime coordinate system is introduced in a way that is independent of the evolution of the gauge-invariant variables $(\delta Q_{I}(t), \delta P^I(t))$.

Now let us consider the case in which the LHS of Eq. \eqref{lapseshift} is non-vanishing. Then $\frac{\delta\tilde{N}^\mu}{N}\big|_{\delta\tilde{C}^{\mu}=0}- \frac{\delta{N}^\mu}{N}\big|_{\delta{C}^{\mu}=0}$ is an arbitrary linear combination of Dirac observables. In this case Eq. \eqref{lapseshift} implies
\begin{equation}
	\begin{aligned}\label{alphabetadot}
		\dot{\alpha}^\mu_{~I}
		=
		-
		\beta^{\mu J}
		\frac{\partial^2\mathcal{H}^{(2)}_{\text{phys}} }{\partial \delta P^J\partial \delta P^I}
		+\alpha^\mu_{~J}
		\frac{\partial^2\mathcal{H}^{(2)}_{\text{phys}} }{\partial \delta Q_J\partial \delta P^I}
		-\lambda_{4\nu}^\mu\alpha^\nu_{~I}
		-\frac{\partial}{\partial \delta P^{I}}\left(\frac{\delta \tilde{N}^\mu-\delta{N}^\mu}{N}\right),
		\\
		\dot{\beta}^{\mu I}
		=
		-\beta^{\mu J}
		\frac{\partial^2\mathcal{H}^{(2)}_{\text{phys}} }{\partial \delta P^J\partial \delta Q_I}
		+\alpha^\mu_{~J}
		\frac{\partial^2\mathcal{H}^{(2)}_{\text{phys}} }{\partial \delta Q_J\partial \delta Q_I}-\lambda_{4\nu}^\mu\beta^{\nu I}
		-\frac{\partial}{\partial \delta Q_{I}}\left(\frac{\delta \tilde{N}^\mu-\delta{N}^\mu}{N}\right).
		\end{aligned}
\end{equation}

where a \textbf{ particular solution} can be obtained by initially assuming $\alpha^\mu_{~I}(t_0)={\beta}^{\mu I}(t_0)=0$. The complete space of solutions is then constructed by combining it with the solutions of Eq. (\ref{kernel}).

\section{Summary}

In this chapter we discussed how to perturb the Hamiltonian up to $n$-th order in perturbation while only considering terms linear in the perturbation of the metric, i.e. $(\delta q)^n$. We used this result to obtain the second order perturbation of the metric, we saw that the 0-th and 1-st order are constraints, while the second, and last, order is non-zero. This allows us to work with a constraint system, although, contrary to the full GR case, the total Hamiltonian is non-vanishing. Using the notion gathered in Ch. \ref{Chapter1}, we were able to study our perturbed system using the \textbf{Dirac method}. This allowed us to \textbf{reduce} our Hamiltonian from the full phase-space, in which redundant degrees of freedom are present, to the constraints surface by solving the constraints. The Hamiltonian in the constraints surface is yet non-physical and has a gauge freedom, which allows us to set the most convenient gauge based on our needs. Once the gauge is fixed, we do get a new Hamiltonian in terms of the ADM variables. The form of this Hamiltonian will then change based on the gauge-fixing chosen. To overcome this issue we introduce the \textbf{Dirac observables} which are defined as those variables which commute with the constraints. These new variables allow us to write our Hamiltonian in a gauge-independent way and thus determine the final \textbf{physical Hamiltonian}. The final issue to be addressed is the spacetime reconstruction, i.e.  finding the value of the lapse and shift, whose values do depend on the gauge choice. This is easily done by considering the Poisson brackets between the gauge-fixing conditions and the total Hamiltonian, which are zero as the gauge-fixing conditions must be dynamically preserved. 


We then introduced an alternative parametrization based on the \textbf{Kucha\v r decomposition}, allowing the separation of variables defining the hypersurface and those defining the coordinates. The new phase space variables are given as two types of canonical pairs: one defined by the Dirac observables and the other by the gauge-fixing functions and Hamiltonian constraints.

When deriving spacetime solutions for arbitrary gauges, we establish a \textbf{gauge frame} by placing a chosen gauge at the point of origin. The respective full spacetime metric for that particular gauge is computed, as explained in Section \ref{ISR}. The usual and convenient choice for the point of origin is the spatially flat (or spatially uniform) gauge. Subsequently, we construct other gauges conveniently by making arbitrary choices of the parameters $\alpha^\mu_{~I}$ and $\beta^{\mu I}$ within a fixed gauge frame. As demonstrated, these choices completely determine the new full spacetime metric: (i) they determine the lapse function and the shift vector on the three-surfaces through the stability equation \eqref{lapseshift}; and (ii) they define the metric of the three-surfaces via Eq. \eqref{ab3g}.

\section{Conclusions}\label{conclusion}
The purpose of this work was to develop a complete Hamiltonian approach to CPT. The basic property of our approach is the separation of the gauge-independent dynamics of perturbations from the problem of gauge-fixing and spacetime reconstruction. We use the \textbf{Dirac procedure} for constrained systems to derive the dynamics of gauge-dependent perturbations and to rewrite it in terms of gauge-independent quantities, the Dirac observables. A key element of our approach is the reconstruction of spacetime based on  gauge-fixing conditions. 

Similar work in which the problem of gauge-fixing is addressed like in \cite{Mukhanov:1990me,Malik:2008im,sasaki-cpt}, do not offer a methodological choice of gauge-fixing conditions. Although some commonly used gauges, their validity and, in some cases, their residual freedom are studied in the mentioned references, no general method for defining a valid gauge and its residual freedom is provided. As such, no clear exposition of the connection between the residual freedom, the lapse and the shift, and the choice of the initial three-surface is presented. 
To overcome this problem, we introduced the \textbf{Kucha\v r decomposition} for the ADM perturbation phase space. The space of all the possible gauge-fixing conditions and the gauge transformations induced by the linear diffeomorphisms of three-surfaces are made explicit via this decomposition. Moreover, it makes the transformations of the lapse and shift manifestly dependent on purely gauge-independent terms of the full Hamiltonian. This simplifies the problem of spacetime reconstruction and provides a tool for studying partial gauge-fixing.


The \textbf{possible applications} of the presented Hamiltonian formalism include addressing key conceptual problems in quantum cosmology such as:
\begin{itemize}
	\item[$\hookrightarrow$] The time problem (see Ch. \ref{Chapter4});
	\item[$\hookrightarrow$] The semiclassical spacetime reconstruction;
	\item[$\hookrightarrow$] The relation between the kinematical and reduced phase space quantization.
\end{itemize}
 The full clarification of the Hamiltonian formalism and its structure is essential for understanding these and similar issues.

\newpage

%% file: Chapters/Chapter3.tex
\allowdisplaybreaks
\chapter{Perturbation theory in anisotropic universes} 

\label{Chapter3} 

The goal of this chapter is to apply the Hamiltonian formalism presented in Ch. \ref{Chapter2} to cosmological perturbations within the context of an anisotropic background, specifically the \textbf{Bianchi I universe}. 
Our result agrees with \cite{Uzan}, where the standard configuration space approach is used. However, we express the physical Hamiltonian in a form that is better suited for the (affine or canonical) quantization of both the background and the perturbations in a consistent manner. We provide full canonical expressions for gauge-invariant variables, including the canonical definition of a gravitational wave, which differs from the isotropic case. We utilize the Dirac procedure to test various gauge-fixing conditions and their isotropic limits. Furthermore, we discuss the reconstruction of the full spacetime metric in terms of physical phase space variables and illustrate it with an example. 

The study of anisotropic backgrounds presents many interesting differences compared to isotropic ones. In particular, the dynamics of cosmological perturbations in anisotropic backgrounds exhibit new and fascinating phenomena, such as \textbf{couplings} between two modes of a gravitational wave or between gravitational waves and scalar fields \cite{Uzan, Cho_1995}.
Moreover, the definition of polarization modes becomes ambiguous in these scenarios \cite{Cho_1995}. 
 Our method relies on \textbf{Fermi-Walker-propagated }vectors tangent to the wavefronts, offering a more comprehensive framework for understanding polarization in anisotropic backgrounds.
 Although this definition may not be optimal for quantization purposes, it provides the simplest form of the dynamics and aligns with the prescription given in \cite{Uzan}.
Another interesting difference arises from the admissible gauge-fixing conditions, which can differ from those in isotropic spacetimes. Indeed, we provide an example of a valid gauge that does not exist within the isotropic limit. Furthermore, the expression of Dirac observables in kinematical phase space variables becomes more intricate due to the mixing of scalar, vector, and tensor modes. This mixing introduces a complexity such that gravitational waves can no longer be unequivocally associated with the transverse and traceless metric perturbations. As a result, the geometrical meaning of physical variables can vary significantly when switching from one gauge-fixing surface to another.

We approach the issue of \textbf{spacetime reconstruction} using both the \textbf{Dirac method} and \textbf{Kucha\v r decomposition}. With the Dirac method, once a gauge choice has been made, it is necessary to return to the kinematical phase space to determine the values of the Lagrange multipliers, $\delta N$ and $\delta N^i$. This issue has been discussed in \cite{Malkiewicz_2019} in the context of isotropic spacetimes. The spacetime reconstruction can be simplified by using the Kucha\v r decomposition, as discussed in Ch. \ref{Chapter2}.

For the most part, we consider a \textbf{minimally coupled canonical scalar field} as the only matter component. However, in Sec. \ref{multifield}, the formalism is extended to the case of any number of minimally coupled scalar fields in an arbitrary potential.


The canonical formalism obtained in this chapter serves as a starting point for the quantization of all gravitational degrees of freedom, not only for tensor modes, but also scalar ones. The quantization of a perturbed anisotropic universe provides a framework for testing the hypothesis of a primordial anisotropic universe, in which quantum gravity effects play an essential role. These quantum gravity effects, such as singularity resolution, spread, interference, and entanglement, may offer a physically rich and viable alternative, or complement, to the inflationary paradigm models.
For a more comprehensive understanding of the quantum cosmological dynamics up to first order, refer to the detailed explanation provided in \cite{malkiewicz2020dynamics}, particularly Eq. (17).
It must be noted that presently available analogous quantum frameworks (see e.g. \cite{Peter_2008}) are much simpler as they usually assume primordial isotropy and neglect the quantum spread or entanglement. They also generically predict (slightly) blue-tilted primordial amplitude spectrum of density perturbations (scalar modes) in tension with the cosmic microwave background (CMB) data \cite{Aghanim:2018eyx}.


The \textbf{primary motivation} behind this work is the need for new frameworks which include fewer primordial symmetries but still cover all the important quantum effects. This idea is supported both theoretically and by what has been observed. For instance, when data from the CMB is examined, statistical anomalies emerge that standard theories struggle to explain. Hence, exploring anisotropic inflationary models can result in an effort to explain them \cite{pereira}.

For earlier works on the cosmological perturbations in a Bianchi I universe, apart from the already mentioned \cite{Uzan}, see earlier discussions of linearized Einstein's equations in \cite{Cho_1995, Tomita:1985me}, where some approximate solutions were derived. The solutions for the vacuum case (i.e. for the perturbed Kasner universe) were also studied in \cite{Kofman:2011tr}. Recently, an interesting canonical analysis of the theory, by another method, was also considered in \cite{Agullo2020}. It is worth noting that similarly the perturbation theory of anisotropically curved cosmologies, the Bianchi III and the Kantowski-Sachs models, is found to lead to interesting physical effects \cite{Franco:2017pxt}.


\section{Perturbative expansion of the canonical formalism}\label{expansion}
We start by assuming the spacetime topology given by $\mathcal{M}\simeq \mathbb{T}^3\times \mathbb{R}$ in order to have a \textbf{spatially compact} universe. {This choice allows us to avoid the ambiguity in the definition of the symplectic structure for background (homogeneous) variables and to avoid the introduction of a boundary term known as the \textbf{Gibbons–Hawking–York boundary term}
\cite{York:1972sj, Gibbons:1976ue, Kiefer:2021zdq}. Notice that the choice of any 3D compact space gives us discrete eigenvalues for the Laplace operator. This means that, when going to the Fourier space (see Sec. \ref{decomposition}), the wave numbers $k$ are discrete. We choose a torus as it is the simplest 3D compact space. }

For any compact 3D space, the wave numbers are discrete. This is a result of the compactness of the space, which imposes boundary conditions that quantize the possible wavelengths, leading to discrete eigenvalues for the Laplace operator. This quantization reflects the fact that only certain modes of vibration (or wave numbers) can exist in a finite, bounded space.

The line element in the ADM formalism is given in Eq. \eqref{dsadm}.
The spacetime is filled with a \textbf{real scalar field} $\phi$ in a \textbf{potential} $V(\phi)$. The phase space of this model includes the ADM variables. In the \textbf{geometry sector}, these variables include the three-metric $q_{ij}$ and the three-momenta $\pi^{ij}$ (defined in Eq. \eqref{admomento}). In the \textbf{matter sector}, we have the scalar field $\phi$ and its momentum $\pi_{\phi}$. These variables satisfy the following canonical equations:
\begin{equation}
	\{q_{ij}(x),\pi^{kl}(x')\}
	=\delta_{(i}^{~(k}\delta_{j)}^{~l)}\delta^3(x-x')
	,~~~
	\{\phi(x),\pi_{\phi}(x')\}
	=\delta^3(x-x').
\end{equation}
The dynamics is generated by the Hamiltonian in Eq. \eqref{hamgra}, which we rewrite here for convenience with a slight change of notation as,
\begin{equation}\label{fullH}
	{\bf H}=
	\int 
	\left(
	N \mathcal{H}_{0} 
	+N^i\mathcal{H}_{i}
	\right)~\ud^3x.
\end{equation}
The above Hamiltonian comprises the \textbf{gravity} and \textbf{matter} parts:
\begin{equation}
	\mathcal{H}_{0}=\mathcal{H}_{g,0}+\mathcal{H}_{m,0}
	,~~~
	\mathcal{H}_{i}=\mathcal{H}_{g,i}+\mathcal{H}_{m,i}~,
\end{equation}
with $\mathcal{H}_{0}$ and $\mathcal{H}_{g,i}$ defined in Eq. \eqref{hamtemp} and Eq. \eqref{hampspa} respectively. The \textbf{matter Hamiltonian} is defined as follows
\begin{equation}
	\begin{aligned}
		\mathcal{H}_{m,0}
		=\sqrt{q}
		\left[
		\frac{1}{2}q^{-1}\pi_{\phi}^2
		+\frac{1}{2}q^{ij}\phi_{,i}\phi_{,j}
		+V(\phi
		)\right]
		,~~
		\mathcal{H}_{ m,i}
		=\pi_{\phi}\phi_{,i}
		~.
	\end{aligned}
\end{equation}
We expand the above canonical formalism in perturbations around a \textbf{Bianchi I} background, where the three-metric and the three-momentum perturbations are given in Eq. \eqref{3mpert}. The \textbf{lapse} and the \textbf{shift} are perturbed as per Eq. \eqref{expN}, which at first order reads,  $N\mapsto N+\delta N$ and $N^i\mapsto N^i+\delta N^i$, where $N$ and $N^i$ are now understood as zero-order quantities. The total Hamiltonian expanded up to second order was defined in Eq. \eqref{Htot}. By considering $\int_{\mathbb{T}^3}\ud^3x=1$ and separating the vector and the scalar parts of the first-order Hamiltonian, we can write
\begin{equation}\label{hamtot}
	{\bf H}=N\mathcal{H}_{0}^{(0)}+\int_{\mathbb{T}^3} \bigg(N\mathcal{H}_{0}^{(2)}+\delta N\delta\mathcal{H}_{0}+\delta N^i\delta\mathcal{H}_{i}\bigg)~\ud^3x.
\end{equation}
Recall that $\mathcal{H}_{0}^{(0)}$ is zeroth order, $\delta\mathcal{H}_{0}$, $\delta\mathcal{H}_{i}$ are first order and $\mathcal{H}_{0}^{(2)}$ is second order in perturbation.  
The expansion is defined around a \textbf{non-tilted} Bianchi I spacetime, meaning that the flow of the scalar field is orthogonal to the spatial hypersurfaces at any time, and thus $\mathcal{H}_{i}^{(0)}=0$, as seen in Sec. \ref{ADMpert}.
 Moreover, we can simplify our formalism by choosing the simplest value for the shift vector $N^i$, i.e. $N^i=0$. As explained in \cite{Ryan}, fixing both the values of $N$ and $N^i$ generally equates to selecting a particular set of coordinates, which is usually avoided. 


\subsection{The anisotropic model}
In this work, we perturb around the Bianchi I universe, which is one of the possible solutions of the Einstein equations obtained by relaxing the assumption of isotropy while maintaining homogeneity in the Universe.
In this section we will briefly introduce the Bianchi Universes \cite{bianchiIT, bianchiEN} and then focus on the particular case of the Bianchi I solution.

Following the approach presented in \cite{Landau:1975pou}, we define a triad, or frame, for describing spacetime. We introduce the triad vectors $e_i$, which are differential operators independent of the coordinates and provide tangent vectors to the three-surface at each point in the manifold.
The operators $\partial_\mu$ are basis vectors, meaning they are linearly independent, and every vector field $e_i$ can be uniquely expressed as a linear combination of the $\partial_\mu$.
We can thus define a linear differential operator such that
\begin{equation}
	e_i=e^\mu_{i}\frac{\partial}{\partial x^\mu}.
\end{equation}
The Latin indices label the frame vectors, while the Greek indices label the spacetime coordinates.
The line element in this new basis reads
\begin{equation}\label{metricL}
\ud s^2=
\gamma_{ij}
e^{i}_\alpha e^{j}_\beta 
\ud x^\alpha \ud x^\beta
\end{equation} 
which describes a non-Euclidean homogeneous space.
 
\begin{figure}[t]
	\includegraphics[width=0.8\textwidth,center]{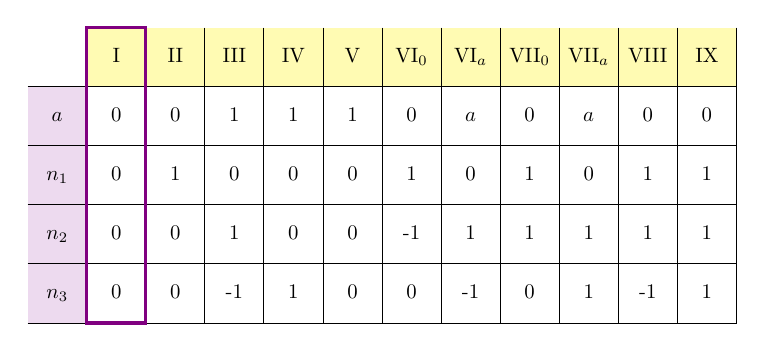}
	\caption{Classification of all the Bianchi universes. We will focus on the Bianchi I universe, highlighted in purple.}
	\label{bianchitab}
\end{figure}

A generic set of vector fields $e_i$ forms a non-coordinate basis if $[e_i,e_j]\neq0$. The commutator is also defined as the Lie derivative of $e_j$ with respect to $e_i$, denoted as \cite{Wald,Landau:1975pou}
\begin{equation}\label{comm}
	\mathcal{L}_{e_i}e_j\equiv [e_i,e_j]
	=
	C^k_{~ij}e_k
\end{equation}
where $C^k_{~ij}$ are the \textbf{structure constants}\footnote{For a more detailed derivation of Eq. \eqref{comm}, refer to \cite{Landau:1975pou}.} of the basis $e_i$, representing the torsion of the vector field \cite{Ryan}. 
Eq. \eqref{comm} represents the required conditions for space homogeneity, 
Introducing the quantities $C^{lk}$ defined by the relation
\begin{equation}
 C^k_{~ij}=\epsilon_{ijl}C^{lk},
\end{equation}
we can have an object behaving like a tensor under linear transformation of the frame vectors. As such, by using tensor-like properties we are able to determine all the non-equivalent structure constants. This allows us to obtain a classification of all the \textbf{3-parameter homogeneous spaces}.
We can write  $C^{lk}$ as the sum of symmetric and antisymmetric components, i.e., using the notation in \cite{Landau:1975pou},
\begin{equation}\label{symm-ant}
C^{lk}=n^{lk}+\epsilon^{lkj}a_j,
\end{equation}
where $n^{lk}$ is the symmetric part and $\epsilon^{lkj}a_j$ the antisymmetric one. 
We can finally introduce the Bianchi classification using the eigenvalues of $n^{lk}$, namely $n_1$, $n_2$ and $n_3$, and considering, without loss of generality \cite{Landau:1975pou},  the simpler case in which the antisymmetric part is  $a_k=(a,0,0)$.
The full list of possible combinations is given in Table \ref{bianchitab}
We are interested in the case in which $a=0$, $n_1=n_2=n_3=0$ describing an Euclidean space (invariant under commuting translations) as all the curvature components vanish and we are left with a flat anisotropic spacetime which we will consider as background for our model.

Let us now focus on the Bianchi I background metric, which reads:
\begin{equation}\label{BImetric}
	\ud s^2=-\ud t^2+\sum_ia_i^2(\ud x^i)^2,~~~a=(a_1a_2a_3)^{\frac{1}{3}},
\end{equation}
where $a$ is the \textbf{average scale factor }and we assume  the coordinates\footnote{This assumption is due to our initial choice of spacelike hypersurfaces (see the beginning of the current section) which we assume to be a torus.} $(x^1,x^2,x^3)\in [0,1)^3$. We will now use the computation presented in Sec. \ref{ADMpert} to obtain the zeroth, first and second order of the Hamiltonian in the BI universe. From \eqref{hamtemp0}, setting the background variables to $\bar{q}_{ij}=a_i^2\delta_{ij}$ and $\bar{\pi}^{ij}=p^i\delta^{ij}$, we have that the background Hamiltonian is
\begin{equation}\label{H0}
	\mathcal{H}_{0}^{(0)}
	=
	a^{-3}
	\bigg(
	\frac{1}{2}
	\sum_i
	(a_i^2p^i)^2
	-\sum_{i> j}
	a_i^2p^ia_j^2p^j
	+\frac{1}{2}p_{\phi}^2
	+a^6V
	\bigg),
	~~\{a_i^2,p^j\}=\delta_{i}^{~j},
	~\{\phi,p_{\phi}\}=1,
\end{equation}
while \eqref{hampspa0} is zero as discussed in Ch. \ref{Chapter2}.
The equations of motion generated by Eq. \eqref{H0} are,
\begin{equation}\label{backeom-pa}
	\begin{aligned}
		\dot{p}^{i}=-\frac{1}{a^3a_{i}^2}\bigg((a_i^2p^i)^2-\sum_{j\neq (i)}a_{i}^2p^{i}a_j^2p^j+a^6V\bigg),
\\
		\dot{a}^2_{i}=a^{-3}\bigg(a_{i}^4p^{i}-\sum_{j\neq (i)}a_{i}^2a_j^2p^j\bigg),
		\end{aligned}
\end{equation}
and
\begin{equation}\label{pdot}
	\dot{p}_{\phi}=-a^3V_{,\phi},~~\dot{\phi}=a^{-3}{p}_{\phi},
\end{equation}
where the dynamics is confined in the constraint surface, $\mathcal{H}_{0}^{(0)}=0$. 
Finding the solution to the above equations in general can be difficult, see \cite{Mohanty:2003bj,Folomeev:2007uw,Rybakov:2010uv,Fadragas:2013ina,Chaubey:2016qtx,Kohli:2017ogi} for some results on the Bianchi I dynamics.

\subsection{First-order constraints}
The canonical perturbation variables of Eq. \eqref{3mpert} are expressed as follows:
\begin{equation}\label{defP0}
	\begin{gathered}
	\delta q_{ij}=q_{ij}-a_i^2\delta_{ij}
	,~~\delta\pi^{ij}=\pi^{ij}-p^i\delta^{ij},
	\\
	\delta\phi=\phi-\bar{\phi},
	~~\delta \pi_{\phi}=\pi_{\phi}-{p}_{\phi},
	\end{gathered}
\end{equation}
and satisfy the following Poisson brackets
\begin{equation}\label{PB1}
		\{\delta \phi(x),\delta \pi_{\phi}(x')\}
		=\delta^3(x-x'),
		~~\{\delta q_{ij}(x),\delta\pi^{kl}(x')\}
		=\delta_{(i}^{~k}\delta_{j)}^{~l}\delta^3(x-x').
\end{equation}
We obtain that the \textbf{first-order scalar constraint }from Eq. \eqref{H0-1}, is now given by
\begin{equation}\label{scx}
	\begin{aligned}
		\delta\mathcal{H}_{0}&=
		a^{-3}
		\bigg(
		2\overline{\pi}_{ij}-\overline{q}_{ij}\overline{\pi}^{k}_{~k}
		\bigg)
		\delta\pi^{ij}
		\\&
		-a^{-3}
		\bigg[
		\frac{1}{2}\overline{q}^{ij}\bigg(\overline{\pi}^{kl}\overline{\pi}_{kl}
		-\frac{1}{2}(\overline{\pi}^{k}_{~k})^2\bigg)
		-\overline{\pi}^{i}_{~k}\overline{\pi}^{kj}
		+\frac{1}{2}\overline{\pi}^{ij}\overline{\pi}^{k}_{~k}\bigg]\delta q_{ij}
		\\
		&
		-a^3\overline{q}^{ij}\overline{q}^{kl}(\delta q_{ik,jl}-\delta q_{ij,kl})+a^{-3}p_{\phi}\delta\pi_{\phi}
		-\frac{a^{-3}}{4}p_{\phi}^2\overline{q}^{ij}\delta q_{ij}
		\\&
		+\frac{a^{3}}{2}V\overline{q}^{ij}\delta q_{ij} 
		+a^{3}V_{,\phi}\delta\phi.
	\end{aligned}
\end{equation}
The\textbf{ first-order vector constraint}, given in Eq. \eqref{Hi1-inizio}, reads
\begin{equation}\label{vcx}
	\delta\mathcal{H}^{i}=
	-2\bigg(\delta\pi^{ij}_{,j}
	+\overline{q}^{ij}\delta q_{kj,l}
	\overline{\pi}^{kl}
	-\frac{1}{2}\overline{q}^{ij}\delta q_{kl,j}\overline{\pi}^{kl}\bigg)+\overline{q}^{ij} p_{\phi}\delta\phi_{,j}~,
\end{equation}
where $\delta q^{ij}$ is given in Eq. \eqref{delta-q-contrav}, and $\delta \pi_{ij}:=\delta \pi^{kl}\overline{q}_{ki}\overline{q}_{lj}$. 

\subsection{Second-order Hamiltonian}
The \textbf{second-order scalar Hamiltonian} \eqref{H0-2} is found to read
\begin{align}\label{2cx}
			\mathcal{H}_{0}^{(2)}
			&=
			a^{-3}\bigg[\delta\pi_{ij}\delta\pi^{ij}
			-\frac{1}{2}(\delta\pi^{i}_{~i})^2
			+4\delta\pi^{j}_{~k}\overline{\pi}^{ki}\delta q_{ij}
			-\overline{\pi}^{k}_k(\delta\pi^{ij}\delta q_{ij})
			-\delta\pi^{i}_{~i}(\overline{\pi}^{ij}\delta q_{ij})
			\nonumber
			\\
			&
			-(\delta\pi^{ij}\overline{\pi}_{ij})(\delta q_{ij}\overline{q}^{ij})
			+\frac{1}{2}\overline{\pi}^{k}_{~k}\delta\pi^l_{~l}(\delta q_{ij}\overline{q}^{ij})
			+\overline{\pi}^{jl}\overline{\pi}^{in}\delta q_{il}\delta q_{jn}
			\nonumber
			\\
			&
			-\frac{1}{2}(\overline{\pi}^{ij}\delta q_{ij})^2
			-(\overline{q}^{ij}\delta q_{ij})(\delta q_{ln}\overline{\pi}^{n}_{~m}\overline{\pi}^{ml})
			+\frac{1}{2}\overline{\pi}^{k}_{~k}(\delta q_{ij}\overline{q}^{ij})(\delta q_{ij}\overline{\pi}^{ij})\bigg]
			\nonumber
			\\
			&
			+a^{-3}\left[\frac{1}{8}(\delta q_{ij}\overline{q}^{ij})^2
			+\frac{1}{4}(\delta q_{ij}\delta q^{ij})\right]
			\left[(\overline{\pi}^{ij}\overline{\pi}_{ij})-\frac{1}{2}(\overline{\pi}^{k}_{~k})^2\right]
			\\
			&
			+\frac{1}{4}a^{3}(\delta q_{ij}\overline{q}^{ij})(\overline{q}^{ij}\overline{q}^{kl}\delta q_{ij,kl})
			+\frac{1}{4}a^{3}(\delta q^{ij}\overline{q}^{kl})(2\delta q_{ik,jl}-\delta q_{ij,kl}
			-2\delta q_{kl,ij})
			\nonumber
			\\
			&
			+\frac{1}{16}a^{-3}(\delta q_{ij}\overline{q}^{ij})^2p_{\phi}^2
			+\frac{1}{8}a^{-3}(\delta q^{ij}\delta q_{ij})p^2_{\phi}-\frac{1}{2}a^{-3}(\delta q_{ij}\overline{q}^{ij})p_{\phi}\delta\pi_{\phi}+\frac{1}{2}a^{-3}(\delta\pi_{\phi})^2
			\nonumber
			\\
			&
			+\frac{1}{2}a^3\overline{q}^{ij}\delta\phi_{,i}\delta\phi_{,j}
			+\frac{1}{2}a^3V_{,\phi\phi}(\delta\phi)^2
			+\frac{1}{2}a^3(\delta q_{ij}\overline{q}^{ij})V_{,\phi}\delta\phi
					\nonumber
			\\
			&
			+\frac{1}{8}a^3(\delta q_{ij}\overline{q}^{ij})^2V
			-\frac{1}{4}a^3(\delta q_{ij}\delta q^{ij})V.\nonumber
	\end{align}

\section{Mode decomposition}\label{decomposition}

It is useful to introduce the \textbf{rescaled spatial metric tensor} $\gamma_{ij}=a^{-2}\overline{q}_{ij}$. From now on, we shall use $\gamma$ to define the \textbf{duals of spatial tensors}. In particular, the indices of the basic perturbation variables (\ref{defP0}) are raised and lowered with  $\gamma^{ij}=(\gamma^{-1})^{ij}$ and $\gamma_{ij}$, respectively.

Let us consider the \textbf{Fourier transform} of a perturbation variable, denoted by $\delta X$,
\begin{equation}\label{Fourier}
	\delta\check{X}(\underline{k})
	=\int \delta X(\overline{x})e^{-ik_ix^i}\ud^3x,
\end{equation}
and its inverse
\begin{equation*}\label{Fourierinv}
	\delta\check{X}(\overline{x})
	=\frac{1}{(2\pi)^3}\int \delta X(\underline{k})e^{ik_ix^i}\ud^3k.
\end{equation*}
The components $k_i$ of a fixed spatial co-vector $\underline{k}$ determine the respective Fourier mode by introducing wavefronts into the coordinate space. The components of the dual vector $\overline{k}$ read $k^i=k_j\gamma^{ji}$. Note that $k^i$ are in general \textbf{time-dependent }as $\gamma^{ij}$ evolves. It can be shown by using the integral representation of Kronecker delta, that the Fourier transform of the basic perturbation variables (\ref{defP0}) yields the following expression for the Poisson brackets (\ref{PB1}),
\begin{equation}\label{PB2}
		\{
		\delta \check{\phi}(\underline{k}),
		\delta \check{\pi}^{\phi}(\underline{k}')
		\}
		=
		\delta_{{\underline{k}},-{\underline{k}}'},
		~~
		\{
		\delta \check{q}_{ij}(\underline{k}),
		\delta\check{\pi}^{lm}(\underline{k}')
		\}
		=
		\delta_{(i}^{~l}\delta_{j)}^{~m}
		\delta_{\underline{k},-\underline{k}'}.
\end{equation}
The Kronecker delta arises as a consequence of the compact (toroidal) topology of the spatial sheet $\Sigma\simeq \mathbb{T}^3$, for which we have that $k_i=2\pi n_i$, where $n_i\in\mathbb{Z}$.
We introduce an \textbf{orthonormal spatial triad} consisting of the normalized vector $\hat{k}=\overline{k}/k$, where $k=\sqrt{k_i k_j \gamma^{ij}}$, along with the remaining vectors denoted as $\hat{v}$ and $\hat{w}$. We postpone the discussion regarding the ambiguity in defining the vectors $\hat{v}$ and $\hat{w}$ for a given $\hat{k}$  to Sec. \ref{Fermi}. Note that the components of $\hat{v}$ and $\hat{w}$ are in general \textbf{time-dependent} due to the presence of the metric in their definition.

The introduced fixed triad $(\hat{k},\hat{v},\hat{w})$, can be used to decompose any spatial symmetric $2$-rank covariant tensor in the $\gamma$-orthogonal basis $(A^1,A^2,A^3,A^4,A^5,A^6)$ \cite{Dapor_2013}, where
\begin{equation}\label{baseA}
	\begin{aligned}
		A_{ij}^1=\gamma_{ij}
		&,
		\quad
		A_{ij}^2=\hat{k}_i\hat{k}_j-\frac{1}{3}\gamma_{ij}
		,
		\\
		A_{ij}^3=
		\frac{1}{\sqrt{2}}
		\Big(
		\hat{k}_i\hat{v}_j+\hat{v}_i\hat{k}_j
		\Big)
		&,
		\quad
		A_{ij}^4=
		\frac{1}{\sqrt{2}}
		\Big(
		\hat{k}_i\hat{w}_j+\hat{w}_i\hat{k}_j
		\Big)
		,
		\\
		A_{ij}^5=
		\frac{1}{\sqrt{2}}
		\Big(
		\hat{v}_i\hat{w}_j+\hat{w}_i\hat{v}_j
		\Big) 
		&,
		\quad
		A_{ij}^6=
		\frac{1}{\sqrt{2}}
		\Big(
		\hat{v}_i\hat{v}_j-\hat{w}_i\hat{w}_j
		\Big).
	\end{aligned}
\end{equation}
The dual basis $(A_1,A_2,A_3,A_4,A_5,A_6)$ is defined such that $A_n^{ij}A^m_{ij}=\delta_{n}^m$. The new basis splits the perturbations into \textbf{scalar} ($A^1$, $A^2$), \textbf{vector} ($A^3$, $A^4$) and \textbf{tensor} ($A^5$, $A^6$) modes. Note that both $A^n_{ij}$'s and $A_n^{ij}$'s are in general \textbf{time-dependent} as $\gamma_{ij}$ and $(\hat{k}_i,\hat{v}_i,\hat{w}_i)$ evolve. The perturbations of the three-metric and the three-momentum may be expressed
  via the following\textbf{ time-dependent linear transformations},
\begin{equation}\label{nperts}
	\delta {q}_{n}=\delta \check{q}_{ij}A_n^{ij},
	~~~\delta {\pi}^{n}=\delta \check{\pi}^{ij}A^n_{ij},
\end{equation}
which define the new basis $(\delta {q}_{n},\delta {\pi}^{m})$.
Using the orthonormality of the triad $(\hat{k},\hat{v},\hat{w})$, and the properties of the metric $\gamma$, it can be shown that the Poisson brackets (\ref{PB2}) now read
\begin{align}\label{PB3}
	\begin{split}
		\{\delta \check{\phi}(\underline{k}),\delta \check{\pi}^{\phi}(\underline{k}')\}=\delta_{\underline{k},-\underline{k}'},
		~~\{\delta {q}_{n}(\underline{k}),\delta{\pi}^{m}(\underline{k}')\}=\delta^m_n\delta_{\underline{k},-\underline{k}'}.
	\end{split}
\end{align}
As previously discussed, the presence of a time-dependent transformation introduces an additional term in the Hamiltonian. In Section \ref{Fermi}, we will explore how the transformations in Eq. (\ref{nperts}) contribute an extra term to the second-order Hamiltonian.

The variables $\delta {q}_5$ and $\delta {q}_6$ describe the \textbf{transverse-traceless} metric perturbations. As we will discuss in Sec. \ref{diracobs}, it is important to note that they are \textbf{not} gauge-invariant quantities. This lack of gauge invariance led to difficulties in defining gravitational waves solely as transverse and traceless metric perturbations \cite{Cho_1995}. In isotropic universes, the transverse-traceless metric perturbations are commonly referred to as tensor modes and are associated with gravitational waves.
However, as shown in \cite{Uzan}, in anisotropic universes, it is also possible to define tensor modes as certain linear combinations of \textbf{transverse-traceless} ($\delta {q}_5$, $\delta {q}_6$) and \textbf{scalar trace and traceless} ($\delta {q}_1$, $\delta {q}_2$) metric perturbations that are gauge-invariant. 

\subsection{Zeroth-order revisited}\label{0ord}

Let us define a mapping $P:\mathrm{T}\Sigma\rightarrow \mathrm{T}\Sigma$ in the space of spatial vectors,
\begin{equation*}
P^i_{~j}:=\bar{\pi}^{ik}\bar{q}_{kj}=a^2\bar{\pi}^{ik}\gamma_{kj}.
\end{equation*} 
 We make use of the introduced frame to define the components of $P$ such as 
\begin{equation}
	P_{nm}
	=\hat{n}_iP^i_{~j}\hat{m}^j
	=a^2\bar{\pi}^{ij}\hat{n}_i\hat{m}_j
	=a^2\sum_ip^i\hat{n}_i\hat{m}_i,
\end{equation}
where $\hat{n},\hat{m}\in (\hat{k},\hat{v},\hat{w})$. Using Eq. \eqref{admomento}, one can show that $P$ is related to the \textbf{shear tensor} as follows
\begin{equation}\label{Pshear}
	P_{{n}{m}}=a^2\sigma_{nm}-2a^2\mathcal{H}\delta_{{n}{m}},
\end{equation}
where $\mathcal{H}=-\frac{1}{6a^2}\Tr P$ is the \textbf{mean conformal Hubble rate}. The shear is defined as $\sigma_{{n}{m}}=\sigma_{ij}\hat{n}^i\hat{m}^j$, where
\begin{equation}\label{shear}
	\sigma_{ij}
	=\frac{1}{2}\frac{\ud}{\ud\eta}\bigg(\frac{a_i^2}{a^2}\bigg)\delta_{ij}
	=
	\frac{a_i^2}{a^2}\bigg(\mathcal{H}_i-\mathcal{H}\bigg)\delta_{ij},
\end{equation}
and $\mathcal{H}_i=a_i^{-1}\frac{\ud a_i}{\ud \eta}$ are the \textbf{directional conformal Hubble rates}. Using the new frame, the Hamiltonian constraint \eqref{H0} becomes
\begin{equation*}
	\mathcal{H}_0^{(0)}
	=a^{-3}
	\left[
	(\Tr P^2)
	-\frac{1}{2}(\Tr P)^2
	+\frac{1}{2}p_{\phi}^2
	+a^6V
	\right].
\end{equation*}
The\textbf{ isotropic limit} is obtained by taking $P_{kk}=P_{vv}=P_{ww}=\frac{\Tr P}{3}$,  and $P_{kv}=P_{kw}=P_{vw}=0$, from which we obtain $(\Tr P^2)=\frac{(\Tr P)^2}{3}$, in agreement with \cite{Malkiewicz_2019}.
\subsection{First-order constraints revisited}

We express first-order constraints in the new basis 
\eqref{nperts}. The \textbf{gravity scalar constraint} \eqref{scx} can thus we written as follows
\begin{align}\label{linS}
			\delta\mathcal{H}_{g,0}
			&=
			-\frac{1}{3}a^{-1}(\Tr P)\delta\pi^1
			+a^{-1}
			\left[
			3P_{kk}
			-(\Tr P)
			\right]\delta\pi^2
			+a^{-1}2\sqrt{2}P_{kv}\delta\pi^3
			\nonumber
						\\
			&\quad
			+a^{-1}2\sqrt{2}P_{kw}\delta\pi^4
			+a^{-1}2\sqrt{2}P_{vw}\delta\pi^5
			+a^{-1}\sqrt{2}(P_{vv}
			-P_{ww})\delta\pi^6
			\nonumber
						\\
			&\quad
			+\frac{1}{2}a^{-5}[(\Tr P^2)
			-\frac{1}{2}(\Tr P)^2]\delta q_1
			\nonumber
			\\
			&\quad
			+\frac{1}{3}a^{-5}
			\big\{
			-2(\Tr P^2)
			+6(P_{kk}^2+P_{kv}^2+P_{kw}^2)
			-(\Tr P)
			\left[
			3P_{kk}-(\Tr P)
			\right]
			\big\}\delta q_2
						\nonumber
			\\
			&\quad
			+\sqrt{2}a^{-5}
			\left[
			2P_{kw} P_{vw}
			+P_{kv} 
			\left(P_{vv}-P_{ww}\right)
			+P_{kk} P_{kv}
			\right]\delta q_3
			\nonumber
			\\
			&\quad
			+\sqrt{2}a^{-5}
			\left[
			2P_{kv}P_{vw}
			-P_{kw}
			\left(P_{vv}-P_{ww}\right)
			+P_{kk}P_{kw}
			\right]\delta q_4
			\nonumber
			\\
			&\quad
			+\sqrt{2}a^{-5}
			\left[
			2P_{kv}P_{kw}
			-2P_{kk}P_{vw}
			+(\Tr P)P_{vw}
			\right]\delta q_5
			\nonumber
			\\
			&\quad
			+\frac{1}{\sqrt{2}}a^{-5}
			\left[
			2(P_{kv}^2
			-P_{kw}^2)
			-2P_{kk}
			\left(P_{vv}-P_{ww}\right)
			+(\Tr P)(P_{vv}-P_{ww})
			\right]\delta q_6
			\nonumber
			\\
			&\quad
			-2a^{-1}k^2
			\left(\delta q_1
			-\frac{1}{3}\delta q_2\right),
\end{align}
and the \textbf{gravity vector constraint} \eqref{vcx}:
		\begin{align}\label{linD}
				\delta\mathcal{H}^{~i}_g=&
				-2ik\hat{k}^i
				\bigg[\frac{1}{3}\delta\pi^1
				+\delta\pi^2+a^{-4}
				\left(
				P_{kk}
				-\frac{1}{2}(\Tr P)
				\right)\delta q_1
								\nonumber
				\\
				&\qquad
				+\frac{a^{-4}}{6}
				\left(
				P_{kk}
				+(\Tr P)
				\right)\delta q_2
				-\frac{a^{-4}}{\sqrt{2}}P_{vw}\delta q_5
				-\frac{a^{-4}}{\sqrt{2}}\frac{P_{vv}-P_{ww}}{2}\delta q_6
				\bigg]
				\nonumber
				\\
				&
				-2ik\hat{v}^i
				\bigg[\frac{1}{\sqrt{2}}\delta\pi^3
				+a^{-4}P_{kv}
				\left(
				\delta q_1
				-\frac{1}{3}\delta q_2
				\right)
				+\frac{a^{-4}}{\sqrt{2}}P_{kk}\delta q_3
				\nonumber
				\\
				&\qquad
				+\frac{a^{-4}}{\sqrt{2}}P_{kw}\delta q_5
				+\frac{a^{-4}}{\sqrt{2}}P_{kv}\delta q_6
				\bigg]
				\nonumber
				\\
				&
				-2ik\hat{w}^i
				\bigg[\frac{1}{\sqrt{2}}\delta\pi^4
				+a^{-4}P_{kw}
				\left(
				\delta q_1
				-\frac{1}{3}\delta q_2
				\right)
				+\frac{a^{-4}}{\sqrt{2}}P_{kk}\delta q_4
				\nonumber
				\\
				&\qquad
				+\frac{a^{-4}}{\sqrt{2}}P_{kv}\delta q_5
				-\frac{a^{-4}}{\sqrt{2}}P_{kw}\delta q_6
				\bigg].
\end{align}
The matter scalar and vector constraints read
\begin{equation}\label{eqmatter}
\begin{gathered}
	\delta\mathcal{H}_{m,0} = a^{-3}p_{\phi}\delta\pi_{\phi}-\frac{3}{4}a^{-5}p_{\phi}^2\delta q_{1}+\frac{3}{2}aV\delta q_{1} +a^{3}V_{,\phi}\delta\phi, \\
	\delta\mathcal{H}_{m}^{i} = ia^{-2}k^ip_{\phi}\delta\phi.
\end{gathered}
\end{equation}
We make use of the identity $\hat{k}^i \hat{k}_j+\hat{v}^i \hat{v}_j+\hat{w}^i \hat{w}_j=\delta^{i}_{~j}$ to introduce 
\begin{equation}\label{1-class-const}
	\begin{aligned}
	\delta \mathcal{H}_k&=\delta\mathcal{H}^{~i}_g\hat{k}_i+\delta\mathcal{H}_{m}^{i}\hat{k}_i,
	\\
	\delta \mathcal{H}_v&=\delta\mathcal{H}^{~i}_g\hat{v}_i+\delta\mathcal{H}_{m}^{i}\hat{v}_i,
	\\
	\delta \mathcal{H}_w&=\delta\mathcal{H}^{~i}_g\hat{w}_i+\delta\mathcal{H}_{m}^{i}\hat{w}_i,
	\end{aligned}
\end{equation}  
as well as
\begin{equation*}
	\begin{aligned}
\delta N^{k}&=\delta N_i\hat{k}^i,
\\
\delta N^{v}&=\delta N_i\hat{v}^i,
\\
\delta N^{w}&=\delta N_i\hat{w}^i.
	\end{aligned}
\end{equation*} 

\subsection{The Fermi-Walker basis}\label{Fermi}

The frame $(\hat{k},\hat{v},\hat{w})$ can be rotated around the $\hat{k}$-axis, making its definition ambiguous. Moreover, the pair $(\hat{v},\hat{w})$ determines the definitions of the polarization modes for vector and tensor perturbations. In an \textbf{isotropic universe}, such as Minkowski spacetime, the definition of the two polarization modes is naturally guided by the requirement that the modes must dynamically decouple from each other. However, decoupling is generally not possible in an \textbf{anisotropic universe}. Nonetheless, we can still aim to find the simplest possible form of dynamical law.

The Fourier transform \eqref{Fourier} fixes a foliation of the spatial coordinate space $(x^1,x^2,x^3)$ with the wavefronts of plane waves. In the physical space, due to the anisotropic dynamics, the wavefronts are not fixed but are being continuously tilted and anisotropically contracted or expanded. 
To address this issue, the natural choice is to assume that the vectors $(\hat{v},\hat{w})$ are\textbf{ Fermi-Walker-transported} along the future-oriented null vector field $\vec{p}$. The spatial component of $\vec{p}$  is dual to the wavefront $\underline{k}$ of the gravitational wave, that is,
\begin{equation*}
	\vec{p}=\overline{k}+|\overline{k}|\partial_{\eta}~,
\end{equation*}
where $\nabla_{\vec{p}}\vec{p}=0$ and we work with $\ud s^2=-\ud\eta^2+\gamma_{ij}\ud x^i\ud x^j$. Note that $\vec{p}$ may be identified with a tangent to a bundle of null geodesics that in the \textbf{eikonal approximation}, i.e., for large wavenumbers, are associated with rays of gravitational waves. This construction is the so-called \textbf{Sachs basis} \cite{sachsbasis} that, in the context of the propagation of light in the Bianchi I universe, was also considered in \cite{Fleury:2014rea}.
A \textbf{Fermi-Walker-propagated vector field }$\vec{E}$ is defined by $\nabla_{\vec{p}}\vec{E}=0$,
\begin{equation*}
	\frac{\ud E^0}{\ud\lambda}
	=
	-k^i\sigma_{ij}E^j,~~\frac{\ud E^j}{\ud\lambda}=-|\overline{k}|\sigma^j_{~i}E^i,
\end{equation*}
where $\lambda$ is an affine parameter. Even if $\vec{E}$ initially lies within the plane $(\hat{v},\hat{w})$, it will eventually develop longitudinal and temporal components.
 Therefore, we project the covariant derivative onto the plane. We assume $^{\bot}\nabla_{\vec{p}}\vec{E}=0$, where $^{\bot}$ denotes the orthogonal projection onto the plane $(\hat{v},\hat{w})$,
\begin{equation*}
	\frac{\ud E^j}{\ud\eta}=
	-\sigma^j_{~i}E^i
	+\hat{k}^j\sigma_{k i}E^i,
\end{equation*}
where the right-hand side\footnote{We identify $\sigma_{ki}=\hat{k}^j\sigma_{ji}$. Analogously defined are the quantities $\sigma_{kk}$, $\sigma_{vw}$, etc.} is projected onto the plane and  the affine parameter is replaced with conformal time via the relation $|\overline{k}|\ud\lambda=\ud \eta$. Let us assume that $(\hat{v},\hat{w})$ form a pair of Fermi-Walker-propagated vectors. Then making use of the fact that $(\hat{k},\hat{v},\hat{w})$ is an orthogonal spatial basis we obtain,
\begin{equation}\label{fermilaw}
	\frac{\ud \hat{v}^j}{\ud\eta}=-\sigma_{{v}{v}}\hat{v}^j-\sigma_{{v}{w}}\hat{w}^j,~~\frac{\ud \hat{w}^j}{\ud\eta}=-\sigma_{{w}{w}}\hat{w}^j-\sigma_{{w}{v}}\hat{v}^j.
\end{equation}
The above equations can be rewritten using Eq. \eqref{Pshear} and using the relation between cosmic and conformal time $\ud\eta=a \ud t$, which leads to:
\begin{equation}\label{fermilaw2}
	\begin{aligned}
		\frac{\ud \hat{v}^j}{\ud t}
		&=
		-a^{-3}
		\left(
		P_{{v}{v}}
		-\frac{1}{3}\Tr P
		\right)
		\hat{v}^j
		-a^{-3}P_{{v}{w}}\hat{w}^j,
		\\
		\frac{\ud \hat{w}^j}{\ud t}
		&=
		-a^{-3}
		\left(
		P_{{w}{w}}
		-\frac{1}{3}\Tr P
		\right)\hat{w}^j
		-a^{-3}P_{{w}{v}}\hat{v}^j.
\end{aligned}
\end{equation}
The derivatives of the dual vectors $\hat{v}_j$ and $\hat{w}_j$ are easily determined from the time derivatives of the metric components $\gamma_{ij}$ used to lower and raise the indices.


Equations \eqref{fermilaw2} are necessary to derive the extra 
Hamiltonian resulting from the time-dependent canonical transformations 
\eqref{nperts}, used to express the Hamiltonian 
from coordinate-based to triad-based perturbation variables.
 The \textbf{symplectic form} in the new variables reads:
\begin{equation}
	\ud  \check{q}_{ij}\wedge\ud \check{\pi}^{ij}
	=
	\ud  \delta {q}_{n}\wedge\ud \delta {\pi}^{n}
	+
	\ud t \wedge\ud 
	\left(
	\frac{\ud A^n_{ij}}{\ud t} A_m^{ij}
	\delta {q}_{n}\delta {\pi}^{m}
	\right).
\end{equation}
leading to the \textbf{extra Hamiltonian} $H_{\text{ext}}=-\frac{\ud A^n_{ij}}{\ud t} A_m^{ij}\delta {q}_{n}\delta {\pi}^{m}$. The full expression for $H_{\text{ext}}$ and its reduced form in the flat slicing gauge\footnote{This is the gauge that we will use in the next section, i.e., $\delta q_1=\delta q_2=\delta q_3=\delta q_4=0$ (see Sec. \ref{kucharpert}).} are given in Appendix \ref{NRHam0}.
The full second-order Hamiltonian \eqref{2cx} written in terms of $(\delta {q}_{n},\delta {\pi}^{n})$, $n=1,\dots,6$, is given in Appendix \ref{NRHam}. In the following computation we will not need the full expression, as we intend to impose gauge-fixing conditions that considerably simplify Eq. \eqref{2cx}.

\section{The physical Hamiltonian}\label{physical}

In deriving the physical Hamiltonian we shall follow the Dirac procedure for constrained systems in perturbation theory, presented in Sec. \ref{sec-Dirac-metodo}. Let us briefly recall the main steps in this procedure:
\begin{itemize}
\item[$\hookrightarrow$] We set \textbf{4 gauge-fixing conditions}. In this case we have $\delta c_{1}=0$, $\delta c_{2}=0$, $\delta c_{3}=0$, $\delta c_{4}=0$. Together with the initial \textbf{4 first-class constraints}  $\delta\mathcal{H}_{0}=0$, $\delta\mathcal{H}_{k}=0$, $\delta\mathcal{H}_{v}=0$, $\delta\mathcal{H}_{w}=0$, these form a set of \textbf{8 second-class constraints}. We denote them collectively by $\delta C_{\rho}=0$, where
$
	\delta C_{\rho}
	=
	\left\{
	\delta c_{1}
	, \delta c_{2}
	, \delta c_{3}
	, \delta c_{4}
	, \delta\mathcal{H}_{0}
	, \delta\mathcal{H}_{k}
	, \delta\mathcal{H}_{v}
	, \delta\mathcal{H}_{w}
	\right\}.
$

\item[$\hookrightarrow$]  Provided that Eq. \eqref{det}, here written as 
\begin{equation}\label{gaugecondition}
\text{Det} \{\delta C_{\rho},\delta C_{\sigma}\}\neq 0,
\end{equation}
 is satisfied, we introduce the Dirac bracket \eqref{DB}:
\begin{equation}\label{dirbrack}
\{A, B \}_D = \{A, B \} - \{A, \delta C_\rho\}\{\delta C_\rho, \delta C_\sigma\}^{-1}\{\delta C_\sigma, B\}.
\end{equation}

\item[$\hookrightarrow$]  We obtain the physical Hamiltonian by \textbf{strongly} imposing the second-class constraints on the second-order Hamiltonian, which is done by removing the redundant
dynamical variables. This yields the physical Hamiltonian, 
\begin{equation}\label{physred}
	\begin{aligned}
	H_{phys}
	&=
	\left(
	N\mathcal{H}_{0}^{(2)}
	+\delta N\delta\mathcal{H}_{0}+\delta N^k\delta\mathcal{H}_{k}+\delta N^v\delta\mathcal{H}_{v}+\delta N^w\delta\mathcal{H}_{w}
	\right)\bigg|_{\delta C_{\rho}= 0}
	\\
	&= N\mathcal{H}_{0}^{(2)}\bigg|_{\delta C_{\rho}= 0}.
	\end{aligned}
\end{equation}
\end{itemize}
The\textbf{ Hamilton equations} in the gauge-fixing surface for any basic observable $\mathcal{O}$ are generated by the physical Hamiltonian via the Dirac bracket,
\begin{align}
	\dot{\mathcal{O}}=\left\{\mathcal{O},N\mathcal{H}_{0}^{(2)}\big|_{\delta C_{\rho}= 0}\right\}_D.
\end{align}
The basic property of the above dynamics is that $\delta \dot{C}_{\rho}=0$ for all $\rho$.

\subsection{Gauge-fixing conditions}\label{kucharpert}
In what follows we impose a set of gauge-fixing conditions called the \textbf{spatially flat slicing} gauge \cite{Malik:2008im},
\begin{align}\label{gauge}
	\delta c_1:=\delta q_1,~~\delta c_2:=\delta q_2,~~\delta c_3:=\delta q_3,~~\delta c_4:=\delta q_4.
\end{align}
Considering the perturbation of the three-curvature $\delta({}^3R)$, as given in Eq. \eqref{deltaRapp} in Appendix \ref{geoquant}, it becomes evident that in the above gauge, $\delta({}^3R)=0$ on constant-time slices. As discussed in Sec. \ref{geomconst}, once the gauge conditions are fixed, we have the flexibility to decide which degrees of freedom to remove. In this case, the best choice is to solve the constraints for the conjugate momenta to the variables in Eq. \eqref{gauge}. This is accomplished using the first-class constraints \eqref{1-class-const}, from which we obtain:
	\begin{align}\label{firtclassconst}
			\delta\pi^1&=
			\frac{2\sqrt{2}P_{vw}}{P_{kk}}\delta\pi^5
			+\frac{\sqrt{2}(P_{vv}-P_{ww})}{P_{kk}}\delta\pi^6
							\nonumber
			\\
			&
			\quad
			+\frac{\sqrt{2}a^{-4}}{P_{kk}}
			\bigg(
			\frac{1}{2}P_{vw}[(\Tr P)-P_{kk}]
			-2P_{kv} P_{kw}
			\bigg)\delta q_5
				\nonumber
			\\
			&
			\quad
			+\frac{a^{-4}}{\sqrt{2}P_{kk}}\bigg[
			\bigg(\frac{\Tr P}{2}-\frac{7}{2}P_{kk}\bigg)(P_{vv}-P_{ww})-2(P_{kv}^2-P_{kw}^2)
			\bigg]\delta q_6
			\nonumber
			\\
			&
			\quad
			+\frac{a}{P_{kk}}\delta\mathcal{H}_{m,0}
			+\sum_i \frac{1}{2ik_i P_{kk}}[3P_{kk}-(\Tr P)]\delta\mathcal{H}_{m,i}
			~~,
				\nonumber
			\\
			\delta\pi^2&=
			-\frac{2\sqrt{2}P_{vw}}{3P_{kk}}\delta\pi^5
			-\frac{\sqrt{2}}{3P_{kk}}(P_{vv}-P_{ww})\delta\pi^6
					\nonumber
			\\
			&
			\quad
			-\frac{\sqrt{2}a^{-4}}{3P_{kk}}
			\bigg[
			\left(
			\frac{\Tr P}{2}
			-2P_{kk}\right)P_{vw}
			-2P_{kv} P_{kw}
			\bigg]\delta q_5
				\nonumber
			\\
			&
			\quad
			-\frac{a^{-4}}{3\sqrt{2}P_{kk}}
			\bigg[
			\bigg(\frac{\Tr P}{2}-2P_{kk}\bigg)(P_{vv}-P_{ww})
			-2(P_{kv}^2
			-P_{kw}^2)
			\bigg]\delta q_6
					\nonumber
			\\
			&
			\quad
			-\frac{a}{3P_{kk}}\delta\mathcal{H}_{m,0}
			+\sum_i \frac{(\Tr P)}{6ik_i P_{kk}}\delta\mathcal{H}_{m,i}
			~~,
				\nonumber
			\\
			&
			\delta\pi^3=-a^{-4}P_{kw}\delta q_5
			-a^{-4}P_{kv}\delta q_6
			~~,
				\nonumber
			\\
			&
			\delta\pi^4=-a^{-4}P_{kv}\delta q_5+a^{-4}P_{kw}\delta q_6
			~~,
\end{align}
where $P_{kk}\neq 0$. Therefore, imposing the \textbf{8 second-class constraints}, $\delta C_{\rho}=0$, naturally leads to the removal of\textbf{ 4 canonical pairs} $(\delta q_1,\delta\pi^1,\delta q_2,\delta\pi^2,\delta q_3,\delta\pi^3,\delta q_4,\delta\pi^4)$ from the phase space. The remaining variables $(\delta q_5,\delta\pi^5,\delta q_6,\delta\pi^6,\delta {\phi},\delta {\pi}^{\phi})$ are considered \textbf{physical}, see Sec. \ref{sec-Dirac-metodo}. It is easy to see that they must form 3 canonical pairs with respect to the Dirac bracket  (\ref{DB}).

As mentioned in Ch. \ref{Chapter2}, the above procedure shares similarities with the \textbf{Kucha\v r decomposition} discussed in Sec. \ref{kuchar}. In the Kucha\v r decomposition, one selects \textbf{four scalar fields} from the kinematical variables to linearize the canonical conjugate momenta of the four constraints, effectively serving as \textbf{internal spacetime coordinates} with respect to which the unconstrained, "physical" degrees of freedom and their dynamics are expressed. The Brown-Kucha\v r dust model serves as a well-known example \cite{Brown:1994py}. However, this decomposition becomes impossible when the spacetime model possesses spatial symmetries or approximate spatial symmetries, as discussed in \cite{Torre:1992rg}.

In cosmology, where no internal spatial coordinates exist, the role of gauge-fixing conditions differs slightly. Rather than setting internal coordinates, the first-order gauge-fixing conditions determine the embedding of the fixed background spacetime in the perturbed spacetime. This is known as "\textbf{gauge-fixing of the second kind}" \cite{DeWitt,Nakamura}. These conditions implicitly determine the embedding by imposing constraints on the difference between the perturbed and background spacetime, using the three-metric and three-momentum components.
 In this way the space and time coordinates for the perturbed spacetime are unambiguously provided by the space and time coordinates of the background spacetime.  The time coordinate of the background model is an internal variable made of the kinematical background variables such as the scalar field or the scale factor, whereas the spatial coordinates of the background model are external. The latter come from a natural parametrization of the homogeneous three-space. For instance, in a toroidal Bianchi I universe, the spatial coordinates are determined only up to a constant spatial shift, that is, modulo the action of the given homogeneity group, if the condition $\int_{\mathbb{S}_i}\ud x^i=1$ holds for all $i$. 

A more general notion than the gauge-fixing of second kind is that of covariant gauge-fixing \cite{Hajicek:1999ht}, which is a coordinate-independent notion. It involves identifying spacetime points belonging to different spacetime solutions and assumes an underlying background manifold with no preferred coordinates. The covariant gauge-fixing is invariant with respect to the group of diffeomorphisms of the background manifold. The arbitrary choice of particular coordinates on the background manifold is called the gauge-fixing of first kind. Both the point-by-point identification and the background coordinates are provided by the Kucha\v r decomposition. In the case of perturbation theory, two extra restrictions hold. First, both space and time coordinates are already fixed (modulo the spatial shifts) on the background manifold. Second, the identification of the spacetime points belonging to different spacetime solutions has to respect the condition of the smallness of the three-metric and three-momentum perturbations.

\subsection{Physical Hamiltonian}
The \textbf{physical Hamiltonian} \eqref{physred} is obtained from the \textbf{second-order Hamiltonian} \eqref{2cx} by:
\begin{enumerate}[i)]
	\item  Expressing the perturbation variables in the Fermi-Walker-propagated basis;
	\item Adding the extra Hamiltonian \eqref{ext} yielded by the time-dependent basis transformation;
	\item Reducing it into the flat slicing gauge \eqref{gauge}, and introducing the Dirac bracket.
\end{enumerate}	
The explicit formula for the physical Hamiltonian is, thus, found to read:
\begin{equation}
	\begin{aligned}\label{hfin}
		H_{phys}&=
		\frac{\delta \pi_\phi ^2}{2 a^3}
		+
		a \delta \pi_5^2
		+a
		\delta\pi_6^2
		\\&
		+\left(
		\frac{k^2	a}{2}+\tilde{U}_\phi
		\right) 
		\delta \phi ^2
		+\left(
		\frac{k^2}{4 a^3}+\tilde{U}_5
		\right)
		\delta q_5^2
		+\left(
		\frac{k^2}{4 a^3}+\tilde{U}_6
		\right) 
		\delta q_6^2
		\\&
		+
		C_{\phi\phi}
		\delta\phi\delta\pi_\phi
		+C_{55}\delta q_5
		\delta \pi_5
		+C_{66}\delta q_6 \delta \pi_6
		\\&
		+C_{5\phi}
		\left(
		\sqrt{2}a^2\delta \pi_5 \delta \phi
		+\frac{1}{\sqrt{2} a^2}
		\delta q_5\delta \pi_\phi
		\right)
		\\&
		+C_{6\phi}
		\left(
		\sqrt{2}a^2\delta\pi_6\delta \phi 
		+\frac{1}{\sqrt{2} a^2}\delta q_6 \delta \pi_{\phi} 
		\right)
		\\&
		+C_{56}
		\left(
		\delta q_5\delta \pi_6+\delta q_6\delta \pi_5
		\right)
		+ 
		\tilde{C}_1 \delta q_5
		\delta q_6
		+\tilde{C}_2 \delta \phi\delta q_5
		+\tilde{C}_3 \delta \phi\delta q_6,
	\end{aligned}
\end{equation}
where the zero-order coefficients are given in Appendix \ref{Hfin}. The formula for the full, unconstrained second-order Hamiltonian in the Fermi-Walker basis is provided in Appendix \ref{NRHam}.

\subsection{Mukhanov-Sasaki variables}\label{mukha}

By Mukhanov-Sasaki variables, \cite{SasakiMS, MukhanovMS} we mean \textbf{gauge-invariant} perturbation variables that, in the limit of a flat spacetime, satisfy the equation of motion for the \textbf{harmonic oscillator}. The physical Hamiltonian  \eqref{hfin}
can be written as a quasi-harmonic oscillator by considering an adequate change of coordinates. In particular, we \textbf{rescale} the kinetic terms and perform a\textbf{ linear canonical transformation} on the momenta to decouple them from the positions.
 From Eq. \eqref{hfin} it is easy to see that the \textbf{rescaling} should read:
\begin{equation}\label{res}
	\begin{aligned}
		\delta q_{5,6}\rightarrow \delta \tilde{q}_{5,6}=\frac{1}{\sqrt{2}a}\delta q_{5,6}&,~~~\delta \pi_{5,6}\rightarrow \delta \tilde{\pi}_{5,6}=\sqrt{2}a\delta \pi_{5,6},\\
		~\delta\phi\rightarrow\delta\tilde{\phi}=a\delta\phi&,
		~~~\delta\pi_{\phi}\rightarrow\delta\tilde{\pi}_{\phi}=a^{-1}\delta\pi_{\phi}.
	\end{aligned}
\end{equation}
As a time-dependent transformation, it produces an \textbf{extra term} in the Hamiltonian: \begin{equation}
H_{\text{ext}}=
\frac{\Tr P}{6a^3}
\left(
 \delta \tilde{q}_{5}\delta \tilde{\pi}_{5}
+\delta \tilde{q}_{6}\delta \tilde{\pi}_{6}
-\delta\tilde{\phi}\delta\tilde{\pi}
\right).
\end{equation}
The \textbf{canonical transformation} that removes the momentum-position couplings follows from the general prescription. Given the Hamiltonian,
\begin{align}
	H=\frac{1}{a}\bigg(\sum_i \frac{1}{2}p_i^2+\sum_{i,j}C_{ij}p_iq_j+\dots\bigg),~~~C_{ij}=C_{ji},
\end{align}
the transformation
\begin{equation}\label{linear-tra}
	p_i\rightarrow\tilde{p}_i=p_i+C_{ij}q_j ~,
\end{equation}
leads to the new form
\begin{equation}
	\label{Hshift}
	H=\frac{1}{a}\bigg(\sum_i \frac{1}{2}\tilde{p}_i^2-\frac{a}{2}\sum_{i,j}\dot{C}_{ij}q_iq_j+\dots\bigg),
\end{equation}
where $~\dot{}~$ denotes the derivative with respect to cosmological time. The second term in \eqref{Hshift} arises from the time dependence of the transformation \eqref{linear-tra}. We apply this transformation to the Hamiltonian (\ref{hfin}) for which the coefficients $C_{ij}$ are given in Eq. (\ref{HfinCoeff}). The complete form of the Hamiltonian \eqref{Hshift} is obtained using the background variables (\ref{backeom-pa}), which are also needed for the components of $P$ in the new basis, which are given in Appendix \ref{Appbackeom}.

We arrive at the\textbf{ final Hamiltonian} expressed in terms of the \textbf{anisotropic Mukhanov-Sasaki variables}:
\begin{equation}\label{HBI}
	\begin{aligned}
		H_{BI}=&
		\frac{N}{2a}
		\bigg[
		\delta\tilde{\pi}_{\phi}^2
		+\delta\tilde{\pi}_5^2
		+\delta\tilde{\pi}_6^2
		+\left(k^2+U_{\phi}\right)\delta\tilde{\phi}^2
		+\left(k^2+U_5\right)\delta \tilde{q}_5^2
		+\left(k^2+U_6\right)\delta\tilde{q}_6^2
		\\
		&
		+C_{1}\delta \tilde{q}_5\delta \tilde{q}_6
		+C_{2}\delta \tilde{q}_5\delta\tilde{\phi}
		+C_{3}\delta\tilde{q}_6\delta\tilde{\phi}
		\bigg],
	\end{aligned}
\end{equation}
where the coefficients are given in Appendix \ref{MSAppgen}.  Notice that in Eq. \eqref{HBI}, the tensor modes are coupled to each other with the coupling $C_1$. These modes decouple when the shear of the wavefront vanishes. Analogously, the scalar field is coupled to the tensor modes with the couplings $C_2$ and $C_3$. In this case, decoupling can happen with the vanishing of the shear of the wavefront and of the planes perpendicular to it, given by the normal $\hat{k}\times\hat{v}$ or $\hat{k}\times\hat{w}$. The isotropic limit can be obtained as described at the end of Sec. \ref{0ord}. As expected, in this limit $C_1=C_2=C_3=0$, and the two polarization modes of the gravitational wave and the scalar mode all decouple from each other, giving the results obtained in \cite{Malkiewicz_2019}. 

The coefficients in \eqref{HBI} are combinations of the background phase space variables and the vectors $\hat{v}$ and $\hat{w}$. To make the Hamiltonian suitable for quantization, the vectors should be expressed in terms of the background phase space variables. This can be achieved by solving the dynamical equations \eqref{fermilaw} or, equivalently, Eqs \eqref{fermilaw2}, which can be very difficult. Alternatively, it is also possible to fix $\hat{v}$ and $\hat{w}$ explicitly. However, in general, this procedure yields vectors that are not Fermi-Walker-propagated and therefore results in a correction to the Hamiltonian \eqref{ext}. For instance, we can define:
\begin{equation}
	\hat{v}^i=\frac{v^i}{\sqrt{\gamma_{lj}v^lv^j}},~~\hat{w}^i=\epsilon^i_{~jk}\hat{v}^j\hat{k}^k,
\end{equation}
where $v^i$ is a constant such that $v^ik_i=0$, and $\epsilon_{ijk}$ is a totally antisymmetric tensor. We find the extra Hamiltonian to read:
\begin{equation}
	H_{\text{ext}}'=
	2a^{-2}P_{vw}
	\left(
	\delta \tilde{q}_5\delta\tilde{\pi}^6
	-\delta \tilde{q}_6\delta\tilde{\pi}^5
	\right).
\end{equation}
Now the vectors $\hat{v}$ and $\hat{w}$ are given explicitly in terms of the phase space variables, and the Hamiltonian \eqref{HBI} supplemented with the above term can now be quantized.

\subsection{Dirac observables}\label{diracobs}

The Dirac observables first defined in Sec. \ref{secdirac} and Sec. \ref{gid}, are defined to be \textbf{first-order kinematical phase space observables},  which weakly commute with the first-class constraints,
\begin{equation}\label{defdir}
	\forall_{\delta{\xi}^{\rho}}
	~\left\{
	\delta D_i~,
	\int\delta{\xi}^{\rho}
	\delta\mathcal{H}_{\rho}
	\right\}\approx 0.
\end{equation} 
We denote the complete set of solutions to Eq. \eqref{defdir} in the kinematical phase space by ${\delta D_i}$, where $i=1,\dots,10$. The explicit formulae are given in Appendix \ref{Dext}. There are \textbf{10 independent solutions} in the kinematical phase space. Hence there must be six independent Dirac observables in the constraint surface. To choose them conveniently it is useful to impose the\textbf{ flat slicing gauge-fixing} conditions. that is $\delta q_1=\delta q_2=\delta q_3=\delta q_4=0$, which largely simplifies the expressions for ${\delta D_i}$:
\begin{equation}
	\begin{aligned}
		&{\delta D_1}
		\big|_{_{GF}}
		=
		{\delta q _5},
		\quad
		{\delta D_2}
		\big|_{_{GF}}
		=	
		{\delta q _6},
		\quad
		{\delta D_3}
		\big|_{_{GF}}
		=
		{	\delta \pi_1},
		\quad
		{\delta D_4}
		\big|_{_{GF}}
		=
		{	\delta \pi_2},
		\quad
		\\&
		{\delta D_5}
		\big|_{_{GF}}
		=	
		{\delta \pi_3},	
		\quad
		{\delta D_6}
		\big|_{_{GF}}
		=	
		{\delta \pi_4},
		\quad
		{\delta D_7}
		\big|_{_{GF}}
		=
		{\delta \pi_5},
		\quad
		{\delta D_8}
		\big|_{_{_{GF}}}
		=
		{\delta \pi_6},
		\quad
				\\&
		{\delta D_9}
		\big|_{_{GF}}
		=
		{\delta \phi},
		\quad
		{\delta D_{10}}
		\big|_{_{GF}}
		=	
		{\delta \pi_\phi},
	\end{aligned}
\end{equation}
where the label $|_{_{GF}}$ means ``in the gauge-fixing surface". It follows from Eq. (\ref{firtclassconst}) that the observables ${\delta D_{3}}|_{_{GF}}$, ${\delta D_{4}}|_{_{GF}}$, ${\delta D_{5}}|_{_{GF}}$, ${\delta D_{6}}|_{_{GF}}$ are not independent and may be expressed in terms of the remaining solutions. Hence we discard them. On the other hand, ${\delta D_{1}}|_{_{GF}}$, ${\delta D_{2}}|_{_{GF}}$, ${\delta D_{7}}|_{_{GF}}$, ${\delta D_{8}}|_{_{GF}}$, ${\delta D_{9}}|_{_{GF}}$ and ${\delta D_{10}}|_{_{GF}}$ can be easily related to the tilded variables used in the final Hamiltonian (\ref{HBI}). Since any combination of Dirac observables is a Dirac observable, we introduce a \textbf{new basis} in the space of Dirac observables,
\begin{equation}
	\begin{aligned}
		\delta{Q}_1&=
		\frac{1}{\sqrt{2}a}{\delta D_{1}},
		~~
		\delta{Q}_2=
		\frac{1}{\sqrt{2}a}{\delta D_{2}},
		~~
		\delta{Q}_{3}=
		a{\delta D_{9}},
		\\
		\delta{P}_1&=
		\sqrt{2}a{\delta D_{7}}
		+\frac{C_{55}+\frac{\Tr P}{6a^3}}{\sqrt{2}}{\delta D_{1}}
		+\frac{C_{56}}{\sqrt{2}}{\delta D_{2}}
		+C_{5\phi}a^2 {\delta D_{9}},
		\\
		\delta{P}_2&=
		\sqrt{2}a{\delta D_{8}}
		+\frac{C_{56}}{\sqrt{2}}{\delta D_{1}}
		+\frac{C_{66}+\frac{\Tr P}{6a^3}}{\sqrt{2}}{\delta D_{2}}
		+C_{6\phi}a^2{\delta D_{9}},
		\\
		\delta{P}_3&=
		a^{-1}{\delta D_{10}}
		+\frac{C_{\phi5}}{\sqrt{2}}{\delta D_{1}}
		+\frac{C_{\phi6}}{\sqrt{2}}{\delta D_{2}}
		+(C_{\phi\phi}-\frac{\Tr P}{6a^3})a^2{\delta D_{9}},
	\end{aligned}
\end{equation}
where we used the coefficients of the Hamiltonian (\ref{hfin}). It can be verified that in the spatially flat slicing gauge the following identifications hold:
\begin{align*}
	\delta{Q}_1\big|_{_{GF}}=\delta\tilde{q}_5,\quad \delta{Q}_2\big|_{_{GF}}=\delta\tilde{q}_6,\quad \delta{Q}_3\big|_{_{GF}}=\delta\tilde{\phi},
	\\
	 \delta{P}_1\big|_{_{GF}}=\delta\tilde{\pi}_5,\quad \delta{P}_2\big|_{_{GF}}=\delta\tilde{\pi}_6,  \quad\delta{P}_3\big|_{_{GF}}=\delta\tilde{\pi}_{\phi}.
\end{align*}
Making use of the respective Dirac bracket we find the canonical commutation relations,
\[
\{\delta{Q}_i(\underline{k}),\delta{P}_i(\underline{l})\}_D=\delta_{\underline{k},-\underline{l}}.
\] 
Finally, we \textbf{pull-back} the Hamiltonian \eqref{HBI} to the space of Dirac observables with the mapping \eqref{iso} and obtain
\begin{align}\label{BIn}
		H_{BI}=&
		\frac{N}{2a}
		\bigg[
		\delta{P}_{1}^2
		+\delta{P}_2^2
		+\delta{P}_3^2
		+\left(k^2+U_{\phi}\right)\delta{Q}_3^2
		+\left(k^2+U_5\right)\delta{Q}_1^2
		+\left(k^2+U_6\right)\delta{Q}_2^2
		\nonumber
		\\&
		+C_{1}\delta{Q}_1\delta{Q}_2
		+C_{2}\delta{Q}_1\delta{Q}_3
		+C_{3}\delta{Q}_2\delta{Q}_3
		\bigg]. 
\end{align}
The coefficients are given in Appendix \ref{MSAppgen}. 

The above Hamiltonian generates dynamical equations for all Fourier modes. Any solution is uniquely determined by specifying the \textbf{positions}, which describe the three-surface, and the \textbf{momenta}, which describe the extrinsic curvature, at the initial moment of time.
 Alternatively, any solution can be determined by specifying only the positions, i.e., the three-surfaces, at any two fixed moments of time and fixing the gauge at the boundary. This follows from the fact that we can derive the respective reduced Lagrangian (and the respective action) from the Hamiltonian \eqref{BIn}, allowing us to obtain any solution from the \textbf{principle of least action}.   This is in complete agreement with the full GR case \cite{Baierlein:1962zz}. The choice of the gauge at the intermediate three-surfaces is arbitrary and so is the coordinate system of the intermediate spacetime.

The basic Dirac observables expressed in terms of the kinematical phase space variables read
\begin{align}\label{inddir}
			\delta Q_1&=
			\boxed{\frac{1}{\sqrt{2} a}\delta q_5}
			+{\frac{2P_{vw}}{a P_{kk}}}
			{(\delta q_1-\frac{1}{3}\delta q_2)},
			\nonumber
			\\
			\delta Q_2&=
			\boxed{\frac{1}{\sqrt{2} a}\delta q_6}
			+{\frac{P_{vv}-P_{ww}}{a P_{kk}}}
			{(\delta q_1-\frac{1}{3}\delta q_2)},
			\nonumber
			\\
			\delta Q_3&=
			\boxed{a\delta\phi+\frac{p_{\phi}}{aP_{kk}}(\delta q_1-\frac{1}{3}\delta q_2)},
			\nonumber
			\\
			\delta P_1&=
			\boxed{\sqrt{2} a\delta\pi_5+\frac{\frac{5}{6} (\Tr P)- P_{kk}}{\sqrt{2} a^3}\delta q_5}
			\nonumber
			\\
			&\quad
			-{\frac{2P_{vw}}{\sqrt{2} a^3 P_{kk}}}
			{\left(\frac{P_{vv}-P_{ww}}{2}\delta q_6+P_{vw}\delta q_5\right)}
			\nonumber
			\\
			&\quad
			+{\mathcal{F}\left(P_{vw},P_{kv} P_{kw}\right)}
			\left(\delta q_1-\frac{1}{3}\delta q_2\right)
				\nonumber
			\\
			&\quad
			-\frac{P_{vw}}{a^3P_{kk}}
			\left(3P_{kk}\delta q_1 +a^2 p_{\phi}\delta\phi\right)
			+\frac{\sqrt{2}}{a^3}\left(P_{kw} \delta q_3+P_{kv}\delta q_4\right),
			\nonumber
			\\
			\delta P_2&=
			\boxed{\sqrt{2} a\delta\pi_6+\frac{\frac{5}{6} (\Tr P)- P_{kk}}{\sqrt{2} a^3}\delta q_6}
				\nonumber
			\\
			&\quad
			-\frac{P_{vv}-P_{ww}}{\sqrt{2} a^3 P_{kk}}\left(\frac{P_{vv}-P_{ww}}{2}\delta q_6+P_{vw}\delta q_5\right)
			\nonumber
			\\
			&
			\quad
			+{\mathcal{F}\left(\frac{P_{vv} - P_{ww}}{2},\frac{P_{kv}^2-P_{kw}^2}{2}\right)}
			\left(\delta q_1-\frac{1}{3}\delta q_2
			\right)
				\nonumber
			\\
			&\quad
			-{\frac{P_{vv}-P_{ww}}{2a^3P_{kk}}}\left(3P_{kk}\delta q_1 +a^2 p_{\phi}\delta\phi\right)
			+\frac{\sqrt{2}}{a^3}\left(P_{kv} \delta q_3-P_{kw}\delta q_4\right),
			\nonumber
			\\
			\delta P_3&=
			\boxed{\frac{1}{a}\delta\pi_{\phi}
				-\frac{(\Tr P) P_{kk}+3 p_{\phi}^2}{6 a P_{kk}}\delta\phi
				-\frac{3p_{\phi}}{2 a^3}\delta q_1
			}
				\nonumber
			\\
			&\quad
			\boxed{
				+\frac{2 (\Tr P) P_{kk} p_{\phi}-6 a^6 P_{kk} V_{,\phi}-3 p_{\phi}^3}{6 a^3 P_{kk}^2}
				\left(\delta q_1-\frac{1}{3}\delta q_2\right)}
				\nonumber
			\\
			&
			\quad
			-\frac{ p_{\phi}}{\sqrt{2} a^3 P_{kk}}\left(\frac{P_{vv}-P_{ww}}{2}\delta q_6+P_{vw}\delta q_5\right)
				\nonumber
			\\
			&\quad
			-\frac{p_{\phi} \left[
				\left(P_{vv}-P_{ww}\right)^2+4 P_{vw}^2\right]}{2 a^3 P_{kk}^2}(\delta q_1-\frac{1}{3}\delta q_2),
\end{align}
where 
\[
\mathcal{F}(X,Y)=
\frac{4}{a^3 P_{kk}}Y
-\frac{4 \left[
	P_{vw}^2+ \left(\frac{P_{vv} - P_{ww}}{2}\right)^2\right] + p_{\phi}^2-2P_{kk}\left(P_{kk}+\frac{\Tr P}{3}\right)}{a^3 P_{kk}^2}X
.
\]
 The boxed terms are also present in\textbf{ isotropic spacetimes}, whereas the remaining ones are peculiar to \textbf{anisotropic spacetimes}. In Appendix \ref{geodirac} we give the geometric meaning of the Dirac observables, which can be used when imposing gauge-fixing conditions on geometrical quantities. 

Let us make a few observations:
\begin{enumerate}[i)]
\item Given a rotation $R_{\hat{k}}({\theta})$ around the $\hat{k}=\hat{v}\times\hat{w}$ axis by the angle $\theta$, we have that the observables $\delta Q_1$, $\delta Q_2$, $\delta P_1$, $\delta P_2$ transform as
\begin{align*}
R_{\hat{k}}({\theta})\delta Q_1
&=\cos(2\theta)\delta Q_1-\sin(2\theta)\delta Q_2,
\\ 
R_{\hat{k}}({\theta})\delta P_1
&=\cos(2\theta)\delta P_1-\sin(2\theta)\delta P_2,
\\
R_{\hat{k}}({\theta})\delta Q_2
&=\cos(2\theta)\delta Q_2+\sin(2\theta)\delta Q_1,
\\
R_{\hat{k}}({\theta})\delta P_2
&=\cos(2\theta)\delta P_2+\sin(2\theta)\delta P_1.
\end{align*}
That is they transform as tensors and are thus considered to be \textbf{tensor modes}. In contrast, the observables $\delta Q_3$, $\delta P_3$ transform as scalars under rotation, therefore are considered to be \textbf{ scalar modes }.


\item  In the \textbf{isotropic limit} only the boxed terms are surviving. In this limit the number of  background variables is reduced: $P_{vv} - P_{ww}=0$, $P_{vw}=0$, $P_{kv}=0$, $P_{kw}=0$, $3P_{kk}=Tr P$.  The tensor perturbations of the metric and its momentum become gauge-invariant, unambiguously describing the gravitational waves. Similarly, the scalar Dirac observables, $\delta Q_3$ and $\delta P_3$, exclusively consist of scalar perturbations of the field, its momentum, and the metric.

\item  

In the \textbf{anisotropic case}, the tensor modes $\delta Q_1$, $\delta Q_2$, $\delta P_1$, and $\delta P_2$ consist of traceless-transverse perturbations of the metric and its momentum, along with vector and scalar perturbations of the metric. Consequently, the traceless-transverse perturbations of the metric and momentum alone lose their gauge-invariance. The tensorial nature of the "new" terms arises from the zeroth-order coefficients, which can themselves transform as scalars, vectors, or tensors. Specifically, vector perturbations are combined with the vector zeroth-order coefficients to yield tensors, while scalar perturbations are simply multiplied by tensor zeroth-order coefficients. Additionally, tensor perturbations may be multiplied by tensor zeroth-order coefficients to yield scalars, which are again multiplied by tensor zeroth-order coefficients. These various contributions to the tensor modes were not emphasized in \cite{Uzan}, where different definitions were used. In Appendix \ref{Uzantensor}, we compare the two formalisms.

\item  In the anisotropic case the\textbf{ scalar Dirac observable} $\delta P_3$ contains tensor metric perturbations that are contracted with tensorial zeroth-order coefficients to yield a scalar quantity. 
\end{enumerate}

To summarize, in anisotropic models with various matter contents, the Dirac observables are generally defined as a combination of scalar, vector, and tensor three-metric and three-momentum perturbations, as well as scalar, vector, and tensor zeroth-order coefficients. The introduction of the \textit{A-basis} \eqref{baseA} is convenient for both the perturbations and the background quantities.  In this basis, we can define a background contravariant tensor as $X_n=X^{ij}A_{ij}^n$ and the perturbation of a covariant tensor as $Y_n=\delta Y_{ij}A^{ij}_n$, where $n = 1, \dots, 6$. 
The \textbf{quadratic scalar} quantities arise as products of scalar quantities, such as $X_1Y_1$, $X_1Y_2$, and $X_2Y_2$, as well as norms of vector quantities like $X_3Y_3$ and $X_4Y_4$, or norms of tensor quantities, such as $X_5Y_5$ and $X_6Y_6$. \textbf{Quadratic vector }quantities emerge as products of scalar and vector quantities, for example, $X_3Y_5$, $X_1Y_4$, $X_2Y_3$, and $X_2Y_4$, or products of vector and tensor quantities like $X_3Y_5$, $X_3Y_6$, $X_4Y_5$, and $X_4Y_6$. \textbf{Quadratic tensor }quantities manifest as products of vector quantities, such as $X_3Y_4$ and $X_3Y_3-X_4Y_4$, or products of scalar and tensor quantities like $X_1Y_5$, $X_1Y_6$, $X_2Y_5$, and $X_2Y_6$. The ordering of these products and whether the factors are zeroth-order, first-order, or both do not matter. These rules extend straightforwardly to cubic and higher-order products.

\subsection{Canonical isomorphism between gauge-fixing surfaces}\label{canism}
 In the previous section, the Hamiltonian was computed using a specific gauge choice which allowed us to write it in terms of Dirac observables $(\delta{Q}_i,\delta{P}_i)$, thus facilitating the physical interpretation of the dynamical variables. In the following we present other choices of gauge-fixing conditions by making use of the canonical isomorphism \eqref{iso}. It is useful to notice that the condition \eqref{gaugecondition} for the validity of any gauge can be reduced to the following:
\begin{equation}\label{lambda}
	\textrm{Det}~ \Lambda\neq 0,~\textrm{where}~\Lambda_{\mu\nu}=\{\delta c_{\mu},\delta\mathcal{H}_{\nu}\}.
\end{equation}
In order to get a deeper understanding, we will use physical and geometrical quantities given in Appendix \ref{geoquant} to define particular gauges. 

\subsubsection{Uniform density gauge}

As a first example let us consider a gauge known in the isotropic limit as the \textbf{uniform density gauge}.  This gauge assumes the vanishing of the \textbf{metric density}, the \textbf{energy density} (defined in Appendix \ref{geoquant}), and the \textbf{vector metric perturbations}. The gauge conditions are as follows:
\begin{equation}
	\delta c_1:=\delta q_1,
	~~\delta c_2:=\delta \rho,
	~~\delta c_3:=\delta q_3,
	~~\delta c_4:=\delta q_4.
\end{equation}
The determinant of the Poisson brackets \eqref{lambda} is non-vanishing for non-zero scalar field momentum $p_{\phi}\neq 0$,
\begin{equation}
	\textrm{Det}~ \Lambda = -\frac{2 i k^2 p_{\phi}^2 (\Tr P)}{3 a^9},
\end{equation}
proving the validity of the gauge. The \textbf{physical variables}, corresponding to the Mukhanov-Sasaki variables and its conjugate, read:
\begin{equation}
	\begin{aligned}
		\delta Q_3\big|_{_{GF}}&
		=a\delta\phi
		-\frac{p_{\phi}}{3aP_{kk}}\delta q_2,
		\\
		\delta P_3\big|_{_{GF}}&=-
		\frac{P_{vw} p_{\phi}}{\sqrt{2} a^3 P_{kk}}\delta q_5
		-\frac{(P_{vv}-P_{ww}) p_{\phi}}{2\sqrt{2} a^3 P_{kk}}\delta q_6
		\\&
		\quad
		-\frac{
			(Tr P) p_{\phi}P_{kk}+3 p_{\phi}^3+6 a^6 P_{kk}V_{,\phi}
			}
			{6 a p_{\phi}P_{kk}}
			\delta\phi
		\\
		&
		+\frac{
			-2 (Tr P) P_{kk} p_{\phi}+6 a^6 P_{kk} V_{,\phi}+3 p_{\phi} \left[\left(P_{vv}-P_{ww}\right)^2+4 P_{vw}^2+p_{\phi}^2\right]
			}
			{18 a^3 P_{kk}^2}
			\delta q_2.
	\end{aligned}
\end{equation}
In the anisotropic universe they contain tensor modes whereas in the isotropic universe they are combinations of the scalar field and the transverse scalar metric perturbations.

\subsubsection{Longitudinal gauge}

In this example, we consider the \textbf{longitudinal gauge}, another well-known gauge in the isotropic case. This gauge assumes the absence of a scalar shear perturbation (defined in Appendix \ref{geoquant}) and requires that the shift vector vanishes. The first condition reads:
\begin{equation}
	\begin{aligned}
		\delta c_1&:=
		\delta\pi^2
		+\frac{a^{-4}}{2}\bigg(P_{kk}-\frac{\Tr P}{3}\bigg)\delta q_1
		-\frac{2\sqrt{2}a^{-4}}{9}P_{kv}\delta q_3
			\\
		&
		-\frac{2\sqrt{2}a^{-4}}{9}P_{kw}\delta q_4
		-\frac{4\sqrt{2}a^{-4}}{9}P_{vw}\delta q_5
		-\frac{2\sqrt{2}a^{-4}}{9}\left(P_{vv}-P_{ww}\right)\delta q_6.
	\end{aligned}
\end{equation}
In the isotropic limit this condition simply reads $\delta c_1=\delta\pi^2$.
We fix the second condition in agreement with the isotropic limit \cite{Malik:2008im}, i.e.,
\begin{align}
	\delta c_2:=\delta q_2.
\end{align}
When combined with $\delta c_3:=\delta q_3$ and $\delta c_4:=\delta q_4$, this set of conditions yields:
\begin{equation}
	\begin{aligned}
	\textrm{Det}~ \Lambda =
	-2 i k^2 &
	\Bigg\{
	\frac{ 2 
		\left[
		6 a^4 k^2
		+9 P_{kk}^2
		+18
		\left(
		 P_{kv}^2
		 +P_{kw}^2
		 \right)
		 +
		 \left(
		 P_{vv}-P_{ww}
		 \right)^2
		 +4 P_{vw}^2
		\right]
	}{9 a^5}
	\\&\quad
	+
	\frac{	-15 P_{kk} \Tr P+3(\Tr P)^2}{9 a^5}
	\Bigg\},
	\end{aligned}
\end{equation}
which proves it to be a valid gauge.

\subsubsection{Scalar gravity-wave gauge}

Our final example is chosen to demonstrate that anisotropic spacetimes can accommodate gauges that \textbf{do not exist }in isotropic spacetimes. We shall call it the \textbf{scalar gravity-wave gauge}. It assumes the vanishing of a scalar, a vector and a tensor metric perturbation:
\begin{equation}
	\delta c_1=\delta q_2,~~\delta c_2:=\delta q_3,~~\delta c_3:=\delta q_4,~~\delta c_4:=\delta q_5.
\end{equation}
The determinant of the Poisson brackets \eqref{lambda} reads:
\begin{equation}
	\textrm{Det}~ \Lambda = -\frac{8 i \sqrt{2} k^2 P_{vw}}{ a}
\end{equation}
The determinant is non-vanishing as long as $P_{vw}\neq 0$, i.e. the $A_5$-component of the wavefront shear does not vanish\footnote{
The reasoning remains valid if we were to choose $\delta c_4:=\delta q_6$. However, in this scenario, $\textrm{Det}~  \Lambda =- \frac{4 i \sqrt{2} k^2(P_{vv}-P_{ww})}{a}$, that is the $A_6$-component of the wavefront shear is assumed not to vanish. It is worth noting that the simultaneous vanishing of both tensor metric perturbations is impossible, as it would result in $\textrm{Det}~ \Lambda=0$
}. It is clear that this is not  a valid gauge in the isotropic limit, where $P_{vw}\rightarrow 0$. Notice that in this gauge one of the polarization modes of the gravitational wave is entirely carried by the metric density perturbation:
\begin{equation}
	\delta Q_1\big|_{_{GF}}=\frac{2P_{vw}}{a P_{kk}}\delta q_1.
\end{equation}
This gauge corresponds to a coordinate system in which a given tensor metric perturbation vanishes, and the gravity wave is induced by a scalar perturbation that perturbs the wavefronts with non-vanishing shear. It demonstrates that the transverse and traceless metric perturbations cannot be unambiguously identified with gravitational waves. 

We will now apply the Kucha\v r decomposition of Sec. \ref{kuchar} within this gauge. To make use of the gauge frame based on the flat  slicing gauge (FS) we cast both sets of gauge-fixing functions into canonical form \eqref{can}.
 For the flat slicing gauge we have
 	\begin{align}
			\delta C_{FS}=
			\Bigg\{
			&
			\frac{a (3 \delta q_1-\delta q_2)}{3 P_{kk}},
			\nonumber
			\\&
			\frac{i 
				\left[
				\left(P_{vv}+P_{ww}\right)
				\left(3 \delta q_1-\delta q_2\right)
				 -P_{kk}
				 \left(6\delta q_1+\delta q_2\right)
				 \right]
				 }
				 {6 P_{kk}}
			,
			\nonumber
			\\
			&
			-\frac{i 
				\left(12 P_{kv}\delta q_1-4 P_{kv}\delta q_2
				+3 \sqrt{2} P_{kk}\delta q_3
				\right)}{6 P_{kk}},
				\nonumber
				\\&
			-\frac{i \left(12 P_{kw}\delta q_1-4 P_{kw}\delta q_2+3 \sqrt{2} P_{kk}\delta q_4\right)}{6 P_{kk}}
			\Bigg\},
\end{align}
while for the scalar gravity-wave gauge (SG) we find:
 \begin{align}
			\delta C_{SG}=
			\Bigg\{&
			-\frac{a\delta q_5}{2 \sqrt{2}P_{vw}}
			,
			-\frac{i 
				\left[
				4 P_{vw}\delta q_2
				+\sqrt{2}\left(\Tr P-3P_{kk}\right)\delta q_5
				\right]}{8 P_{vw}}
			,\nonumber
			\\&
			-\frac{i 
				\left(
				 P_{vw}\delta q_3
				 - P_{kv}\delta q_5
				 \right)}{\sqrt{2} P_{vw}}
			,
			-\frac{i \left( P_{vw}\delta q_4- P_{kw}\delta q_5\right)}{\sqrt{2} P_{vw}}
			\Bigg\}.
\end{align}
The difference between those two gauge-fixing conditions expressed in terms of $\alpha$ and $\beta$, defined in Eq. \eqref{alphabeta}, reads:
\begin{equation}\label{betaSGFS}
	\begin{gathered}
		\alpha^\mu_{~I}=0,
		\quad
		\beta^{\mu I}=
		\begin{pmatrix}
			-\frac{a^2}{2 P_{vw}}&0\\[0.7em]
			\frac{i (3P_{kk}-\Tr P)}{4 a P_{vw}}&0\\[0.7em]
			\frac{i P_{kv}}{a P_{vw}}&0\\[0.7em]
			\frac{i P_{kw}}{a P_{vw}}&0
		\end{pmatrix}.
\end{gathered}
\end{equation}
Furthermore, we find $\lambda^\mu_{4\nu}$,  defined at the end of Sec. \ref{sec-decomp}, in Eq. \eqref{lambda4}, to be
\begin{equation}
	\lambda_4
	=
	\begin{pmatrix}
		0&0&0&0\\[0.7em]
		i a^{-2}k&\frac{\Tr P-3P_{kk}}{3a^3}&0&0\\[0.7em]
		0&-\frac{2P_{kv}}{a^3}&\frac{\Tr P-3P_{vv}}{3 a^3}&-\frac{P_{vw}}{a^3}\\[0.7em]
		0&-\frac{2P_{kw}}{a^3}&-\frac{P_{vw}}{a^3}&\frac{\Tr P-3P_{ww}}{3 a^3}
	\end{pmatrix}.
\end{equation}
Hence, using Eq. \eqref{lapseshift}, the lapse and shift transform as follows:
\begin{equation}
	\begin{split}
		\frac{\delta{N}_{SG}^\mu}{N}\bigg|_{_{\delta\widetilde{C}_{SG}^{\mu}=0}}
		- \frac{\delta{N}_{FS}^\mu}{N}\bigg|_{_{\delta{C}_{FS}^{\mu}=0}}
		&\approx
		\left(
		\lambda_{4\nu}^\mu\beta^{\nu 1}
		+\dot{\beta}^{\mu 1}
		\right)\delta{Q}_1
		+
		\frac{1}{2a}\beta^{\mu 1}\delta{P}^1.
\end{split}\end{equation}
To reconstruct the three-surfaces we apply the formula \eqref{ab3g} with the matrix $\mathbf{M}^{-1}$ which maps the ADM pertrubation variables into the constraint functions \eqref{linS}, \eqref{linD}, the gauge-fixing functions in the spatially flat slicing gauge \eqref{gauge} and the Dirac observables \eqref{inddir}.
\\
\subsubsection{Synchronous gauge}
The synchronous gauge is given by\textbf{ partial gauge-fixing}, $\delta N^\mu=0$.
The gauge-fixing conditions for the synchronous gauge are obtained in terms of $\alpha^\mu_{~I}$ and $\beta^{\mu I}$ as solutions to Eqs \eqref{alphabetadot}. In the gauge frame based on SF, we find: 
\begin{align}\begin{split}
		\dot{\alpha}^\mu_{~I}
		&	=
		-\frac{1}{2a}
		\beta^{\mu I}
		+\frac{\partial}{\partial \delta P^{I}}\left(\frac{\delta {N}_{FS}^\mu}{N}\right),\\
		\dot{\beta}^{\mu I}
		&	=
		U_I\alpha^\mu_{~I}+C_{IJ}\alpha^\mu_{~J}
		-\lambda_{4\nu}^\mu\beta^{\nu I}
		+\frac{\partial}{\partial \delta Q_{I}}\left(\frac{\delta{N}_{FS}^\mu}{N}\right),\end{split}
\end{align}
where $\frac{\delta{N}_{FS}^\mu}{N}$ are given by Eqs \eqref{NB}. The choice of the initial data ${\alpha}^\mu_{~I}(t_0)$, ${\beta}^{\mu I}(t_0)$ determines unambiguously the synchronous gauge-fixing conditions. Then the three-surfaces can be reconstructed with the use of the matrix $\mathbf{M}$ as in the previous example. 

\subsection{Spacetime reconstruction}
 In this section we proceed as discussed in Sec. \ref{ISR},
in order to obtain a \textbf{complete spacetime }picture as in Eq. \eqref{dsadm}.  We need to determine the values of $\delta N$, $\delta N^k$, $\delta N^v$ and $\delta N^w$ from the consistency equations \eqref{stabilityEq},
where ${\mathbb H}$ is the \textbf{full Hamiltonian} of Eq. \eqref{hamtot}. For each mode $\underline{k}$ and each index $\rho$, we obtain a linear algebraic equation,
\begin{equation}
	N
	\left\{
	\delta c_{\rho},
	\mathcal{H}_{0}^{(0)}+\mathcal{H}_{0}^{(2)}
	\right\}
	+\delta N^{\mu}\Lambda_{\rho\mu}
	 \approx 0,
\end{equation}
where $\mathcal{H}_{0}^{(2)}$ should include the extra Hamiltonian if the perturbation variables are expressed in the $A$-basis. Given a complete set of gauge-fixing conditions, the above equation is easily solved finding \eqref{coneq}, which in this case reads
\begin{equation}
	\frac{\delta N^{\mu}}{N}\approx-(\Lambda^{-1})^{\mu\rho}
	\left\{
	\delta c_{\rho},\mathcal{H}_{0}^{(0)}+\mathcal{H}_{0}^{(2)}
	\right\}.
\end{equation}
For the particular case of the flat slicing gauge \eqref{gauge}, the consistency equations in the Fermi-Walker-propagated basis yield, in terms of the Dirac observables,
\begin{align}
		\frac{\delta N}{N}&=
		-\frac{P_{vw} }{ a P_{kk}}\delta Q_1
		-\frac{P_{vv}-P_{ww}}{2 a P_{kk}}\delta Q_2
		-\frac{p_\phi  }{2a P_{kk}}\delta Q_3,
		\nonumber
		\\
				\nonumber
		\\
		\frac{\delta N^k}{N}&=
		\delta Q_1
		\bigg(
		\frac{ P_{vw}}{2 a^2} 
		-\frac{ 2 P_{kv} P_{kw} }{a^2 P_{kk}}
		\bigg)
		+\delta Q_2
		\bigg(
		\frac{ P_{kw}^2}{a^2 P_{kk}} 
		-\frac{ P_{kv}^2 }{a^2 P_{kk}}
		+\frac{ P_{vv}-P_{ww} }{4  a^2}
		\bigg)
				\nonumber
		\\&
		+\delta Q_3
		\bigg(
		\frac{ a^4 V_{,\phi} }{2 P_{kk}}
		-\frac{p_\phi (\Tr P)  }{2 a^2 P_{kk}}
		+\frac{3  p_\phi  }{4 a^2}
		\bigg)
		+\frac{ P_{vw}}{P_{kk}}\delta P_1
		+\frac{ P_{vv}-P_{ww}}{2 P_{kk}}\delta P_2
		+\frac{ p_\phi  }{2 P_{kk}}\delta P_3
				\nonumber
		,\\
				\nonumber
		\\
		\frac{\delta N^v}{N}&=
		\delta Q_1
		\bigg(
		\frac{ 2 P_{kv} P_{vw} }{a^2 k P_{kk}}
		+\frac{ 2 P_{kw}}{a^2 k}
		\bigg)
		+
		\delta Q_2
		\bigg(
		\frac{ P_{kv} (P_{vv}-P_{ww}) }{a^2 k P_{kk}}
		+\frac{ 2 P_{kv} }{a^2 k}
		\bigg)
		+\frac{ P_{kv} p_\phi  }{a^2 k P_{kk}}\delta Q_3,
				\nonumber
		\\
				\nonumber
		\\
		\frac{\delta N^w}{N}&=
		\delta Q_1
		\bigg(
		\frac{2 P_{kw} P_{vw}}{a^2 k P_{kk}}
		+\frac{ 2 P_{kv}}{a^2 k}
		\bigg)
		+\delta Q_2
		\bigg(
		\frac{ P_{kw} (P_{vv}- P_{ww})}{a^2 k P_{kk}}
		-\frac{2 P_{kw} }{a^2 k}
		\bigg)
		+\frac{P_{kw} p_\phi  }{a^2 k P_{kk}}\delta Q_3
		.\end{align}
Note that both the lapse function $\delta N$ and the shift vector $\delta N^k$ are \textbf{scalars} under rotations around $\hat{k}$. This follows from the fact that the tensor Dirac observables $\delta Q_1$, $\delta Q_2$, $\delta P_1$ and $\delta P_2$ suitably combined with the tensor zeroth-order coefficients (i.e. forming the products $X_5Y_5$ and $X_6Y_6$, see our discussion below \eqref{inddir}) become scalars. 
The shift vectors $\delta N^v$ and $\delta N^w$, under a rotation $R_{\hat{k}}({\theta})$ around $\hat{k}=\hat{v}\times\hat{w}$ by the angle $\theta$, that is
\begin{equation}
\begin{aligned}
	R_{\hat{k}}({\theta})\delta N^v&=\cos(\theta)\delta N^v+\sin(\theta)\delta N^w
	\\R_{\hat{k}}({\theta})\delta N^w&=\cos(\theta)\delta N^w-\sin(\theta)\delta N^v,
	\end{aligned}
\end{equation}
  transform as vectors, as they include quadratic and cubic terms $X_3Y_5$, $X_3Y_6$, $X_4Y_5$, $X_4Y_6$, $X_3Y_5Z_5$, $X_4Y_5Z_5$, $X_3Y_6Z_6$, $X_4Y_6Z_6$. 
  
  The knowledge of $\frac{\delta N}{N}$, $\frac{\delta N^k}{N}$,  $\frac{\delta N^v}{N}$,  $\frac{\delta N^w}{N}$ allows one to reconstruct the full spacetime metric tensor of Eq. \eqref{dsadm} as a function of Dirac observables. However, the metric components \eqref{dsadm} themselves are not Dirac observables as the above relations are tied to the particular choice of gauge-fixing conditions, i.e. they depend on the employed coordinate system.

An alternative approach to fixing gauge conditions is to begin by specifying $\frac{\delta N}{N}$, $\frac{\delta N^k}{N}$, $\frac{\delta N^v}{N}$, and $\frac{\delta N^w}{N}$. Then, we solve the equation:
\begin{equation}
	\{\delta c_{\rho},\mathcal{H}_{0}^{(0)}+\mathcal{H}_{0}^{(2)}\}=-\Lambda_{\rho\mu}\frac{\delta N^{\mu}}{N},
\end{equation}
which is a first-order linear partial differential equation from which we find:
\begin{align}\label{NB}
		\frac{\delta N}{N}&=
		-\frac{P_{vw} }{ a P_{kk}}\delta Q_1
		-\frac{P_{vv}-P_{ww}}{2 a P_{kk}}\delta Q_2,
				\nonumber
		\\
				\nonumber
		\\
		\frac{1}{i}\frac{\delta N^k}{N}&=
		\bigg[
		\frac{ 2P_{vw}}{3 a^4} 
		-\frac{ 2 P_{kv} P_{kw} }{a^4 P_{kk}}
		+\frac{ 2P_{vw}^3}{a^4P_{kk}^2}
		+\frac{ P_{vw}(P_{vv}-P_{ww})^2}{2a^4P_{kk}^2}
				\nonumber
		\\&\quad
		-\frac{ 5P_{vw}(P_{vv}+P_{ww})}{6a^4P_{kk}}
		\bigg]\delta Q_1
				\nonumber
		\\&\quad
		+\bigg[
		\frac{ P_{kw}^2}{a^4 P_{kk}} 
		-\frac{ P_{kv}^2 }{a^4 P_{kk}}
		+\frac{ P_{vv}-P_{ww} }{3 a^4}+\frac{ 5(P_{ww}^2 -P_{vv}^2)}{12 a^4P_{kk}}
				\nonumber
		\\&\quad
		+\frac{(P_{vv}-P_{ww})P_{vw}^2}{a^4P_{kk}^2}
		+\frac{(P_{vv}-P_{ww})^3}{4a^4P_{kk}^2}
		\bigg]\delta Q_2
				\nonumber
		\\&\quad
		+\frac{ P_{vw}}{a^2 P_{kk}}\delta P^1
		+\frac{ P_{vv}-P_{ww}}{2a^2 P_{kk}}\delta P^2
				\nonumber
		,\\
				\nonumber
		\\
		\frac{1}{i}\frac{\delta N^v}{N}&=
		\bigg(
		\frac{ 2 P_{kv} P_{vw} }{a^4 P_{kk}}
		+\frac{ 2 P_{kw}}{a^4}
		\bigg)\delta Q_1
		+
		\bigg[
		\frac{ P_{kv} (P_{vv}-P_{ww}) }{a^4 P_{kk}}
		+\frac{ 2 P_{kv} }{a^4}
		\bigg]\delta Q_2
				\nonumber
		,\\
				\nonumber
		\\
		\frac{1}{i}\frac{\delta N^w}{N}&=
		\bigg(
		\frac{2 P_{kw} P_{vw}}{a^4 P_{kk}}
		+\frac{ 2 P_{kv}}{a^4}
		\bigg)\delta Q_1
		+
		\bigg[
		\frac{ P_{kw} (P_{vv}- P_{ww})}{a^4 P_{kk}}
		-\frac{2 P_{kw} }{a^4}
		\bigg]\delta Q_2.
\end{align}
{Note that the imaginary units in front of $\delta N^k$, $\delta N^v$, $\delta N^w$ cancel out in Eq. \eqref{KHw} when multiplied by the respective constraints \eqref{linD}.}

 Generally, there may not be a unique solution, leading to what is known as the \textbf{residual gauge freedom}. Geometrically, choosing $\frac{\delta N^{\mu}}{N}$ uniquely fixes the coordinates once the initial Cauchy surface is determined. Selecting a specific solution to this equation is equivalent to defining the initial Cauchy surface in the perturbed spacetime.
\section{Multi-field case}\label{multifield}

Particle physics models allow for the possible presence of more than one scalar field in the primordial universe. The multi-field scenario offers new and attractive features in particular to the theory of inflation \cite{2009pdpL,Bassett_2006,Schutz:2013fua}. It is quite straightforward to extend the above single-field framework to the case when the matter is made of a collection of scalar fields. The matter constraints read:
\begin{equation}
	\begin{gathered}
		\mathcal{H}_{m,0}=
		\sqrt{q}
		\bigg[
		\sum_I\frac{1}{2}q^{-1}\pi_{\phi^I}^2
		+\sum_I\frac{1}{2}q^{ij}\phi_{~,i}^I\phi_{,j}^I
		+V(\dots,\phi^I,\dots)
		\bigg],
		\\
		\mathcal{H}^{~i}_m=-\sum_I\pi_{\phi^I}\phi_{,}^{I~i}~,
	\end{gathered}
\end{equation}
where the index $I$ labels the \textbf{real scalar fields}. The Fourier components of the linearized scalar and vector matter constraints \eqref{eqmatter}, in the multi-field case, are found to be:
\begin{equation}
	\begin{gathered}
		\delta\mathcal{H}_{m,0}=
		\sum_I
		\left[
		a^{-3}p_{\phi^I}\delta\pi_{\phi^I}
		-\frac{3}{4}a^{-5}(p_{\phi^I})^2\delta q_{1}
		+\frac{3}{2}aV\delta q_{1}
		 +a^{3}V_{,\phi^I}\delta\phi^I
		\right],\\
		\delta\mathcal{H}_{m}^{i}=\sum_Iia^{-2}k^ip_{\phi^I}\delta\phi^I.
	\end{gathered}
\end{equation}
Repeating all the steps made in the single-field case, as well as setting the spatially flat gauge-fixing conditions \eqref{gauge}, we arrive at the following generalization of the gauge-invariant Hamiltonian \eqref{HBI}:
\begin{equation}\label{HBIgen}
	\begin{aligned}
		H_{BI}=&
		\frac{N}{2a}\bigg[
		\delta{P}_{1}^2+\delta{P}_2^2+\sum_I\delta{P}_{3I}^2
		+\sum_I(k^2+U_{\phi})\delta{Q}_{3I}^2+(k^2+U_5)\delta{Q}_1^2+(k^2+U_6)\delta{Q}_2^2
		\\&
		+C_{1}\delta{Q}_1\delta{Q}_2+\sum_IC_{2I}\delta{Q}_1\delta{Q}_{3I}+\sum_IC_{3I}\delta{Q}_2\delta{Q}_{3I}+\sum_{I>J}C_{IJ}\delta Q_{3I}\delta Q_{3J}\bigg], 
		\end{aligned}
\end{equation}
where the coefficients are given in Appendix \ref{MSAppgen}. 
The Hamiltonian presented above shares a similar structure with \eqref{HBI}, but with the addition of more matter perturbations denoted by $\delta Q_{3I}$ and the inclusion of new coupling terms $C_{IJ}$. It is worth noting that, under the spatially flat gauge, the following equivalences hold:
\begin{equation*}
	\delta{Q}_1=\delta\tilde{q}_5,\quad \delta{Q}_2=\delta\tilde{q}_6,\quad \delta{Q}_{3I}=\delta\tilde{\phi^I},\quad \delta{P}_1=\delta\tilde{\pi}_5,\quad \delta{P}_2=\delta\tilde{\pi}_6 \quad \text{and} \quad\delta{P}_{3I}=\delta\tilde{\pi}_{\phi^I}.
\end{equation*}

The complete set of Dirac observables is given in Appendix \ref{Dext}. The physical variables are related to them as follows:
\begin{equation}
	\begin{aligned}
		\delta{Q}_1&=
		\frac{1}{\sqrt{2}a}{\delta D_{1}},
		~~\delta{P}_1=
		\sqrt{2}a{\delta D_{7}}
		+\frac{C_{55}}{\sqrt{2}a}{\delta D_{1}}
		+\frac{C_{56}}{\sqrt{2}a}{\delta D_{2}}
		+C_{5\phi^J}a {\delta D_{9J}},\\
		\delta{Q}_2&=
		\frac{1}{\sqrt{2}a}{\delta D_{2}}
		,~~\delta{P}_2=
		\sqrt{2}a{\delta D_{8}}
		+\frac{C_{56}}{\sqrt{2}a}{\delta D_{1}}
		+\frac{C_{66}}{\sqrt{2}a}{\delta D_{2}}
		+C_{6\phi^J}a{\delta D_{9J}},\\
		\delta{Q}_{3I}&=
		a{\delta D_{9I}},
		~~~~~~\delta{P}_{3I}=
		a^{-1}{\delta D_{10I}}
		+\frac{C_{I 5}}{\sqrt{2}a}{\delta D_{1}}
		+\frac{C_{ I6}}{\sqrt{2}a}{\delta D_{2}}
		+C_{\phi^I\phi^J}a{\delta D_{9J}}.
	\end{aligned}
\end{equation}

\section{Summary}

In the present work we have derived the physical phase space and the physical Hamiltonian for anisotropic cosmological theory by means of the \textbf{Dirac procedure}. We introduced the \textbf{Fermi-Walker-propagated basis} in which we expressed the tensor and vector modes and computed the extra Hamiltonian generated by the choice of that basis, we also gave and alternative definition of the frame which is more suitable for quantization. We chose convenient gauge-fixing conditions and obtained the physical Hamiltonian in terms of background and perturbation canonical variables. We showed that the obtained result is valid in\textbf{ any gauge} if the physical variables are replaced by the respective\textbf{ Dirac observables}. We also reconstructed the full spacetime by means of canonical variables in the flat slicing gauge. Finally, we extended the obtained result to the multi-field case which may be relevant for models of the primordial universe.

The Dirac method relies on the existence of the \textbf{canonical isomorphism} between different gauge-fixing surfaces and provides a useful framework for studying gauge-fixing conditions. We considered a few examples of gauge-fixing conditions, where most were extensions of the gauge-fixing conditions used in \textbf{isotropic spacetimes}, while one was specific to the anisotropic spacetime and does not constitute a valid gauge in the isotropic spacetimes. For any choice of gauge-fixing conditions, the Dirac observables acquire a distinct physical interpretation. In particular, it turns out that a gravitational wave could be seen as a \textbf{scalar perturbation} of the metric.
It is worth mentioning that the scalar mode of gravitational waves is also predicted in extended or modified GR theories \cite{capozziello}. However, it differs from our model's "scalar" wave, which maintains tensorial transformation properties due to induced anisotropy in the wavefronts by the global dynamics.

%% file: Chapters/Chapter4.tex

\chapter{Time problem in perturbation theory} 

\label{Chapter4} 

In the previous chapters we presented the Dirac method to obtain the physical perturbed Hamiltonian. We applied this method in Chapter \ref{Chapter3} to obtain the gauge-invariant Hamiltonian in an anisotropic Bianchi I universe. The aim of this work is to prepare for the quantization of our system. Although in this thesis we are not going to quantize the obtained Hamiltonian \eqref{BIn}, in the following chapter we are going to make use, once again, of the Dirac observables in order to tackle one of the main issues related to the quantization of relativistic systems, i.e. the time problem.

	
	\section{Introduction}
	
	The \textbf{time problem} present in quantum gravity models \cite{Kuchar:1991qf, Isham:1992ms, Anderson:2010xm, Kiefer:2021zdq} is due to the different definitions of time in non-relativistic physics, in which an external and \textbf{absolute time} exists, and in Einstein's theory of gravity, where one has
	to rely on largely arbitrary physical variables, known as \textbf{internal
	time} variables or internal clocks, to follow changes occurring in gravitational systems. By virtue of the principle of general relativity, the	choice of internal time variable has no physical consequence in the classical theory. Upon passing to quantum theory, however, different choices of internal time variables are known to produce unitarily inequivalent quantum models \cite{Deser:1961zza, Beluardi:1994jb, Hajicek:1999ti, Catren:2000br,
		Hajicek:2000dy,Malkiewicz:2014fja,
		Malkiewicz:2017cuw,Bojowald_2018}. The problem of finding the	correct interpretation of these non-equivalent models is commonly	referred to as \textbf{the time problem}.
	
	In this chapter, we aim to find an interpretation for the non-equivalent clocks in these models. For this purpose we consider the model of primordial gravitational waves propagating across the \textbf{Friedmann universe}. It is important to note that similar models were previously used for making physical predictions for the primordial amplitude spectrum of gravity waves and of density perturbations, and are constrained by observations (see, e.g., Refs.~\cite{Peter:2006hx, Pinto-Neto:2008gdo,
		Peter_2008}). 
	
It was shown that dynamical observables defined in different clocks are \textbf{unitary inequivalent}, see, for instance, Refs.~\cite{Malkiewicz:2016hjr, Malkiewicz2020, Malkiewicz:2022szx}. In quantum gravity, only gauge-invariant variables are measurable quantities \cite{Rovelli:2001bz, Dittrich:2004cb}. These variables are constants of motion, yet they encode the so-called relational dynamics. From this perspective, dynamical quantities can be ambiguously expressed as one-parameter families of gauge-invariant quantities, with each family representing motion with respect to a specific internal time. Alternatively, the dynamical variables can be viewed as fundamental variables. The differences in their dynamics should be carefully studied before proposing their physical interpretation. This interpretation would influence the predictions derived from quantum gravity models.
	
	
	The cosmological system considered in this chapter, shows dynamical
	discrepancies when based on different clocks, both in the background and perturbation variables. This leads
	us to focus on the fundamental question: \textit{what are the dynamical
	predictions of quantum cosmological models,
	which do not depend on the employed time variable?}
	
	We address this question within the\textbf{ reduced phase space
	quantization}. Namely, we solve the Hamiltonian constraint and choose the internal time variable prior to
	quantization. An alternative approach would be to first quantize and
	then solve the constraint quantum mechanically while promoting one of
	the variables as internal time. Both approaches lead to
	the same time problem \cite{Bojowald_2023} and therefore using the
	technically less involved reduced phase approach seems well-justified. Most significantly, within the reduced phase space approach,
	there exists a theory of clock transformations, which is completely crucial for the purpose of this work \cite{Malkiewicz:2015fqa}. Thanks to these precisely defined
	transformations, we are able to explore all possible clocks and
	quantize them with an assumption of fixed operator ordering. Hence, any quantum ambiguities found, arise from the differences between clocks rather than the differences between quantization prescriptions.
	\section{Clock transformations in totally constrained systems}
	\label{sec2}
	As discussed in Ch. \ref{Chapter1}, in canonical relativity the presence of the Hamiltonian constraint is a consequence of the fact that
	the dynamics of three-surfaces is generated by infinitesimal time-like diffeomorphisms, and the latter leave the full four-dimensional spacetime invariant. The Hamiltonian
	constraint dynamics is a feature of any canonical relativistic theory of gravity, be it Einstein's or any modified gravity theory, although their dynamics differ. Canonical
	relativity assumes the lack of an absolute, external time in which three-surfaces evolve, and replaces it with internal variables that serve as clocks with which the dynamics of three-surfaces can be described. None of the internal clocks can play a privileged role as the principle of relativity states. This picture is certainly true in the classical theory. At the quantum level, no spacetime exists and, as we will see later, the principle of	relativity must take a somewhat altered form. In order to find it, we need to extend the canonical formalism by including clock transformations	that transform a canonical description from one internal clock to
	another; only then can we move to the quantum level
	where these new transformations become a key to unlock the \textbf{principle	of quantum relativity}.
	
	\subsection{Internal time}
	
	We will now select the internal variable which will serve as cock in our system. 
	Let us consider a system consisting of a set of $N+1$ canonical
	pairs $\{q_\alpha,p^\alpha\}_{\alpha= 0,\cdots, N}$ and assume a
	Hamiltonian constraint
\begin{equation*}
	C(q_\alpha,p^\alpha)\approx 0.
\end{equation*}
 Suppose that one of the positions, say $q_0$, varies monotonically with the evolution
	generated by the constraint, i.e. $\forall q_0, \{q_0,C\}_\textsc{pb}
	\neq 0$. It is then possible to assign to $q_0$ the role of an \textbf{internal clock} in which the evolution of the remaining variables occurs. This evolution is then governed by a Hamiltonian that is not a constraint.
	
	At this stage, it may seem that the time variable is fixed
	once and for all, which would contradict the principle of relativity;
	we discuss below in what sense this is not the case.
	
Let us briefly recall how the \textbf{reduced Hamiltonian formalism} is obtained from the initial
	symplectic form $\Sigma = \dd q_\alpha \wedge \dd p^\alpha$ (Einstein
	convention assumed), evaluated on the constraining surface, namely
	\begin{equation}
		\begin{gathered}
			\Sigma\big|_{C=0} = \left(\ud q_I\wedge \ud p^I+\ud q_0\wedge \ud
			p^0\right)\big|_{C=0}\\
			\null =\ud q_I \wedge \ud p^I-\ud t\wedge \ud H,
			\label{OmegaC}
		\end{gathered}
	\end{equation}
	where $I=1,\cdots, N$, and $H=H\left( q_0,q_I,p^I\right)$ is the
	nonvanishing reduced Hamiltonian such that $p_0+H\approx 0$. Note that
	both $q_0$ (denoted by $t$ from now on to emphasize its role as a time
	variable) and $p^0$ are removed from the phase space and the remaining
	dynamical variables are no longer constrained, as seen in \ref{geomconst}. Indeed, their dynamics
	read
	\begin{equation*}
		\frac{\ud q_I}{\ud t}
		=
		\frac{\partial H}{\partial p^I}
		\qquad
		\hbox{and} 
		\qquad
		 \frac{\ud p^I}{\ud t}=
		 -\frac{\partial H}{\partial q_I},
	\end{equation*}
	 which is entirely solved once an arbitrary initial condition
	$(q_I^\text{ini},p^I_\text{ini}, q_0^\text{ini})$ is provided.
	
	In order to restore the principle of relativity, we need to allow for	\textbf{any clock}, denoted by $\tilde{t}$, which monotonically varies with the
	evolution generated by the constraint, $\{\tilde{t},C\}_\textsc{pb}
	\neq 0$. This new clock must be a function of the old clock and the
	old canonical variables, $\tilde{t}=\tilde{t}(q_I,p^I, t)$. Thus, it
	must satisfy
	\begin{equation}
		\frac{\dd \tilde{t}}{\dd t} = \frac{\partial \tilde{t}}{\partial t} +
		\underset{\{\tilde{t},H \}_\textsc{pb}}{\boxed{\frac{\partial \tilde{t}}{\partial q_I} \frac{\partial
				H}{\partial p^I} - \frac{\partial \tilde{t}}{\partial p^I}
			\frac{\partial H}{\partial q_I}}}
		\neq 0.
	\end{equation}
For the new formalism to be canonical, the two-form\footnote{} \eqref{OmegaC} induced on the constraint surface $C=0$ in some new canonical variables
	must read:
	\begin{equation*}
	\Sigma\big|_{C=0}
	=\ud \tilde{q}_I\wedge\ud \tilde{p}^I
	 - \ud
	 \tilde{t}
	\wedge\ud \tilde{H},
	\end{equation*}
	 This implies that there must
	exist an \textbf{invertible map} between the old and the new variables:
	\begin{equation}
		\tilde{t}=\tilde{t}(q_I,p^I, t),\ 
		\tilde{q}_I=\tilde{q}_I(q_J,p^J,t), \
		\tilde{p}^I=\tilde{p}^I(q_J,p^J, t).
		\label{trans}
	\end{equation}
	These transformations in principle, and in all the relevant cases, are not canonical. It can be shown that clock transformations form a	group of \textbf{generally noncanonical transformations} with canonical
	transformations as its normal subgroup \cite{Malkiewicz:2016hjr};
	finding them is in general a difficult task. However, for an
	integrable dynamical system, the problem can be reduced to that of
	solving a set of algebraic equations.
	
	If a dynamical system is integrable, then we may find a complete set
	of canonical constants of motion, denoted by $D_I$. Let them be
	functions of the old internal time and old canonical variables,
	$D_I=D_I(q_J,p^J, t)$. Note that substituting back $t\rightarrow q_0$,
	they must commute with the original constraint,
	$\{D_I,C\left(q_\alpha,p^\alpha\right)\}_\textsc{pb} = 0$.  They are
	therefore genuine \textbf{Dirac observables }. The new
	internal time $\tilde{t}=\tilde{t}(q_I,p^I, t)$ and new canonical
	variables can then be found according to the algebraic relations
	\begin{equation}
		\tilde{t}=\tilde{t}(q_I,p^I, t),~~D_I(q_J,p^J,
		t)=D_I(\tilde{q}_J,\tilde{p}^J, \tilde{t}),\label{TF}
	\end{equation} 
	where we formally substitute the canonical variables in the
	expressions for Dirac observables $D_I$, i.e., we assume the same
	functional dependence of $D_I$ in both sets of variables. As discussed in the previous chapters, the number
	of $D_I$ is equal to the number of the new canonical variables
	$\tilde{q}_J$ and $\tilde{p}^J$, and thus, leaving aside singular
	cases, the above relations determine $\tilde{q}_J$ and $\tilde{p}^J$
	completely. The result is a new canonical formalism based on a new
	internal clock. Let us note that, by virtue of Eq.~\eqref{TF}, if a
	solution to the dynamics is known in one clock, i.e. $t\rightarrow
	\left[ q_I\left(D_J,t\right),p^I\left(D_J,t\right) \right]$, then it is readily known for all other
	clocks and reads $\tilde{t}\rightarrow
	\left[\tilde{q}_I=q_I\left(D_J,\tilde{t}\right),
	\tilde{p}^I=p^I\left(D_J,\tilde{t}\right)\right]$. This
	makes the choice of the new canonical variables $\tilde{q}_I$ and
	$\tilde{p}^I$ via Eq.~\eqref{TF} very convenient: the formal
	description of the system is the same in all clocks, only the physical
	meaning of the clock and basic variables changes, which is emphasized
	by the use of a tilde (~$\tilde{}$~) over the variable names.
	
	The use of Dirac observables in the derivation of clock
	transformations gives an invaluable advantage when passing to quantum
	theory. Our goal is to make a comparison between quantum theories
	based on different internal clocks of a single physical
	system. Therefore, it is of uttermost importance to make sure that the
	theories are different only insofar as their clocks differ, and not
	due to other quantization ambiguities such as the well-known factor
	ordering. This state of affairs can be achieved by fixing a quantum
	representation of the Dirac observables and then defining basic and
	compound observables as functions of the quantum Dirac observables,
	both in the original
	\begin{equation*}
		\widehat{q}_I=q_I(\widehat{D}_J,t), \qquad
		\widehat{p}^I=p^I(\widehat{D}_J,t),
	\end{equation*}
	and the new variables
	\begin{equation*}
		\widehat{\tilde{q}}_I={q}_I(\widehat{D}_J,\tilde{t}),
		\qquad \widehat{\tilde{p}}^I = {p}^I(\widehat{D}_J,\tilde{t}).
	\end{equation*}
	These definitions imply that ${q}_I$ and ${p}^I$ are promoted to the
	same operators as ${\tilde{q}}_I$ and ${\tilde{p}}^I$,
	respectively. We invert this reasoning and start by assuming the same
	operators for ${q}_I$ and ${\tilde{q}}_I$ as well as ${p}^I$ and
	${\tilde{p}}^I$. This implies that the Dirac observables being the
	same functions in both sets of basic variables are promoted to the
	same operators irrespectively of the choice of clock. Hence, the
	quantum descriptions in different clocks are formally the same, only
	the physical meaning of the basic operators changes from one clock to
	another, which is emphasized by the use of tilde.  Obviously, a unique
	ordering prescription has to be used in all the above formulas. In
	principle, after this step, any physically interesting aspect of the
	quantum theories can be compared. In the following section, we
	introduce the model on which we discuss such comparisons.
	
	\section{Canonical cosmological model}
	\label{sec3}
	
	We consider a \textbf{flat Friedmann-Lema\^{\i}tre-Robertson-Walker} (FLRW)
	universe filled with radiation and perturbed by gravitational waves;
	the line element of the model reads (in units such that $c=1$)
	\[
	\ud s^2=
	-N^2(t)\ud t^2
	+a^2(t)
	\left[
	 \delta_{ij}
	 +h_{ij}(\bm{x},t)
	 \right]
	\ud x^i\ud x^j.
	\] 
	The matrix $h_{ij}$ represents the
	gravitational waves, which in FLRW are associated to tensor perturbations; it satisfies
	$h_{ij}\delta^{ij}=0$ and $\partial^j h_{ij}=0$.
	Finally, as in Ch. \ref{Chapter3}, we assume a \textbf{toroidal spatial topology}
	with each comoving coordinate $x^i\in [0,1)$. Setting
	$N\to a$ means one considers the conformal time which we denote by $\eta$.
	
	\subsection{Perturbative Hamiltonian}
	Let us now build the canonical description of these gravitational
	waves in an FLRW universe. The\textbf{ background variables}, as in Ch. \ref{Chapter3}, are denoted by $(\bar{q},\bar{\pi})$. For the sake of a more coherent notation with the quantization in Sec. \ref{sec4}, we are going to rename them as the
	\textbf{scale factor} $\bar{q}=a$ and its\textbf{ conjugate momentum} $\bar{\pi}=p_a$. The \textbf{tensor perturbations} are represented by the
	\textbf{gravitational wave amplitude} $\mu^{(\lambda)} = a h^{(\lambda)}$ and
	its \textbf{conjugate momentum }$\pi^{(\lambda)}$, with $\lambda \in \{ +,
	\times\}$ and $h_{ij} = \sum_\lambda h^{(\lambda)}
	\varepsilon_{ij(\lambda)}$ (see, e.g., Refs.~\cite{Peter:2013avv, Micheli:2022tld} for details on the helicity expansion).
	
	The\textbf{ matter component} is assumed to be a radiation fluid with energy
	density $p_0$ conjugate to a timelike variable $q_0$. The
	gravitational constraint is expanded to second order through
	\[
	H_\text{tot} = H^\text{(b)} + \sum_{\bm{k}}
	H^\text{(p)}_{\bm{k}},
	\]
where no boundary terms are present as the spatial sections are compact. The quantity $H^\text{(b)}$ is the background
	Hamiltonian and $H^\text{(p)}_{\bm{k}}$ is the perturbation Hamiltonian.
	The background Hamiltonian is given by
	\begin{equation}
		H^\text{(b)} = -\frac12 p_a^2-p_0.
	\end{equation}
	At this stage, one can identify the internal time $q_0$ with the
	conformal time $\eta$ as it reduces the zeroth order Hamiltonian into 
	\begin{equation}
		\begin{gathered}
			\Sigma\big|_{H^{(0)}=0}
			=\left(\ud a\wedge\ud p_a+\ud q_0\wedge \ud
			p_0\right)\big|_{H^\text{(b)}=0}
			\\
			=\ud a\wedge\ud p_a-\ud \eta\wedge\ud
			\left( \frac{1}{2}p_a^2 \right),
			\label{OmegaCbckg}
		\end{gathered}
	\end{equation}
	leading to the\textbf{ physical zeroth-order Hamiltonian} 
	\begin{equation}
		H^{(0)} = \frac12 p_a^2,
		\label{H0phys}
	\end{equation}
	while preserving the form of the perturbation Hamiltonian
	$H^\text{(p)}_{\bm{k}}$. The latter reads, at \textbf{second order}
	\begin{equation}
		H^\text{(p)}_{\bm{k}} \to H^{(2)}_{\bm{k}} = 
		- \sum_{\lambda=+,\times} H^{(2)}_{\bm{k},\lambda}
		\label{Hpert}
	\end{equation}
	with
	\begin{equation}
		H^{(2)}_{\bm{k},\lambda} = 
		\frac12 \left| \pi^{(\lambda)}_{\bm{k}} \right|^2
		+ \frac12 \left( k^2 - \frac{a''}{a} \right)
		\left| \mu^{(\lambda)}_{\bm{k}} \right|^2,
		\label{H2kini}
	\end{equation}
	where a prime stands for a derivative with respect to the conformal
	time. Since the tensor perturbations are real, one has $\mu^{(\lambda)
		*}_{\bm{k}} = \mu^{(\lambda)}_{-\bm{k}}$.  Moreover, since the
	background is isotropic, one can restrict attention to upward directed
	wavevectors $\bm{k}$ by merely cancelling the factor $\frac12$ in
	$H^{(2)}_{\bm{k},\lambda}$.  This permits to write the \textbf{final
	second-order Hamiltonian} as
	\begin{equation}
		H^{(2)}_{\bm{k},\lambda} = 
		\pi^{(\lambda)}_{\bm{k}} \pi^{(\lambda)}_{-\bm{k}}
		+ \left( k^2 - \frac{a''}{a} \right)
		\mu^{(\lambda)}_{\bm{k}} \mu^{(\lambda)}_{-\bm{k}}.
		\label{H2k}
	\end{equation}
	Note that for the radiation fluid we are concerned with here, the
	Hamiltonian \eqref{H0phys} yields as equations of motion $p_a = a'$
	and $p_a'=0$, thus leading to $a''=0$: the potential for producing
	gravitational waves is indeed \textbf{classically} vanishing if the universe is
	radiation dominated. 
	
 Determining the solution to the dynamics of gravitational waves is straightforward in the radiation case. While it is possible to consider a general fluid with the barotropic index $w>0$ (this case can be solved analytically in terms of Bessel functions, see, e.g., \cite{Mukhanov:1990me}), such consideration is not relevant to the objectives of this work as the examined clock effects are not specific to any matter content but must be present whenever quantum uncertainties in the background geometry are taken into account.
	
	\subsection{Dirac observables}
	
	Now we shall find the constants of motion that form canonical
	pairs. To this end, we need to solve the partial differential
	equations
	\begin{equation}
		\frac{\ud D}{\ud \eta}=
		\frac{\partial D}{\partial
			\eta}
			+
			\left\{
			D,H^{(0)}+H^{(2)}
			\right\}_\textsc{pb}=0.
		\label{Diracs}
	\end{equation}
	At zeroth order, this is
	\[
	\frac{\partial D}{\partial \eta} + p_a
	\frac{\partial D}{\partial a} = 0,
	\]
	with solutions
	\begin{equation}
		D_1=a - p_a \eta 
		\qquad \hbox{and} \qquad
		D_2 = p_a.
		\label{D1D2}
	\end{equation}
	At first order, Eq.~\eqref{Diracs} reads
	\begin{equation}\label{eqdeD}
	\frac{\partial \delta D}{\partial \eta} + p_a
	\frac{\partial \delta D}{\partial a} = 
	\pi^{(\lambda)}_{\bm{k}} \frac{\partial \delta D}
	{\partial \mu^{(\lambda)}_{\bm{k}}}
	-k^2 \mu^{(\lambda)}_{\bm{k}} 
	\frac{\partial \delta D}{\partial \pi^{(\lambda)}_{\bm{k}}},
	\end{equation}
	where we considered the classical solution $a''=0$. Since we are
	considering only first order perturbations, we demand that $\delta D$
	be linear in the perturbation variables $\mu^{(\lambda)}_{\bm{k}}$ and
	$\pi^{(\lambda)}_{\bm{k}}$.  The l.h.s. of Eq. \eqref{eqdeD} is
	greatly simplified if $\delta D$ depends only on the variable $y=\eta
	+ a/p_a$, so we look for a solution of the form $\delta D^{(\lambda)} =
	\mu^{(\lambda)}_{\bm{k}} \alpha (y) + \pi^{(\lambda)}_{\bm{k}}
	\beta(y)$, leading to
	\[
	2\frac{\dd \alpha}{\dd y} \mu^{(\lambda)}_{\bm{k}} +
	2\frac{\dd \beta}{\dd y} \pi^{(\lambda)}_{\bm{k}}
	= \alpha \pi^{(\lambda)}_{\bm{k}} 
	- k^2 \beta \mu^{(\lambda)}_{\bm{k}}.
	\]
	Assuming independent variations of $\mu^{(\lambda)}_{\bm{k}}$ and
	$\pi^{(\lambda)}_{\bm{k}}$, one gets $2\dd\alpha/\dd y=-k^2\beta$ and
	$2\dd\beta/\dd y=\alpha$, and finally $4\dd^2\alpha/\dd y^2 =k^2
	\alpha$, so that, setting $$\Omega_k = \frac{k}{2} \left( \eta +
	\frac{a}{p_a} \right),$$ one gets\textbf{ two independent solutions} for each
	polarization, or in other words\textbf{ four first-order constants}, reading
	\begin{equation}
		\begin{gathered}
			\delta D^{(\lambda)}_{1,\bm{k}} = \sqrt{k}
			\sin\Omega_k \, \mu^{(\lambda)}_{\bm{k}}
			-\frac{\cos\Omega_k}{\sqrt{k}}\,
			\pi^{(\lambda)}_{\bm{k}},\\
			\delta D^{(\lambda)}_{2,\bm{k}} = \sqrt{k} \cos\Omega_k\,
			\mu^{(\lambda)}_{\bm{k}} + \frac{\sin\Omega_k}{\sqrt{k}} \,
			\pi^{(\lambda)}_{\bm{k}}.
			\label{deltaD}
		\end{gathered}
	\end{equation}
	In Eq.~\eqref{deltaD}, the normalisation has been chosen so as to
	ensure that all these Dirac observables indeed form canonical pairs,
	namely
	\[
	\left\{ D_1 , D_2 \right\}_\textsc{pb} = 1\qquad \hbox{and}\qquad
	\left\{ \delta D^{(\lambda)}_{1,\bm{k}}, \delta
	D^{(\bar{\lambda})}_{2,\bm{k}} \right\}_\textsc{pb} =
	\delta_{\lambda\bar{\lambda}}.
	\]
	From now on, we drop the index $\lambda$ and consider just a single
	polarization mode $(\mu_{\bm{k}},\pi_{\bm{k}})$.
	
	\subsection{Clock transformations}\label{clocktra}
	
	Having set the full model, and before moving on to its quantum
	counterpart, let us first consider a \textbf{general clock transformation}
	\begin{equation}
		\eta \rightarrow \tilde{\eta} = \eta + \Delta(a,p_a,\eta),
		\label{CT}
	\end{equation}
	where $\Delta$ is a \textbf{delay function} that in general varies between the
	trajectories as well as along them.  At the background level,
	implementing the recipe given by Eq.~\eqref{trans},
	i.e., $D_{1,2}(a,p_a,\eta) = D_{1,2}
	(\tilde{a},\tilde{p}_a,\tilde{\eta})$ to the transformation~\eqref{CT},
	yields
	\[
	a-p_a \eta = \tilde{a}-\tilde{p}_a \tilde{\eta} \quad
	\hbox{and}\quad p_a = \tilde{p}_a,
	\]
	leading to
	\begin{equation}
		\tilde{a} = a + p_a \Delta  \quad
		\hbox{and} \quad p_a = \tilde{p}_a.
		\label{CTbckg}
	\end{equation}
	In order that the clock transformation \eqref{CT} actually defines a
	new and physically acceptable clock, the delay function $\Delta$ must
	be subject to two conditions. The \textbf{first condition} is that the new clock must run forward,
	that is
	\begin{equation}
		\frac{\ud \tilde{\eta}}{\ud \eta} = 1 + 
		\frac{\ud \Delta}{\ud \eta} = 1 + 
		\frac{\partial \Delta}{\partial \eta} + 
		p_a \frac{\partial \Delta}{\partial a}>0,
		\label{Cond1}
	\end{equation}
	where in the second equality we used the zeroth-order Hamiltonian
	$H^{(0)}$ given by Eq.~\eqref{H0phys} and the associated equations of
	motion.
	
	The \textbf{second condition} that a clock transformation must satisfy is that
	the ranges of the basic variables $a$ and $p_a$ must be preserved,
	thereby preventing the appearance of non-trivial ranges that may
	induce new and potentially unsolvable quantization issues. This second
	condition implies
	\begin{subequations}\label{Cond2}
		\begin{align}
			\lim_{p_a\rightarrow\pm\infty} \tilde{p}_a(a,p_a,\eta) &= \pm\infty,
			\label{Cond2a}
			\\
			\tilde{a}(a,p_a,\eta)\big|_{a=0}&=0.
			\label{Cond2b}
		\end{align}
	\end{subequations}
	The first equality \eqref{Cond2a} is trivially satisfied in the
	present case because of \eqref{CTbckg}.  For $\Delta=\Delta(a,p_a)$,
	to which we shall restrict attention in what follows, the second
	equality \eqref{Cond2b} is identical to demanding that the delay
	function at vanishing scale factor should also vanish,
	$\Delta(0,p_a)=0$. This condition also ensures that the slow-gauge
	clock is transformed into another slow-gauge clock, that is, the
	boundary is reached within a finite amount of time (see
	Ref.~\cite{Malkiewicz2020}). Such a condition \eqref{Cond2b}, although
	irrelevant in the classical theory, is crucial for the existence of a
	bounce at the quantum level where the clock must smoothly connect
	contracting and expanding trajectories. Were \eqref{Cond2} violated,
	the clock transformations would break the bouncing trajectories.
	
	It turns out that the condition \eqref{Cond1} is equivalent to the
	existence of a \textbf{one-to-one map} between the reduced phase spaces
	$(a,p_a)$ and $(\tilde{a},\tilde{p}_a)$, i.e. the determinant
	\begin{equation}
		\frac{\partial \left( \tilde{a},\tilde{p}_a\right)}
		{\partial\left( a,p_a \right)} = \left|
		\begin{array}{cc} \displaystyle\frac{\partial\tilde{a}}{\partial a} & 
			\displaystyle\frac{\partial\tilde{a}}{\partial p_a} \\ & \\
			\displaystyle\frac{\partial\tilde{p}_a}{\partial a} & 
			\displaystyle\frac{\partial\tilde{p}_a}{\partial p_a}
		\end{array}\right|> 0,
		\label{Jac}
	\end{equation}
	which is indeed Eq.~\eqref{Cond1} when $\partial\Delta/\partial\eta
	=0$.
	
	At first order, one must solve
	\begin{equation*}
		\delta D_1 (a, p_a, \mu_{\bm{k}}, \pi_{\bm{k}}) = 
		\delta D_1 (\tilde{a}, \tilde{p}_a, \tilde{\mu}_{\bm{k}},
		\tilde{\pi}_{\bm{k}})
	\end{equation*}
	and
	\begin{equation*}
		\delta D_2 (a, p_a, \mu_{\bm{k}}, \pi_{\bm{k}}) = 
		\delta D_2 (\tilde{a}, \tilde{p}_a, \tilde{\mu}_{\bm{k}},
		\tilde{\pi}_{\bm{k}})
	\end{equation*}
	in order to determine the clock-transformed perturbation
	variables. Explicitly, using \eqref{deltaD}, one gets
	\begin{equation}
		\begin{gathered}
			\sqrt{k} \sin\Omega_k\,\mu_{\bm{k}}
			- \frac{\cos\Omega_k}{\sqrt{k}}\,\pi_{\bm{k}} 
			= \sqrt{k} \sin\tilde{\Omega}_k\,\tilde{\mu}_{\bm{k}}
			- \frac{\cos\tilde{\Omega}_k}{\sqrt{k}}\,\tilde{\pi}_{\bm{k}},\\ 
			\sqrt{k} \cos\Omega_k \,\mu_{\bm{k}}
			+ \frac{\sin\Omega_k}{\sqrt{k}}\,\pi_{\bm{k}} 
			= \sqrt{k} \cos\tilde{\Omega}_k\,\tilde{\mu}_{\bm{k}}
			+ \frac{\sin\tilde{\Omega}_k}{\sqrt{k}}\,\tilde{\pi}_{\bm{k}} 
			,
			\label{tildemupi}
		\end{gathered}
	\end{equation}
	where $\tilde{\Omega}_k=\frac12 k (\tilde{\eta} +
	\tilde{a}/\tilde{p}_a) = \Omega_k + k\Delta$.  The algebraic
	equations \eqref{tildemupi} can easily be inverted to yield the new
	canonical perturbation variables, namely
	\begin{equation}
		\left(
		\begin{array}{c}
			\tilde{\mu}_{\bm{k}} \\ \\
			\displaystyle\frac{\tilde{\pi}_{\bm{k}}}{k}
		\end{array}
		\right)
		=
		\left(
		\begin{array}{cc}
			\cos k\Delta & -\sin k\Delta \\ \\
			\sin k\Delta &  \cos k\Delta
		\end{array}
		\right)
		\left(
		\begin{array}{c}
			\mu_{\bm{k}} \\ \\ 
			\displaystyle\frac{\pi_{\bm{k}}}{k}
		\end{array}
		\right).
		\label{funfor}
	\end{equation}
	It is important to note that the above are \textbf{classical relations} between
	canonical variables belonging to distinct canonical frameworks based
	on distinct internal clocks. Although they are canonically
	inequivalent, these two frameworks generate the\textbf{ same physical dynamics}
	of the system, which is required by the principle of relativity.
	
Finally, we observe that the clock transformations described in our framework preserve the foliation of cosmological spacetimes by homogeneous spatial leaves with small perturbations. Given that the initial clock $\eta$ corresponds to the conformal time, the new lapse function of the background foliation implied by the new clock $\tilde{\eta}$ in Eq. \eqref{CT}, reads:
		\[
		\tilde{N}=\frac{a}{1 + 
			p_a \frac{\partial \Delta}{\partial a}}>0,
			\]
		where ${\partial \Delta}/{\partial \eta}=0$ was assumed. Note, however, that the idea of clock transformation involves modifying temporal relations between events belonging also to different spacetimes. This aspect of clock transformations is not reflected in the lapse function $\tilde{N}$, which expresses the temporal relation between points within a single spacetime.
	
	\section{Quantization}
	\label{sec4}
	
	Having completed the classical treatment of our system, we now move to
	the investigation of the possible differences between the respective
	quantum dynamics obtained from the quantization of these two different
	frameworks.
	
	\subsection{Semi-classical background}
	
	Since, by definition, the scale factor is positive definite ($a>0$),
	one needs to quantize our previous system on the half line. Although
	the position operator $\widehat{Q}=a$ is self-adjoint on the half
	line, this is not the case for the momentum operator $\widehat{P} = i
	\hbar\partial_a$, so we instead use the\textbf{ symmetric dilation
	operator} 
	\[
	\widehat{D} =\{ \widehat{P},\widehat{Q}\} = \frac12 \left(
	\widehat{P} \widehat{Q} + \widehat{Q} \widehat{P}\right) = \frac12
	i\hbar \left( a\partial_a + \partial_a a\right).
	\]
	 Classically the
	dilation variable is $d = a p_a$, so that the Hamiltonian, expressed
	in term of $d$, is $H^{(0)} = \frac12 p_a^2 = \frac12 d^2/a^2$, and
	one can define its quantum counterpart as a symmetric ordering of
	$\frac12 \widehat{Q}^{-2} \widehat{D}^2$.  Expanding on the basis
	($\widehat{Q},\widehat{P}$), this yields
	$$
	\widehat{H}^{(0)} = - \frac{1}{2}\frac{\partial^2}{\partial a^2} + 
	\frac{\hbar^2K}{a^2},
	$$
	where the value of $K>0$ depends on the ordering; fixing one ordering such that
	$K>\frac34$ ensures $\widehat{H}^{(0)}$ is self-adjoint on the half
	line~\cite{Vilenkin:1987kf}.
	
	We can find some approximate solutions to the Schr\"odinger equation
	with a family of \textbf{coherent states} (see, e.g.,
	Refs.~\cite{Klauder:2015ifa,Klauder:2012vh,Martin:2021dbz}
	for the specific case under study here). We choose the coherent states
	to read
	\begin{equation}
		|a(\eta),p_a(\eta)\rangle = \ex^{i p_a(\eta) \widehat{Q}/\hbar}
		\ex^{-i \ln [a(\eta)] \widehat{D}/\hbar}|\xi\rangle,
		\label{CohSta}
	\end{equation}
	where $|\xi\rangle$ is such that the expectation values of
	$\widehat{Q}$ and $\widehat{P}$ in $|a(\eta),p_a(\eta)\rangle$ are
	respectively $a(\eta)$ and $p_a(\eta)$, and otherwise arbitrary
	(see also Ref.~\cite{Bergeron:2023zzo}).
	
\begin{figure}[t]
		\centering
		\includegraphics[width=0.6\textwidth]{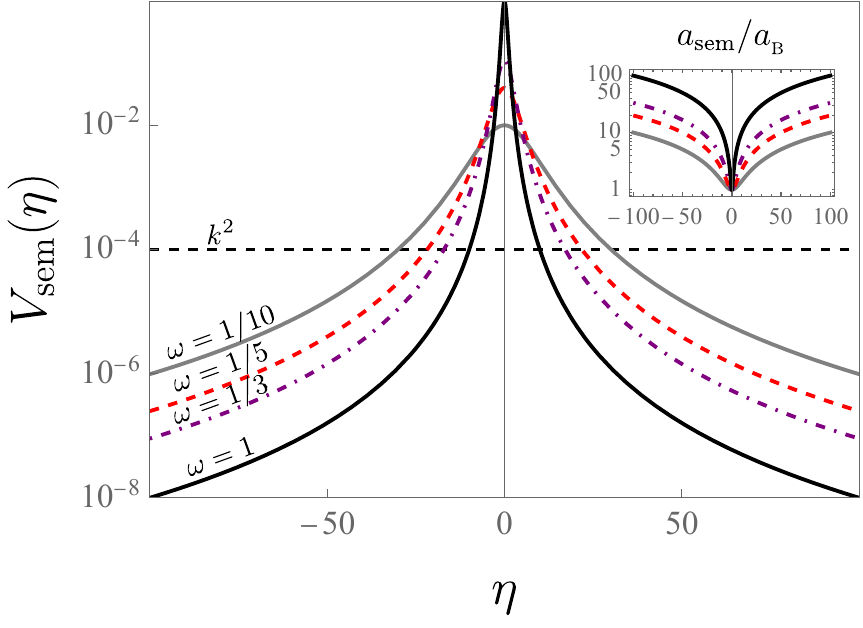}
		
		\caption{The semi-classical potential $V_\text{sem}$ given by
			Eq.~\eqref{Vsem} as a function of the conformal time $\eta$ for
			various values of the inverse bounce duration $\omega$. The
			potential has to be compared with the relevant value of $k^2$ ($k=0.01)$,
			indicated as a straight line. The corresponding scale factor time
			evolution is shown in the insert.}
		
		\label{VsemFig}
	\end{figure}

	The dynamics confined to the coherent states can be deduced from the
	quantum action
	\[
		\mathcal{S}_\textsc{q} =\int \left\{ a'(\eta) p_a(\eta) -
		H_\text{sem}\left[a(\eta),p_a(\eta)\right] \right\}\dd \eta,
	\]
	with the semiclassical Hamiltonian given by
	\[
		H_\text{sem} =\langle a,p_a| \widehat{H}^{(0)} |a,p_a\rangle,
	\]
	from which one derives the ordinary \textbf{Hamilton equations}
	\begin{equation}
		a' = \frac{\partial {H}_\text{sem}}{\partial p_a}
		\quad \text{and}
		\quad p'_a = - \frac{\partial {H}_\text{sem}}{\partial a}. \label{eomsem}
	\end{equation}
	We find that the \textbf{semiclassical background Hamiltonian} reads
	\cite{Martin:2021dbz}
	\begin{equation}		\label{HKt}
		H_\text{sem}=\frac12 \left( p_a^2 + 
		\frac{\hbar^2\mathfrak{K}}{a^2}\right),
	\end{equation}
	where the new constant $\mathfrak{K}$ is positive ($\mathfrak{K}>0$).
	Its specific value is related with both $K$ and the fiducial state
	$|\xi\rangle$. We find the solution to \eqref{eomsem} to read
	\[
	a^2(\eta) = a_0 + a_1 \eta + a^2(\eta),
	\]
	 with $a_0 a_2 -a_1^2/4	=\hbar^2\mathfrak{K} >0$, so that the equation $a(\eta) =0$
	has no longer any real solution;\textit{ the singularity is indeed
	quantum mechanically avoided}. Choosing the origin of time
	such that $a'=0$ for $\eta=0$ permits to rewrite this
	solution in full generality as
	\begin{subequations}
		\begin{align}
			a(\eta) & = a_\textsc{b} \sqrt{1+(\omega \eta)^2},
			\label{aeta}\\
			p_a(\eta) & = \frac{a_\textsc{b} \omega^2 \eta}
			{\sqrt{1+(\omega\eta)^2}},
			\label{paeta}
		\end{align}
		\label{solsem}%
	\end{subequations}
	where $a_\textsc{b}^4 \omega^2 = \hbar^2\mathfrak{K}$, which in turn
	implies $H_\text{sem} = \frac12 a_\textsc{b}^2 \omega =\frac12 \hbar
	\sqrt{\frak{K}}$; it is clear that the model contains one and only one
	free parameter, namely $\frak{K}$.  From now on, we assume that the
	background evolution is given by Eqs.~\eqref{solsem}: this means the
	\textbf{semi-classical potential}
	\begin{equation}
		V_\text{sem} = \frac{a''}{a} = \frac{\hbar^2 \frak{K}}{a^4}
		=\left[ \frac{\omega}{1+\left(\omega\eta\right)^2} \right]^2,
		\label{Vsem}
	\end{equation}
	shown in Fig.~\ref{VsemFig}, never vanishes except in the large scale
	factor limit ($a\gg 1 \Longrightarrow \eta \gg \omega^{-1}$). This is
	appropriate as this is also the classical limit for which $a''\to
	0$. A classical radiation-dominated universe begins or ends with a
	singularity and produces no gravitational waves, whereas our quantum
	radiation-dominated universe naturally connects the contracting and
	expanding phases through a bounce which is subsequently responsible
	for a non-vacuum spectrum of tensor perturbations, to which we now
	turn.
	
	\subsection{Quantum perturbations}
	
	For a given mode $\bm{k}$, the Hamiltonian $H^{(2)}_{\bm{k}}$, given
	by Eq.~\eqref{H2k}, is easily quantized using the usual
	prescriptions. We assume that the background follows the
	semi-classical approximation described above, so that the potential
	for the perturbation is given by $V_\text{sem}$, defined in Eq.~\eqref{Vsem}. The basic variables are replaced by a set of
	operators
	\begin{equation}
		\begin{gathered}
			\mu_{\bm{k}} \mapsto \widehat{\mu}_{\bm{k}} =
			\sqrt{\frac{\hbar}{2}}\left[\widehat{a}_{\bm{k}} \mu^*_k(\eta)+
			\widehat{a}^{\dagger}_{-\bm{k}}
			\mu_k(\eta)\right],\\ 
			\pi_{\bm{k}} \mapsto
			\widehat{\pi}_{\bm{k}} = \sqrt{\frac{\hbar}{2}}
			\left[\widehat{a}_{\bm{k}}
			\mu^{*\prime}_k(\eta)+ \widehat{a}^{\dagger}_{-\bm{k}}
			\mu'_k(\eta)\right],
		\end{gathered}
		\label{operators}
	\end{equation}
	where we assume the Wronskian normalization condition $\mu'_k \mu^*_k
	- \mu_k \mu^{*\prime}_k = 2i$ for the complex mode functions
	$\mu_k$. The creation $\widehat{a}^\dagger_{\bm{k}}$ and annihilation
	$\widehat{a}_{\bm{k}}$ operators satisfy the commutation relations
	$\left[ \widehat{a}_{\bm{k}}, \widehat{a}^\dagger_{\bm{p}} \right] =
	\delta_{\bm{k}, \bm{p}}$ stemming from the canonical ones between the
	field operators $\left[\widehat{\mu}_{\bm{k}}, \widehat{\pi}_{-\bm{p}}
	\right] = i\hbar \delta_{\bm{k}, \bm{p}}$.
	
	In the \textbf{Heisenberg picture}, the equations of motion take the form
	\[
	i\hbar\frac{\dd \widehat{\mu}_{\bm{k}}}{\dd\eta} = 
	\left[ H^{(2)}_{\bm{k}}, \widehat{\mu}_{\bm{k}}\right]
	\quad \hbox{and} \quad
	i\hbar\frac{\dd \widehat{\pi}_{\bm{k}}}{\dd\eta} = 
	\left[ H^{(2)}_{\bm{k}}, \widehat{\pi}_{\bm{k}}\right].
	\]
	The above equations imply that the mode function $\mu_k(\eta)$
	satisfies
	\begin{equation}
		\frac{\ud^2\mu_k}{\ud\eta^2}+\left(k^2 -
		\frac{\hbar^2\mathfrak{K}}{a^4}\right)\mu_k=0,
		\label{mode1}
	\end{equation}
	where $a(\eta)$ is given by the semi-classical solution
	\eqref{aeta}. Using \eqref{Vsem}, this transforms into
	\begin{equation}
		\frac{\ud^2\mu_k}{\ud\eta^2}+\left\{ k^2 -
		\left[ \frac{\omega}{1+\left(\omega\eta\right)^2} \right]^2
		\right\}
		\mu_k=0,
		\label{modemu}
	\end{equation}
	which can be integrated numerically once\textbf{ initial conditions} are fixed. We assume that far in the contracting branch, with
	$\eta_\text{ini} <0$ and $V_\text{sem}(\eta_\text{ini}) \ll k^2$,
	there was no gravitational wave, so the field was in a vacuum
	state. This implies the mode function satisfies $\mu_k
	(\eta_\text{ini}) = \ex^{-ik\eta_\text{ini}}/\sqrt{2k}$ and $\mu'_k
	(\eta_\text{ini}) =-i\sqrt{k/2} \, \ex^{-ik\eta_\text{ini}}$.
	
	\begin{figure}[t]
		\centering
		\includegraphics[width=0.6\textwidth]{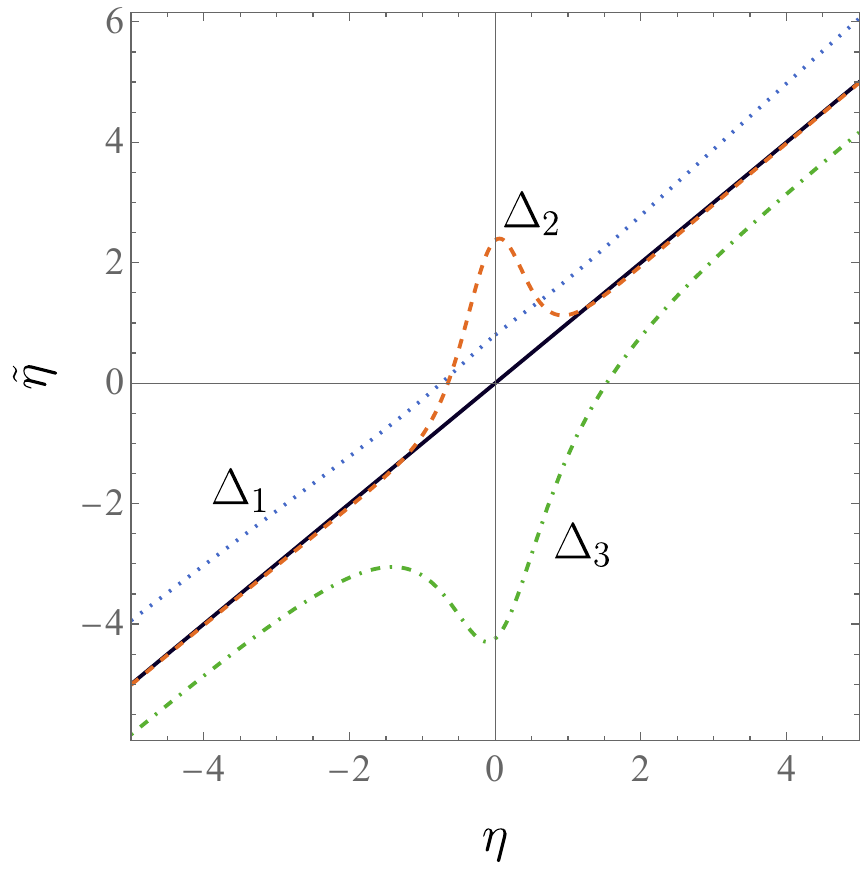}
		
		\caption{The new time $\tilde{\eta}$ as a function 
			of the original one $\eta$ for three different shapes
			of delay functions $\Delta_1$, $\Delta_2$ and
			$\Delta_3$ defined through Eq.~\eqref{delfun} along
			the original fixed bouncing trajectory \eqref{solsem}.
			The parameters are chosen as $A=B=D=1$, $C=4$ and $E=2$
			for $\Delta_1$, while we set $A=2$, $B=0.2$, $C=0.5$,
			$D=3$ and $E=4$ for $\Delta_2$, and finally the set 
			$A=-1$, $B=C=1$, $D=0.5$ and $E=3$ defines $\Delta_3$.}
		\label{deltas}
	\end{figure}
	
	\section{Quantum "clocks"}
	\label{sec5}
	
	We study the effect of clocks on the quantum and
	semi-classical dynamics of selected dynamical variables by considering the following steps:
	\begin{enumerate}[1.] 
\item First, we
obtain the \textbf{dynamical trajectories }in the reduced phase space
$(a,{p},{\mu}_k,\pi_k)$ that is associated with the initial clock
${\eta}$; note that from that point on, since there is no risk of confusion, we shall replace what was previously denoted as $p_a$ simply by $p$.

\item 	 Next, we choose a set of \textbf{delay functions} $\Delta(a,p)$ to
define new clocks $\tilde{\eta}$ and obtain the new reduced phase spaces $(\tilde{a},
\tilde{p},\tilde{\mu}_k,\tilde{\pi}_k)$ associated with the new
clocks. 

\item Then, we make use of Eqs \eqref{CT}, \eqref{CTbckg} and \eqref{funfor} to \textbf{transport the dynamical trajectories }to these new phase spaces. We assume that the latter admit a unique physical interpretation, and so the trajectories can be meaningfully compared in these new variables. 
In other words, there are many clocks denoted by $\tilde{\eta}$ and only one denoted by ${\eta}$. Note that for $\Delta=0$ the clocks ${\eta}$ and $\tilde{\eta}$ coincide. For this case, we assume that ${\eta}$ and $(a,{p},{\mu}_k,\pi_k)$ are the variables of Sec. \ref{sec2}, which sets the physical meaning of the phase space $(\tilde{a}, \tilde{p},\tilde{\mu}_k,\tilde{\pi}_k)$ and the clock $\tilde{\eta}$.
	\end{enumerate}

	\begin{figure}[t]
		\centering
		\includegraphics[width=0.6\textwidth]{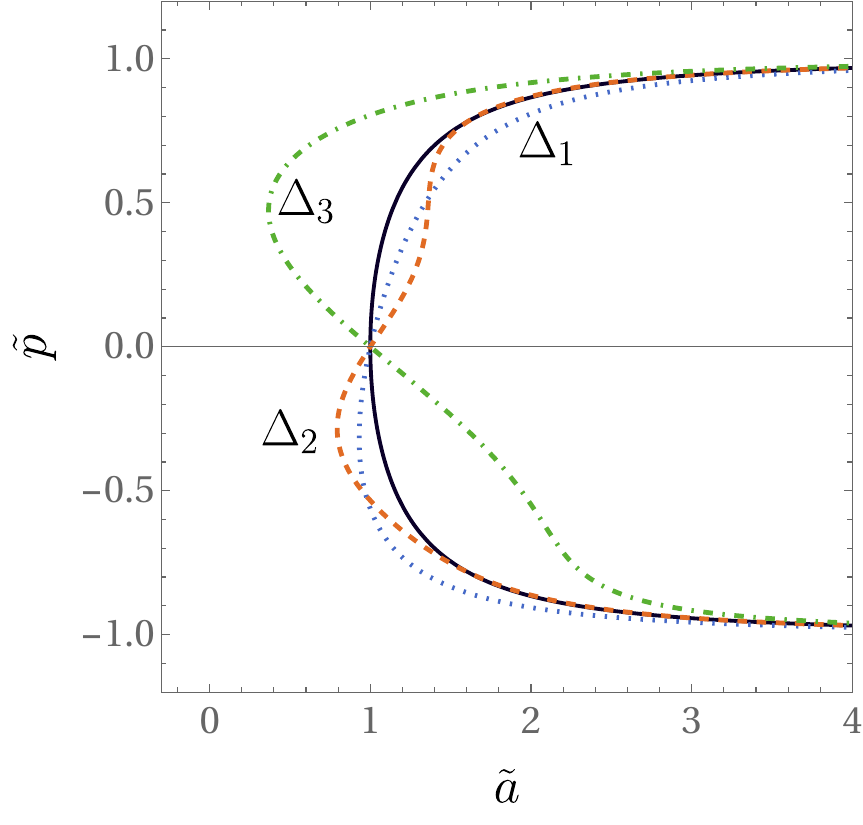}
		
		\caption{Semi-classical trajectories obtained in different clocks and
			mapped into the initial reduced phase space $(\tilde{a},\tilde{p})$ to compare with
			the original trajectory represented by the full black line.}
		
		\label{qpbackground}
	\end{figure}

	\begin{figure}[t]
		\includegraphics[width=0.5\textwidth]{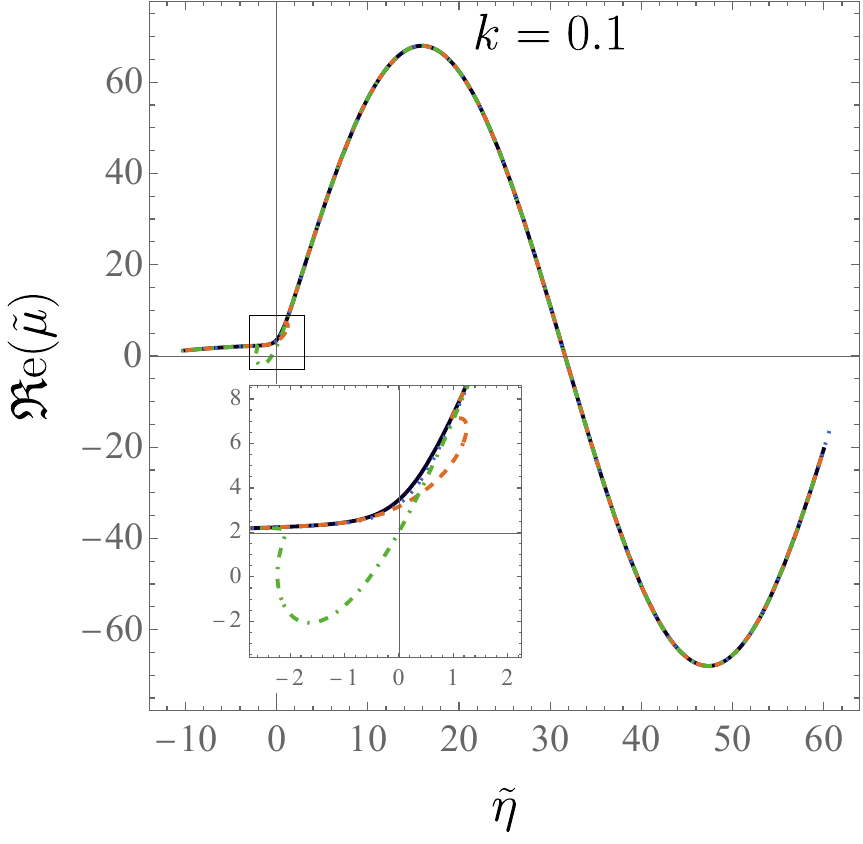}
		\includegraphics[width=0.5\textwidth]{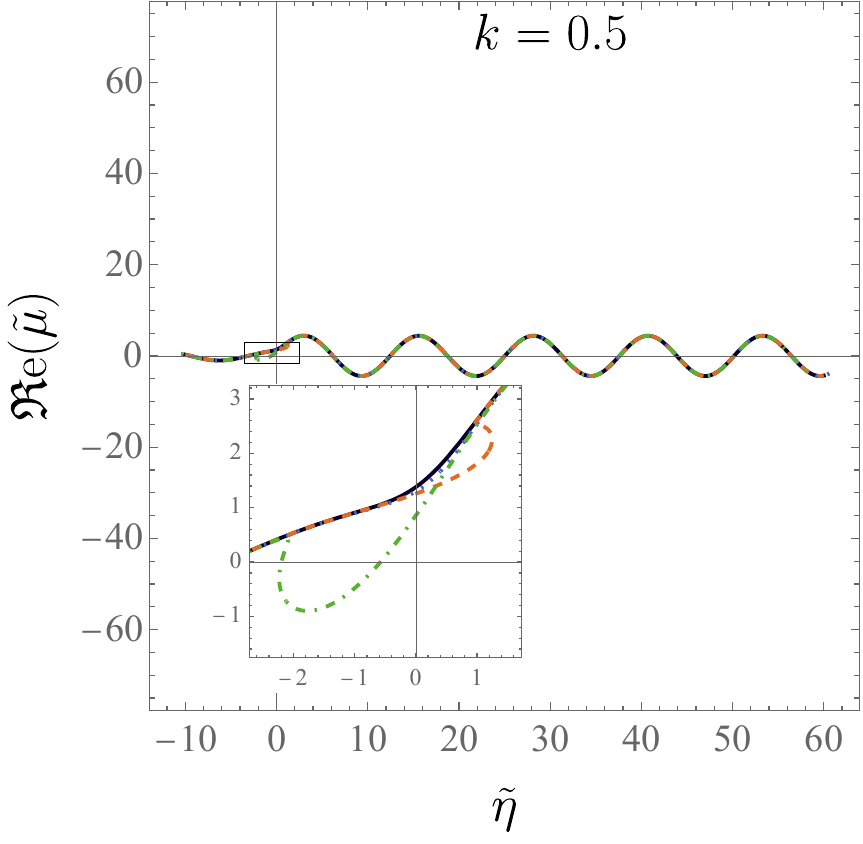}
		
		\caption{Evolution of the primordial gravity-wave
			$\mathfrak{R}\mathrm{e}(\tilde{\mu})$ for two different
			wavenumbers, $k=0.1$ (left panel) and $k=0.5$ (right panel),
			and for different clocks based
			on the first class of delay functions, $\Delta_1$, $\Delta_2$ and
			$\Delta_3$, represented by the dotted blue line, dashed red line
			and dashed-dotted green line respectively. The original trajectory
			is represented by the full black line. In Fig. \ref{mu-k-clock2}
			the same plot for the second class of delay functions is depicted
			to show how the choice of delay function affects the time of
			convergence.}
		
		\label{mu-k-clock}
	\end{figure}
	
	\begin{figure}[ht]
		\centering
		\includegraphics[width=0.6\textwidth]{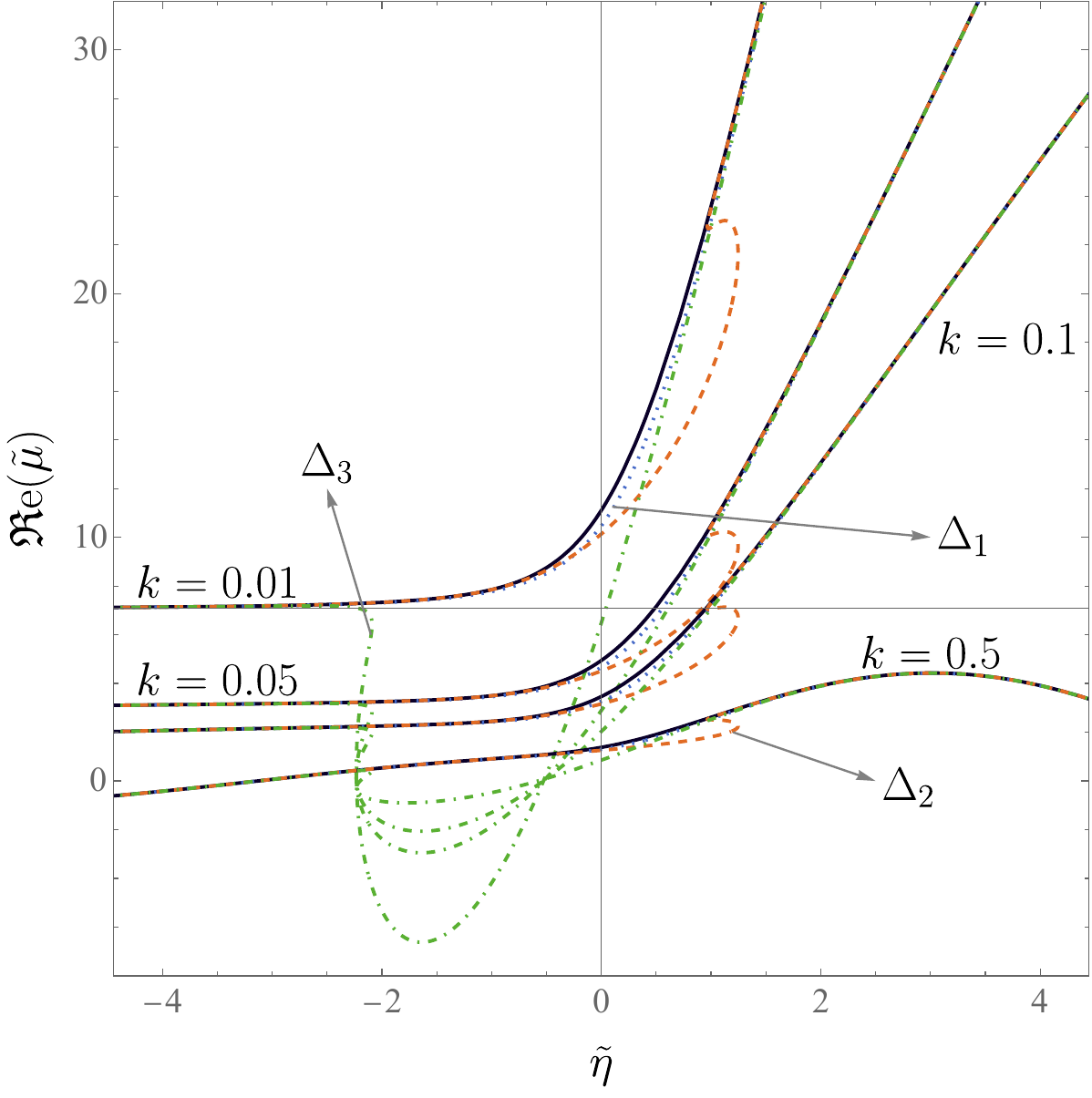}
		
		\caption{Evolution of the primordial gravity-wave
			$\mathfrak{R}\mathrm{e}(\tilde{\mu}_k)$ plotted for 4 different
			wavenumber $k$ values. For each fixed $k$ we changed the clock
			considering the family of delay functions $\Delta$, whose value
			is the same as in Fig. \ref{qpbackground}.}
		
		\label{mu-all-k}
	\end{figure}

	\subsection{Clock choices and background}
	
	In order to illustrate the clock choice issue, we consider
	a \textbf{family of delay functions}, namely
	\begin{equation}\label{delfun}
		\Delta(a,p)=A \frac{a^B}{(a+C)^D}\frac{\sin(E p)}{p},
	\end{equation}
	where $A$, $B$, $C$, $D$ and $E$, are arbitrary coefficients, whose
	values are limited to ensure that the conditions presented in
	Sec.~\ref{clocktra} hold. In Appendix \ref{Appendix-time},
	we consider another set of acceptable delay functions to show
	that our conclusions are not restricted to the choice \eqref{delfun}. 
	
	A few "clocks" corresponding to the
	delay function $\Delta(a,p)$ are
	represented along a semi-classical dynamical trajectory for different
	choices of the free parameters in Fig.~\ref{deltas}. It shows that, contrary to the classical case where the condition \eqref{Cond1} holds, the new clocks, in general, are no longer monotonic due to quantum corrections.
	
	Applying the clock transformation of Fig.~\ref{deltas} to the
	background solution \eqref{solsem} yields Fig.~\ref{qpbackground} once
	mapped into the reduced phase space, with the original trajectory
	superimposed for comparison.
	All the trajectories originate in the same
	classical regime at large $\tilde{a}$ and negative $\tilde{p}$, i.e. at a time at
	which the universe is large and contracting. Close to the $\tilde{a}=0$
	boundary, where the quantum behavior dominates, they all somehow
	bounce in the variables $\tilde{a}$ and $\tilde{p}$, diverging from one another and
	providing different accounts of the bounce. Finally, they reach the
	region of large $\tilde{a}$ and positive $\tilde{p}$ where they converge again to the
	unique classical behavior representing a large and expanding universe.
	
	Possible differences between the trajectories include the values of
	$\tilde{a}$ and $\tilde{p}$ at which the bounce occurs, the level of asymmetry between
	contracting and expanding branches, or even the number of
	bounces. These semi-classical trajectories illustrate the non-unitary
	relation between different clocks.  Nevertheless, they all originate
	from a unique contracting classical universe and end towards a
	similarly unique expanding classical universe. Therefore, the
	semi-classical trajectories in different clocks yield the same outcome
	for large and classical universes.  Notice that the trajectories'
	convergence before and after the bounce can be delayed as much as one
	wants by making use of appropriate delay functions,
	such as that discussed in Appendix \ref{Appendix-time}, i.e. Eq.~\eqref{delfun2},
	whose effects on both background and
	perturbation trajectories can be seen in Appendix \ref{Appendix-time}. 
	
	Let us now move to the perturbation of these homogeneous solutions and
	compare the different evolution that can result from using different
	clocks.
	
	\subsection{Clocks and perturbations}
	
	In Fig.~\ref{mu-k-clock}, we plot the dynamics of the real part of the
	perturbation variable $\tilde\mu_k$ against the delayed time
	$\tilde{\eta}$ for the three different functions of Eq.~\eqref{delfun}
	displayed in Fig.~\ref{deltas}, and for two values of the comoving
	wavenumber $k$. The figure illustrates our general finding that the
	absolute clock effect is more or less equally strong and lasting
	roughly equally long for all-wavelength perturbations. This is
	shown more convincingly in Fig.~\ref{mu-all-k} in which the evolution
	of 4 different modes is shown as a close-up in the quantum-dominated
	bouncing region. This means that
	the larger the wavelength of the perturbation, the larger the
	relative clock effect, and the longer it lasts in units of its
	oscillation period. Thus, the clock effect is more important for
	phenomena occurring at \textbf{small timescales} and over \textbf{short
	distances}.
	Moreover, the evolving amplitude $\tilde\mu_k$, 
	in general, is not a function of the clock $\tilde{\eta}$ due to 
	quantum effects that disrupt the monotonicity relation between quantized clocks.
	
	Given that both the background and the perturbation modes
	evolve in such a way as to reach a unique configuration, 
	the primordial gravity-wave amplitude $\tilde\mu_k/\tilde{a}$,
	which is the quantity one expects to measure in 
	practice~\cite{Micheli:2022tld}, also converges to a
	unique solution, making the model predictive.
	
	All the plots above illustrate the\textbf{ non-unitary} relation between
	different clocks, as well as the spoiling of the clock monotonicity at
	the quantum level, which is illustrated in Fig.~\ref{deltas}.
	Nevertheless,
	similarly to the semi-classical background trajectories, the perturbation
	variable $\mathfrak{R}\mathrm{e}(\tilde{\mu})$ visibly converges to a \textbf{unique classical solution} from a well-defined asymptotic past
	initial condition to the
	asymptotic future. Therefore, one can safely extend the background
	conclusion to the perturbations: the time development of the mode
	$\mathfrak{R}\mathrm{e}(\tilde{\mu})$ using different clocks yields
	the same predictions in the large and classical universe regime.
	The delay of the convergence due to different choices of delay
	functions can be seen in Fig.~\ref{mu-all-k}.
	
\begin{figure}[h]
		\centering
		\includegraphics[width=0.8\textwidth]{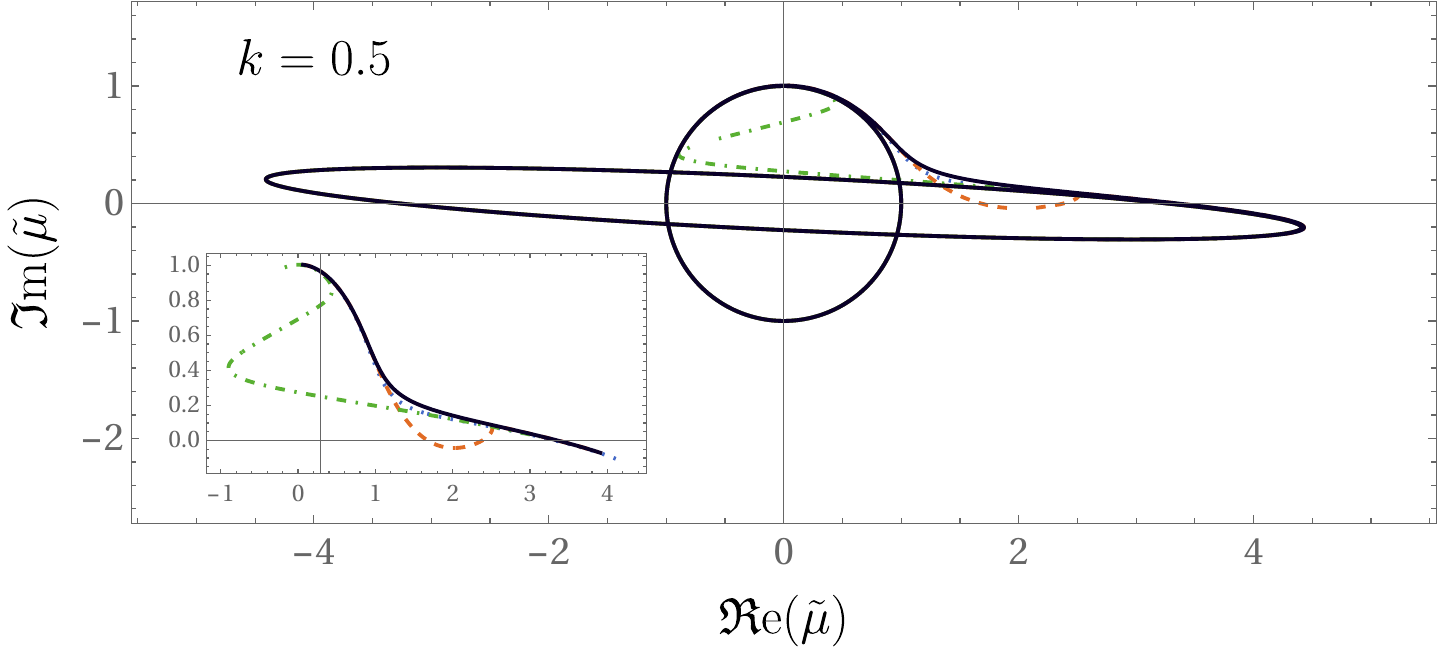}
		
		\caption{Evolution of the real versus imaginary part of $\tilde{\mu}_k$
			for a wavenumber $k=0.5$ and a bounce parameter $\omega=1$.
			The initial circle represents the initial vacuum state of the
			perturbation, while the ellipse shows the final squeezed state,
			which happens, in the case at hand, to have a slight phase shift
			with respect to the real axis. The transition between these two
			asymptotic cases differs for the different delay functions
			$\Delta_1$, $\Delta_2$ and $\Delta_3$, whose trajectories are
			represented by the dotted blue line, dashed red line and 
			dashed-dotted green line respectively, the original trajectory
			being represented by the full black line.}
		
		\label{RealIMmu}
	\end{figure}

	As a final illustration of the perturbation behavior through the
	quantum bounce, we find it useful to inspect the \textbf{phase space
	trajectories }in the plane
	$[\mathfrak{R}\mathrm{e}(\tilde\mu_k),\mathfrak{I}\mathrm{m}(\tilde\mu_k)]$ as displayed in Fig. \ref{RealIMmu}. The initial vacuum state is
	represented by a circle that is squeezed into an ellipse
	during the contraction and bounce, squeezing that represents
	the amplification of the amplitude of
	the perturbation. From the point of view of the time problem, the
	initial circle and the final ellipse respectively represent the
	asymptotic past and future of the amplitude: from the point of view
	of physical prediction, the indeterminacy occurring near the
	bounce as may develop through various different times 
	disappears in the asymptotic regimes, so that the existence
	of a classical approximation in our trajectory approach ensures
	the standard procedure of treating the perturbations leads
	to physically meaningful predictions.
	\section{Summary}\label{sec6}
	
	In this chapter, we explored the time problem in the framework of quantum
	fields on quantum spacetimes. We considered the specific
	example of primordial gravitational waves propagating through a
	\textbf{bouncing quantum Friedmann universe}. 
	
We showed that the dynamical variables, such as the scale
	factor or the amplitude of a gravitational wave, obtained from
	different internal clocks, evolve differently when compared in a clock-independent
	manner. These expectation values (background evolution)
	and mode functions of operators (perturbations), irrespective of the
	clock chosen, converge to a \textbf{unique evolution} for large 
	classically-behaving universes. This is the phase space
	domain in which unambiguous predictions can be made. Then we showed that for
	different clocks, the dynamics converges to the classical behavior at
	different times. In principle, there is no restriction on how far from
	the bounce the system must be in order to display the classical
	behavior. In practice, however, all the clocks considered were found
	to converge very quickly, allowing for unambiguous predictions shortly
	after the bounce.
	
	\section{Conclusions}
	
	Based on the above findings, we postulate that
	the physical predictions are only those predictions provided by any
	clock, which are not altered upon the clock's transformation. The fact
	that for large universes the semi-classical background dynamics
	and the quantum perturbation dynamics do not depend
	on the clock, implies the following: {\it Despite the fact
		that the dynamical variables are not Dirac observables, they provide
		physical predictions for large universes, which is precisely the regime 
		in which we observe the actual Universe}.
	
	Note, however, that the word {\it large} is never precisely
	defined. One could expect that, at least in principle, some clocks
	require times larger than the present age of the Universe to converge
	to the classical behavior. This, however, poses no problem to our
	interpretation as we simply exclude such clocks and retain only those
	that behave classically in the domain for which we make
	predictions. This may seem arbitrary and unjustified. We must,
	however, remember that, as a matter of fact, any semi-classical
	description of ordinary quantum mechanics is necessarily restricted
	to a limited set of observables, usually the simple ones, while more
	compound observables often display classically incompatible
	behavior (e.g. $\langle x \rangle^2 \not= \langle x^2 \rangle$).
	For similar reasons, we are allowed to choose only those
	clocks in which the dynamics of the relevant observables is
	classically consistent.
	
	On the one hand, we proved that the evolution of the expectation
	values of
	some observables constitute {\it physical} predictions of quantum
	cosmological models. On the other hand, the expectation values are not
	all that is measured in the large Universe. In other words, not all objects are classical in the large Universe. For instance, the position
	of an electron is a dynamical variable that can be measured in a
	lab. So, \textit{could the outcomes of such a measurement also be unambiguously predicted by a
	quantum cosmological model}? The answer is affirmative. Note that the mode function $\mu_k$, whose dynamics becomes unambiguous in large universe, determines the evolution of the operator  $\widehat{\mu}_k$ via Eq.~\eqref{operators}. This implies that the Heisenberg equation of motion encoded in Eq.~\eqref{modemu} becomes unambiguous too. Obviously, the evolution of perturbation in the Schr\"odinger picture must consequently become unique as well. Hence, ordinary quantum mechanics of perturbation modes is recovered in a large universe. This conclusions must also apply to electrons and, in general, 
	to all non-gravitational degrees of freedom.
	
	To better understand the origin of the emergence of ordinary quantum mechanics, notice that any clock transformation \eqref{CT} involves, by definition, only background variables. If the latter behave classically, the clock transformation is completely classical and amounts to a mere (in general, nonlinear) change of units of time. In Ref.~\cite{Malkiewicz:2017cuw}, it was demonstrated that the
	relational dynamics of {\it a quantum variable in a classical clock}
	is unambiguous in the sense that switching to another classical
	clock does not induce any clock effect. 
	
	Let us put to test our approach and our result by addressing a set of questions that were proposed in Ref.~\cite{Isham:1992ms} for assessing the
	completeness of any potential solution to the time problem.
	\begin{enumerate}[1.]
\item \textit{How should the notion of time be
		re-introduced into the quantum theory of gravity?} 
	
	Our approach relies on evolving internal variables called clocks. We
	express the dynamics of the dynamical variables in terms of these
	clocks.
	
\item\textit{ In particular, should attempts to identify
		time be made at the classical level, i.e., before quantization, or
		should the theory be quantized first?}
	
	
	In our approach, we first reduce the Hamiltonian formalism based on a
	selected clock, next we quantize the reduced formalism as if the clock
	was an external and absolute time. However, it is neither external nor
	absolute. The instantaneous value of the clock determines the
	instantaneous physical state of the system. Switching to another clock
	entails a change in the physical interpretation of the clock and the
	entire state of the system.
	
\item \textit{Can "time" still be regarded as a
		fundamental concept in a quantum theory of gravity, or is its status
		purely phenomenological? [...] }

In our approach, there is no fundamental "time". The fundamental concept is
"change" or "evolution", meaning we merely need to assume that the
3+1 split of the underlying geometry imposes an ordered set of
hypersurfaces. As we showed in this paper, extracting
dynamical predictions from such a formalism is a subtle issue. The
clocks serve as tools for deriving the predictions. Once a class of
clocks converges to a unique dynamics, any one of them can be treated
analogously and deserves the qualification of "time", and any quantum
dynamical variable becomes described in them by a unique Schr\"odinger 
equation. This is how ordinary quantum mechanics emerges.

\item \textit{If "time" is only an approximate concept,
	how reliable is the rest of the quantum-mechanical formalism in
	those regimes where the normal notion of time is not applicable? In
	particular, how closely tied to the concept of time is the idea of
	probability? [...] }

The quantum-mechanical description in the regime where different
clocks exhibit different dynamics is an essential part of our
theory. It describes the deterministic evolution of the
system. However, this regime does not seem to allow for any meaningful
dynamical interpretation in terms of relational change.
Although we have not explicitly addressed
this question in the present work, our approach permits to do
it.
\end{enumerate}
\bigskip

To conclude, one can mention that the chosen "clock" degrees
of freedom, although perfectly acceptable as such in the 
classical framework of general relativity, are arguably not
in the quantum regime. They do not qualify as actual
clock since, along the quantum trajectory, they yield
a non monotonic change of time variable, in other words,
they provide different hypersurface orderings. This might
be cured by adding to the classical clock transformation
\eqref{CT} a quantum term that need be identified. One may
also argue that we are insisting upon using a trajectory
to define the background evolution, while some might
insist upon the fact that there is no such thing as a
trajectory in quantum mechanics.

In any case, it is interesting to note that whichever
of the possibilities above happens to be valid, the
critical point that is made here is that even though
the quantum-dominated phase is indeed ill-defined both
at the background and perturbation levels from the
point of view of time development, the asymptotic
regimes end up being unique. As a result, setting
well-motivated initial conditions in the classical
past, one gets unambiguous physical predictions for
the classical future in which we happen to perform
the measurements. In other words, we have
shown that the lack of predictability in the quantum
regime does not exclude the fact that the theory permits
meaningful physical predictions that can be tested
with observations.

%% file: Chapters/Chapter5.tex
\chapter*{Discussion and future prospects}

\label{Chapter5}
\addcontentsline{toc}{chapter}{Discussion and future prospects}
\markboth{Discussion and future prospects}{}

In this work, we studied in detail the applications of the Dirac method for constrained systems. As explained in Ch. \ref{Chapter3}, this method allows us to choose an initial gauge, usually one that simplifies our computations, and then compute the physical Hamiltonian. However, the spacetime reconstruction  involves the full Hamiltonian, i.e. before setting any gauge, thus leading to heavy computations. 
In Ch. \ref{Chapter2} we were able to refine this approach. We derive the full space of gauge-fixing conditions and formulated a new and simpler method to reconstruct the spacetime for any gauge.
This technical result allows us to get a deeper physical insight into the relation between the abstract gauge invariant quantities and the full spacetime.

For instance, the application of this method to cosmological perturbation theory around the anisotropic Bianchi I universe, showed that, unlike in the isotropic universes where gravitational waves are unambiguously given by tensor modes, in anisotropic settings, they can be given by scalar modes in some gauges. In the future we would like to use this finding to rethink the nature of the gravitational waves {in broader terms}. This could be of both theoretical and experimental interest.

This leads us to wonder what would happen if we were to apply this method to a more involved metric such as a Bianchi IX universe. This metric, although still homogeneous and anisotropic, describes a curved spacetime with non-linear interactions between the scale factors.  The application to a Bianchi IX model presents technical challenges, however could allow us to study promising new gauges and  bring us insight on both the system under investigation and the procedure. A first step towards this would be to consider a Bianchi II universe, which constitutes a more complex spacetime than Bianchi I by introducing a non-null torque, but still a far simpler universe than Bianchi IX. We would like to pursue this research in the near future.

The next step is to quantize the obtained Hamiltonian for a perturbed Bianchi I universe. This step seems to be even more intriguing considering the interesting results obtained by the quantization of the Hamiltonian in a FLRW background. The use of Dirac observables, combined with the fact that they are not dynamical, has allowed us to study the evolution of both background and perturbation trajectories for different clocks. This has been possible since the form of the Dirac observables remains invariant for different clocks, allowing us to depict the behaviour of such trajectories in different clocks. We found out that the discrepancies due to the change of clocks are only present around the bounce, i.e. at a time in which the universe is in a quantum regime. At the classical limit, i.e. for large scale factors, all the trajectories were eventually converging.

Further analyses of this method can be done by assuming a full quantum background instead of a semi-classical one \cite{Bergeron:2024art}. It could be also interesting to see if there is any relation between the quantum states and the clock, that is, studying if there are any preferred internal clocks for a given quantum state. Also, we would like to investigate if the matter content can affect the clock effect, e.g. if stiffer matter can suppress the clock effect. Moreover it would be interesting to investigate the time of convergence of these trajectories, from which many questions arise naturally: Can the time of convergence for the scale factor be arbitrarily large, or is there a limit? If there is no limit, and we find that it is always possible to find a delay function such that the convergence is infinitely delayed, does this mean that quantum behaviour persists into the classical regime? If so, should we be able to detect such discrepancies in the clocks with sufficient sensitivity?

Additionally, if the discrepancies due to the clock change are visible only in a quantum regime, can we estimate at what value scale factor the universe transitions from being quantum to classical?

The last question seems to have a possible answer as the limit between the two regimes should be found around the Planck scale. This is indeed what was found in \cite{Ashtekar:2005qt}, where, similarly, the Dirac observables were employed to construct  a semi-classical space which was consequently studied with numerical method. In this case the authors considered a semi-classical state for large universes which was then evolved back in time near the classical singularity, i.e. around the Planck regime where the quantum effects dominate.
They found that the semi-classical solution extends beyond the singularity, meaning that the singularity is not a final stop for the physical space. Similar interesting results have been obtained in the context of loop quantum cosmology in \cite{Mielczarek:2008zv,Husain:2011tm}.

The model presented in \cite{Ashtekar:2005qt} was a toy model based on minisuperspaces \cite{Klauder:1972je,Misner:1972js} with no matter content, aiming to prove that the singularity could be avoided. In our case, the singularity has been avoided by assuming a bounce scenario. This implies that, naturally, the background and perturbed trajectories are free to go from the time before the bounce to the one after the bounce, i.e. from one classical regime to another without encountering a singularity.  Following \cite{Ashtekar:2005qt}, it could be interesting to see if, by changing the potential in the work in Ch. \ref{Chapter4} so as to include a singularity, the trajectory were to still follow a path across the singularity, which would act like "a bridge" between the two classical regimes, to cite the words used in  \cite{Ashtekar:2005qt}. 

The applications of the Dirac method and the quantization model introduced in this context, therefore, offer numerous opportunities for gaining valuable insights into the dynamics of the universe, particularly in the vicinity of singularities that one expects to smooth out within the quantum gravity regime. 

%% file: Appendices/AppendixBianchi.tex

\chapter{Bianchi I in perturbation theory} 

\label{Appendix-Bianchi} 

\section{Extra Hamiltonian}\label{NRHam0}

Since the canonical transformation \eqref{nperts} depends on time through the new tensorial basis \eqref{baseA}, it generates an extra term in the full Hamiltonian (\ref{hamtot}). The extra term depends on the gravitational variables only, that is, $(\delta {q}_{n},\delta {\pi}^{n})$, $n=1,\dots,6$, and reads

\begin{align}\label{fullext}
	&H_{ext}=
	\delta q_1 \bigg(\frac{2  (3 P_{kk}	-(TrP))}{3 a^3}\delta \pi_2
	+\frac{2 \sqrt{2}  P_{kv}}{a^3}\delta \pi_3
	+\frac{2 \sqrt{2}  P_{kw}}{a^3}\delta \pi_4
	+\frac{\sqrt{2}\left( P_{vv}- P_{ww}\right)}{ a^3}\delta \pi_6 
	+\frac{2 \sqrt{2}P_{vw}}{a^3} \delta \pi_5
	\bigg)
	\nonumber
	\\
	\nonumber
	&
	\quad
	+\delta q_2 \bigg(\frac{ 2(3 P_{kk}	- (TrP))}{9 a^3}\delta \pi_1
	+\frac{ (3 P_{kk}	-(TrP))}{3 a^3}\delta \pi_2
	-\frac{2 \sqrt{2}  P_{kv}}{3 a^3}\delta \pi_3
	-\frac{2 \sqrt{2}  P_{kw}}{3 a^3}\delta \pi_4
	\nonumber
	\\
	\nonumber
	&
	\qquad
	-\frac{2 \sqrt{2} P_{vw}}{3 	a^3}\delta \pi_5
	+\frac{ \sqrt{2} \left( P_{ww}	- P_{vv}\right)}{3 a^3}\delta \pi_6
	\bigg)
	\\
	\nonumber
	&
	\quad
	+\delta q_3\bigg(\frac{2 \sqrt{2} P_{kv}}{3 a^3} \delta \pi_1
	+\frac{2 \sqrt{2}  P_{kv}}{a^3}\delta \pi_2
	+\frac{ ( (TrP)	-3 P_{ww})}{3 a^3}\delta \pi_3
	+\frac{P_{vw}}{a^3}\delta \pi_4
	\bigg)
	\\
	&
	\quad
	+\delta q_4 \bigg(\frac{2 \sqrt{2}  P_{kw}}{3 a^3}\delta \pi_1
	+\frac{2 \sqrt{2}  P_{kw}}{a^3}\delta \pi_2
	+\frac{P_{vw}}{a^3}\delta \pi_3
	+\frac{ ( (TrP)-3 P_{vv})}{3 a^3}\delta \pi_4
	\bigg)
	\\
	\nonumber
	&
	\quad
	+\delta q_5 \bigg(
	+\frac{2 \sqrt{2} P_{vw}}{3 a^3}\delta \pi_1
	-\frac{\sqrt{2} P_{vw}}{a^3}\delta \pi_2
	+\frac{2  P_{kw}}{a^3}\delta \pi_3
	+\frac{2  P_{kv}}{a^3}\delta \pi_4
	+\frac{ ( (TrP)-3 P_{kk})}{3 a^3}\delta \pi_5
	\bigg)
	\\
	\nonumber
	&
	\quad
	+\delta q_6 \bigg(
	\frac{\sqrt{2} \left( P_{vv}- P_{ww}\right)}{3 a^3}\delta \pi_1
	-\frac{ \left( P_{vv}	- P_{ww}\right)}{\sqrt{2} a^3}\delta \pi_2
	+\frac{2  P_{kv}}{a^3}\delta \pi_3
	-\frac{2  P_{kw}}{a^3}\delta \pi_4
	+\frac{ ((TrP)-3 P_{kk})}{3 a^3}\delta \pi_6
	\bigg).
\end{align}

In the spatially flat slicing gauge, the extra Hamiltonian takes the form
\small{\begin{align}\label{ext}
		\begin{aligned}
			H_{ext}=-a^{-3}\delta q_5\left(\frac{2\sqrt{2}}{3}P_{vw}\delta\pi^1-\sqrt{2}P_{vw}\delta\pi^2+2P_{kw}\delta\pi^3+2P_{vk}\delta\pi^4+\left(\frac{Tr P}{3}-P_{kk}\right)\delta\pi^5\right)\\
			-a^{-3}\delta q_6\left(\frac{\sqrt{2}}{3}(P_{vv}-P_{ww})\delta\pi^1-\frac{P_{vv}-P_{ww}}{\sqrt{2}}\delta\pi^2+2P_{kv}\delta\pi^3-2P_{kw}\delta\pi^4+\left(\frac{Tr P}{3}-P_{kk}\right)\delta\pi^6\right).
		\end{aligned}
\end{align}}

\section{Second-order constraint}\label{NRHam}

The second order constraint (\ref{2cx}) expressed in the new tensorial basis (\ref{baseA}) and supplemented with the extra term (\ref{fullext}) generated by the corresponding time-dependent canonical transformation reads

{\small
	\begin{align}
		&\mathcal{H}_{0}^{(2)}+H_{ext}=
		-\frac{a \delta \pi_1^2}{6}
		+\frac{3 a \delta\pi_2^2}{2}
		+a \delta \pi_3^2
		+a \delta \pi_4^2
		+a \delta \pi_5^2
		+a \delta \pi_6^2
		+\frac{\delta\pi_\phi ^2}{2 a^3}
		\nonumber
		\\
		&
		+\delta\phi ^2\bigg(
		\frac{ a^3V_{,\phi\phi}}{2}
		+\frac{k^2a}{2}
		\bigg) 
		+
		\delta q_1^2\bigg(\frac{9 V a^7}{8}
		-\frac{3 V}{4 a}
		-\frac{k^2}{2 a^3}
		+\frac{15p_\phi^2}{16 a^7}
		-\frac{47 (Tr P)^2}{16a^7}
		-\frac{(Tr P^2)}{8 a^7}\bigg) 
		\nonumber
		\\
		&
		+\delta q_2^2
		\bigg(-\frac{k^2}{18 a^3}
		-\frac{P_{kk}^2}{6 a^7}
		-\frac{2 P_{kv}^2}{3a^7}
		-\frac{2 P_{kw}^2}{3 a^7}
		+\frac{p_\phi^2}{12 a^7}
		-\frac{V}{6 a}
		+
		\frac{P_{kk} (Tr P)}{3	a^7}
		-\frac{5(Tr P)^2}{36 a^7}
		+\frac{5 (Tr P^2)}{18 a^7}\bigg)
		\nonumber
		\\
		&
		+\delta q_3^2
		\bigg(
		\frac{p_\phi^2}{8 a^7}
		-\frac{(Tr P)^2}{8a^7}
		+\frac{P_{kk} P_{vv}}{a^7}
		-\frac{V}{4 a}
		+\frac{(Tr P^2)}{4a^7}
		\bigg)
		+\delta q_4^2 
		\bigg(
		\frac{p_\phi^2}{8 a^7}
		-\frac{(Tr P)^2}{8 a^7}
		+\frac{P_{kk} P_{ww}}{a^7}
		-\frac{V}{4 a}
		+\frac{(Tr P^2)}{4 a^7}
		\bigg)
		\nonumber
		\\
		&
		+\delta q_5^2 
		\bigg(
		\frac{k^2}{4 a^3}
		+\frac{p_\phi^2}{8a^7}
		-\frac{(Tr P)^2}{8 a^7}
		+\frac{P_{vv}P_{ww}}{a^7}
		-\frac{V}{4 a}
		+\frac{(Tr P^2)}{4 a^7}
		\bigg)
		\nonumber
		\\
		&
		+\delta q_6^2
		\bigg(
		\frac{k^2}{4 a^3}
		+\frac{(P_{vv}+P_{ww})^2}{4a^7}
		-\frac{P_{vw}^2}{a^7}
		+\frac{p_\phi^2}{8 a^7}
		-\frac{(Tr P)^2}{8 a^7}
		-\frac{V}{4 a}
		+\frac{(Tr P^2)}{4 a^7}
		\bigg)
		\nonumber
		\\
		&
		+
		\delta q_1\bigg[
		-\frac{\sqrt{2}	P_{kv} \delta \pi_3}{a^3}
		-\frac{\sqrt{2} P_{kw} \delta \pi_4}{a^3}
		-\frac{\sqrt{2} P_{vw} \delta \pi_5}{a^3}
		+\left(\frac{P_{ww}}{\sqrt{2} a^3}
		-\frac{P_{vv}}{\sqrt{2}	a^3}\right) \delta \pi_6
		-\frac{3 p_\phi \delta\pi_\phi }{2a^5}
		-\frac{\delta \pi_1 (Tr P)}{6 a^3}
		\nonumber
		\\
		&
		+\delta \pi_2
		\bigg(
		\frac{(Tr P)}{6 a^3}
		-\frac{P_{kk}}{2 a^3}
		\bigg)
		+\delta q_3
		\bigg(
		\frac{\sqrt{2} P_{ww}P_{kv}}{a^7}
		-\frac{7 (Tr P) P_{kv}}{\sqrt{2}	a^7}
		-\frac{\sqrt{2} P_{kw} P_{vw}}{a^7}
		\bigg)
		\nonumber
		\\
		&
		+\delta q_4
		\bigg(
		\frac{\sqrt{2} P_{vv}P_{kw}}{a^7}
		-\frac{7 (Tr P) P_{kw}}{\sqrt{2}	a^7}
		-\frac{\sqrt{2} P_{kv} P_{vw}}{a^7}
		\bigg)
		\nonumber
		\\
		&
		+\delta q_5
		\bigg(
		\frac{\sqrt{2} P_{kk}	P_{vw}}{a^7}
		-\frac{7 	(Tr P)P_{vw}}{\sqrt{2} a^7}
		-\frac{\sqrt{2} P_{kv} P_{kw}}{a^7}
		\bigg)
		\nonumber
		\\
		&
		+\delta q_6
		\bigg(
		-\frac{P_{kv}^2}{\sqrt{2} a^7}
		+\frac{P_{kw}^2}{\sqrt{2}a^7}
		-\frac{(P_{vv}^2-P_{ww}^2)}{\sqrt{2} a^7}
		-\frac{5(P_{vv}-P_{ww}) (Tr P)}{2 \sqrt{2} a^7}
		\bigg)
		\nonumber
		\\
		&
		+\delta q_2 
		\bigg(
		\frac{k^2}{3a^3}
		-\frac{P_{kk}^2}{a^7}
		-\frac{P_{kv}^2}{a^7}
		-\frac{P_{kw}^2}{a^7}
		+\frac{5 (Tr P)^2}{6 a^7}
		-\frac{5 P_{kk} (Tr P)}{2a^7}
		+\frac{(Tr P^2)}{3 a^7}
		\bigg)
		+\frac{3 a V_{,\phi} \phi}{2}
		\bigg]
		\nonumber
		\\
		&
		+\delta q_3 \bigg[
		\frac{2 P_{kk} P_{vw}\delta q_4}{a^7}
		+\frac{2 P_{kw} P_{vv}\delta q_5}{a^7}
		+\bigg(\frac{P_{kv} [(TrP)-P_{kk}]}{a^7}
		-\frac{2 P_{kw}	P_{vw}}{a^7}
		\bigg) \delta q_6
		\nonumber
		\\
		&
		-\frac{\sqrt{2}	P_{kv} \delta \pi_1}{3 a^3}
		-\frac{\sqrt{2} P_{kv} \delta\pi_2}{a^3}
		+\frac{P_{vw} \delta \pi_4}{a^3}
		+\frac{2 P_{kw} \delta \pi_5}{a^3}
		+\frac{2 P_{kv} \delta \pi_6}{a^3}
		+\delta \pi_3 \left(\frac{2(Tr P)}{3 a^3}
		-\frac{P_{ww}}{a^3}
		\right)
		\bigg]
		\nonumber
		\\
		&
		+\delta q_4 \bigg[
		\frac{2 P_{kv}P_{ww} \delta q_5}{a^7}
		+\bigg(
		\frac{2P_{kv} P_{vw}}{a^7}
		-\frac{P_{kw} [(TrP)-P_{kk}]}{a^7}
		\bigg)
		\delta q_6
		-\frac{\sqrt{2} P_{kw} \delta \pi_1}{3 a^3}
		-\frac{\sqrt{2}	P_{kw} \delta \pi_2}{a^3}
		+\frac{P_{vw} \delta \pi_3}{a^3}
		\nonumber
		\\
		&
		+\frac{2P_{kv} \delta \pi_5}{a^3}
		-\frac{2 P_{kw} \delta \pi_6}{a^3}
		+\delta\pi_4 \bigg(
		\frac{2 (Tr P)}{3 a^3}
		-\frac{P_{vv}}{a^3}
		\bigg)\bigg]
		\nonumber
		\\
		&
		+\delta q_5
		\bigg[
		\frac{(P_{vv}-P_{ww}) P_{vw}}{a^7}
		\delta q_6
		-\frac{\sqrt{2} P_{vw} \delta \pi_1}{3a^3}
		-\frac{\sqrt{2} P_{vw} \delta \pi_2}{a^3}
		+\delta \pi_5
		\bigg(
		\frac{2 (Tr P)}{3 a^3}
		-\frac{P_{kk}}{a^3}\bigg)\bigg]
		\nonumber
		\\
		&
		+\delta q_6
		\bigg[
		-\frac{P_{vv}-P_{ww}}{3 \sqrt{2} a^3}
		\delta \pi_1
		-\frac{P_{vv}-P_{ww}}{\sqrt{2}a^3}
		\delta \pi_2
		+\delta \pi_6	\bigg(\frac{2 (Tr P)}{3 a^3}
		-\frac{P_{kk}}{a^3}
		\bigg)\bigg]
		\nonumber
		\\
		&
		+\delta q_2
		\bigg[
		\frac{P_{kk} \delta \pi_2}{a^3}
		+\frac{4 \sqrt{2} P_{kv} \delta \pi_3}{3 a^3}
		+\frac{4 \sqrt{2} P_{kw} \delta \pi_4}{3 a^3}
		-\frac{2 \sqrt{2}P_{vw} \delta \pi_5}{3 a^3}
		-\frac{\sqrt{2} (P_{vv}-P_{ww})}{3	a^3}
		\delta \pi_6
		+\delta \pi_1	\left(\frac{(Tr P)}{9 a^3}
		-\frac{P_{kk}}{3 a^3}\right)
		\nonumber
		\\
		&
		+\delta q_3\bigg(\frac{\sqrt{2} P_{kk} P_{kv}}{3 a^7}
		-\frac{2 \sqrt{2} P_{vv}P_{kv}}{3 a^7}
		+\frac{\sqrt{2} (Tr P) P_{kv}}{3 a^7}
		-\frac{2	\sqrt{2} P_{kw} P_{vw}}{3 a^7}\bigg)
		\nonumber
		\\
		&
		+\delta q_4 \bigg(\frac{\sqrt{2}
			P_{kk} P_{kw}}{3 a^7}-\frac{2 \sqrt{2} P_{ww} P_{kw}}{3
			a^7}+\frac{\sqrt{2} (Tr P) P_{kw}}{3 a^7}-\frac{2 \sqrt{2}
			P_{kv} P_{vw}}{3 a^7}\bigg)
		\nonumber
		\\
		&
		+\delta q_5 
		\bigg(
		\frac{4 \sqrt{2} P_{kv}	P_{kw}}{3 a^7}
		+\frac{\sqrt{2} P_{kk} P_{vw}}{a^7}
		-\frac{5\sqrt{2} P_{vw} (Tr P)}{3 a^7}
		\bigg)
		\nonumber
		\\
		&
		+\delta q_6
		\bigg(
		\frac{2 \sqrt{2} P_{kv}^2}{3 a^7}
		-\frac{2 \sqrt{2} P_{kw}^2}{3a^7}
		-\frac{\sqrt{2} (P_{vv}^2-P_{ww}^2)}{3 a^7}
		\nonumber
		\\
		&
		-\frac{P_{kk} (P_{vv}-P_{ww})}{\sqrt{2} a^7}
		+\frac{(P_{vv}-P_{ww}) (Tr P)}{3 \sqrt{2}	a^7}
		\bigg)\bigg]
		.
	\end{align}
	\par
}

\section{Geometric quantities}\label{geoquant}

The canonical perturbation variables can be used to express geometric quantities. The most useful ones are the Ricci scalar,
\begin{equation}\label{deltaRapp}
	\delta({^3}R)
	=
	2a^{-4}
	k^2
	\left(
	\delta q_1
	-\frac{1}{3} \delta q_2
	\right),
\end{equation}
the energy density of the scalar field,
\begin{align}
	\delta \rho=\frac{\mathcal{H}_{m,0}}{\sqrt{q}}\bigg|^{(1)}=- \frac{5 p_\phi^2}{4 a^8} \delta q_1+V_{,\phi} \delta \phi+\frac{p_\phi}{a^6} \delta \pi_\phi,
\end{align}
and the two scalar modes of the shear,
\begin{align}\begin{split}
		\delta\sigma_1&
		=	
		\frac{2}{9}a^{-2} K_2\delta q_2
		+\frac{1}{3}a^{-2} K_3\delta q_3
		+\frac{1}{3}a^{-2} K_4\delta q_4
		+\frac{1}{3}a^{-2} K_5\delta q_5
		+\frac{1}{3}a^{-2} K_6\delta q_6,\\
		\delta\sigma_{2}&
		=
		\frac{3}{2}a \delta \pi_2
		+\frac{3}{4} a^{-3}\delta q_1 
		\left(
		P_{kk} 
		-
		\frac{1}{3}(Tr P)
		\right)
		+ a^{-3}
		P_{kk} \delta q_2
		+\frac{1}{\sqrt{2}}a^{-3} P_{kv} \delta q_3
		\\&\qquad
		+\frac{1}{\sqrt{2}}a^{-3} P_{kw} \delta q_4
		+\sqrt{2}a^{-3} P_{vw} \delta q_5
		-\frac{1}{\sqrt{2}}a^{-3}( P_{vv}-P_{ww}) \delta q_6,
	\end{split}
\end{align}
where $K_n=K_{ab}A^{ab}_n=a^{-3}\left(\bar{\pi}_{ab}A^{ab}_n-\frac{1}{2}\bar{q}_{ab}A^{ab}_n (Tr P) \right)$ are the components of the zeroth-order extrinsic curvature. 

\section{Physical Hamiltonian}\label{Hfin}

The coefficients in the physical Hamiltonian \eqref{hfin} read:

{\footnotesize \begin{align}\label{HfinCoeff}
		\tilde{U}_\phi&=
		\frac{(Tr P) p_\phi^2}{4 P_{kk} a^3}
		-\frac{3 p_\phi^2}{8a^3}
		-\frac{p_\phi V_{,\phi} a^3}{2 P_{kk}}
		+\frac{V_{,\phi\phi} a^3}{2};
		\nonumber
		\\
		\tilde{U}_5
		&=
		\frac{(Tr P)^2}{8a^7}
		+\frac{(Tr P^2)}{4 a^7}
		-\frac{( Tr P) P_{kk}}{2 a^7}
		+\frac{P_{kk}^2}{4a^7}
		+\frac{P_{kv}^2}{a^7}
		+\frac{P_{kw}^2}{a^7}
		-\frac{(P_{vv}-P_{ww})^2}{4a^7}
		\nonumber
		\\&
		\nonumber
		\quad
		+\frac{2P_{kv}P_{kw}P_{vw}}{a^7P_{kk}}
		-\frac{( Tr P) P_{vw}^2}{2 a^7P_{kk}}
		+\frac{P_{vw}^2}{4 a^7}
		+\frac{p_\phi^2}{8 a^7}
		-\frac{V}{4 a};
		\\
		\nonumber
		\tilde{U}_6
		&=
		\frac{(Tr P^2)}{4 a^7}
		+\frac{(Tr P)^2}{8a^7}
		-\frac{ (Tr P)P_{kk}}{2 a^7}
		+\frac{P_{kk}^2}{4 a^7}
		+\frac{P_{kv}^2}{a^7}
		+\frac{P_{kw}^2}{a^7}
		-\frac{P_{vw}^2}{a^7}
		\\&
		\nonumber
		\quad
		+\frac{(P_{kv}^2-P_{kw}^2)(P_{vv}-P_{ww})}{2a^7P_{kk}}
		-\frac{(Tr P)(P_{vv}-P_{ww})^2}{8a^7P_{kk}}
		+\frac{(P_{vv}-P_{ww})^2}{16a^7}
		+\frac{p_\phi^2}{8 a^7}
		-\frac{V}{4 a};
		\\
		\nonumber
		\tilde{C_1}&=
		-\frac{(Tr P)P_{vw}(P_{vv}-P_{ww})}{2a^7 P_{kk}}+\frac{5P_{vw}(P_{vv}-P_{ww})}{4a^7}+\frac{P_{vw}(P_{kv}^2-P_{kw}^2)}{a^7 P_{kk}}
		\\&
		\nonumber
		\quad
		+\frac{P_{kv} P_{kw}(P_{vv}-P_{ww})}{a^7P_{kk}};
		\\
		\tilde{C}_2&=
		\frac{\sqrt{2} P_{kv} P_{kw} p_\phi}{a^5	P_{kk}}
		-\frac{ P_{vw} p_\phi}{2 \sqrt{2} a^5}
		-\frac{a P_{vw} V_{,\phi}}{\sqrt{2}	P_{kk}};
		\\
		\nonumber
		\tilde{C}_3&=
		\frac{ (P_{kv}^2-P_{kw}^2)p_\phi}{\sqrt{2} a^5 P_{kk}}
		-\frac{(P_{vv}-P_{ww})p_\phi}{4 \sqrt{2} a^5}
		-\frac{a (P_{vv}-P_{ww})V_{,\phi}}{2 \sqrt{2} P_{kk}};
		\\
		\nonumber
		C_{\phi\phi}&=-\frac{p_{\phi}^2}{2a^3 P_{kk}} ;
		\\
		\nonumber
		C_{55}&=\frac{1}{a^3}\bigg[\frac{2}{3}(Tr P)-P_{kk}-\frac{2P_{vw}^2}{P_{kk}}\bigg];
		\\
		\nonumber
		C_{66}&=\frac{1}{a^3}\bigg[\frac{2}{3}(TrP)-P_{kk}-\frac{(P_{vv}-P_{ww})^2}{2P_{kk}}\bigg];
		\\
		\nonumber
		C_{5\phi}&=-\frac{P_{vw} p_{\phi}}{a^3 P_{kk}};
		\\
		\nonumber
		C_{6\phi}&=-\frac{(P_{vv}-P_{ww})p_\phi}{2 a^3 P_{kk}};
		\\
		\nonumber
		C_{56}&=-\frac{P_{vw}(P_{vv}-P_{ww})}{a^3P_{kk}}.
\end{align}}

\section{Dynamics of the operator $P$}\label{Appbackeom}
Making use of Eqs (\ref{backeom-pa}) and (\ref{fermilaw}) we find the dynamics of the components of the operator $P$ in the Fermi-Walker basis:
\begin{align}\begin{aligned}
		\frac{\ud}{\ud t}P_{kk}=a^{-3}\left(-a^6V-2P_{kv}^2-2P_{kw}^2\right),
		&~~\frac{\ud}{\ud t}P_{vv}=a^{-3}\left(-a^6V+2P_{kv}^2\right),
		\\
		\frac{\ud}{\ud t}P_{ww}=a^{-3}\left(-a^6V+2P_{kw}^2\right),
		&~~\frac{\ud}{\ud t}P_{vw}=2a^{-3}P_{kw}P_{kv}
		,
		\\
		\frac{\ud}{\ud t}P_{kv}=a^{-3}\left(P_{kk} P_{kv}-P_{kv}P_{vv}-P_{kw}P_{vw}\right),
		&~~
		\frac{\ud}{\ud t}P_{kw}=a^{-3}\left(P_{kk} P_{kw}-P_{kw}P_{ww}-P_{kv}P_{vw}\right).
	\end{aligned}
\end{align}

\section{Final Hamiltonian in the M-S variables}\label{MSAppgen}

Below the coefficients in the final Hamiltonians (\ref{HBI}), (\ref{BIn}) and (\ref{HBIgen}) are given. In case of a single scalar field there is a unique field label $I=J$ which is omitted in the Hamiltonian.

\begin{align}
	&U_{\phi^I}=\frac{p_{\phi^I}^2(P_{kv}^2 + P_{kw}^2 +P_{kv}P_{kw}+P_{vw}(P_{kv}+P_{kw})-P_{vw}^2+a^6V)}{a^4 P_{kk}^2}-\frac{2a^6p_{\phi^I}V_{,\phi^I}}{a^4 P_{kk}}
	\nonumber
	\\
	&+\frac{(Tr P)^2-18a^6(V-2V_{,\phi^I\phi^I})}{36 a^4};\nonumber\\
	&U_5=\frac{(Tr P)^2+72 \left(P_{kv}^2+P_{kv} P_{kw}+P_{kw}^2+ P_{vw} (P_{kv}+P_{kw})-P_{vw}^2\right)+18 \sum_I p_{\phi^I}^2+18 a^6 V}{36a^4}\nonumber\\
	&+\frac{P_{vw}^2 (4 P_{kv}^2 + 4 P_{kw}^2 - 4 P_{vw}^2 - (P_{vv} - P_{ww})^2 -P_{kk}^2- \sum_Ip_{\phi^I}^2 + 
		2 a^6 V)}{a^4P_{kk}^2}-2\frac{\frac{(P_{vv}-P_{ww})^2}{4}}{a^4}\nonumber\\
	&+\frac{54 P_{kk}^4 - 36 P_{kk}^3 (Tr P)}{36a^4P_{kk}^2}+\frac{2 P_{vw} (8 P_{kv} P_{kw} + P_{vw} (Tr P))}{a^4P_{kk}};\nonumber\\
	&U_6=\frac{(Tr P)^2 + 72 (P_{kv}^2 + P_{kv} P_{kw} +P_{kw}^2+ P_{vw}( P_{kv}+P_{kw} )-P_{vw}^2) + 
		18 \sum_Ip_{\phi^I}^2 + 18 a^6 V}{36a^4}\nonumber\\
	&+\frac{\frac{(P_{vv} - P_{ww})^2}{4} (4 P_{kv}^2 + 4 P_{kw}^2 - 4 P_{vw}^2- (P_{vv} - P_{ww})^2-P_{kk}^2 -
		\sum_Ip_{\phi^I}^2 + 2 a^6 V)}{a^4P_{kk}^2}-2\frac{P_{vw}^2}{a^4}\nonumber\\
	&+\frac{54 P_{kk}^4 -36 P_{kk}^3(Tr P) }{36a^4P_{kk}^2}+\frac{ 2\frac{(P_{vv} - P_{ww})}{2} (4 P_{kv}^2 - 4 P_{kw}^2 +\frac{(P_{vv} - P_{ww})}{2}(Tr P))}{a^4P_{kk}};\nonumber\\
	&
	C_1=\frac{ 
		P_{vw} (P_{vv} - P_{ww}) (4 P_{kv}^2 + 4 P_{kw}^2 - 4 P_{vw}^2 - (P_{vv} - P_{ww})^2 - 
		\sum_Ip_{\phi^I}^2 + 2 a^6 V)}{a^4P_{kk}^2}+
	\nonumber
	\\
	&
	\frac{8(P_{kv}^2 - P_{kw}^2)P_{vw}+ 
		8 P_{kv} P_{kw} (P_{vv} - P_{ww}) + 2(Tr P) P_{vw} (P_{vv} - P_{ww})}{a^4P_{kk}}+\frac{P_{vw} (P_{vv} - P_{ww}) }{a^4};\nonumber\\
	&
	C_{2I}=\frac{P_{vw} p_{\phi^I} (4 P_{kv}^2 + 4 P_{kw}^2 - 4 P_{vw}^2 - (P_{vv} - P_{ww})^2 - 
		\sum_Ip_{\phi^I}^2 + 2 a^6 V) }{a^4P_{kk}^2}-\frac{3 P_{vw} p_{\phi^I}}{a^4}
	\nonumber \\
	&
	+\frac{8 P_{kv} P_{kw}p_{\phi^I} + 2(Tr P) P_{vw} p_{\phi^I} -4 a^6 P_{vw} V_{,\phi^I}}{a^4P_{kk}};\nonumber\\
	&
	C_{3I}=\frac{\frac{(P_{vv}-P_{ww})}{2} p_{\phi^I} (4 P_{kv}^2 + 4 P_{kw}^2 - 4 P_{vw}^2 - (P_{vv} - P_{ww})^2 - 
		\sum_Ip_{\phi^I}^2 + 2 a^6 V) }{a^4P_{kk}^2}-\frac{3 \frac{(P_{vv} - P_{ww})}{2} p_{\phi^I}}{a^4}
	\nonumber \\
	&
	+\frac{ p_{\phi^I}(4 P_{kv}^2- 4P_{kw}^2 + (Tr P) (P_{vv}-P_{ww})) -2a^6 (P_{vv}-P_{ww}) V_{,\phi^I}}{a^4P_{kk}};\nonumber\\
	&
	C_{IJ} (I\neq J)=\frac{ \left(4 a^6 V_{,\phi^I\phi^J}-3 p_{\phi^I} p_{\phi^J}\right)}{2 a^4}-\frac{2 a^6 (p_{\phi^J} V_{\phi^I}+p_{\phi^I} V_{\phi^J})}{ a^4 P_{kk}}-\frac{(Tr P) p_{\phi^I} p_{\phi^J}}{a^4 P_{kk}}
	\nonumber \\
	&
	-\frac{p_{\phi^J} p_{\phi^I} \left(-2 a^6 V-4 P_{kv}^2-4 P_{kw}^2+(P_{vv}-P_{ww})^2+4 P_{vw}^2+\sum_Ip_{\phi^I}^2\right)}{2 a^4 P_{kk}^2}.
\end{align}

\section{Dirac observables}\label{Dext}
This is a complete set of solutions to Eq. (\ref{defdir}). In case of a single scalar field the label $I$ is unique and can be omitted.
\begin{align}
	&{\delta D_1}
	=
	{\delta q _5}
	+
	2\sqrt{2}
	\frac{P_{vw}}{P_{kk}}
	{\delta q_1}
	-
	\frac{2\sqrt{2}}{3}
	\frac{P_{vw}}{P_{kk}}
	{\delta q _2}
	;
	\\&
	{\delta D_2}
	=	
	{\delta q _6}
	+
	\frac{\sqrt{2}}{P_{kk}}
	\big(
	P_{vv}
	-
	P_{ww}
	\big)
	{	\delta q_1}
	-
	\frac{\sqrt{2}}{3}
	\frac{1}{P_{kk}}
	\big(
	P_{vv}
	-
	P_{ww}
	\big)
	{	\delta q _2}
	;
	\\&
	{\delta D_3}
	=	
	{	\delta \pi_1}
	+
	\bigg\{
	\frac{1}{P_{kk}}
	\bigg[
	2
	k^2
	-
	\frac{1}{2}
	a^{-4}
	(Tr P^2)
	+
	\frac{3}{4}
	a^{-4}
	(Tr P)^2
	+4	a^{-4}(P_{kv}^2+ P_{kw}^2)
	+\frac{3}{4}
	a^{-4}
	\sum_Ip^2_{\phi^I}
	-
	\frac{3}{2}
	a^2
	V(\phi)	
	\bigg]
	\nonumber\\
	&
	\qquad\quad
	+3
	a^{-4}
	P_{kk} 
	-\frac{5}{2}
	a^{-4}
	(Tr P) 
	\bigg\}
	{	\delta q_1}
	\nonumber\\
	&
	\qquad\quad
	+
	\frac{1}{3P_{kk}}
	\bigg[
	-
	2
	k^2
	+
	\frac{1}{2}
	a^{-4}
	(Tr P^2)
	-
	\frac{3}{4}
	a^{-4}
	(Tr P )^2
	+
	a^{-4}
	(Tr P)
	P_{kk}
	-\frac{3}{4}
	a^{-4}
	\sum_Ip^2_{\phi^I}
	+
	\frac{3}{2}
	a^2
	V(\phi)
	\nonumber\\
	&
	\qquad\quad
	-
	4
	a^{-4}
	(
	P_{kv}^2
	+
	P_{kw}^2)
	\bigg]
	{	\delta q _2}
	+
	{\sqrt{2}}
	a^{-4}
	P_{kv}
	{	\delta q _3}
	+
	{\sqrt{2}}
	a^{-4}
	P_{kw}
	{\delta q _4};
	\\
	&
	{\delta D_4}
	=
	{	\delta \pi_2}
	+
	\bigg\{
	\frac{1}{P_{kk}}
	\bigg[
	-\frac{2}{3}
	k^2
	+
	\frac{2}{3}
	a^{-4}
	(Tr P^2)
	-
	\frac{1}{2}
	a^{-4}
	(Tr P)^2
	-
	2a^{-4}
	P_k^2
	-
	\frac{4}{3}
	a^{-4}
	(	P_{kv}^2+P_{kw}^2)
	\bigg]
	\nonumber\\
	&
	\qquad\quad
	+
	a^{-4}
	\frac{1}{2} 
	P_{kk}
	+
	a^{-4} \frac{4}{3} 
	(Tr P )
	\bigg\}
	{	\delta q_1}
	\nonumber\\
	&
	\qquad\quad
	+
	\frac{1}{3P_{kk}}
	\bigg[
	\frac{2}{3}
	k^2
	+
	2a^{-4}
	P_{kk}^2
	-
	\frac{2}{3}
	a^{-4}
	(Tr P^2)
	-\frac{5}{6}
	a^{-4}
	(Tr P)
	P_{kk}
	\nonumber\\
	&
	\qquad\quad
	+
	\frac{1}{2}
	a^{-4}
	(Tr P)^2
	+	\frac{10}{3}
	a^{-4}
	(	P_{kw}^2
	+
	P_{kv}^2)
	\bigg]
	{	\delta q _2}
	-\frac{\sqrt{2}}{3}a^{-4}
	P_{kv}	
	{	\delta q _3}
	-\frac{\sqrt{2}}{3}a^{-4}
	P_{kw}
	{	\delta q _4};
	\\
	&
	{\delta D_5}=	
	{\delta \pi_3}
	-
	\frac{1}{P_{kk}}
	\bigg(
	2{\sqrt{2}
		a^{-4}
		(P_{kv}P_{vv}+P_{kw}P_{vw})
	}
	-
	\sqrt{2}
	a^{-4}
	(Tr P)
	P_{kv}
	\bigg)
	{	\delta q_1}
	\nonumber\\
	&
	\qquad\quad
	+
	\frac{1}{3P_{kk}}
	\bigg(
	2{\sqrt{2}}
	a^{-4}
	(P_{kv}P_{vv}+P_{kw}P_{vw})
	-
	\sqrt{2}
	a^{-4}
	(Tr P)
	P_{kv}
	\bigg)
	{	\delta q _2}
	+
	a^{-4}
	P_{kk} 
	{	\delta q _3};
	\\
	&
	{\delta D_6}=
	{\delta \pi_4}	
	-
	\frac{1}{P_{kk}}
	\bigg(
	2\sqrt{2}
	a^{-4}
	(P_{kv}P_{vw}+P_{kw}P_{ww})
	-
	\sqrt{2}
	a^{-4}
	(Tr P)
	P_{kw}
	\bigg)
	{	\delta q_1}
	\nonumber\\
	&
	\qquad\quad
	+
	\frac{1}{3P_{kk}}
	\bigg(
	2
	\sqrt{2}
	a^{-4}
	(P_{kv}P_{vw}+P_{kw}P_{ww})
	-
	\sqrt{2}
	a^{-4}
	(Tr P)
	P_{kw}
	\bigg)
	{	\delta q _2}
	+
	a^{-4}
	P_{kk} 
	{	\delta q _4};
	\\
	&
	{\delta D_7}
	=
	{\delta \pi_5}
	+
	\bigg[
	\frac{1}{P_{kk}}
	\bigg(
	2\sqrt{2}a^{-4}	
	P_{kw}P_{kv}	
	-\frac{1}{\sqrt{2}} 
	a^{-4}(Tr P)
	P_{vw} 	
	\bigg)
	+\frac{1}{\sqrt{2}} 
	a^{-4}
	P_{vw} 
	\bigg]
	{	\delta q_1}
	\nonumber\\
	&
	\qquad\quad
	+
	\frac{1}{3P_{kk}}
	\bigg(
	\frac{1}{\sqrt{2}} 
	a^{-4}(Tr P)
	P_{vw}
	-2
	\sqrt{2}
	a^{-4}
	(P_{kk}P_{vw}
	-
	P_{kw}P_{kv})
	\bigg)
	{	\delta q _2}
	\nonumber\\
	&
	\qquad\quad
	+
	a^{-4}
	P_{kw}
	{	\delta q _3}
	+
	a^{-4}
	P_{kv}
	{	\delta q _4};
	\\
	&
	{\delta D_8}
	=	
	{\delta \pi_6}
	+
	\bigg\{
	\frac{1}{P_{kk}}
	\bigg[
	-\sqrt{2}
	a^{-4}
	(P_{vv}^2-P_{ww}^2)
	+
	\frac{3}{2\sqrt{2}}
	a^{-4}
	(Tr P)
	(P_{vv}-P_{ww})
	+
	\sqrt{2}	a^{-4}	
	(P_{kv}^2-P_{kw}^2)
	\bigg]
	\nonumber\\
	&
	\qquad\quad
	-\frac{3}{2\sqrt{2}} a^{-4}
	(	P_{vv}
	-
	P_{ww})
	\bigg\}
	{	\delta q_1}
	+
	\frac{1}{3P_{kk}}
	\bigg[
	\sqrt{2}
	a^{-4}
	(P_{vv}^2-P_{ww}^2)
	-
	\frac{3}{2\sqrt{2}}
	a^{-4}
	(Tr P)
	(P_{vv}-P_{ww})
	\nonumber\\
	&
	\qquad\quad
	-
	\sqrt{2}	a^{-4}	
	(P_{kv}^2-P_{kw}^2)	
	\bigg]
	{	\delta q _2}
	+
	a^{-4}
	P_{kv}	
	{	\delta q _3}
	-
	a^{-4}
	P_{kw}
	{	\delta q _4};
	\\
	&
	{\delta D_{9I}}
	=
	{\delta \phi^I}
	+
	\frac{1}{P_{kk}}
	a^{-2}
	p_{\phi^I}
	{	\delta q_1}
	-
	\frac{1}{3P_{kk}}
	a^{-2}
	p_{\phi^I}
	{	\delta q _2};
	\\
	&
	{\delta D_{10I}}
	=	
	{\delta \pi_{\phi^I}}
	+
	\frac{1}{a^2}\bigg[
	\frac{1}{ P_{kk}}
	\bigg(
	\frac{1}{2}(Tr P)
	p_{\phi^I}
	-a^6
	V_{,\phi^I}
	\bigg)	
	-\frac{3}{2}
	p_{\phi^I}	
	\bigg]
	{	\delta q_1}
	+
	\frac{1}{3a^2P_{kk}}
	\bigg(a^6
	V_{,\phi^I}
	-\frac{1}{2}
	(Tr P)
	p_{\phi^I}\bigg)	
	{	\delta q _2};
\end{align}

\section{Geometric expressions for the Dirac observables}\label{geodirac}

The independent Dirac observables (\ref{inddir}) may be also given in terms of geometric quantities,
\begin{align}
	&\delta Q_1=
	\frac{1}{\sqrt{2} a}\delta q_5
	-
	\frac{a^3K_5}{2\sqrt{2}k^2\left(K_1-\frac{1}{3}K_2\right)}
	\delta R,
	\nonumber\\
	& \delta Q_2=
	\frac{1}{\sqrt{2} a}\delta q_6
	-\frac{a^3K_6}{2\sqrt{2}k^2\left(K_1-\frac{1}{3}K_2\right)}
	\delta R,
	\nonumber\\
	& \delta Q_3=a\delta\phi
	-\frac{a^2p_{\phi}}{4k^2\left(K_1-\frac{1}{3}K_2\right)}\delta R,
	\nonumber\\
	&\delta P_1=\sqrt{2} \delta K_5
	+\frac{K_5 K_6}{4\sqrt{2} a^2\left(K_1-\frac{1}{3}K_2\right)}\delta q_6
	+\frac{K_1\left(K_1-\frac{1}{3}K_2\right)
		-2\left(K_1-\frac{1}{3}K_2\right)^2 
		+\frac{1}{2}K_5^2}
	{ \sqrt{2} a^2\left(K_1-\frac{1}{3}K_2\right)}\delta q_5,
	\nonumber\\
	&+\mathcal{G}\left(\frac{aK_5}{\sqrt{2}},\frac{a^2K_3K_4}{2} \right)\delta R
	+\frac{K_5 p_{\phi}}{2a\sqrt{2} \left(K_1-\frac{1}{3}K_2\right)}\delta\phi
	\\
	& \delta P_2=\sqrt{2} \delta K_6
	+\frac{K_5 K_6}{4\sqrt{2} a^2\left(K_1-\frac{1}{3}K_2\right)}\delta q_5
	+\frac{ 4K_1\left(K_1-\frac{1}{3}K_2\right)
		-8\left(K_1-\frac{1}{3}K_2\right) ^2
		+\frac{1}{2}K_6^2}
	{4 \sqrt{2} a^2 \left(K_1-\frac{1}{3}K_2\right)}\delta q_6
	\nonumber\\
	&+\mathcal{G}\left(
	\frac{aK_6}{2\sqrt{2}},
	\frac{a^2(K_3^2-K_4^2)}{4}\right)
	\delta R+\frac{p_{\phi} K_6}{4\sqrt{2} a \left(K_1-\frac{1}{3}K_2\right)}\delta\phi
	\nonumber\\
	& \delta P_3=
	\frac{a^5}{p_{\phi}}\delta\rho
	+\frac{K_5 p_{\phi}}{2 a^3\left(K_1-\frac{1}{3}K_2\right)}\delta q_5
	+\frac{K_6 p_{\phi}}{8 a^3 \left(K_1-\frac{1}{3}K_2\right)}\delta q_6
	\nonumber\\&
	+\frac{12a^2K_1\left(K_1-\frac{1}{3}K_2\right)p_{\phi}
		+3 p_{\phi}^3
		-12a^7\left(K_1-\frac{1}{3}K_2\right)V_{,\phi}}
	{12a^2 p_{\phi}\left(K_1-\frac{1}{3}K_2\right)}\delta\phi
	\nonumber\\
	&+\frac{8a^3p_{\phi} K_1\left(K_1-\frac{1}{3}K_2\right)-
		a p_{\phi} \left(\frac{a^2}{2}K_6^2
		+2a^2K_5^2+p_{\phi}^2\right)
		+4 a^8 \left(K_1-\frac{1}{3}K_2\right) V_{,\phi}}{16 k^2a^2 \left(K_1-\frac{1}{3}K_2\right)^2}\delta R,
	\nonumber
\end{align}
where $\mathcal{G}(X,Y)=a\frac{-24a^2\left(K_1-\frac{1}{3}K_2\right)^2X -4a\left(K_1-\frac{1}{3}K_2\right) (6 Y -6a K_1 X) - 
	3 X (2 a^2K_5^2+ \frac{a^2}{2}K_6^2 + p_{\phi}^2)}{6k^2 P_{kk}^2}$. In the isotropic limit we obtain:

\begin{equation}
	\begin{gathered}
		\delta Q_1\rightarrow\frac{1}{\sqrt{2} a}\delta q_5,
		\quad
		\delta Q_2\rightarrow\frac{1}{\sqrt{2} a}\delta q_6,
		\quad
		\delta Q_3\rightarrow 
		a\delta\phi-\frac{a^2p_{\phi}}{4k^2K_1}\delta R,
		\\
		\delta P_1\rightarrow\sqrt{2} \delta K_5 -
		\frac{K_1}{ \sqrt{2} a^2}\delta q_5,
		\quad
		\delta P_2\rightarrow\sqrt{2} \delta K_6
		-\frac{K_1}{ \sqrt{2} a^2}\delta q_6,
		\\
		\delta P_3\rightarrow\frac{a^5}{p_{\phi}}\delta\rho
		+\frac{36a^2K_1^2p_{\phi}+9 p_{\phi}^3-36a^7K_1V_{,\phi}}
		{36 a^2 p_{\phi}K_1}\delta\phi
		+\frac{72a^3p_{\phi} K_1^2-9a p_{\phi}^3+36 a^8 K_1V_{,\phi}}{144 k^2 a^2K_1^2}\delta R.
	\end{gathered}
\end{equation} 

\section{Comparison of metric decompositions}\label{Uzantensor}

The perturbed metric decomposition used in \cite{Uzan} is given by

\begin{align}\label{hij}
	h_{ij}=2C\left(\gamma_{ij}+\frac{\sigma_{ij}}{\mathcal{H}}\right)+2\partial_i\partial_j E+2\partial_{(i}E_{j)}+2E_{ij},
\end{align}
whereas our decomposition reads
\begin{align}\label{deltagamma}
	h_{ij}=a^{-2}\delta q_{ij}=a^{-2}\delta q_n A^n_{ij}.
\end{align}
Expanding \eqref{hij} in the $A$ basis,
\begin{align}\label{hijcoeff}
	h_{ij}&
	=
	A^1_{ij}\left(2C-\frac{2}{3}E\right)
	+
	A^2_{ij}\left(2C\frac{\sigma_2}{\mathcal{H}}-2E\right)
	+
	A^3_{ij}\left(2C\frac{\sigma_3}{\mathcal{H}}+i\sqrt{2}E_v\right)
	+
	A^4_{ij}\left(2C\frac{\sigma_4}{\mathcal{H}}+i\sqrt{2}E_w\right)
	\nonumber\\&\qquad\qquad
	+
	A^5_{ij}\left(2C\frac{\sigma_5}{\mathcal{H}}+2E_5\right)
	+
	A^6_{ij}\left(2C\frac{\sigma_6}{\mathcal{H}}+2E_6\right),
\end{align}
and comparing with \eqref{deltagamma} we obtain the relation between the tensor modes:
\begin{align}
	E_5&
	=\frac{a^{-2}}{2} \delta q_5
	+
	a^{-2}\sqrt{2}
	\frac{P_{vw}}{P_{kk}}
	\left(\delta q_1 -\frac{1}{3}\delta q_2\right)
	,\quad
	E_6
	=\frac{a^{-2}}{2}\delta q_6
	+\frac{a^{-2}(P_{vv}-P_{ww})}{\sqrt{2}P_{kk}}\left(\delta q_1 -\frac{1}{3}\delta q_2\right),
\end{align}
which correspond to the gauge-invariant quantities $\delta Q_1$ and $\delta Q_2$ given in Eq. \eqref{inddir}. For the vector and scalar modes we find the following relations:
\begin{align}\begin{split}
		E_v&=-i\frac{2P_{kv}}{P_{kk}}a^{-2}
		\left(\delta q_1 -\frac{1}{3}\delta q_2\right)
		-i \frac{a^{-2}}{\sqrt{2}}\delta q_3
		,\quad
		E_w
		=-i\frac{2	P_{kw}}{P_{kk}}a^{-2}
		\left(\delta q_1 -\frac{1}{3}\delta q_2\right)
		-i\frac{1}{\sqrt{2}}a^{-2}\delta q_4,\\
		E
		&=
		\frac{a^{-2}(TrP)}{2P_{kk}}
		\left(
		\delta q_1
		-\frac{1}{3}\delta q_2
		\right)
		-\delta q_1a^{-2}\frac{3}{2}
		,\quad
		C
		=
		\frac{a^{-2}(TrP)}{6P_{kk}}
		\left(\delta q_1 -\frac{1}{3}\delta q_2\right).\end{split}
\end{align}

%% file: Appendices/Appendix-time.tex

\chapter{Time problem} 
\renewcommand{\thefigure}{A\arabic{figure}}
\setcounter{figure}{0}
\label{Appendix-time} 

\section{Alternative delay function}\label{appro}

\begin{figure}[h]
	    \centering
	\includegraphics[width=0.6\textwidth]{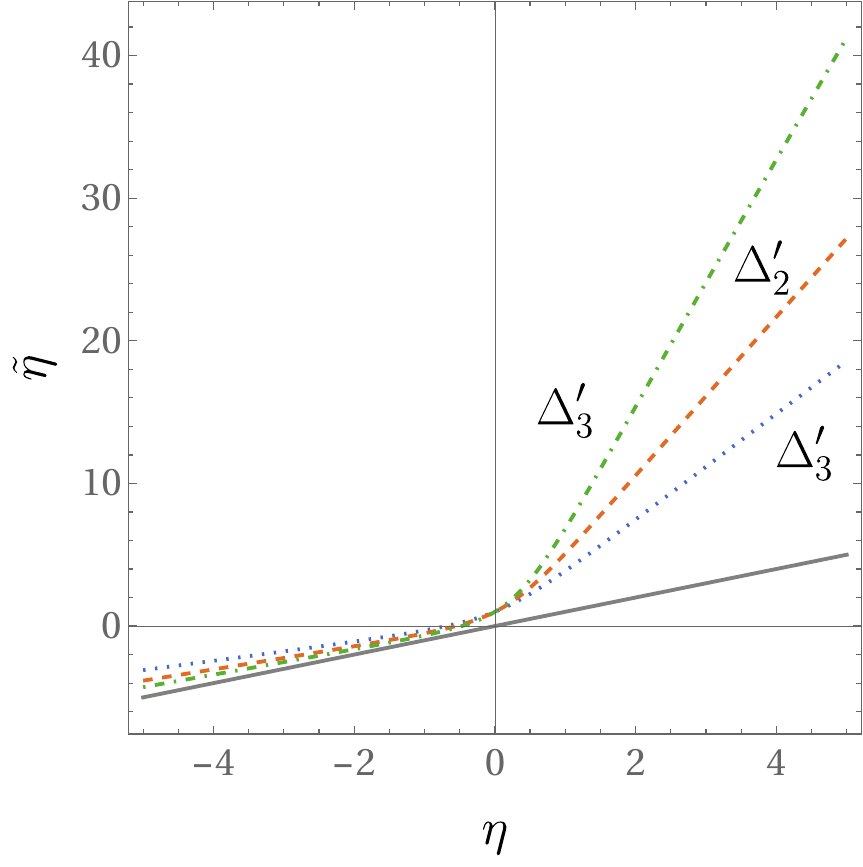}

	\caption{Changes in the time variable $\eta$ for the second family of delay functions
		$\Delta'_1$, $\Delta'_2$ and $\Delta'_3$ given by Eq.~\eqref{delfun2} 
		along a fixed bouncing trajectory, with parameters chosen such that
		$\Delta_1=a e^{p_a}$, $\Delta_2=a e^{3p_a/2}$ and
		$\Delta_3=a e^{2p_a}$. }
	
	\label{deltasprime}
\end{figure}

\begin{figure}[h]
	    \centering
	\includegraphics[width=0.6\textwidth]{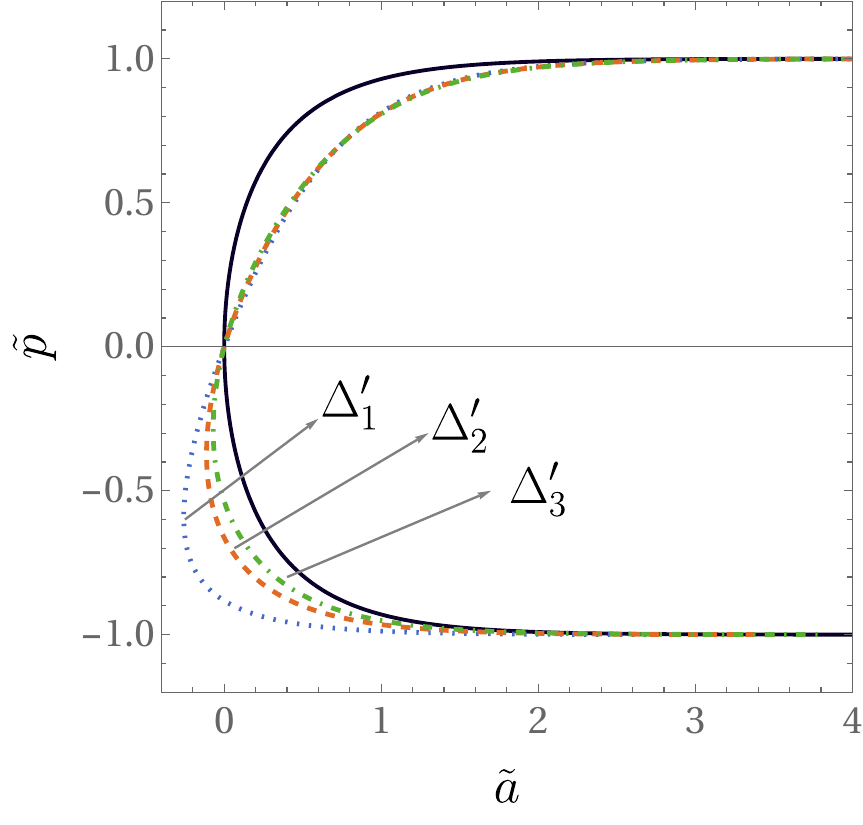}
	
	\caption{Semi-classical trajectories mapped into the initial reduced
		phase space $(a, p)$ for the second class of delay function
		Eq. \eqref{delfun2}, with the same parameters as in 
		Fig.~\ref{deltasprime}.}
	
	\label{qpdeltaprime}
\end{figure}

In this appendix, we consider the alternative choice
of a family of two-parameter delay functions, namely
\begin{equation}\label{delfun2}
	\Delta'(a,p)=a^A e^{B p},
\end{equation}
which define a new set of clocks plotted in Fig.~\ref{deltasprime} for a few
relevant values of the parameters $A$ and $B$. 
Fig.~\ref{qpdeltaprime} depicts the trajectories with
different clocks obtained from $\Delta'(a,p)$ for which the
convergence happens much later than in the case discussed
in the core of this paper, as can be seen by comparing
with Fig.~\ref{qpbackground}. The extent to which this delay
can be increased, and how the matter content of the Universe
can affect this limit, is not dealt with in the present
article and will be the subject of a future work.

One can note that the delay functions \eqref{deltasprime} tend
to diverge in time from one another, all of them growing 
exponentially with the momentum, the phase space trajectories
however do converge to the undelayed one, but at scales that are 
increasingly larger with the amplitude of the exponential
behavior of the relevant delay function.

\begin{figure}[h]
	    \centering
	\includegraphics[width=0.45\textwidth]{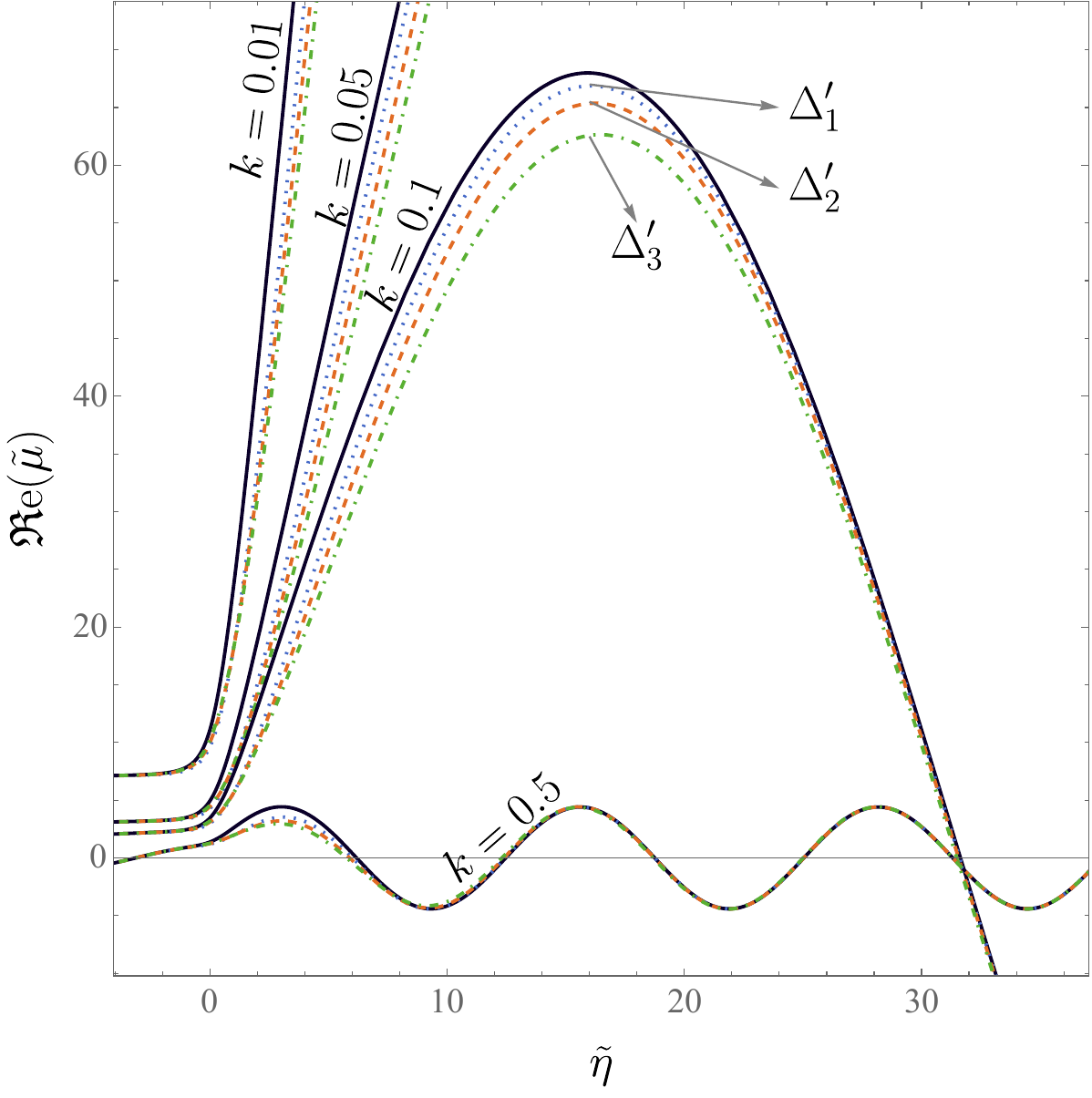}
	
	\caption{Evolution of the primordial gravitational amplitude for
		different clocks obtained from the second class of delay function
		\ref{delfun2}. Convergence happens at a latter time with respect
		to the first class of delay functions \eqref{delfun}, as can be seen
		by comparison with Fig.~\ref{mu-all-k}.}
	
	\label{mu-k-prime}
\end{figure}

\begin{figure}[t]
	\includegraphics[width=0.5\textwidth]{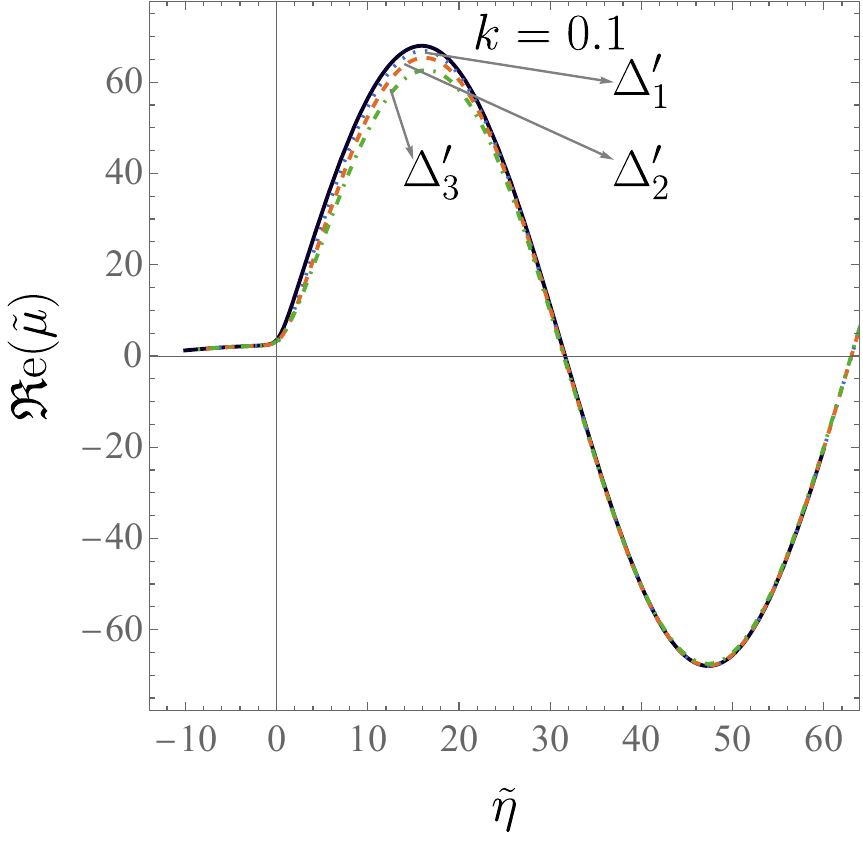}
	\includegraphics[width=0.5\textwidth]{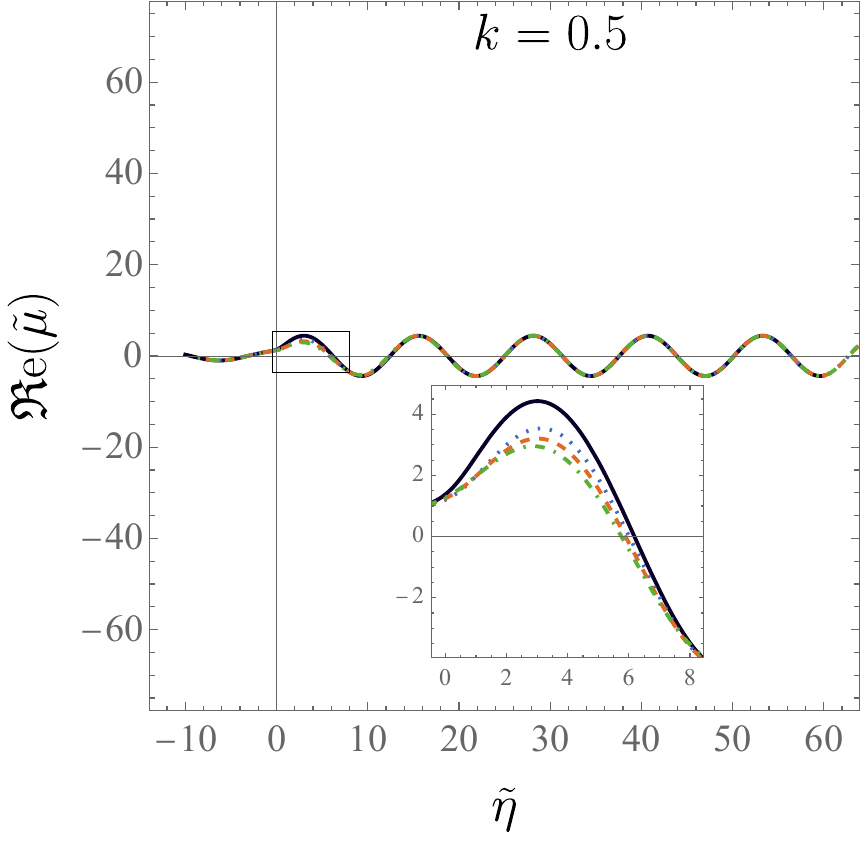}
	\caption{Evolution of the real part of the primordial gravity-wave
		$\mathfrak{R}\mathrm{e}(\tilde{\mu})$ for two different wavenumbers,
		$k=0.1$ and $k=0.5$, and for different clocks for the second class
		of delay functions, $\Delta'_1$, $\Delta'_2$ and $\Delta'_3$,
		respectively represented by the dotted blue line, dashed red line
		and dashed-dotted green line. The original trajectory is represented
		by the full black line.}
	
	\label{mu-k-clock2}
\end{figure}

Moving to the perturbations, we performed the same analysis as
in the core of this paper and show the time development of
the real part of the mode function for different values of the
wavenumber in Fig.~\ref{mu-k-clock2}, with a special emphasis
at the near-bounce regime in Fig.~\ref{mu-k-prime}. As for 
the other family of delay functions, we find that whenever
the classical approximation for the background holds, one
recovers a unique prediction.

